\documentclass[%
superscriptaddress,
preprintnumbers,
amsmath,amssymb,
aps,
prd,
floatfix,
]{revtex4-2}

\usepackage{graphicx}% Include figure files
\graphicspath{{./figures/}}
\usepackage{xcolor}
\usepackage{ulem}
\usepackage{dcolumn}% Align table columns on decimal point
\usepackage{bm}% bold math
\usepackage{hyperref}
\hypersetup{colorlinks, linkcolor = [rgb]{0, 0, 0.5}, citecolor = [rgb]{0,0.0,0.5}, urlcolor = [rgb]{0,0.0,0.5}}
\usepackage[caption=false]{subfig}
\usepackage{siunitx}
\usepackage[capitalise]{cleveref}
\usepackage{subfloat}
\usepackage{float}
\usepackage{amsthm,amsmath,amssymb}
\usepackage{mathrsfs}
\usepackage{amsmath}

\usepackage{ulem}
\usepackage{xcolor}
\newcommand{\new}[1]{{\color{black}{#1}}}

\begin{document}

\preprint{Published 
%as P.~He \& B.-Q.~Ma, 
in \href{https://doi.org/10.1140/epjc/s10052-025-14216-8}{Euro.Phys.J. C 85 (2025) 504}}

\title{Six-dimensional light-front Wigner distributions of the proton}

\newcommand*{\PKU}{School of Physics, Peking University, Beijing 100871, China}\affiliation{\PKU}
\newcommand*{\MOE}{Key Laboratory of Particle Physics and Particle Irradiation (MOE), Institute of Frontier and Interdisciplinary Science, Shandong University, Qingdao, Shandong 266237, China}\affiliation{\MOE}
\newcommand*{\SCNT}{Southern Center for Nuclear-Science Theory (SCNT), Institute of Modern Physics, Chinese Academy of Sciences, Huizhou 516000, China}\affiliation{\SCNT}
\newcommand*{\CHEP}{Center for High Energy Physics, Peking University, Beijing 100871, China}\affiliation{\CHEP}
\newcommand*{\ZZU}{School of Physics, Zhengzhou University, Zhengzhou 450001, China}\affiliation{\ZZU}

\author{Yirui~Yang}\affiliation{\PKU}

\author{Tianbo~Liu}\email{liutb@sdu.edu.cn}\affiliation{\MOE}\affiliation{\SCNT}

\author{Bo-Qiang~Ma}\email{mabq@pku.edu.cn}\affiliation{\PKU}\affiliation{\CHEP}\affiliation{\ZZU}
%\affiliation{\CICQM}


\begin{abstract}
	
We investigate six-dimensional quark Wigner distributions of the proton in a light-front quark spectator-diquark model. Benefiting from the light-front boost-invariant longitudinal variable $\tilde{z}$, these light-front Wigner distributions provide complete information of parton distribution in the phase space and the correlation with spins. At the leading twist, one can define 16 independent distribution functions in according to different combinations of quark and proton polarizations. Numerical results of all these Wigner distributions are presented, unraveling rich structures of the proton, which may potentially provide new observables to be explored at future experiments.
	
\end{abstract}

\maketitle

\section{Introduction}

Quantum chromodynamics (QCD) is the fundamental theory that describes strong interactions, wherein nucleons are composed of quarks and gluons—collectively referred to as partons—within strongly interacting and relativistic bound states. Unlike static particles, nucleons exhibit a complex and dynamic internal structure, and the study of hadron structure has yielded significant insights into the dynamics of quarks and gluons~\cite{Feynman:1969ej,Mulders:1995dh,Goeke:2005hb,Bacchetta:2006tn,Brambilla:2014jmp}. Historically, research has predominantly focused on one-dimensional information, particularly the longitudinal momentum distributions of quarks and gluons within colliding nucleons~\cite{Gross:1973id,Politzer:1973fx}. In the framework of collinear QCD factorization~\cite{Collins:1989gx}, the non-perturbative aspects are encapsulated by the parton distribution function (PDF)~\cite{Collins:1981uw,Martin:1998sq,Gluck:1994uf,Gluck:1998xa}, which provides the probability distribution of longitudinal momentum fraction 
$x$ within the hadron. As research into hadron structure has progressed, parton distribution functions have been extended to higher dimensions, leading to the development of transverse momentum dependent parton distributions (TMDs) and generalized parton distributions (GPDs). TMDs are associated with semi-inclusive deep inelastic scattering (SIDIS) or the Drell-Yan process~\cite{Bacchetta:2006tn,Mulders:1995dh,Barone:2001sp,Brodsky:2002cx}, while GPDs are linked to deep virtual Compton scattering (DVCS) and deep virtual meson production processes (DVMP)~\cite{Ji:1996nm,Diehl:2003ny,Belitsky:2005qn,Goeke:2001tz}. These advancements allow us to extract valuable information regarding the three-dimensional motions of quarks and gluons in colliding nucleons, which are intimately connected to quark orbital angular momentum~\cite{Ji:1996ek,Avakian:2008dz,Avakian:2009jt,She:2009jq,Avakian:2010br}. Today, with the advent of high-energy scattering experiments and the upcoming electron-ion colliders (EIC~\cite{Accardi:2012qut} and EicC~\cite{Anderle:2021wcy}), we have the opportunity to utilize advanced phenomenological tools to further investigate these dynamics.

However, the aforementioned distribution functions primarily describe momentum space distributions and lack information in coordinate space, thereby failing to fully characterize the internal structure of hadrons. A promising approach to overcome this limitation is the utilization of phase space distribution functions, which represent the joint distribution in both coordinate and momentum spaces~\cite{Boussarie:2023izj}. The Wigner distribution function serves as a typical phase space distribution, originally introduced to derive quantum corrections to classical statistical mechanics~\cite{Wigner:1932eb,Hillery:1983ms}. It has since been established that the Wigner distribution can be interpreted as the kernel of the density matrix~\cite{Hiley2015}. In the context of QCD, the Wigner distribution of quarks and gluons in hadrons has been defined, highlighting its connection to GPDs~\cite{Ji:2003ak,Belitsky:2003nz}. These distributions describe the joint distribution of three-dimensional coordinates and momentum, although they have the notable limitation of not accounting for relativistic effects. To address this issue, a five-dimensional Wigner distribution was introduced in the infinite momentum reference frame~\cite{Lorce:2011kd,Lorce:2015sqe}. This formulation allows for the integration over transverse momentum or position coordinates, yielding other distribution functions, such as TMDs or impact parameter dependent parton distributions (IPDs)~\cite{Burkardt:2002hr,Burkardt:2000za}. Additionally, various phenomenological models have been employed to calculate Wigner distributions for both spin-0~\cite{Ma:2018ysi,Kaur:2019jow,Kaur:2019kpi,Zhang:2021tnr} and spin-1/2 hadrons~\cite{Liu:2015eqa,Mukherjee:2014nya,Mukherjee:2015aja,More:2017zqq,Chakrabarti:2016yuw,Chakrabarti:2017teq,Chakrabarti:2019wjx,Kaur:2019lox,Kumar:2017xcm}. However, since the longitudinal space coordinate $x^{-}\rightarrow 0$, this dimension is effectively excluded, resulting in an incomplete characterization of the internal structure of hadrons. Recently, a boost-invariant six-dimensional light-front Wigner distribution function was proposed~\cite{Han:2022tlh}, based on the boost-invariant longitudinal coordinate 
$\tilde{z}$ introduced in earlier works~\cite{Brodsky:2006ku,Brodsky:2006in,Miller:2019ysh}. This six-dimensional light-front Wigner distribution can recover other known distribution functions through integration~\cite{Han:2022tlh,Han:2023fav}.

The light-front (or light-cone) quark spectator-diquark model is a widely used tool to study hadron structures, originally proposed in order to investigate deep inelastic lepton nucleon scattering (DIS)~\cite{Close:1973422,Field:1976ve}. Within the framework of the quark-parton model~\cite{PhysRev.179.1547}, the nucleon can be viewed as consisting of an active quark that is struck by a virtual photon, and a spectator diquark that does not interact directly with the virtual photon. This diquark acts as a quasi-particle in the model. The model has been successful in structure function calculations~\cite{feynman2018photon}. By incorporating relativistic effects~\cite{Ma:1991xq,Ma:1992sj} of Wigner rotations~\cite{Wigner:1939cj} or Melosh transformations~\cite{Melosh:1974cu,Buccella:1974bz}, the model can be extended to study a variety of physical quantities, including quark unpolarized, helicity, and transversity distributions~\cite{Ma:1996np,Ma:1997gy}; form factors~\cite{Ma:2002ir,Liu:2014npa}; TMDs~\cite{Lu:2004au,She:2009jq,Bacchetta:2008af,Lu:2012gu}; and GPDs~\cite{Burkardt:2003je,Chakrabarti:2005zm,Hwang:2007tb}. More recently, a six-dimensional light-front Wigner distribution function, based on the spectator quark model, has been calculated for spin-0 particles~\cite{Han:2023fav}.

\new{In this work, we present a significant extension of  previous studies~\cite{Liu:2015eqa,Han:2022tlh} regarding light-front Wigner distributions. While the earlier works only considered the five-dimensional case~\cite{Liu:2015eqa} and proposed the concept of six-dimensional light-front Wigner distributions and provided theoretical definitions for the unpolarized scenario~\cite{Han:2022tlh}, the current study provides a comprehensive calculation of all 16 leading-twist Wigner distributions for various polarization combinations, offering a complete picture of six-dimensional Wigner distributions, which was not previously available. Moreover, this work delves deeper into the physical interpretations of these distributions, providing new insights into the proton structure.} The structure of this paper is organized as follows. In Sec.~\ref{Sec:model}, we briefly review the light-front spectator model of the proton. In Sec.~\ref{Sec:6DWigner}, we provide the six-dimensional Wigner distribution for all polarization states and explain the phenomenological implications and connection to experimental observables. The numerical results for different polarization states are presented in Sec.~\ref{Sec:Numerical}. Finally, the summary and conclusions are provided in Sec.~\ref{Sec:SC}.

\section{The light-front spectator model of the proton}
\label{Sec:model}

According to the light-front quark spectator-diquark model, a proton state can be expanded in the base of the Fock state as
\begin{equation}
	\begin{aligned}
		\left|P^{+}, \boldsymbol{P}_{\perp}, \lambda_{N}\right\rangle =\sum_{q,D,\lambda_{q},\lambda_{D}}\int \frac{\mathrm{d}x\mathrm{d}^2\boldsymbol{k}_{\perp}}{2(2\pi)^{3}\sqrt{x(1-x)}}\Psi_{qD}(x,\boldsymbol{k}_{\perp})\left|qD;x,\boldsymbol{k}_{\perp},\lambda_{q},\lambda_{D}\right\rangle,
	\end{aligned}    
\end{equation}    
where $q$ denotes the detected quark, $D$ represents the species of the spectator, $\lambda$ is the light-front helicity of the corresponding particle, and $\Psi_{qD}(x,\boldsymbol{k}_{\perp})$ is the probability amplitude that hadron state $\left|P^{+}, \boldsymbol{P}_{\perp}, \lambda_{N}\right\rangle$ in the Fock state $\Psi_{qD}(x,\boldsymbol{k}_{\perp})\left|qD;x,\boldsymbol{k}_{\perp},\lambda_{q},\lambda_{D}\right\rangle$, which is called the light-front wave function.

Considering the three-quark state $\left|uud\right\rangle$ of a proton in the leading twist, it can be approximated as a decomposition into a quark state and a spectator state with a double-quark quantum number, where non-perturbative effects between quarks and gluons are absorbed into the mass of the spectator~\cite{Liu:2015eqa,Liu:2014vwa}. The proton state, taking into account the quantum numbers of both the proton and the quarks, can involve scalar or axial-vector spectators, with the axial-vector case being essential for flavor separation. Therefore, the proton state can be written as
\begin{equation}
\left|p\right\rangle=\sin{\theta}\phi_{S}\left|qS\right\rangle+\cos{\theta}\phi_{V}\left|qV\right\rangle,
\end{equation}
where $S$ and $V$ represent scalar and axial-vector spectators respectively, and $\phi_{S}$ and $\phi_{V}$ are the corresponding light-front wave functions in momentum space. The angle $\theta$ reflects the spin-flavor $\mathrm{SU}(6)$ symmetry breaking and is chosen as $\theta=\pi/4$ in the calculation, analogous to the analysis of helicity and transversity distributions~\cite{Ma:1996np,Ma:1997gy}, form factors~\cite{Ma:2002ir,Liu:2014npa}, and single-spin asymmetries~\cite{Zhu:2010ewp,Zhu:2011zza,Lu:2011cw}.

The wave function in spin space is derived by applying the Melosh-Wigner rotation to the wave function in the instantaneous form. The instantaneous wave function of the proton is given by~\cite{Ma:1996np}
\begin{equation}
	|p\rangle^{\uparrow, \downarrow}=\frac{1}{\sqrt{2}}|u S(u d)\rangle_T^{\uparrow, \downarrow}-\frac{1}{\sqrt{6}}|u V(u d)\rangle_T^{\uparrow, \downarrow}+\frac{1}{\sqrt{3}}|\mathrm{d} V(u u)\rangle_T^{\uparrow, \downarrow}.
\end{equation}	

To transform the spinor of the quark from the instantaneous form to the light-front form, the corresponding transformation is
\begin{equation}
	\left[\begin{array}{c}
		q_T^{\uparrow} \\
		q_T^{\downarrow}
	\end{array}\right]=W_q\left[\begin{array}{c}
		q_F^{\uparrow} \\
		q_F^{\downarrow}
	\end{array}\right],
\end{equation}  
where $q_T$ and $q_F$ denote the spinors in the instant form and the light-front form, respectively. The results obtained by this transformation method are in agreement with the light-front field theory~\cite{Xiao:2003wf}. And the Melosh-Wigner rotation matrix of spinors $W_q$ is given by~\cite{Wigner:1939cj,Melosh:1974cu,Buccella:1974bz}
\begin{equation}
	W_q=\omega\left[\begin{array}{cc}
		k^{+}+m & -k^R \\
		k^L & k^{+}+m
	\end{array}\right],
\end{equation}   
where $m$ is the mass of the quark and $\omega=1 / \sqrt{\left(k^{+}+m\right)^2+k^L k^R}$ is a normalization factor. This effect plays an important role in understanding the “proton spin puzzle”~\cite{Ma:1991xq,Ma:1992sj}.

For the spectator, since the scalar di-quark has spin zero, this transformation remains invariant. The Melosh-Wigner transformation for the axial-vector di-quark is given by~\cite{Ahluwalia:1993xa}
\begin{equation}
	\left[\begin{array}{c}
		V_T^1 \\
		V_T^0 \\
		V_T^{-1}            
	\end{array}\right]=W_V\left[\begin{array}{c}
		V_F^1 \\
		V_F^0 \\
		V_F^{-1}
	\end{array}\right].
\end{equation}
The corresponding Melosh–Wigner transformation matrix is
\begin{equation}
	W_V=\omega_V^2\left[\begin{array}{ccc}
		\left(k_V^{+}+m_V\right)^2 & -\sqrt{2}\left(k_V^{+}+m_V\right) k_V^R & \left(k_V^R\right)^2 \\
		\sqrt{2}\left(k_V^{+}+m_V\right) k_V^L & \left(k_V^{+}+m_V\right)^2-k_V^L k_V^R & -\sqrt{2}\left(k_V^{+}+m_V\right) k_V^R \\
		\left(k_V^L\right)^2 & \sqrt{2}\left(k_V^{+}+m_V\right) k^L & \left(k_V^{+}+m_V\right)^2
	\end{array}\right],    
\end{equation}
where $m_V$ is the mass of the di-quark and $\omega_V=1 / \sqrt{(k_V^{+}+m_V)^2+k_V^L k_V^R}$ is a normalization factor.

For the momentum space light-front wave function, various phenomenological wave functions for spin-averaged quark-spectator states have been proposed, including the Brodsky-Huang-Lepage (BHL) prescription~\cite{Brodsky:1980vj,brodsky1981slac,Brodsky:1982nx}, the Terentev-Karmanov (TK) prescription~\cite{Terentev:1976jk,Karmanov:1979if}, the Chung-Coester-Polyzou (CCP) prescription~\cite{Chung:1988mu}, and the Vega-Schmidt-Gutsche-Lyubovitskij (VSGL) prescription~\cite{Vega:2013bxa}. Following the BHL form of the light-front wave function~\cite{brodsky1981quantum,Huang:1994dy}, the momentum space light-front wave function is given by
\begin{equation}
	\varphi_{q D}\left(x, \boldsymbol{k}_{\perp}\right)=A_{q D} \exp \left(-\frac{\mathcal{M}^2}{8 \beta_D^2}\right),
\end{equation}
where $D$ represents the spectator, with $D=S$ for the scalar spectator and $D=V$ for the axial-vector spectator. Besides, $\beta_D$ is the scale parameter of the harmonic oscillator, $A_{q D}$ is the normalization factor and $\mathcal{M}$ is the invariant mass as 
\begin{equation}
	\mathcal{M}=\sqrt{\frac{m_q^2+\boldsymbol{k}_{\perp}^2}{x}+\frac{m_D^2+\boldsymbol{k}_{\perp}^2}{1-x}},
\end{equation}     
where $m_q$ and $m_D$ are the mass of quarks and spectators, respectively.

According to the previous discussion, the quark-parton state can be expressed as 
\begin{align}
	|q S\rangle_T^\lambda=\int \frac{\mathrm{d} x \mathrm{d}^2 \boldsymbol{k}_{\perp}}{16 \pi^3} \sum_{i=\{\uparrow, \downarrow\}} A_{q S}^{i, \lambda}\left|q^i S\right\rangle_F, \\
	|q V\rangle_T^\lambda=\int \frac{\mathrm{d} x \mathrm{d}^2 \boldsymbol{k}_{\perp}}{16 \pi^3} \sum_{i=\{\uparrow, \downarrow\}} \sum_{j=\{-1,0,1\}} B_{q V}^{i j, \lambda}\left|q^i V^j\right\rangle_F,
\end{align}
where $\lambda = \uparrow$ or $\downarrow$, and the coefficients $A_{q S}^{i, \lambda}$ and $B_{q V}^{i j, \lambda}$ are defined as
\begin{align}
	& A_{q S}^{i, \lambda}=W_q^{\lambda, i} \varphi_S, \\
	& B_{q V}^{i j, \uparrow}=\left(-\sqrt{\frac{1}{3}} W_q^{\uparrow, i} W_V^{0, j}+\sqrt{\frac{2}{3}} W_q^{\downarrow, i} W_V^{1, j}\right) \varphi_V, \\
	& B_{q V}^{i j, \downarrow}=\left(-\sqrt{\frac{1}{3}} W_q^{\downarrow, i} W_V^{0, j}+\sqrt{\frac{2}{3}} W_q^{\uparrow, i} W_V^{-1, j}\right) \varphi_V.
\end{align}

From these expressions, the light-front wave functions of the proton can be derived as
\begin{align}
	& \psi_{u S(u d)}^\lambda\left(x, \boldsymbol{k}_{\perp}, \lambda_u=i\right)=\frac{1}{\sqrt{2}} A_{u S(u d)}^{i, \lambda}, \\
	& \psi_{u V(u d)}^\lambda\left(x, \boldsymbol{k}_{\perp}, \lambda_u=i, \lambda_V=j\right)=-\frac{1}{\sqrt{6}} B_{u V(u d)}^{i j, \lambda}, \\
	& \psi_{\mathrm{d} V(u u)}^\lambda\left(x, \boldsymbol{k}_{\perp}, \lambda_d=i, \lambda_V=j\right)=\frac{1}{\sqrt{3}} B_{\mathrm{d} V(u u)}^{i j, \lambda} .
\end{align}
These wave functions require summation over the spectator types and spin states. The results are primarily determined by the momentum space wave function, which influences the quantitative aspects, while the spin structure primarily governs the qualitative features.

%%%%%%%%%%%%%%%%%%%%%%%%%%%%%%%%%%%%%%%%%%%%%%%%%%%%%%%%

\section{Six-dimensional light-front Wigner distribution of proton}
\label{Sec:6DWigner}

\subsection{\new{Theoretical Framework and Definitions}}

Considering the dependence of the longitudinal coordinates and fixing the light-front time, the six-dimensional light-front Wigner operator can be expressed as~\cite{Han:2022tlh}
\begin{equation}
	\begin{gathered}
		\widehat{W}^{[\Gamma]}\left(b^{-}, \boldsymbol{b}_{\perp}, k^{+}, \boldsymbol{k}_{\perp}\right)=\int \frac{\mathrm{d} y^{-} \mathrm{d}^2 \boldsymbol{y}_{\perp}}{2(2 \pi)^3} e^{i k^{+} y^{-}-i\boldsymbol{k}_{\perp} \cdot \boldsymbol{y}_{\perp}} \\
		\quad \times\left.\bar{\psi}\left(b-\frac{y}{2}\right) \Gamma \mathcal{L}_{[b-y / 2, b+y / 2]} \Psi\left(b+\frac{y}{2}\right)\right|_{y^{+}=0},
	\end{gathered}
\end{equation}
where $b=\left(0, b^{-}, \boldsymbol{b}_{\perp}\right)$ is the average position for the quark fixed to the light-front time $y^{+}=0$, $\mathcal{L}_{[b-y / 2, b+y / 2]}$ represents the Wilson line connecting the quark operators to ensure gauge invariance, and $\Gamma$ is the Dirac matrix projecting the quarks to different spin states in leading-order twist. The possible choices for $\Gamma$ include $\mathrm{\Gamma}=\gamma^{+}, \gamma^{+} \gamma_{5}$ and $i \sigma^{i+} \gamma_{5}$, corresponding to unpolarized, longitudinal polarized, and transverse polarized states of quarks, respectively.

%%%%%%%%%%%%%%%%%%%%%%%%%%%%%%%%%%%%%%%%%%%%

The six-dimensional light-front Wigner distribution of the hadron can be defined by placing the Wigner operator between the initial and final states of the hadron
\begin{equation}
    \label{wignereq1}
	\begin{aligned}
		& \rho^{[\Gamma]}\left(\tilde{z}, x, \boldsymbol{b}_{\perp}, \boldsymbol{k}_{\perp}, S\right)=\int \frac{\mathrm{d} \xi \mathrm{d}^2 \boldsymbol{\Delta}_{\perp}}{4 \pi^3} \\
		& \quad \times\left\langle P+\frac{\Delta}{2}, S\left|\widehat{W}^{[\Gamma]}\left(b^{-}, \boldsymbol{b}_{\perp}, k^{+}, \boldsymbol{k}_{\perp}\right)\right| P-\frac{\Delta}{2}, S
		\right\rangle,
	\end{aligned}
\end{equation}
where $\Delta=\left(\Delta^{+}, \Delta^{-}, \boldsymbol{\Delta}_{\perp}\right)$ is the transferred momentum between the initial and final states of the hadron, $\xi=-\Delta^{+} /\left(2 P^{+}\right)$ is the skewness variable indicating the physical longitudinal momentum transfer during the process, $P^{+}$ is the average light-front momentum of the nucleon, $x=k^{+} / P^{+}$ is the average momentum fraction of the nucleon, and $\tilde{z}=b^{-} P^{+}$ is the boost-invariant longitudinal coordinate defined by Brodsky and Miller \cite{Brodsky:2006ku,Brodsky:2006in,Miller:2019ysh}, and $S$ is the spin state of the nucleon. The integration over $\xi$ ranges from $-x$ to $x$, corresponding to the DGLAP region~\cite{Diehl:2003ny}, ensuring the quark momentum fraction remains non-negative. This equation expresses the quantum phase space distribution of individual quarks within the nucleon.

We clarify that although $\tilde{z}$ is the Fourier conjugate variable of $\xi$, it offers unique information. When considering the boost-invariant longitudinal space, $\tilde{z}$ is related to the longitudinal impact parameter. In the context of GPDs, the variable $\xi$ and its conjugate $\tilde{z}$ play crucial roles in describing the internal structure of the proton. For example, the $\tilde{z}$-dependence can be related to the diffraction-like patterns observed in the DVCS process, which is measurable. This new explanation helps to establish the physical significance of $\tilde{z}$ beyond its relationship with $\xi$.

By combining different polarization states of hadrons and quarks, such as unpolarized (U), longitudinal polarized (L), and transverse polarized (T), we can define 16 independent Wigner distributions in the leading twist. For a hadron with spin $1/2$, the following six-dimensional different light-front Wigner distribution functions can be obtained. The six-dimensional light-front Wigner distributions are defined as
\begin{equation}
	\rho_{\mathrm{UU}}\left(\tilde{z},x,\boldsymbol{b}_{\perp}, \boldsymbol{k}_{\perp}\right)=\frac{1}{2}\left[\rho^{\left[\gamma^{+}\right]}\left(\tilde{z},x,\boldsymbol{b}_{\perp}, \boldsymbol{k}_{\perp}, \hat{\boldsymbol{e}}_z\right)+\rho^{\left[\gamma^{+}\right]}\left(\tilde{z},x,\boldsymbol{b}_{\perp}, \boldsymbol{k}_{\perp}, -\hat{\boldsymbol{e}}_z\right)\right],
\end{equation}
for unpolarized case,
\begin{equation}
	\rho_{\mathrm{UL}}\left(\tilde{z},x,\boldsymbol{b}_{\perp}, \boldsymbol{k}_{\perp}\right) =\frac{1}{2}\left[\rho^{\left[\gamma^{+} \gamma_5\right]}\left(\tilde{z},x,\boldsymbol{b}_{\perp}, \boldsymbol{k}_{\perp},  \hat{\boldsymbol{e}}_z\right)+\rho^{\left[\gamma^{+} \gamma_5\right]}\left(\tilde{z},x,\boldsymbol{b}_{\perp}, \boldsymbol{k}_{\perp},-\hat{\boldsymbol{e}}_z\right)\right],
\end{equation}
for unpolarized-longitudinal case,
\begin{equation}
	\rho_{\mathrm{UT}}^j\left(\tilde{z},x,\boldsymbol{b}_{\perp}, \boldsymbol{k}_{\perp}\right)=\frac{1}{2}\left[\rho^{\left[i\sigma^{j+}\gamma_{5}\right]}\left(\tilde{z},x,\boldsymbol{b}_{\perp}, \boldsymbol{k}_{\perp}, \hat{\boldsymbol{e}}_z\right)+\rho^{\left[i \sigma^{j+} \gamma_5\right]}\left(\tilde{z},x,\boldsymbol{b}_{\perp}, \boldsymbol{k}_{\perp},-\hat{\boldsymbol{e}}_z\right)\right],
\end{equation}
for unpolarized-transverse case,
\begin{equation}
	\rho_{\mathrm{LU}}\left(\tilde{z},x,\boldsymbol{b}_{\perp}, \boldsymbol{k}_{\perp}\right)=\frac{1}{2}\left[\rho^{\left[\gamma^{+}\right]}\left(\tilde{z},x,\boldsymbol{b}_{\perp}, \boldsymbol{k}_{\perp},  \hat{\boldsymbol{e}}_z\right)-\rho^{\left[\gamma^{+}\right]}\left(\tilde{z},x,\boldsymbol{b}_{\perp}, \boldsymbol{k}_{\perp},-\hat{\boldsymbol{e}}_z\right)\right],   
\end{equation}
for longitudinal-unpolarized case,
\begin{equation}
	\rho_{\mathrm{LL}}\left(\tilde{z},x,\boldsymbol{b}_{\perp}, \boldsymbol{k}_{\perp}\right)=\frac{1}{2}\left[\rho^{\left[\gamma^{+} \gamma_5\right]}\left(\tilde{z},x,\boldsymbol{b}_{\perp}, \boldsymbol{k}_{\perp},  \hat{\boldsymbol{e}}_z\right)-\rho^{\left[\gamma^{+} \gamma_5\right]}\left(\tilde{z},x,\boldsymbol{b}_{\perp}, \boldsymbol{k}_{\perp},  -\hat{\boldsymbol{e}}_z\right)\right], 
\end{equation}
for longitudinal case,
\begin{equation}
	\rho_{\mathrm{LT}}^j\left(\tilde{z},x,\boldsymbol{b}_{\perp}, \boldsymbol{k}_{\perp}\right)=\frac{1}{2}\left[\rho^{\left[i \sigma^{j+} \gamma_5\right]}\left(\tilde{z},x,\boldsymbol{b}_{\perp}, \boldsymbol{k}_{\perp},\hat{\boldsymbol{e}}_z\right)-\rho^{\left[i \sigma^{j+} \gamma_5\right]}\left(\tilde{z},x,\boldsymbol{b}_{\perp}, \boldsymbol{k}_{\perp}, -\hat{\boldsymbol{e}}_z\right)\right],    
\end{equation}
for longitudinal-transverse case,
\begin{equation}
	\rho^{i}_{\mathrm{TU}}\left(\tilde{z},x,\boldsymbol{b}_{\perp}, \boldsymbol{k}_{\perp}\right)=\frac{1}{2}\left[\rho^{\left[\gamma^{+}\right]}\left(\tilde{z},x,\boldsymbol{b}_{\perp}, \boldsymbol{k}_{\perp},  \hat{\boldsymbol{e}}_i\right)-\rho^{\left[\gamma^{+}\right]}\left(\tilde{z},x,\boldsymbol{b}_{\perp}, \boldsymbol{k}_{\perp},-\hat{\boldsymbol{e}}_i\right)\right],    
\end{equation}
for transverse-unpolarized case,
\begin{equation}
	\rho^{i}_{\mathrm{TL}}\left(\tilde{z},x,\boldsymbol{b}_{\perp}, \boldsymbol{k}_{\perp}\right)=\frac{1}{2}\left[\rho^{\left[\gamma^{+} \gamma_5\right]}\left(\tilde{z},x,\boldsymbol{b}_{\perp}, \boldsymbol{k}_{\perp},\hat{\boldsymbol{e}}_i\right)-\rho^{\left[\gamma^{+} \gamma_5\right]}\left(\tilde{z},x,\boldsymbol{b}_{\perp}, \boldsymbol{k}_{\perp}, -\hat{\boldsymbol{e}}_i\right)\right],    
\end{equation}
for transverse-longitudinal case,
\begin{equation}
	\rho_{\mathrm{TT}}\left(\tilde{z},x,\boldsymbol{b}_{\perp}, \boldsymbol{k}_{\perp}\right)=\frac{1}{2}\delta_{ij}\left[\rho^{\left[i \sigma^{j+} \gamma_5\right]}\left(\tilde{z},x,\boldsymbol{b}_{\perp}, \boldsymbol{k}_{\perp}, \hat{\boldsymbol{e}}_i\right)-\rho^{\left[i \sigma^{j+} \gamma_5\right]}\left(\tilde{z},x,\boldsymbol{b}_{\perp}, \boldsymbol{k}_{\perp},-\hat{\boldsymbol{e}}_i\right)\right],    
\end{equation}
for transverse case,
\begin{equation}
	\rho_{\mathrm{TT}}^{\perp}\left(\tilde{z},x,\boldsymbol{b}_{\perp}, \boldsymbol{k}_{\perp}\right)=\frac{1}{2}\epsilon_{ij}\left[\rho^{\left[i \sigma^{j+} \gamma_5\right]}\left(\tilde{z},x,\boldsymbol{b}_{\perp}, \boldsymbol{k}_{\perp}, \hat{\boldsymbol{e}}_i\right)-\rho^{\left[i \sigma^{j+} \gamma_5\right]}\left(\tilde{z},x,\boldsymbol{b}_{\perp}, \boldsymbol{k}_{\perp},-\hat{\boldsymbol{e}}_i\right)\right],   
\end{equation}
for pretzelous case. The first subscript denotes the proton polarization, while the second one represents the quark polarization. Subsequently, a prefix is appended to characterize the proton polarization, except when the proton polarization is parallel to that of the quark.

By integrating over all relevant phase-space variables ($\tilde{z}$, $x$, $\boldsymbol{b}_{\perp}$, and $\boldsymbol{k}_{\perp}$), we can obtain the following relations for the six-dimensional Wigner distributions of the proton for different polarization states
\begin{align}
	\int \mathrm{d}\tilde{z}\mathrm{d}x\mathrm{d}^2\boldsymbol{b}_{\perp}\mathrm{d}^2\boldsymbol{k}_{\perp}\rho_{\mathrm{UU}}\left(\tilde{z}, x, \boldsymbol{b}_{\perp}, \boldsymbol{k}_{\perp}\right)&=N_{q},\label{eq:UU}\\
	\int \mathrm{d}\tilde{z}\mathrm{d}x\mathrm{d}^2\boldsymbol{b}_{\perp}\mathrm{d}^2\boldsymbol{k}_{\perp}\rho_{\mathrm{UL}}\left(\tilde{z}, x, \boldsymbol{b}_{\perp}, \boldsymbol{k}_{\perp}\right)&=0,\label{eq:UL}\\
	\int \mathrm{d}\tilde{z}\mathrm{d}x\mathrm{d}^2\boldsymbol{b}_{\perp}\mathrm{d}^2\boldsymbol{k}_{\perp}\rho_{\mathrm{UT}}^{j}\left(\tilde{z}, x, \boldsymbol{b}_{\perp}, \boldsymbol{k}_{\perp}\right)&=0,\label{eq:UT}\\
        \int \mathrm{d}\tilde{z}\mathrm{d}x\mathrm{d}^2\boldsymbol{b}_{\perp}\mathrm{d}^2\boldsymbol{k}_{\perp}\rho_{\mathrm{LU}}\left(\tilde{z}, x, \boldsymbol{b}_{\perp}, \boldsymbol{k}_{\perp}\right)&=0,\label{eq:LU}\\
        \int \mathrm{d}\tilde{z}\mathrm{d}x\mathrm{d}^2\boldsymbol{b}_{\perp}\mathrm{d}^2\boldsymbol{k}_{\perp}\rho_{\mathrm{LL}}\left(\tilde{z}, x, \boldsymbol{b}_{\perp}, \boldsymbol{k}_{\perp}\right)&=\Delta_{q},\label{eq:LL}\\
        \int \mathrm{d}\tilde{z}\mathrm{d}x\mathrm{d}^2\boldsymbol{b}_{\perp}\mathrm{d}^2\boldsymbol{k}_{\perp}\rho_{\mathrm{LT}}^{j}\left(\tilde{z}, x, \boldsymbol{b}_{\perp}, \boldsymbol{k}_{\perp}\right)&=0,\label{eq:LT}\\
        \int \mathrm{d}\tilde{z}\mathrm{d}x\mathrm{d}^2\boldsymbol{b}_{\perp}\mathrm{d}^2\boldsymbol{k}_{\perp}\rho_{\mathrm{TU}}^{i}\left(\tilde{z}, x, \boldsymbol{b}_{\perp}, \boldsymbol{k}_{\perp}\right)&=0,\label{eq:TU}\\
        \int \mathrm{d}\tilde{z}\mathrm{d}x\mathrm{d}^2\boldsymbol{b}_{\perp}\mathrm{d}^2\boldsymbol{k}_{\perp}\rho_{\mathrm{TL}}^{i}\left(\tilde{z}, x, \boldsymbol{b}_{\perp}, \boldsymbol{k}_{\perp}\right)&=0,\label{eq:TL}\\
	\int \mathrm{d}\tilde{z}\mathrm{d}x\mathrm{d}^2\boldsymbol{b}_{\perp}\mathrm{d}^2\boldsymbol{k}_{\perp}\rho_{\mathrm{TT}}\left(\tilde{z}, x, \boldsymbol{b}_{\perp}, \boldsymbol{k}_{\perp}\right)&=\Delta_{Tq},\label{eq:TT}\\
	\int \mathrm{d}\tilde{z}\mathrm{d}x\mathrm{d}^2\boldsymbol{b}_{\perp}\mathrm{d}^2\boldsymbol{k}_{\perp}\rho_{\mathrm{TT}}^{\perp}\left(\tilde{z}, \boldsymbol{b}_{\perp}, \boldsymbol{k}_{\perp}, x\right)&=0,\label{eq:Pretzelous}
\end{align}
where Eq.~\eqref{eq:UU} represents the quark number density in the unpolarized case, e.g., for the proton, $N_{u}=2$ and $N_{d}=1$. Eqs.~\eqref{eq:UL}, \eqref{eq:UT}, \eqref{eq:LU}, \eqref{eq:LT}, \eqref{eq:TU}, and \eqref{eq:TL} yield zero due to the symmetry of the polarization states, indicating that the quarks are polarized along the positive direction with the same probability as the negative direction. Eq.~\eqref{eq:LL} corresponds to the helicity distribution $\Delta q$, representing the contribution of quark spin to the longitudinal polarization of the hadron. Eq.~\eqref{eq:TT} gives the tensor charge $\Delta_T q$, which quantifies the contribution of quark spin to the hadron transverse polarization.  A positive value of $\Delta_T q$ indicates that the quark spin is aligned with the hadron transverse polarization, while a negative value suggests an anti-alignment.

Based on the preceding discussion, the six-dimensional light-front  Wigner distributions of twist-two can be expressed as a wave function overlap representation, which are given by
\begin{equation}
	\begin{aligned}
		\rho^{\left[\gamma^{+}\right]}\left(\tilde{z}, x, \boldsymbol{b}_{\perp}, \boldsymbol{k}_{\perp}\right)=\int \frac{\mathrm{d} \xi \mathrm{d}^2 \boldsymbol{\Delta}_{\perp}}{4 \pi^3} e^{-2 i \xi \tilde{z}-i \boldsymbol{b}_{\perp} \cdot \boldsymbol{\Delta}_{\perp}} \frac{1}{16 \pi^3}
		& \sum_{\lambda_{\bar{q}}}\left(\psi^*\left(x^{\mathrm{out}}, \boldsymbol{k}_{\perp}^{\mathrm{out}},+, \lambda_{\bar{q}}\right) \psi\left(x^{\mathrm{in}}, \boldsymbol{k}_{\perp}^{\mathrm{in}},+, \lambda_{\bar{q}}\right)\right. \\
		& \left.+\psi^*\left(x^{\mathrm{out}}, \boldsymbol{k}_{\perp}^{\mathrm{out}},-, \lambda_{\bar{q}}\right) \psi\left(x^{\mathrm{in}}, \boldsymbol{k}_{\perp}^{\mathrm{in}},-, \lambda_{\bar{q}}\right)\right),
	\end{aligned}    
\end{equation}
\begin{equation}
	\begin{aligned}
		\rho^{\left[\gamma^{+}\gamma_{5}\right]}\left(\tilde{z}, x, \boldsymbol{b}_{\perp}, \boldsymbol{k}_{\perp}\right)=\int \frac{\mathrm{d} \xi \mathrm{d}^2 \boldsymbol{\Delta}_{\perp}}{4 \pi^3} e^{-2 i \xi \tilde{z}-i \boldsymbol{b}_{\perp} \cdot \boldsymbol{\Delta}_{\perp}} \frac{1}{16 \pi^3}
		& \sum_{\lambda_{\bar{q}}}\left(\psi^*\left(x^{\mathrm{out}}, \boldsymbol{k}_{\perp}^{\mathrm{out}},+, \lambda_{\bar{q}}\right) \psi\left(x^{\mathrm{in}}, \boldsymbol{k}_{\perp}^{\mathrm{in}},+, \lambda_{\bar{q}}\right)\right. \\
		& \left.-\psi^*\left(x^{\mathrm{out}}, \boldsymbol{k}_{\perp}^{\mathrm{out}},-, \lambda_{\bar{q}}\right) \psi\left(x^{\mathrm{in}}, \boldsymbol{k}_{\perp}^{\mathrm {in }},-, \lambda_{\bar{q}}\right)\right),
	\end{aligned}    
\end{equation}
\begin{equation}
	\begin{aligned}
		\rho^{\left[i\sigma^{j+}\gamma_{5}\right]}\left(\tilde{z}, x, \boldsymbol{b}_{\perp}, \boldsymbol{k}_{\perp}\right)=\int \frac{\mathrm{d} \xi \mathrm{d}^2 \boldsymbol{\Delta}_{\perp}}{4 \pi^3} e^{-2 i \xi \tilde{z}-i \boldsymbol{b}_{\perp} \cdot \boldsymbol{\Delta}_{\perp}} \frac{1}{16 \pi^3}
		& \sum_{\lambda_{\bar{q}}}\left(\psi^*\left(x^{\mathrm{out}}, \boldsymbol{k}_{\perp}^{\mathrm{out}}, \uparrow, \lambda_{\bar{q}}\right) \psi\left(x^{\mathrm{in}}, \boldsymbol{k}_{\perp}^{\mathrm {in }}, \uparrow, \lambda_{\bar{q}}\right)\right. \\
		& \left.-\psi^*\left(x^{\mathrm{out}}, \boldsymbol{k}_{\perp}^{\mathrm{out}}, \downarrow, \lambda_{\bar{q}}\right) \psi\left(x^{\mathrm{in}}, \boldsymbol{k}_{\perp}^{\mathrm{in}}, \downarrow, \lambda_{\bar{q}}\right)\right),
	\end{aligned}    
\end{equation}
where $\uparrow$ and $\downarrow$ represent the transverse polarization of quarks along the directions of $\hat{\boldsymbol{e}}_x$ and $-\hat{\boldsymbol{e}}_x$, and $\lambda_{q, \bar{q}}$ denotes the helicities of quark and spectator diquark. The longitudinal momentum fractions $x^{\mathrm{out}}$ and $x^{\mathrm{in}}$ correspond to the momentum fractions carried by the struck quark in the final and initial states, expressed as
\begin{equation}
	x^{\mathrm {out }}=\frac{x-\xi}{1-\xi}, \quad x^{\mathrm {in }}=\frac{x+\xi}{1+\xi}.
\end{equation}
The intrinsic transverse momentum of the struck quark in the final and initial states relative to the hadrons are denoted by $\boldsymbol{k}_{\perp}^{\text{out}}$ and $\boldsymbol{k}_{\perp}^{\text{in}}$, which are given by
\begin{equation}
	\boldsymbol{k}_{\perp}^{\text{out }}=\boldsymbol{k}_{\perp}+\frac{\boldsymbol{\Delta}_{\perp}}{2}\frac{1-x}{1-\xi}, \quad \boldsymbol{k}_{\perp}^{\text {in }}=\boldsymbol{k}_{\perp}-\frac{\boldsymbol{\Delta}_{\perp}}{2}\frac{1-x}{1+\xi}.
\end{equation}
The transverse momentum of the final and initial state hadrons are $\boldsymbol{\Delta}_{\perp}/2$ and $-\boldsymbol{\Delta}_{\perp}/2$, respectively. And the transverse momentum of the struck quark can be expressed by the following equation
\begin{align}
	\boldsymbol{p}_{\perp}^{\text {out }}&=\boldsymbol{k}_{\perp}^{\text {out }}+x^{\text {out }} \frac{\boldsymbol{\Delta}_{\perp}}{2}=\boldsymbol{k}_{\perp}+\frac{\boldsymbol{\Delta}_{\perp}}{2}, \\
	\boldsymbol{p}_{\perp}^{\text {in }}&=\boldsymbol{k}_{\perp}^{\text {in }}-x^{\text {in }} \frac{\boldsymbol{\Delta}_{\perp}}{2}=\boldsymbol{k}_{\perp}-\frac{\boldsymbol{\Delta}_{\perp}}{2}.    
\end{align}
The difference between these two expressions gives the transferred transverse momentum
\begin{equation}
	\boldsymbol{p}_{\perp}^{\text {out }}-\boldsymbol{p}_{\perp}^{\text {in }}=\boldsymbol{\Delta}_{\perp}.
\end{equation}

\subsection{\new{Phenomenological Implications and Connection to Experimental Observables}}

\new{Without considering the gauge link effects, the distribution functions of momentum space depend mainly on the following six variables: $x$, $\boldsymbol{k}_{\perp}$, $\xi$ and $\boldsymbol{\Delta}_{\perp}$, which are also the variables on which generalized transverse-momentum dependent parton distributions (GTMDs) rely~\cite{Lorce:2011dv}. By performing Fourier transform of the transverse momentum, one can map the momentum space distributions into coordinate space, yielding the six-dimensional Wigner distribution functions. Since the GTMDs can be considered partonic “mother functions” from which other functions can be derived, this transformation enables the exploration of new distribution functions obtained by taking limits or integrating the Wigner distributions. These derived functions encapsulate valuable information that warrants further investigation.}

\new{It is worth noting that we consider integrating the Wigner function with respect to certain variables related to experimentally measurable quantities. For example, integrating the Wigner function over momentum variables can yield functions related to the spatial distribution of quarks within the proton, similar to the spatial distributions associated with electromagnetic form factors. In the light-front case, we can also view the Wigner function in a similar manner. Through integration, taking limits, or performing Fourier transforms, we can obtain functions that describe different aspects of the nucleon structure in the complete 3+3 dimensional space, such as PDFs, TMDs, GPDs, GTMDs, and other new distribution functions. This approach is consistent with the broader framework of using phase-space distributions to gain insights into hadron properties.}

\new{Prior to obtaining a more comprehensive understanding of the experimentally observable boost-invariant longitudinal coordinate in our research, it is crucial to thoroughly explore the theoretical underpinnings that define the physical significance of this variable. The skewness variable \(\xi\), a dimensionless parameter that characterizes longitudinal momentum transfer, is defined as
\begin{equation}
	\xi = -\frac{\Delta^+}{2P^+},
\end{equation}
where \(\Delta^+\) represents the longitudinal momentum transfer, and \(P^+\) denotes the average light-front momentum of the nucleon. The variable \(\xi\) describes the difference in longitudinal momentum fractions between the incoming and outgoing partons, typically lying within the range \(-1 \leq \xi \leq 1\), with \(\xi = 0\) corresponding to the scenario of no longitudinal momentum transfer.}

\new{The longitudinal coordinate \(\tilde{z}\), on the other hand, is defined as
\begin{equation}
	\tilde{z} = b^- P^+,
\end{equation}
where \(b^-\) represents the longitudinal coordinate, and \(P^+\) is the average light-front momentum of the nucleon. In the light-front formalism, \(\tilde{z}\) characterizes the distribution of partons along the longitudinal axis. Within the Wigner distribution function, \(\tilde{z}\) serves as a fundamental variable to describe the longitudinal spatial distribution of partons.}

\new{The relationship between \(\xi\) and \(\tilde{z}\) can be elucidated through Fourier transformation. Specifically, in the definition of the Wigner distribution function Eq.~\eqref{wignereq1}, \(\xi\) and \(\tilde{z}\) are interconnected via the exponential factor \(e^{-2i\xi\tilde{z}}\), where \(\xi\) acts as the Fourier conjugate variable to \(\tilde{z}\). This factor, \(e^{-2i\xi\tilde{z}}\), represents the Fourier kernel that links the longitudinal momentum transfer \(\xi\) and the longitudinal spatial distribution \(\tilde{z}\). Consequently, \(\xi\) and \(\tilde{z}\) exhibit a dual relationship governed by the uncertainty principle due to their Fourier conjugate nature. The product of \(\xi\) and \(\tilde{z}\) is dimensionless, and their respective ranges are mutually constrained. Therefore, the new boost-invariant longitudinal coordinate \(\tilde{z}\) can be indirectly measured by measuring the skewness variable \(\xi\).}

\new{In momentum space, the most commonly used distribution functions are the PDFs~\cite{Collins:1981uw}, which describe the parton distribution within a hadron as a function of the longitudinal momentum fraction \( x \). These are primarily measured through DIS experiments. Referring to the definition of PDFs in the momentum space, by integrating over the transverse momentum $\boldsymbol{k}_{\perp}$, the transverse position $\boldsymbol{b}_{\perp}$ and longitudinal momentum fraction $x$, the distribution function with respect to the longitudinal coordinates $\tilde{z}$ can be defined as follows
\begin{equation}
	\rho^{[\Gamma]}(\tilde{z}, S) = \int \mathrm{d}^2 \boldsymbol{k}_{\perp} \, \mathrm{d}^2 \boldsymbol{b}_{\perp} \, \mathrm{d} x \, \rho^{[\Gamma]}\left(\tilde{z}, x, \boldsymbol{b}_{\perp}, \boldsymbol{k}_{\perp}, S\right).
\end{equation}
The distribution defined in coordinate space can be viewed as a Fourier transform of the momentum-space distribution functions. It can be interpreted as a kind of thickness distribution of the nucleon in the new spatial $\tilde{z}$-direction, showing diffraction-like patterns similar to the distribution in the DVCS experiment. This provides a new way to understand the internal structure of the nucleon from a longitudinal perspective, which characterizes the shape of hadrons in longitudinal space.}

\new{Furthermore, the form factors (FFs) can be defined as a function of the momentum transfer $\boldsymbol{\Delta}_{\perp}^2$, which describe the charge and magnetic moment distributions of hadrons~\cite{Hofstadter:1953zjy}. Integrating the Wigner function over momentum variables yields spatial-only functions related to charge and magnetic distributions, analogous to electromagnetic form factors.}

\new{By considering the dependence on the longitudinal momentum fraction \( x \), the skewness variable $\xi$ and the momentum transfer $\boldsymbol{\Delta}_{\perp}^2$, the GPDs~\cite{Belitsky:2005qn} can be introduced. Performing the Fourier transform of the longitudinal and transverse coordinates to momentum space and integrating over the transverse momentum yields the expression for GPDs as 
\begin{equation}
	H(x, \xi, t, S) = \int \mathrm{d}^2\boldsymbol{k}_\perp \int \mathrm{d}\tilde{z} \, \mathrm{d}^2\boldsymbol{b}_\perp \, e^{2i\xi\tilde{z} + i\boldsymbol{b}_\perp \cdot \boldsymbol{\Delta}_\perp} \rho(\tilde{z}, x, \boldsymbol{k}_\perp, \boldsymbol{b}_\perp, S),
\end{equation}
where $t = \boldsymbol{\Delta}_{\perp}^2$ is the square of  the transverse momentum transfer and the variable \(\xi\) quantifies the asymmetry in longitudinal momentum. These distributions encode information about the longitudinal momentum transfer and the transverse position of partons, accessible through DVCS and related exclusive processes. Integrating over the longitudinal momentum fraction  \(x\) and setting the transverse momentum transfer \(\boldsymbol{\Delta}_\perp = 0\) yield a distribution function in terms of \(\xi\). This distribution can subsequently be utilized to infer the corresponding distribution in \(\tilde{z}\), establishing an indirect relationship between the longitudinal momentum transfer and the longitudinal spatial coordinate.}

\new{Additionally, the transverse momentum-dependent form factors (TMFFs)~\cite{Pasquini:2013uja} can be defined in the context of transverse momentum $\boldsymbol{k}_{\perp}$ and momentum transfer $\boldsymbol{\Delta}_{\perp}$, which can be probed in high-energy scattering experiments such as electron-proton scattering. These distributions can be derived through the integration of the six-dimensional Wigner distribution, and the newly introduced three-dimensional distributions can be analogously correlated with experimentally observable quantities, following a similar approach as in relevant physical frameworks.}

\new{Moreover, the GTMDs can be defined to describe the dependence of parton distributions on the longitudinal momentum fraction $x$, the longitudinal coordinate $\tilde{z}$, the transverse momentum $\boldsymbol{k}_{\perp}$ and the momentum transfer \(\boldsymbol{\Delta}_\perp\). GTMDs can be probed through exclusive processes like diffractive di-jet production (sensitive to gluon GTMDs)~\cite{Hatta:2016dxp} and double Drell-Yan processes (sensitive to quark GTMDs)~\cite{Bhattacharya:2017bvs}. Since the six-dimensional light-front Wigner distribution functions are related to the GTMDs via a Fourier transform, this relationship allows indirect access to the Wigner function, offering deeper insights into hadron structure.}

\new{When considering transverse momentum dependence, the TMDs can be defined, which describe the parton distribution as a function of both the longitudinal momentum fraction \( x \) and the transverse momentum $\boldsymbol{k}_{\perp}$. TMDs are typically probed through SIDIS and Drell-Yan processes, so the transverse Wigner distributions can provide insights into the single spin asymmetries observed in SIDIS and the Drell-Yan process~\cite{Ji:2004xq}.}

\new{With reference to the introduction of TMDs, we can also consider the dependence of the transverse momentum $\boldsymbol{k}_{\perp}$ on the basis of the previous longitudinal spatial distribution $\rho^{[\Gamma]}(\tilde{z})$. By integrating the longitudinal momentum fraction $x$ and the transverse position $\boldsymbol{b}_{\perp}$, the three-dimensional parton distributions can be obtained as
\begin{equation}
	\rho^{[\Gamma]}\left(\tilde{z}, \boldsymbol{k}_{\perp}, S\right)=\int \mathrm{d} x \, \mathrm{d}^2 \boldsymbol{b}_{\perp} \, \rho^{[\Gamma]}\left(\tilde{z}, x, \boldsymbol{b}_{\perp}, \boldsymbol{k}_{\perp}, S\right),  
\end{equation}
which describes the parton distribution in the joint three-dimensional space of longitudinal coordinate $\tilde{z}$ and transverse momentum $\boldsymbol{k}_{\perp}$, with the dependence on light-front time.}

\new{Additionally, by considering the dependence on the transverse coordinate \( \boldsymbol{b}_\perp \), we can define parton distribution functions that depend on both \( \tilde{z} \) and \( \boldsymbol{b}_\perp \). Integrating over the longitudinal momentum fraction $x$ and the transverse momentum $\boldsymbol{k}_{\perp}$ defines the three-dimensional position-space parton distributions as
\begin{equation}
	\rho^{[\Gamma]}\left(\tilde{z}, \boldsymbol{b}_{\perp}, S\right)=\int \mathrm{d} x \, \mathrm{d}^2 \boldsymbol{k}_{\perp} \, \rho^{[\Gamma]}\left(\tilde{z}, x, \boldsymbol{b}_{\perp}, \boldsymbol{k}_{\perp}, S\right),
\end{equation}
which contains longitudinal position $\tilde{z}$ and transverse position $\boldsymbol{b}_{\perp}$. }

\new{In addition to the correlations between transverse coordinate $\boldsymbol{b}_{\perp}$ and transverse momentum $\boldsymbol{k}_{\perp}$, we are also capable of defining mixed distributions that pertain to the association between transverse coordinates and specific directional components of the transverse momentum. By integrating over different combinations of variables, one can define
\begin{align}
	& \rho^{[\Gamma]}\left(\tilde{z}, b_x, k_y, S\right)=\int \mathrm{d} x \, \mathrm{d} b_y \, \mathrm{d} k_x \, \rho^{[\Gamma]}\left(\tilde{z}, x, \boldsymbol{b}_{\perp}, \boldsymbol{k}_{\perp}, S\right),\\
	& \rho^{[\Gamma]}\left(\tilde{z}, b_y, k_x, S\right)=\int \mathrm{d} x \, \mathrm{d} b_x \, \mathrm{d} k_y \, \rho^{[\Gamma]}\left(\tilde{z}, x, \boldsymbol{b}_{\perp}, \boldsymbol{k}_{\perp}, S\right),\\
	& \rho^{[\Gamma]}\left(\tilde{z}, b_y, k_y, S\right)=\int \mathrm{d} x \, \mathrm{d} b_x \, \mathrm{d} k_x \, \rho^{[\Gamma]}\left(\tilde{z}, x, \boldsymbol{b}_{\perp}, \boldsymbol{k}_{\perp}, S\right),\\
	& \rho^{[\Gamma]}\left(\tilde{z}, b_x, k_x, S\right)=\int \mathrm{d} x \, \mathrm{d} b_y \, \mathrm{d} k_y \, \rho^{[\Gamma]}\left(\tilde{z}, x, \boldsymbol{b}_{\perp}, \boldsymbol{k}_{\perp}, S\right).
\end{align}
Such mixed distributions encode correlations between transverse position and momentum, providing a novel window into the partonic structure of the nucleon.}

\new{Specifically, the six-dimensional Wigner operator can be associated with the intrinsic orbital angular momentum and can be expressed as
\begin{equation}
	\ell_{z}^{q}=\int\mathrm{d}\tilde{z}\,\mathrm{d}x\,\mathrm{d}^2\boldsymbol{k}_{\perp}\,\mathrm{d}^2\boldsymbol{b}_{\perp}(\boldsymbol{b}_{\perp}\times\boldsymbol{k}_{\perp})_z\rho^q_{LU}(\tilde{z},x,\boldsymbol{b}_{\perp},\boldsymbol{k}_{\perp}, S),
\end{equation}	
where the longitudinal intrinsic orbital angular momentum of quarks can be deduced from the longitudinal unpolarized distribution. Similarly, the spin-orbit correlation of quarks can be associated with the unpolarized longitudinal distribution as
\begin{equation}
	\label{OAM}
	C_z^q=\int\mathrm{d}\tilde{z}\,\mathrm{d}x\,\mathrm{d}^2\boldsymbol{k}_{\perp}\,\mathrm{d}^2\boldsymbol{b}_{\perp}(\boldsymbol{b}_{\perp}\times\boldsymbol{k}_{\perp})_z\rho^q_{UL}(\tilde{z},x,\boldsymbol{b}_{\perp},\boldsymbol{k}_{\perp}, S).
\end{equation}	
The distributions provide information on the quark orbital angular momentum, which is a key component of the proton spin. }

\new{Beyond the discussion of conventional distribution functions, we can also investigate quasi-probability distributions, revealing deeper aspects of quantum mechanics. Due to the uncertainty principle, it is not possible to simultaneously measure both coordinate and momentum space variables directly. Therefore, the six-dimensional Wigner distribution functions cannot be measured directly but can be inferred indirectly through experimental measurements of related quantities, such as GTMDs or other high-energy scattering observables. To better demonstrate this quasi-probabilistic nature, we integrate the transverse momentum $\boldsymbol{k}_{\perp}$ and the transverse position $\boldsymbol{b}_{\perp}$, which gives us a reduction as
\begin{equation}
	\rho^{[\Gamma]}(\tilde{z}, x, S) = \int \mathrm{d}^2 \boldsymbol{k}_{\perp} \, \mathrm{d}^2 \boldsymbol{b}_{\perp} \, \rho^{[\Gamma]}\left(\tilde{z}, x, \boldsymbol{b}_{\perp}, \boldsymbol{k}_{\perp}, S\right).
\end{equation}
It is important to note that, due to the non-zero value of $\xi$, there is a non-zero $\Delta^{-}$ depending on $x^{+}$. This resulting distribution is not positive definite, indicating the presence of quantum mechanical interference effects.}

\section{Numerical results and discussion}
\label{Sec:Numerical}

Based on the relevant literatures, the parameters chosen for our calculations are $m = 0.33\,\mathrm{GeV}$ and $\beta_D = 0.33\,\mathrm{GeV}$ ~\cite{Schlumpf:1993rm,Weber:1990fx,Huang:1994dy}. It is important to note that the masses of the spectators are effectively treated as effective parameters in this context. For our calculations, we adopt $m_S = 0.6\,\mathrm{GeV}$ and $m_V = 0.8\,\mathrm{GeV}$ as the effective masses for the scalar and vector diquarks, respectively. Further investigation indicates that the spectator masses exhibit a spectrum that can be extracted through experimental methods. Recent studies focusing on diquark correlations within nucleons have provided valuable insights into this aspect of hadron structure~\cite{Maris:2002yu,Maris:2004bp,Eichmann:2008ef,Eichmann:2009qa,Eichmann:2011ec,Cloet:2013gva}. These findings enhance our understanding of the role of diquarks in nucleon dynamics and their contributions to the overall structure of hadrons. 
%The incorporation of these parameters and insights into our calculations will improve the accuracy and relevance of our theoretical predictions.

The newly defined six-dimensional light-front Wigner distribution is a six-dimensional function with respect to the coordinates $\tilde{z}$, $x$, $b_x$, $b_y$, $k_x$, $k_y$. To analyze how the boost-invariant six-dimensional light-front Wigner distribution function behaves with respect to the newly introduced longitudinal coordinate $\tilde{z}$, we integrate over three dimensions of the transverse coordinates and transverse momentum. Subsequently, we fix the dimension $x$ at specific values and plot the Wigner distribution with respect to the remaining transverse coordinates or transverse momentum. \new{It is worth mentioning that, in contrast to the prior work~\cite{Han:2022tlh}, we optimize key parameters to improve the accuracy and physical relevance of our results. Specifically, we narrow the range of the longitudinal momentum fraction $x$ from $x=0.25,0.5,0.75$ to $x=0.1,0.25,0.4$, better capturing the valence quark distribution peak at $x\approx0.25$~\cite{H1:2012xnw}. Additionally, we adjust the transverse momentum parameter from \( \boldsymbol{k}_\perp = 0.6\,\mathrm{GeV}\boldsymbol{\hat{e}}_x \) to \( \boldsymbol{k}_\perp = 0.3\,\mathrm{GeV}\boldsymbol{\hat{e}}_x \), aligning it more closely with the spatial extent set by the harmonic oscillator scale parameter $\beta_D=0.33\,\mathrm{GeV}$. These adjustments ensure that our results are more consistent to reflect the proton internal structure.}

In this paper, we have performed calculations for all polarization states of the proton. For clarity and conciseness, we present the representative results for the unpolarized, longitudinal-polarized, longitudinal-transverse, and transverse-longitudinal  polarization states in the main text, which are detailed in Sec.~\ref{UU}, Sec.~\ref{LL}, Sec.~\ref{LT} and Sec.~\ref{TL}, respectively. The results for the remaining polarization configurations are provided in the Appendix~\ref{Appendix} for completeness.

\new{Specifically, our six-dimensional Wigner distribution can be reduced to five-dimensional distribution by integrating over the longitudinal coordinate $\tilde{z}$, as demonstrated by the following expression
\begin{equation}
	\rho\left(x, \boldsymbol{b}_{\perp}, \boldsymbol{k}_{\perp}, S\right)=\int \mathrm{d} \tilde{z} \rho\left(\tilde{z}, x, \boldsymbol{b}_{\perp}, \boldsymbol{k}_{\perp}, S\right).
\end{equation}
Substituting the explicit form of the six-dimensional Wigner distribution, we obtain 
\begin{equation}
	\rho\left(x, \boldsymbol{b}_{\perp}, \boldsymbol{k}_{\perp}, S\right)=\int \mathrm{d} \tilde{z} \int \frac{\mathrm{d} \xi \mathrm{d}^2 \boldsymbol{\Delta}_{\perp}}{4 \pi^3} e^{-2 i \xi \tilde{z}-i \boldsymbol{b}_\perp \cdot \boldsymbol{\Delta}_{\perp}} \left\langle P+\frac{\boldsymbol{\Delta}}{2}, S\left| \widehat{W}^{[\Gamma]}\right| P-\frac{\boldsymbol{\Delta}}{2}, S\right\rangle.
\end{equation}
The integration over $\tilde{z}$ produces a Dirac delta function
\begin{equation}
	\int \mathrm{d} \tilde{z} e^{-2 i \xi \tilde{z}}=2 \pi \delta(2 \xi),
\end{equation}
which simplifies the expression to
\begin{equation}
	\rho\left(x, \boldsymbol{b}_{\perp}, \boldsymbol{k}_{\perp}, S\right)=\left.\int \frac{\mathrm{d}^2 \boldsymbol{\Delta}_{\perp}}{4 \pi^2} e^{-i \boldsymbol{b}_{\perp} \cdot \boldsymbol{\Delta}_{\perp}}\left\langle P+\frac{\boldsymbol{\Delta}}{2}, S\right| \widehat{W}^{[\Gamma]}\left|P-\frac{\boldsymbol{\Delta}}{2}, S\right\rangle\right|_{\xi=0}.
\end{equation}
This result indicates that, by integrating over the longitudinal coordinates $\tilde{z}$, the Wigner distribution returns to the $\xi=0$ frame, which depends only on $x$, $\boldsymbol{b}_{\perp}$ and $\boldsymbol{k}_{\perp}$, effectively reducing it to a five-dimensional distribution. Further integration over $x$ allows us to numerically study the behavior in terms of transverse coordinate $\boldsymbol{b}_{\perp}$ or transverse momentum $\boldsymbol{k}_{\perp}$, which aligns with the results presented in the previous research~\cite{Liu:2015eqa}. Our analysis confirms that the reduced distributions are consistent with existing results in the literature, thereby validating the self-consistency of our approach. This comparison also highlights the model-dependence of our conclusions, as the integration process preserves the essential features of the distributions while simplifying their dimensionality.}

\new{Overall, within the framework of the six-dimensional light-front Wigner distributions formalism, the longitudinal coordinate $\tilde{z}$ exhibits a centrosymmetric or dipole-symmetric characteristic when compared to the transverse coordinate $\boldsymbol{b}_{\perp}$ or the transverse momentum $\boldsymbol{k}_{\perp}$. A distinct feature setting it apart from traditional distribution functions is that these six-dimensional Wigner distributions is not positive definite, reflecting its unique role in quantum phase-space analysis rather than serving as a probability density function.}

%\subsection{Unpolarized proton}

\subsection{Unpolarized  Wigner distribution}
\label{UU}
 
\begin{figure}[htbp]
	\centering
	\subfloat{
		\includegraphics[width=0.31\textwidth]{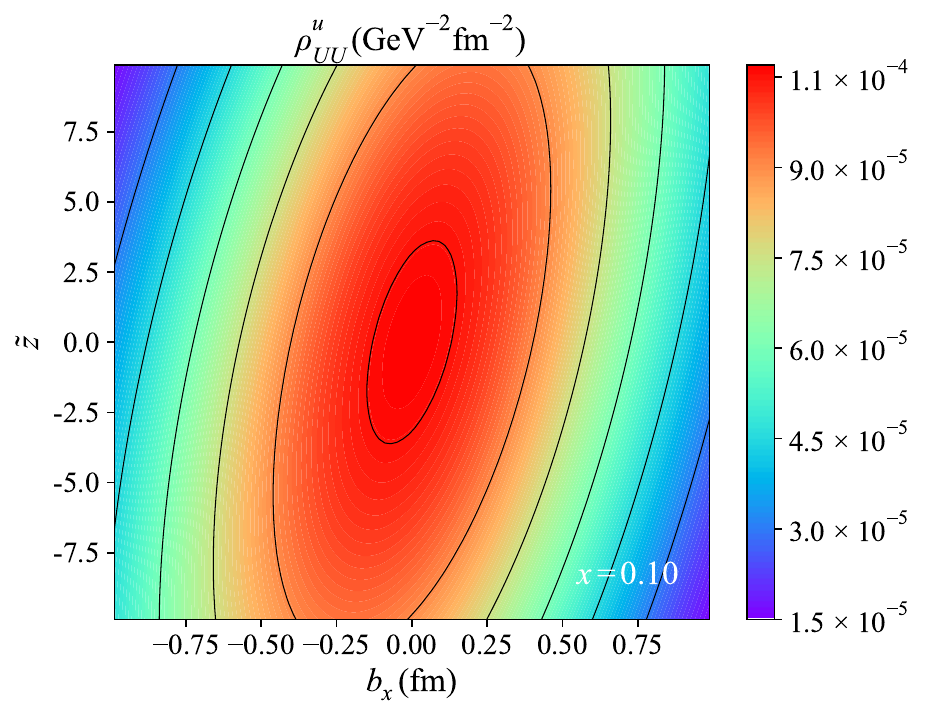}
	}
	\subfloat{
		\includegraphics[width=0.31\textwidth]{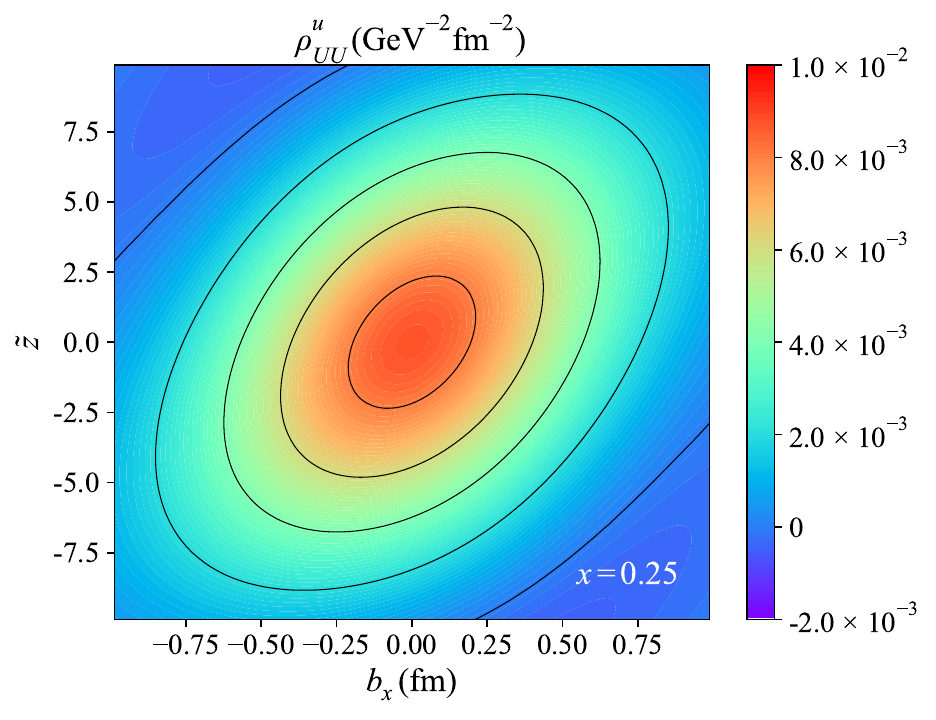}
	}
	\subfloat{
		\includegraphics[width=0.31\textwidth]{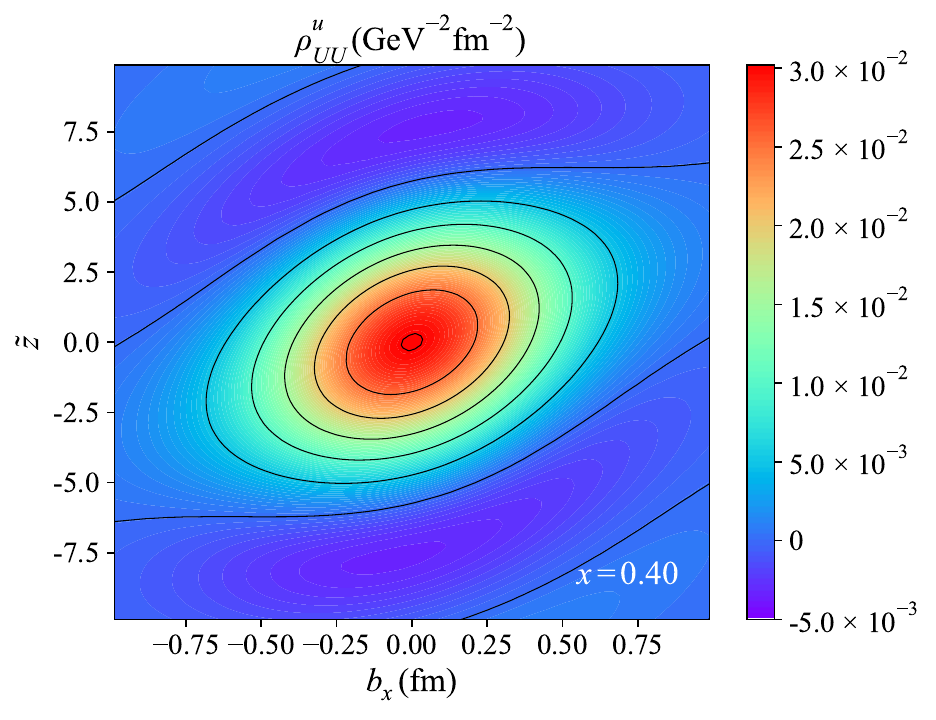}
	}\\
	\subfloat{
		\includegraphics[width=0.31\textwidth]{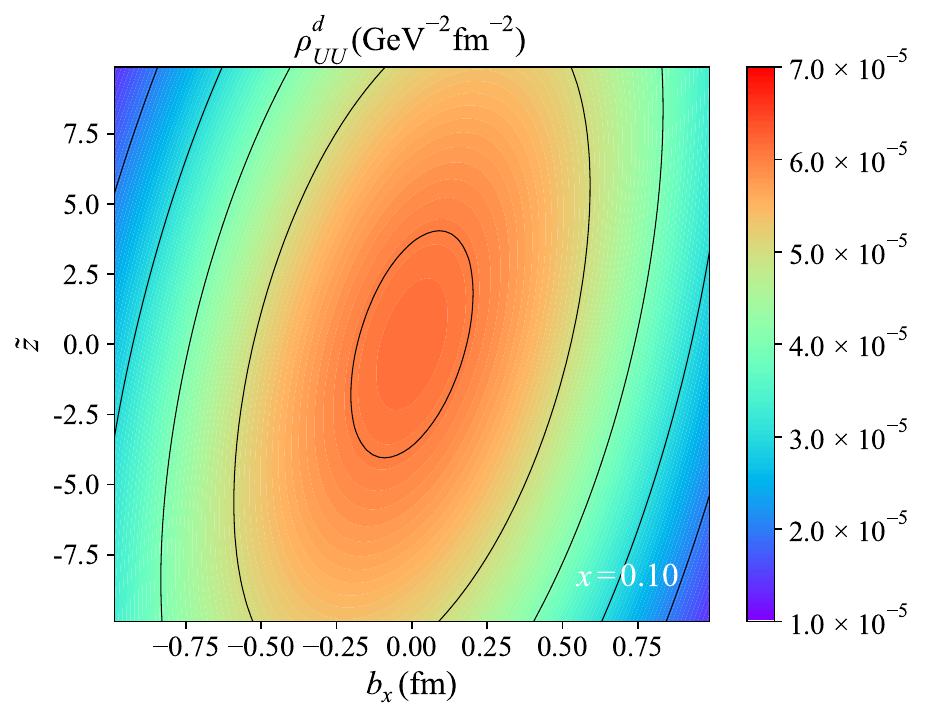}
	}
	\subfloat{
		\includegraphics[width=0.31\textwidth]{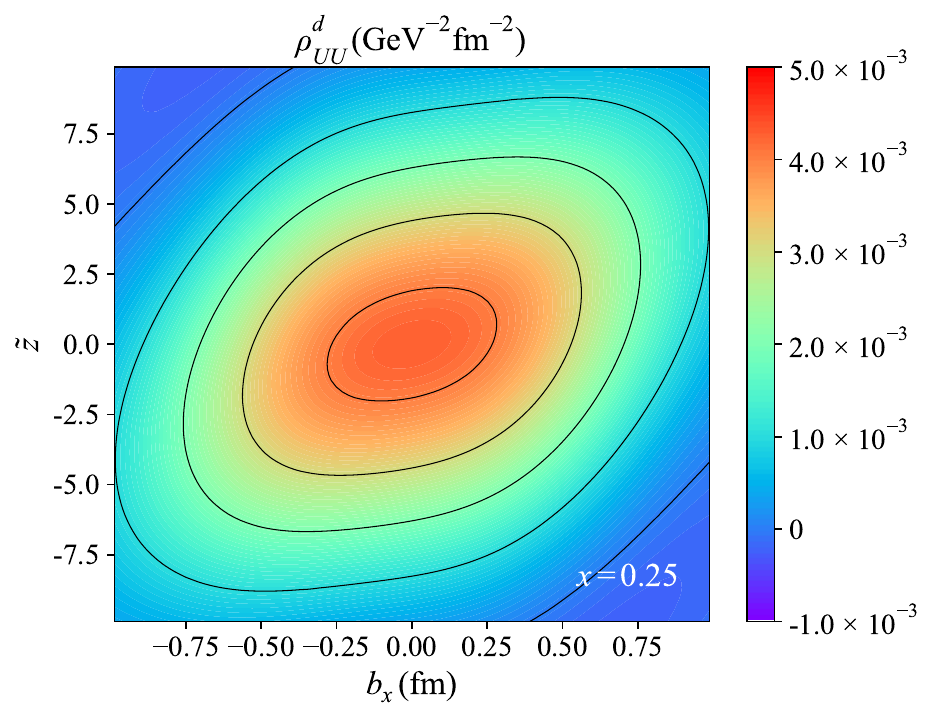}
	}
	\subfloat{
		\includegraphics[width=0.31\textwidth]{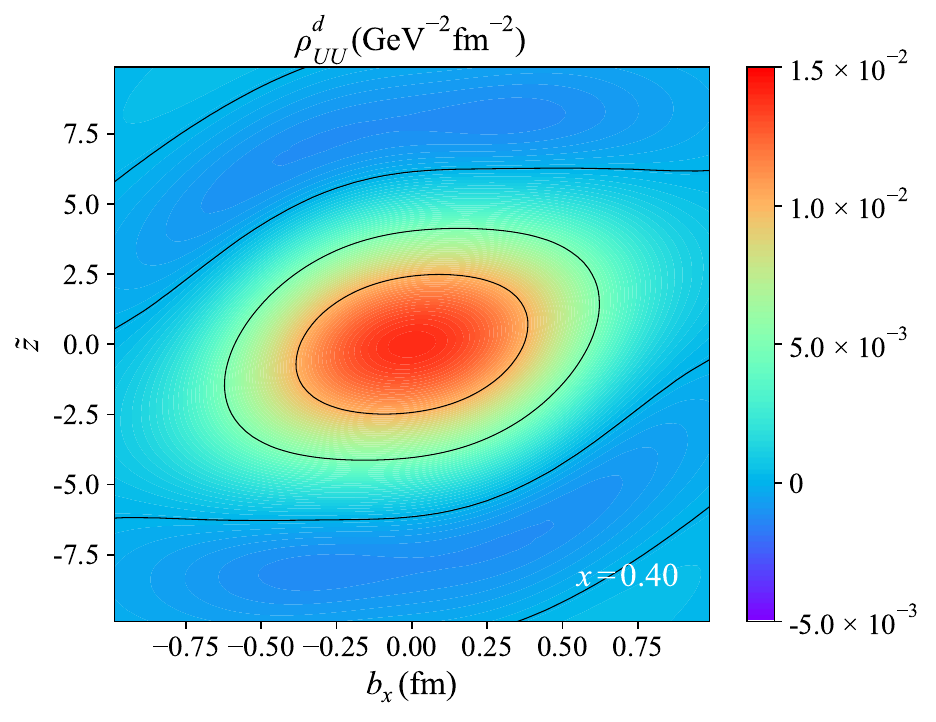}
	}
	\caption{Six-dimensional unpolarized light-front Wigner distribution  $\rho_{\mathrm{UU}}\left(\tilde{z},x,\boldsymbol{b}_{\perp}, \boldsymbol{k}_{\perp}\right)$ for $u$ quark (upper panels) and $d$ quark (lower panels). The figure presents the Wigner distribution in the $\tilde{z}-b_x$ plane, with the transverse momentum fixed at $\boldsymbol{k}_{\perp}=0.3\,\mathrm{GeV}\boldsymbol{\hat{e}}_x$ (where $\boldsymbol{\hat{e}}_x$ is the unit vector in the $x$-direction) and the transverse coordinate component fixed at $b_y=0.4\,\mathrm{GeV}^{-1}$. The three columns correspond to $x=0.10$, $x=0.25$, and $x=0.40$.}
	\label{6DProtonUUudzbx}
\end{figure}

\begin{figure}[htbp]
	\centering
	\subfloat{
		\includegraphics[width=0.31\textwidth]{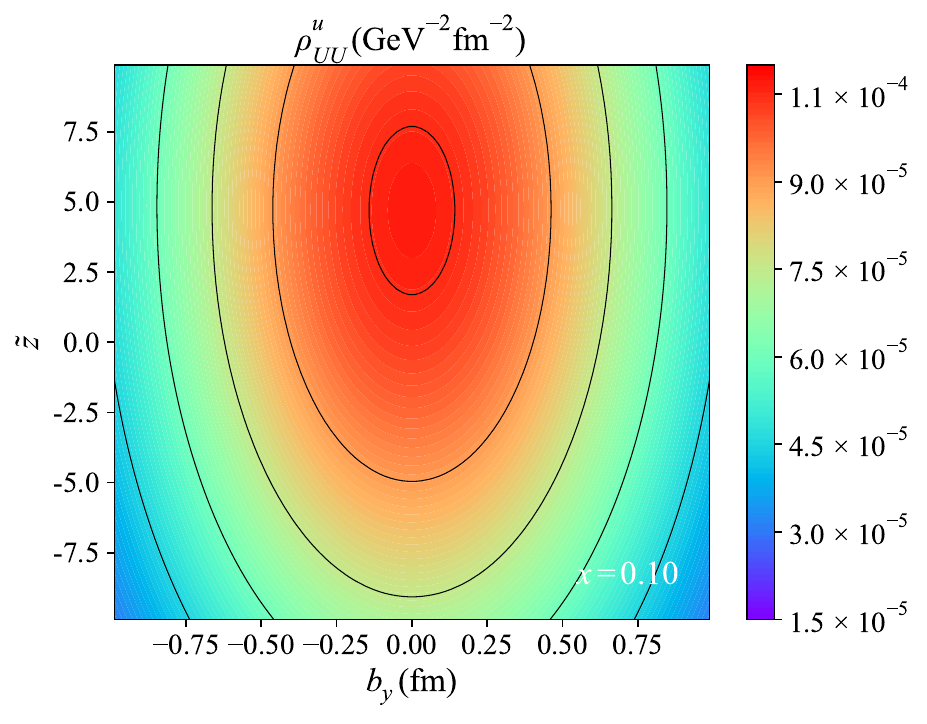}
	}
	\subfloat{
		\includegraphics[width=0.31\textwidth]{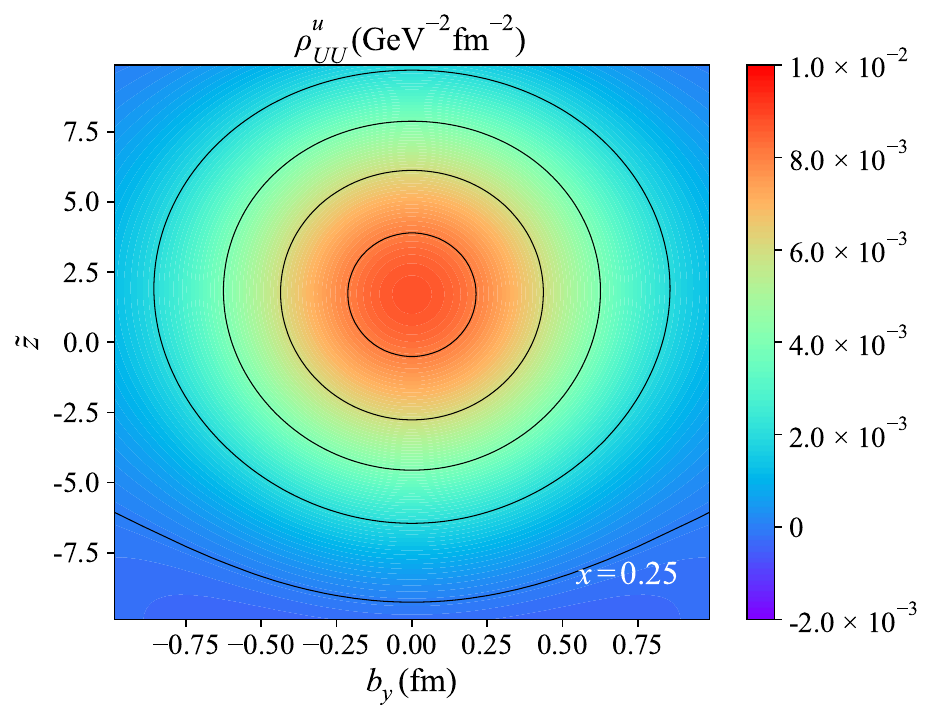}
	}
	\subfloat{
		\includegraphics[width=0.31\textwidth]{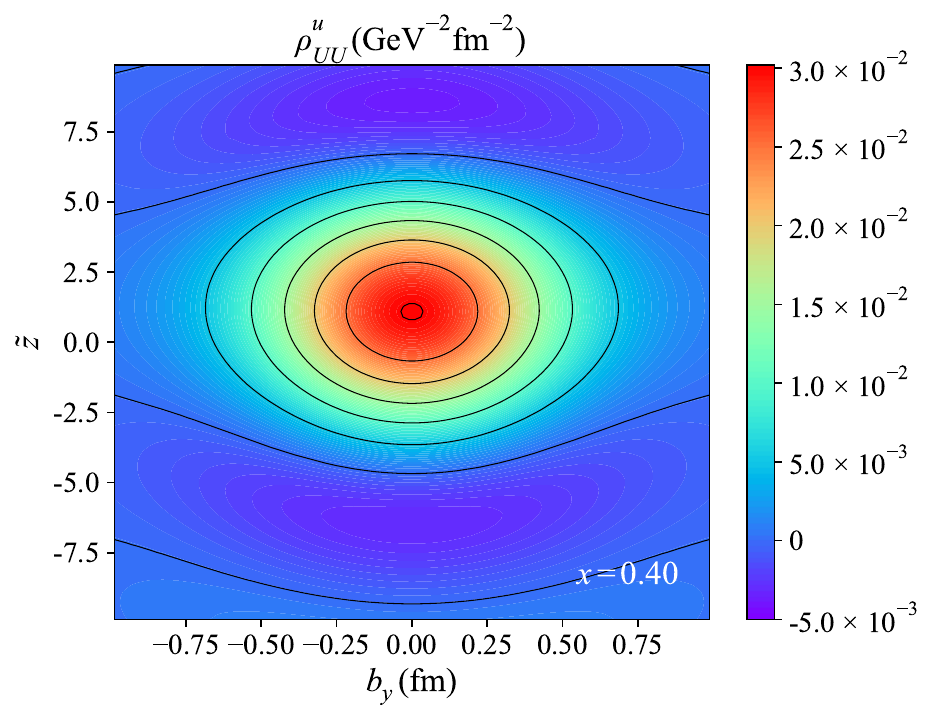}
	}\\
	\subfloat{
		\includegraphics[width=0.31\textwidth]{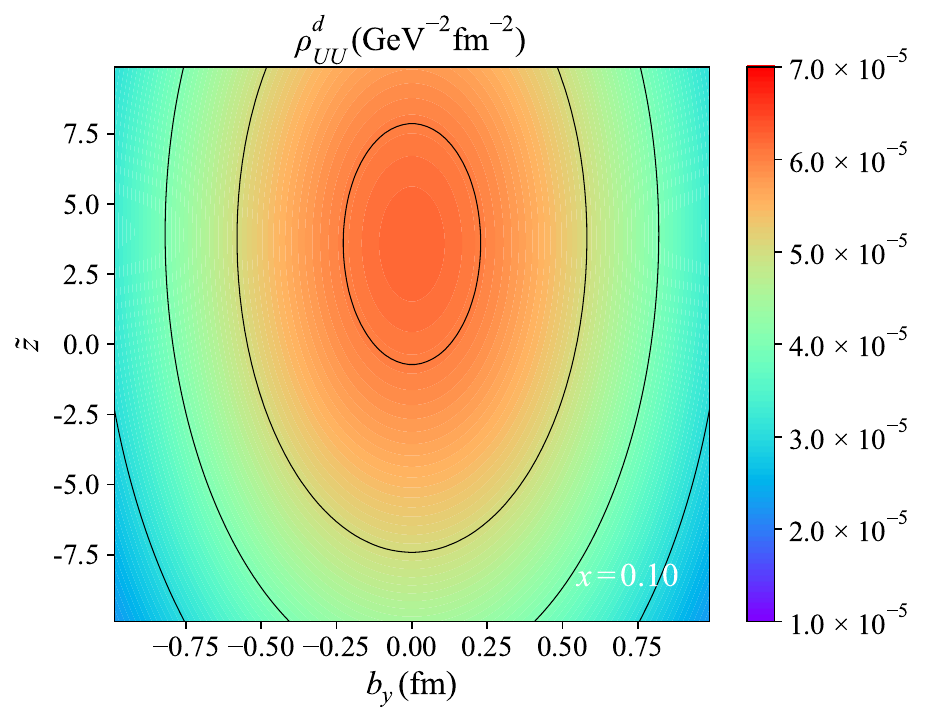}
	}
	\subfloat{
		\includegraphics[width=0.31\textwidth]{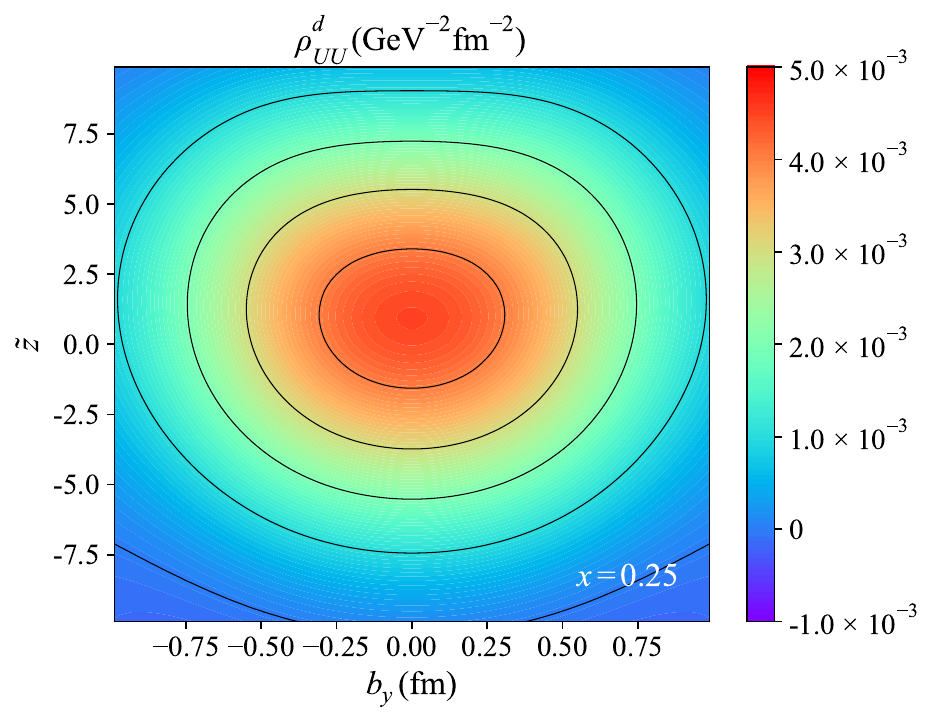}
	}
	\subfloat{
		\includegraphics[width=0.31\textwidth]{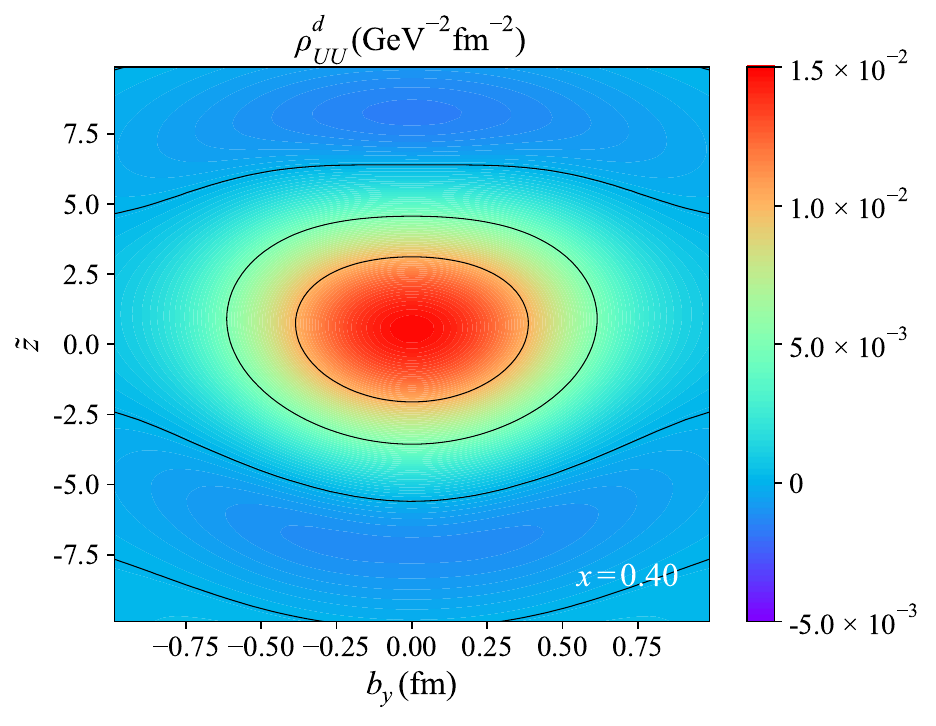}
	}
	\caption{Six-dimensional unpolarized light-front Wigner distribution  $\rho_{\mathrm{UU}}\left(\tilde{z},x,\boldsymbol{b}_{\perp}, \boldsymbol{k}_{\perp}\right)$ for $u$ quark (upper panels) and $d$ quark (lower panels). The figure presents the Wigner distributions in the $\tilde{z}-b_y$ plane, with the transverse momentum fixed at $\boldsymbol{k}_{\perp}=0.3\,\mathrm{GeV}\boldsymbol{\hat{e}}_x$ (where $\boldsymbol{\hat{e}}_x$ is the unit vector in the $x$-direction) and the transverse coordinate component fixed at $b_x=0.4\,\mathrm{GeV}^{-1}$. The three columns correspond to $x=0.10$, $x=0.25$, and $x=0.40$.}
	\label{6DProtonUUudzby}
\end{figure}

\begin{figure}[htbp]
	\centering
	\subfloat{
		\includegraphics[width=0.31\textwidth]{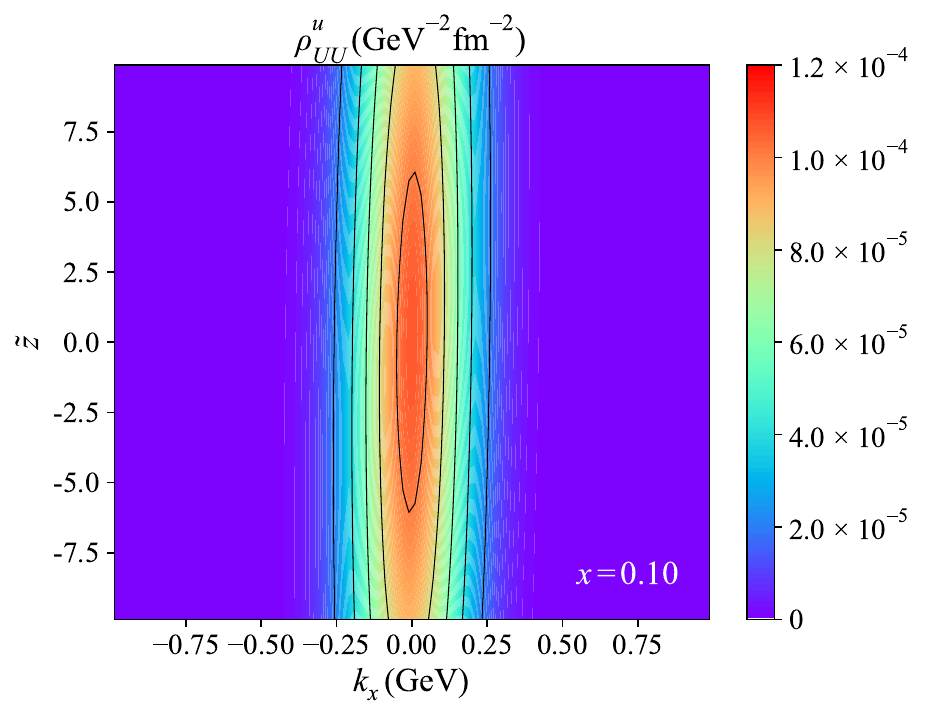}
	}
	\subfloat{
		\includegraphics[width=0.31\textwidth]{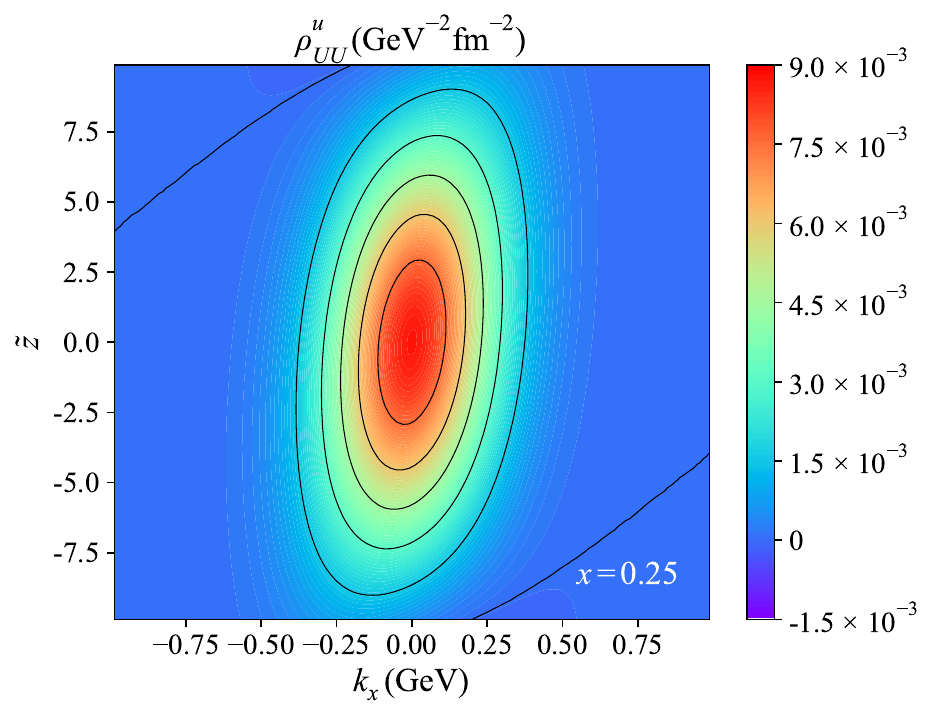}
	}
	\subfloat{
		\includegraphics[width=0.31\textwidth]{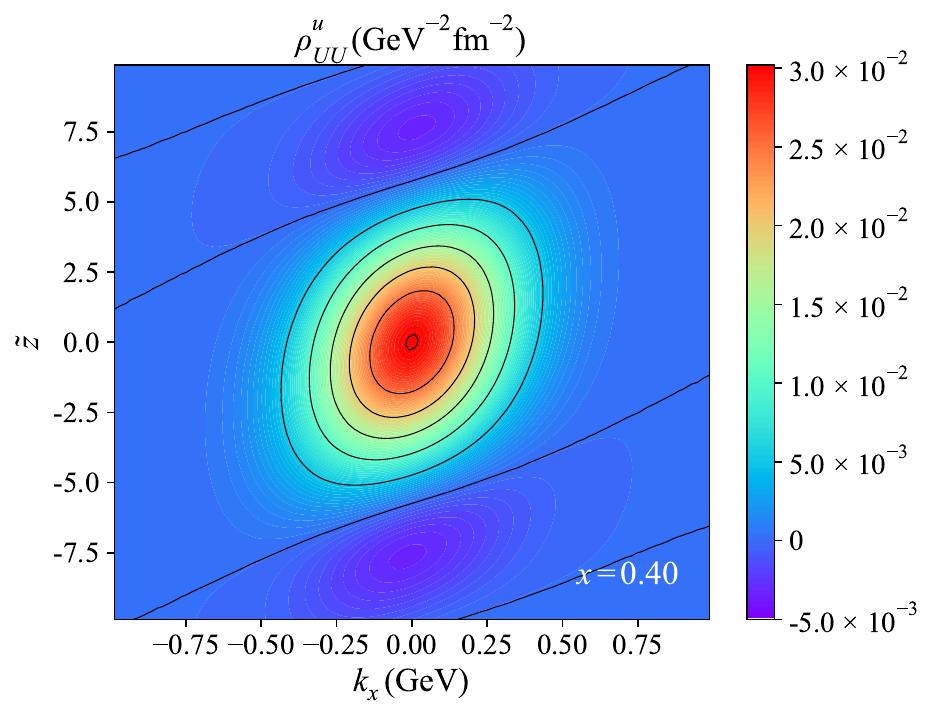}
	}\\
	\subfloat{
		\includegraphics[width=0.31\textwidth]{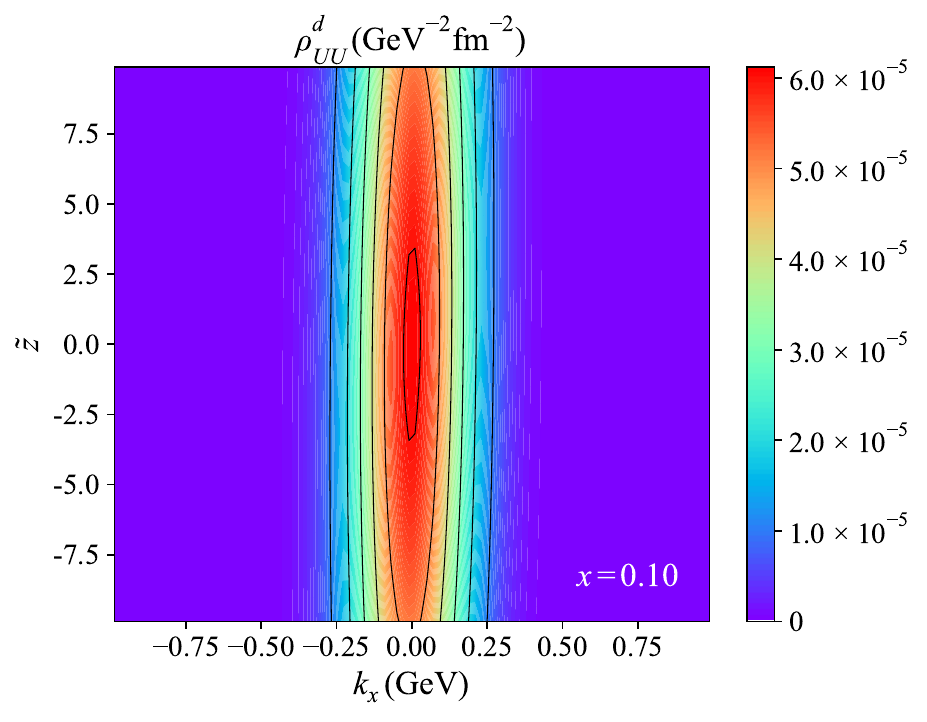}
	}
	\subfloat{
		\includegraphics[width=0.31\textwidth]{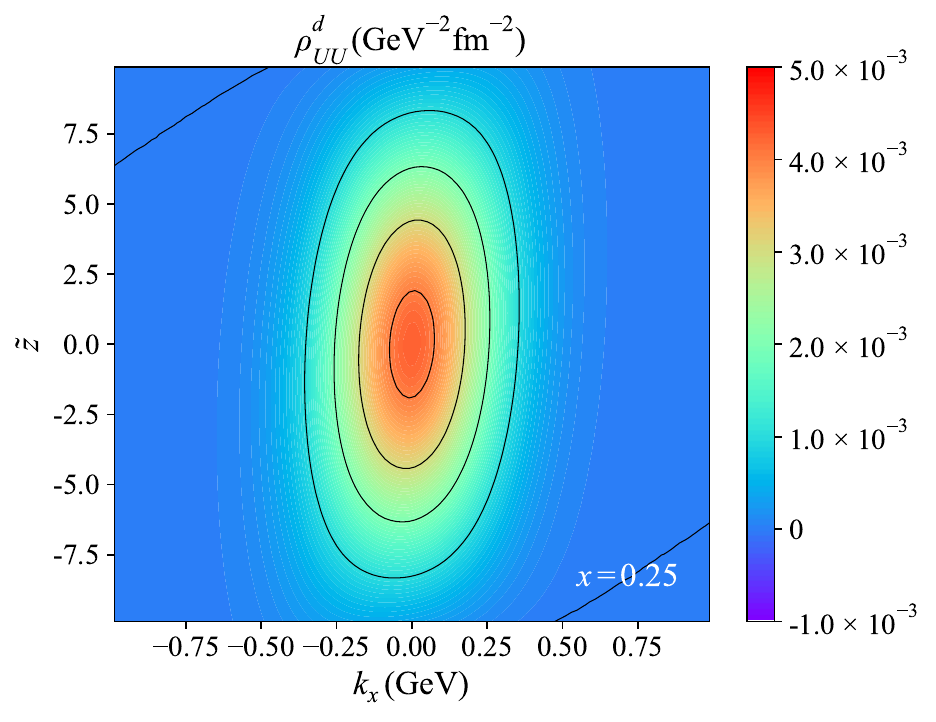}
	}
	\subfloat{
		\includegraphics[width=0.31\textwidth]{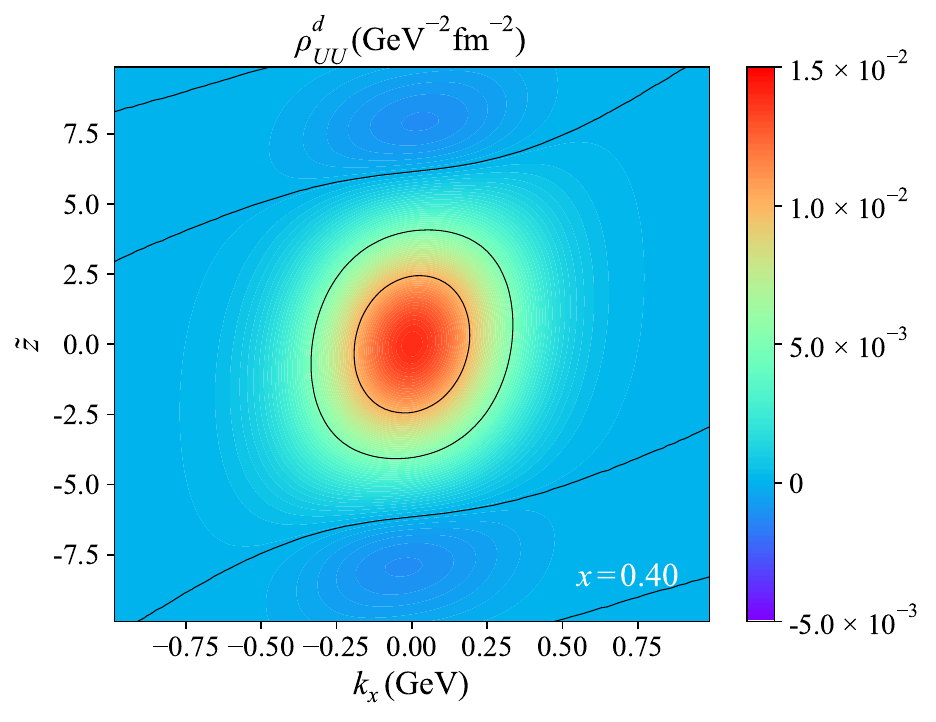}
	}
	\caption{Six-dimensional unpolarized light-front Wigner distribution  $\rho_{\mathrm{UU}}\left(\tilde{z},x,\boldsymbol{b}_{\perp}, \boldsymbol{k}_{\perp}\right)$ for $u$ quark (upper panels) and $d$ quark (lower panels). The figure presents the Wigner distributions in the $\tilde{z}-k_x$ plane, with the transverse coordinate fixed at $\boldsymbol{b}_{\perp}=0.4\,\mathrm{GeV}^{-1}\boldsymbol{\hat{e}}_x$ (where $\boldsymbol{\hat{e}}_x$ is the unit vector along the $x$-axis) and the transverse momentum component fixed at $k_y=0.3\,\mathrm{GeV}$. The three columns correspond to $x=0.10$, $x=0.25$, and $x=0.40$.}
	\label{6DProtonUUudzkx}
\end{figure}

\begin{figure}[htbp]
	\centering
	\subfloat{
		\includegraphics[width=0.31\textwidth]{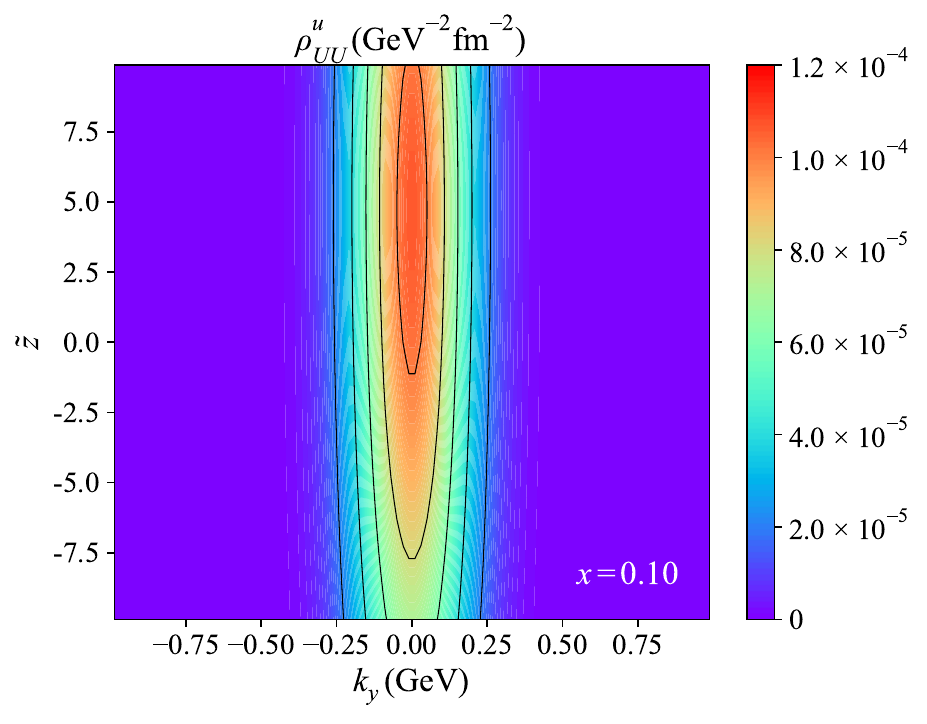}
	}
	\subfloat{
		\includegraphics[width=0.31\textwidth]{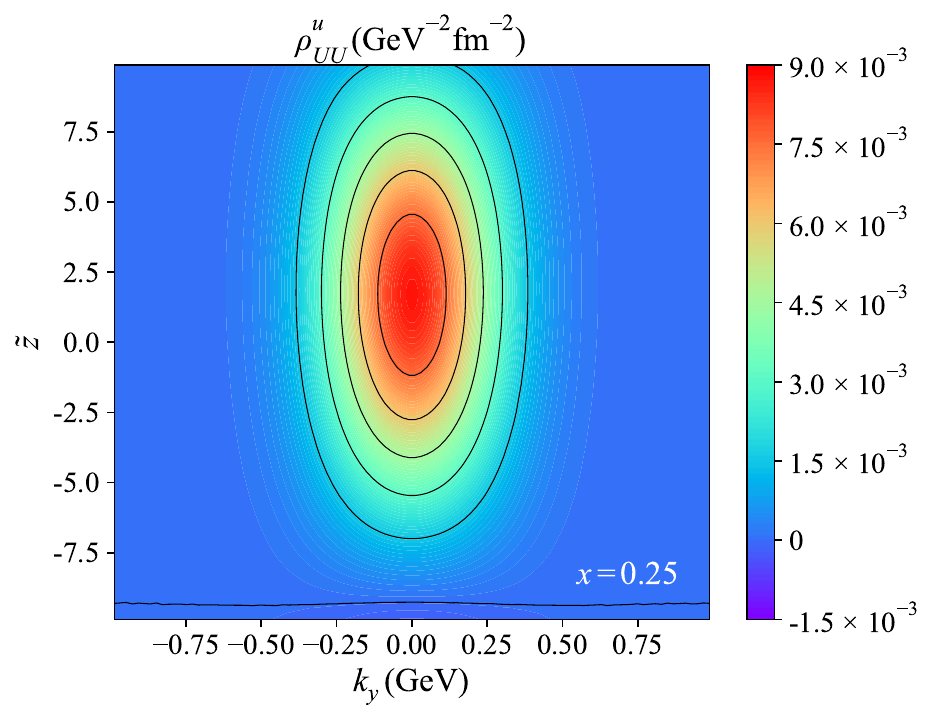}
	}
	\subfloat{
		\includegraphics[width=0.31\textwidth]{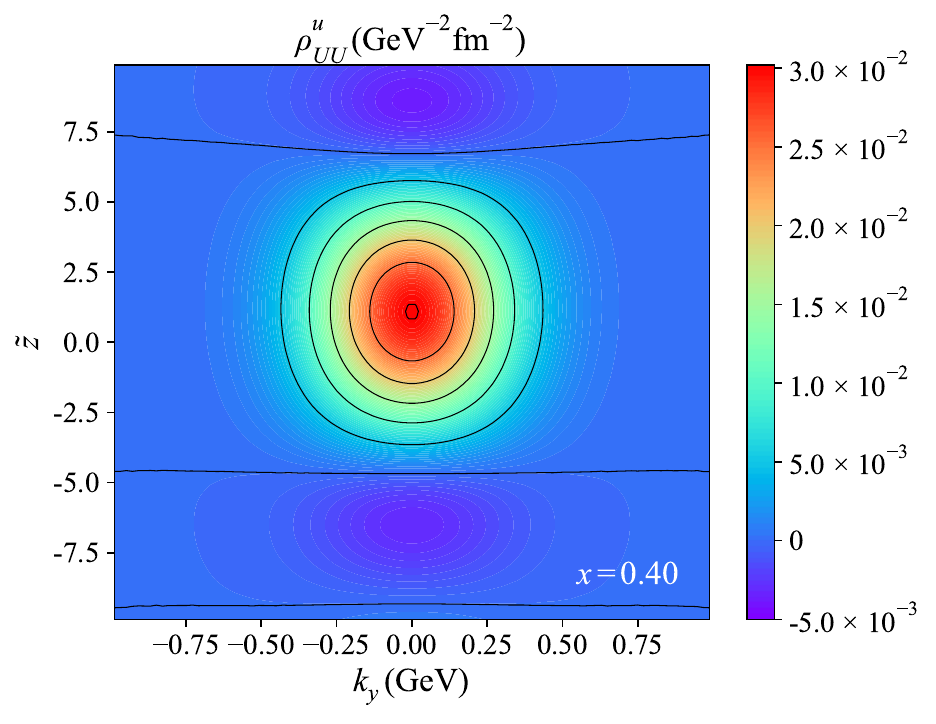}
	}\\
	\subfloat{
		\includegraphics[width=0.31\textwidth]{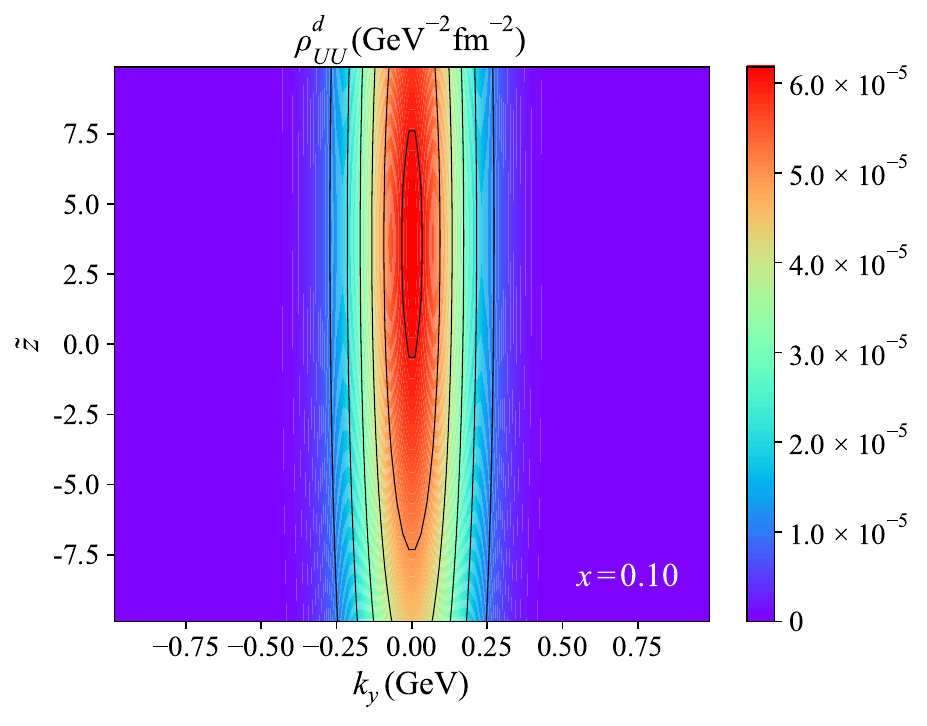}
	}
	\subfloat{
		\includegraphics[width=0.31\textwidth]{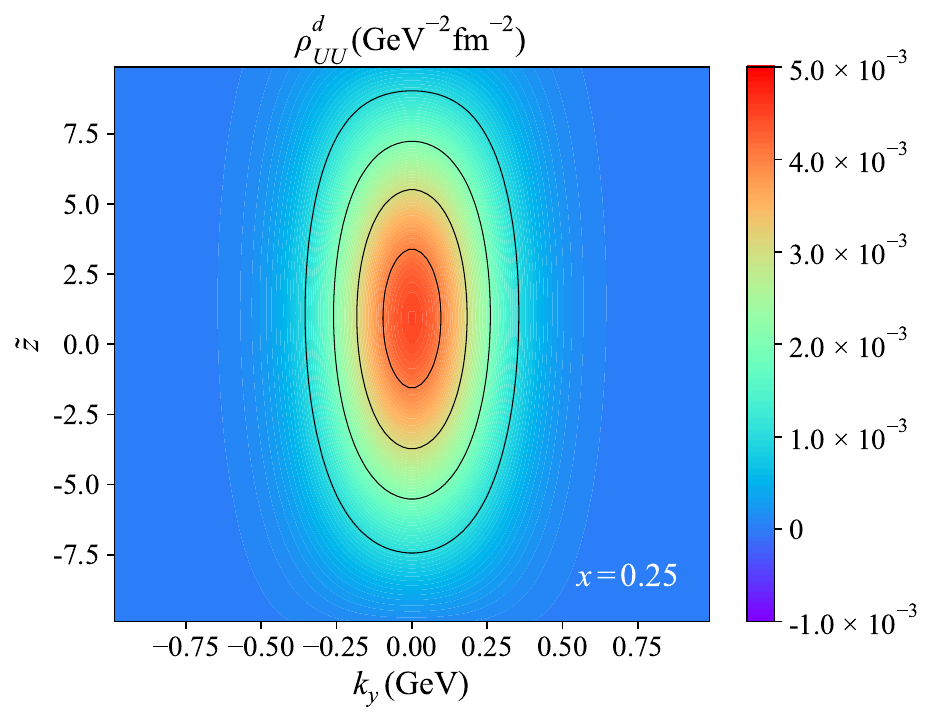}
	}
	\subfloat{
		\includegraphics[width=0.31\textwidth]{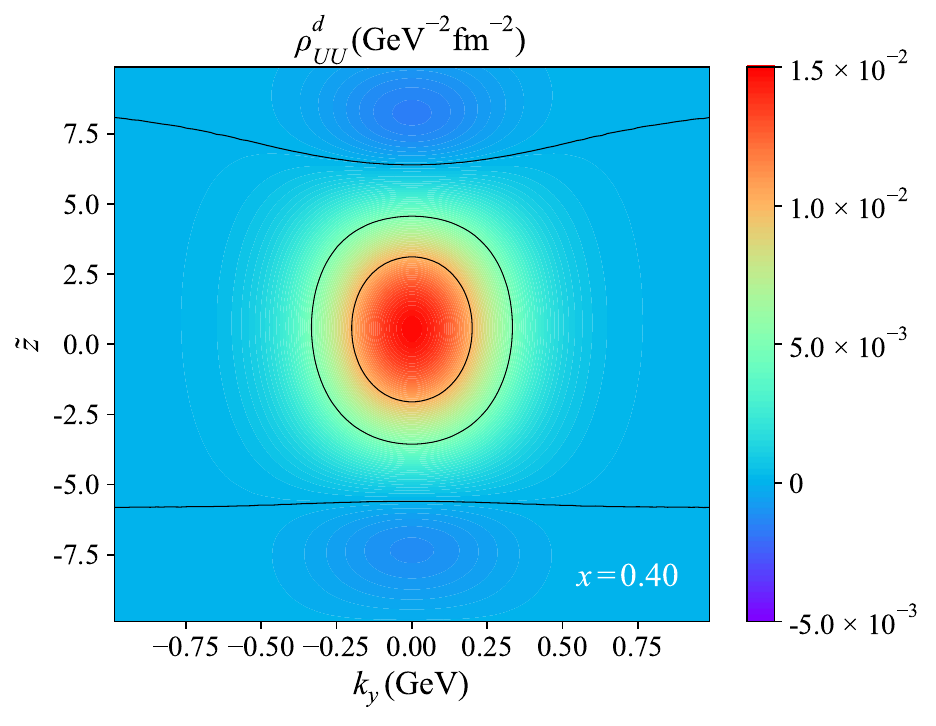}
	}
	\caption{Six-dimensional unpolarized light-front Wigner distribution $\rho_{\mathrm{UU}}\left(\tilde{z},x,\boldsymbol{b}_{\perp}, \boldsymbol{k}_{\perp}\right)$ for $u$ quark (upper panels) and $d$ quark (lower panels). The figure presents the Wigner distributions in the $\tilde{z}-k_y$ plane, with the transverse coordinate fixed at $\boldsymbol{b}_{\perp}=0.4\,\mathrm{GeV}^{-1}\boldsymbol{\hat{e}}_x$ (where $\boldsymbol{\hat{e}}_x$ is the unit vector along the $x$-axis) and the transverse momentum component fixed at $k_x=0.3\,\mathrm{GeV}$. The three columns correspond to $x=0.10$, $x=0.25$, and $x=0.40$.}
	\label{6DProtonUUudzky}
\end{figure}

In Figs.~\ref{6DProtonUUudzbx}--\ref{6DProtonUUudzky}, % Fig.~\ref{6DProtonUUudzby}, Fig.~\ref{6DProtonUUudzkx} and Fig.~\ref{6DProtonUUudzky}, 
we plot the six-dimensional unpolarized light-front Wigner distribution $\rho_{\mathrm{UU}}\left(\tilde{z},x,\boldsymbol{b}_{\perp}, \boldsymbol{k}_{\perp}\right)$ for the $u$ and $d$ quarks of the proton, displayed in the $\tilde{z}-b_x$, $\tilde{z}-b_y$, $\tilde{z}-k_x$, and $\tilde{z}-k_y$ subspaces, respectively. The six-dimensional unpolarized light-front Wigner distributions represent the transverse phase-space distribution of the unpolarized quark in an unpolarized proton. The numerical results shown are obtained by fixing the transverse momentum $\boldsymbol{k}_{\perp}$ or the transverse coordinate $\boldsymbol{b}_{\perp}$ at specific values, and the longitudinal momentum fraction $x$ is set at $x = 0.10$, $x = 0.25$, and $x = 0.40$ in the first, second, and third columns, respectively.

\new{For Fig.~\ref{6DProtonUUudzbx} and Fig.~\ref{6DProtonUUudzkx}, the distribution exhibits exact parity symmetry in the $\tilde{z}-b_x$ and $\tilde{z}-k_x$ subspaces, manifested through central symmetry about the origin that reflects the rotational invariance of unpolarized systems. However, a subtle $\tilde{z}$-shift emerges in the $\tilde{z}-b_y$ and $\tilde{z}-k_y$ projections, originating from the phase factor $e^{-i\boldsymbol{b}_{\perp} \cdot \boldsymbol{\Delta}_{\perp}}$ in the Fourier transform that couples transverse position and momentum degrees of freedom. This shift disappears upon integration over $\boldsymbol{k}_{\perp}$, restoring the full axial symmetry expected from IPD.}

\new{The symmetry $\rho_{\mathrm{UU}}\left(\tilde{z},x,\boldsymbol{b}_{\perp}, \boldsymbol{k}_{\perp}\right)=\rho_{\mathrm{UU}}\left(\tilde{z},x,-\boldsymbol{b}_{\perp}, -\boldsymbol{k}_{\perp}\right)$ reflects the parity invariance of QCD for unpolarized states, while the $\tilde{z}$-dependence reveals longitudinal localization effects beyond the traditional five-dimensional formulation. Notably, the $u$ quark distribution displays stronger spatial localization (peaking at $\boldsymbol{b}_{\perp}=0$) compared to $d$ quarks, consistent with both their dominant role in determining the proton charge radius and their preferential coupling to scalar diquarks in the $\mathrm{SU(6)}$ spin-flavor wavefunction. The momentum-space distributions for both flavors follow the Gaussian form of the BHL wavefunction, with the $d$ quark broader $\boldsymbol{k}_{\perp}$ spread reflecting its higher virtuality through coupling to axial-vector diquarks. The non-positive definiteness of these distributions directly manifests quantum interference effects in phase space, while maintaining the expected $\mathrm{SO(2)}$ rotational symmetry that confirms fundamental QCD constraints on unpolarized systems.}

\new{Through appropriate integration schemes, $\rho_{\mathrm{UU}}$ connects directly to experimentally measurable quantities: integration over $\boldsymbol{b}_{\perp}$ and $\tilde{z}$ yields the TMD $f_1(x,\boldsymbol{k}_{\perp})$ accessible in SIDIS at EIC~\cite{Brodsky:2002cx}, while integration over $\boldsymbol{k}_{\perp}$ and $\tilde{z}$ gives the GPD $H(x,\xi,t=-\Delta_{\perp}^2)$ measured in DVCS~\cite{Belitsky:2005qn}. Future precision data at $Q^2 > 10 \text{ GeV}^2$ could resolve the $\hat{z}$-dependence through $\xi$-dependent analyses~\cite{Accardi:2012qut}. The $x$-dependence at $\boldsymbol{b}_{\perp}=0$ further correlates with electromagnetic form factors, providing multiple pathways for experimental validation. These results, obtained using light-front wavefunctions with parameters constrained by proton structure data, demonstrate how six-dimensional Wigner distributions encode rich information about quark dynamics that transcends conventional one-dimensional parton descriptions, offering new insights into the proton quark-diquark structure and orbital dynamics. Future high-precision measurements at the EIC will further constrain the $\tilde{z}$-dependence through $\xi$-resolved analyses of exclusive processes.}

%\subsubsection{u Quark}

%\subsubsection{d Quark}

% UL

% UT

%\subsubsection{d Quark}

\subsection{Longitudinal-polarized Wigner distribution}
\label{LL}

% \subsubsection{LL}

\begin{figure}[htbp]
	\centering
	\subfloat{
		\includegraphics[width=0.31\textwidth]{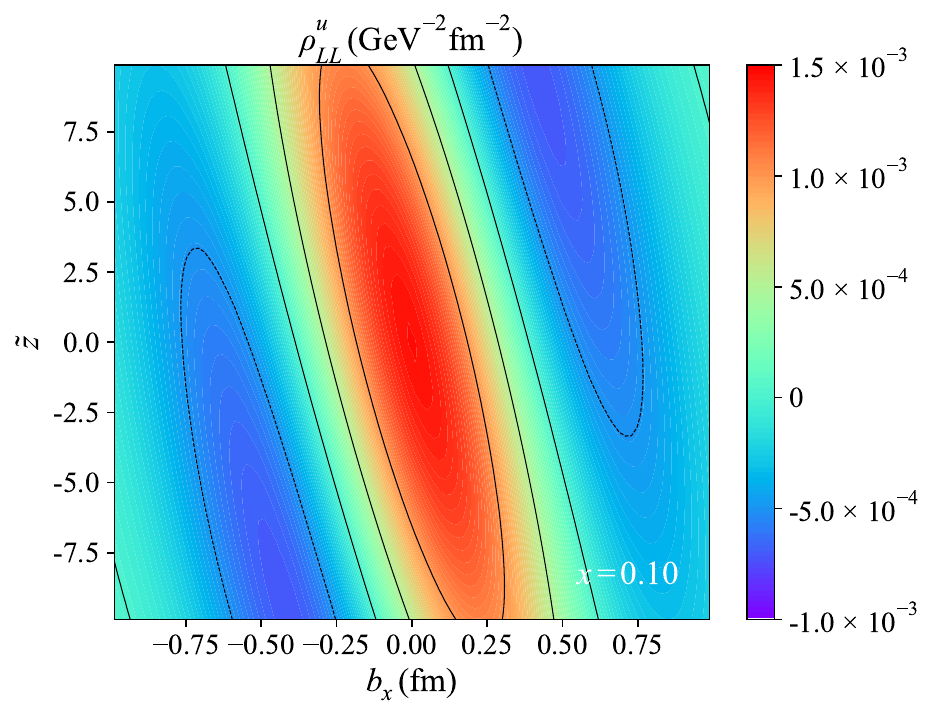}
	}
	\subfloat{
		\includegraphics[width=0.31\textwidth]{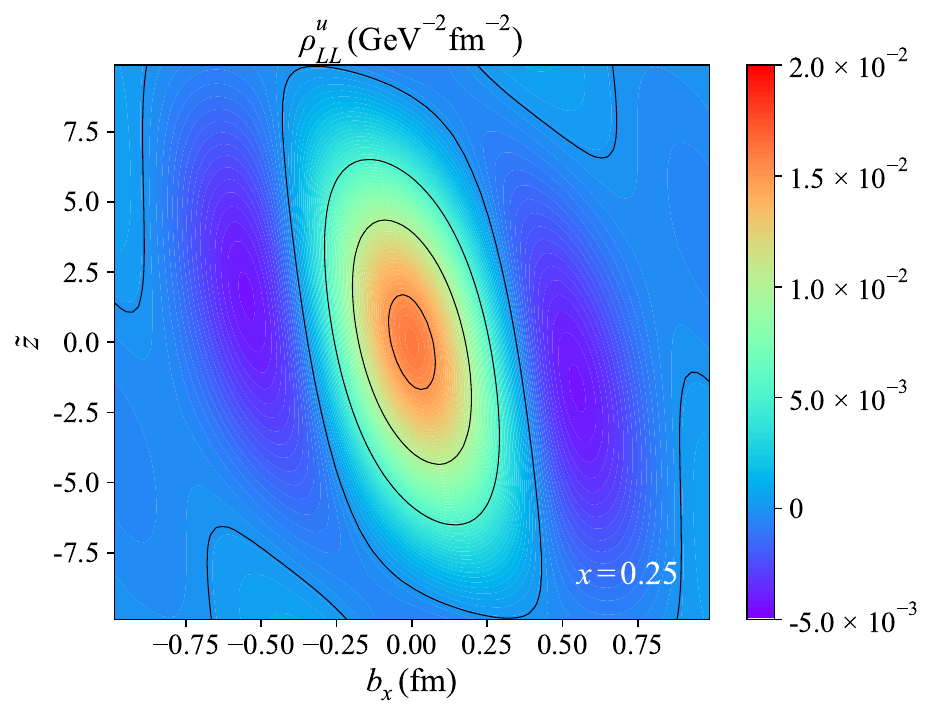}
	}
	\subfloat{
		\includegraphics[width=0.31\textwidth]{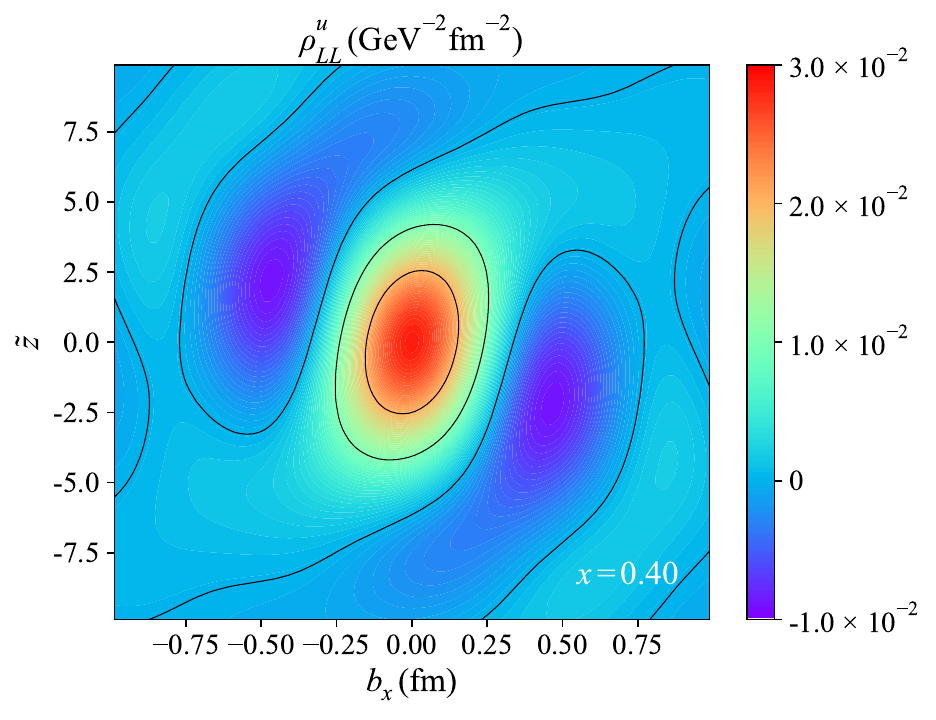}
	}\\
	\subfloat{
		\includegraphics[width=0.31\textwidth]{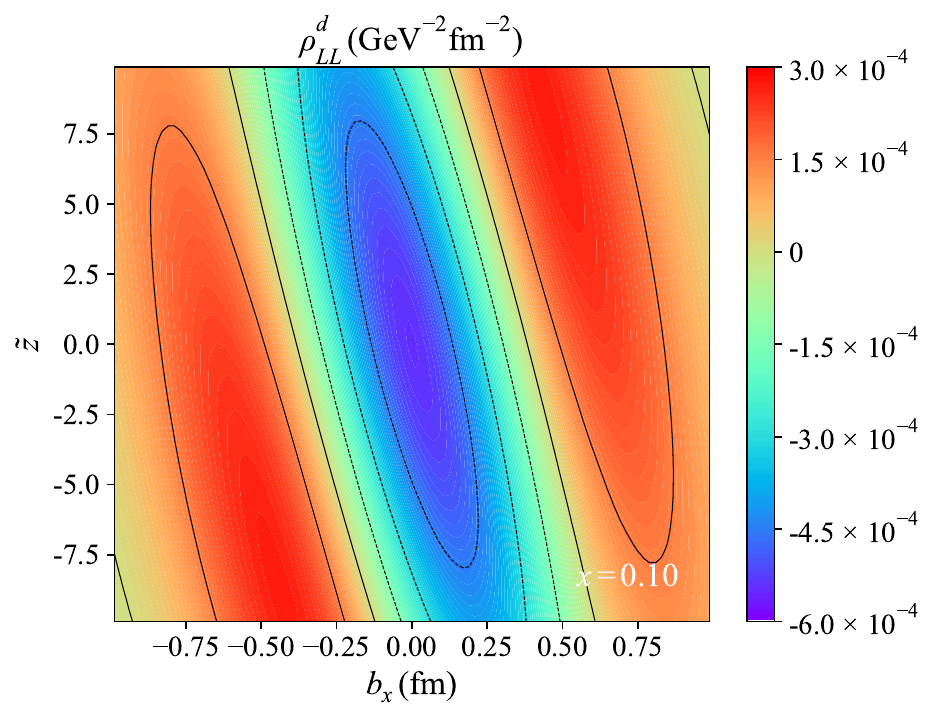}
	}
	\subfloat{
		\includegraphics[width=0.31\textwidth]{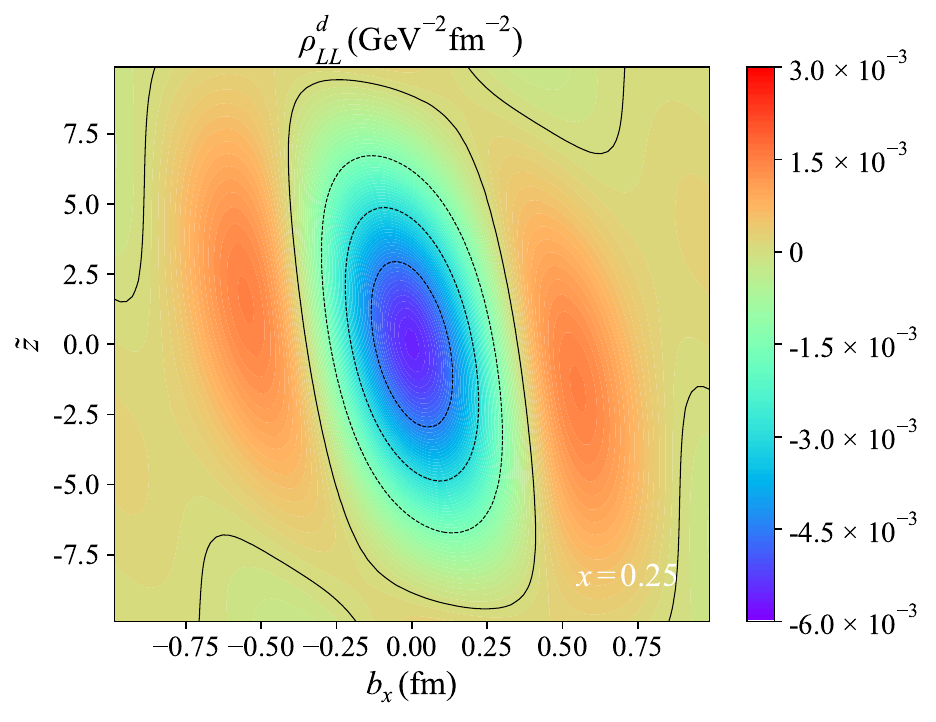}
	}
	\subfloat{
		\includegraphics[width=0.31\textwidth]{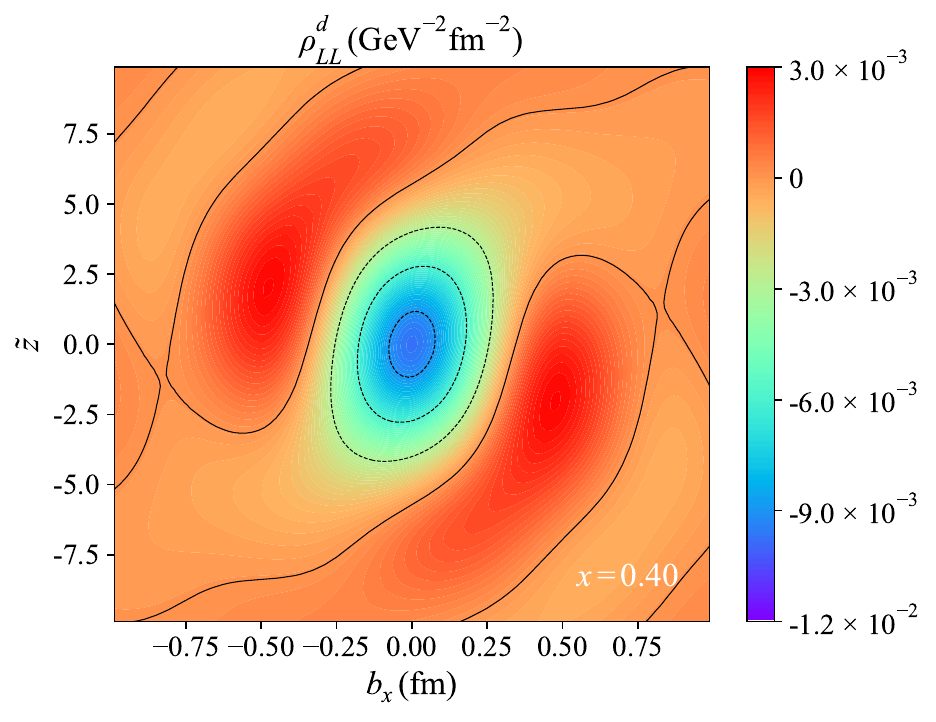}
	}
	\caption{Six-dimensional longitudinal light-front Wigner distribution $\rho_{\mathrm{LL}}\left(\tilde{z},x,\boldsymbol{b}_{\perp}, \boldsymbol{k}_{\perp}\right)$ for $u$ quark (upper panels) and $d$ quark (lower panels). The figure presents the Wigner distribution in the $\tilde{z}-b_x$ plane, with the transverse momentum fixed at $\boldsymbol{k}_{\perp}=0.3\,\mathrm{GeV}\boldsymbol{\hat{e}}_x$ (where $\boldsymbol{\hat{e}}_x$ is the unit vector in the $x$-direction) and the transverse coordinate component fixed at $b_y=0.4\,\mathrm{GeV}^{-1}$. The three columns correspond to $x=0.10$, $x=0.25$, and $x=0.40$.}
	\label{6DProtonLLudzbx}
\end{figure}

\begin{figure}[htbp]
	\centering
	\subfloat{
		\includegraphics[width=0.31\textwidth]{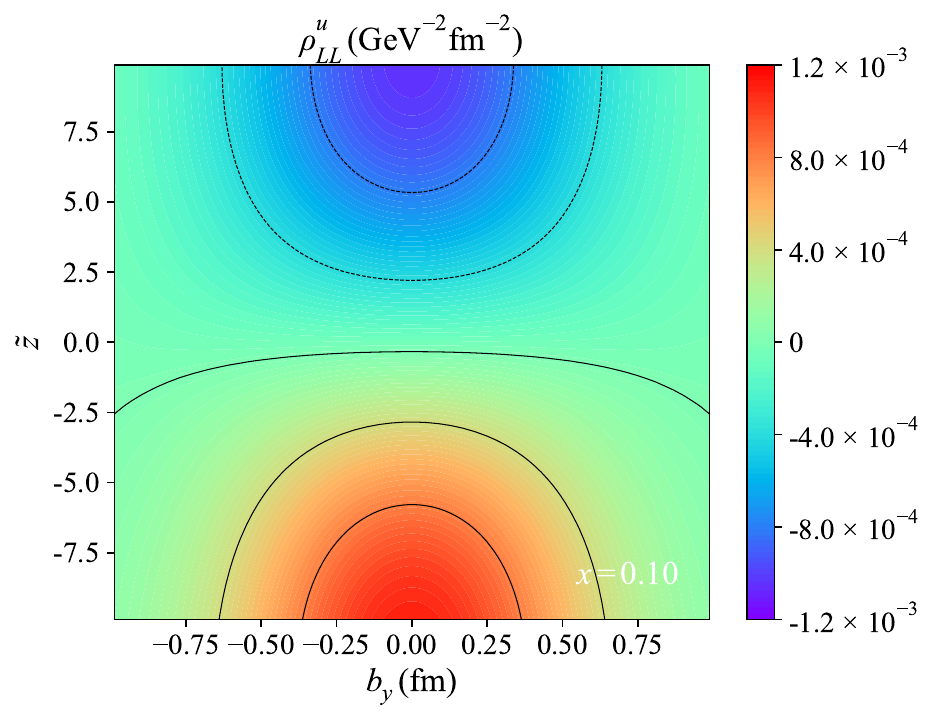}
	}
	\subfloat{
		\includegraphics[width=0.31\textwidth]{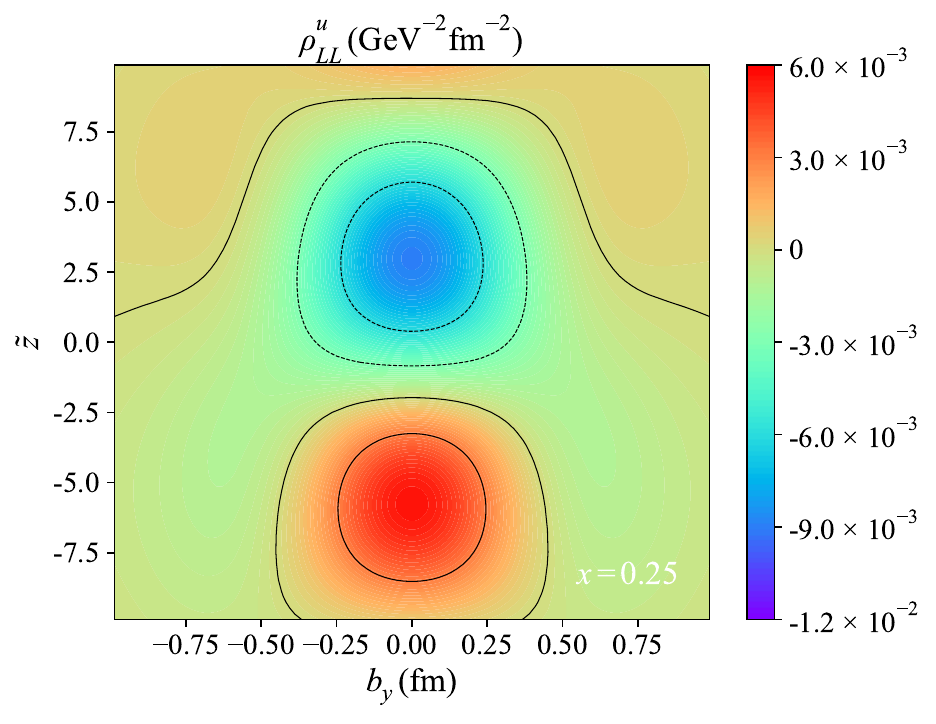}
	}
	\subfloat{
		\includegraphics[width=0.31\textwidth]{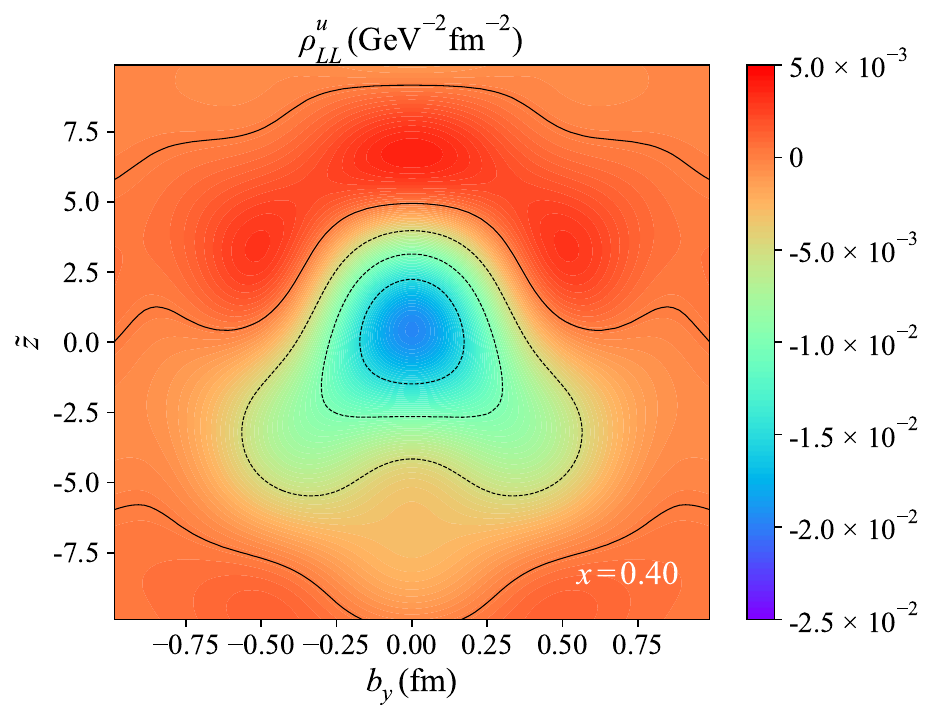}
	}\\
	\subfloat{
		\includegraphics[width=0.31\textwidth]{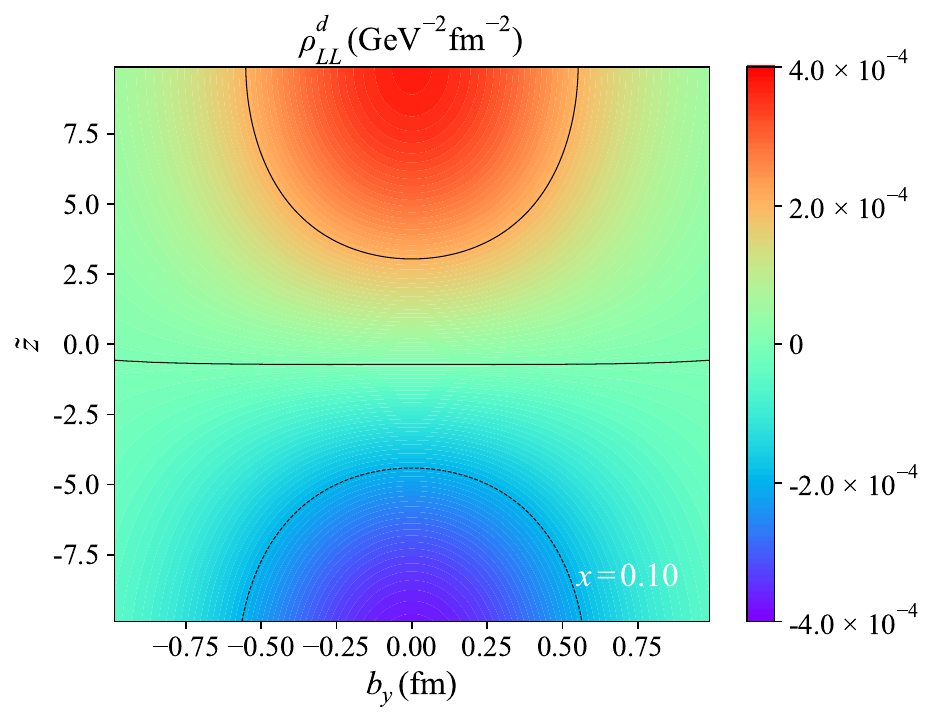}
	}
	\subfloat{
		\includegraphics[width=0.31\textwidth]{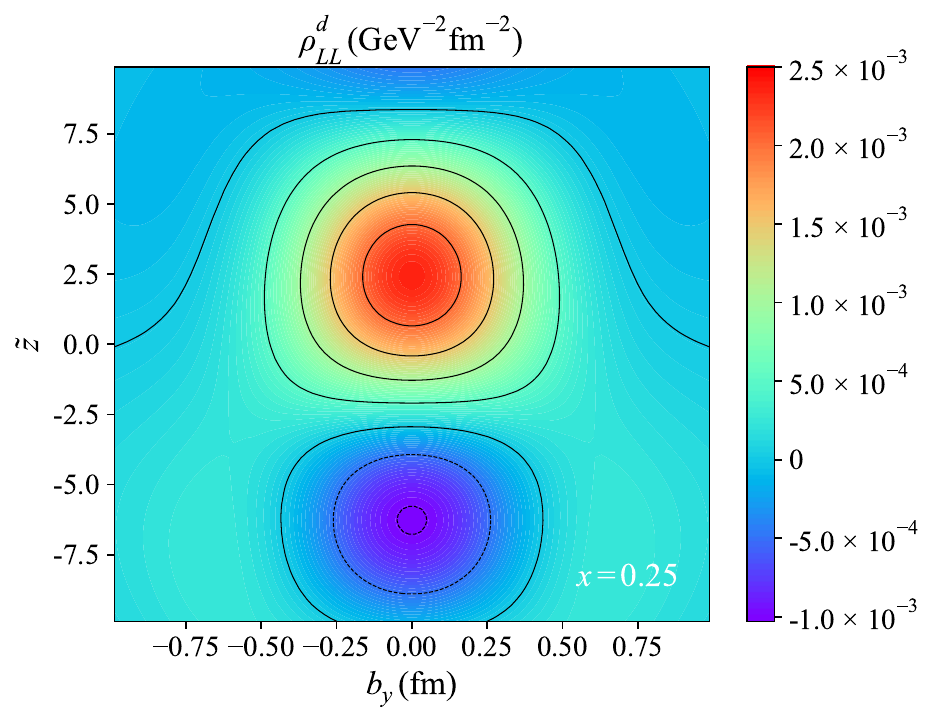}
	}
	\subfloat{
		\includegraphics[width=0.31\textwidth]{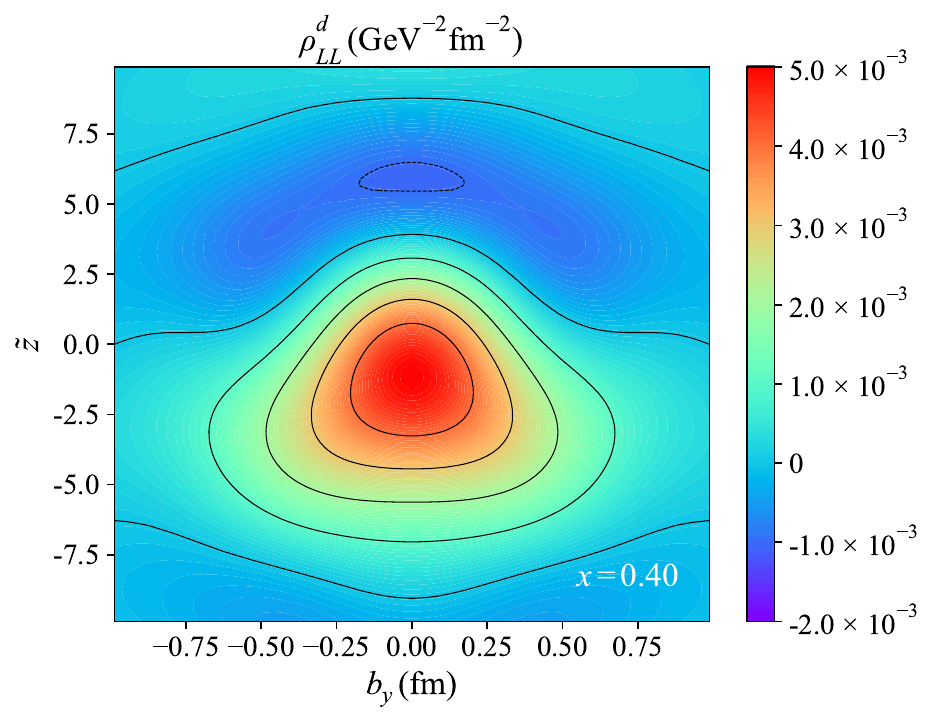}
	}
	\caption{Six-dimensional longitudinal light-front Wigner distribution $\rho_{\mathrm{LL}}\left(\tilde{z},x,\boldsymbol{b}_{\perp}, \boldsymbol{k}_{\perp}\right)$ for $u$ quark (upper panels) and $d$ quark (lower panels). The figure presents the Wigner distributions in the $\tilde{z}-b_y$ plane, with the transverse momentum fixed at $\boldsymbol{k}_{\perp}=0.3\,\mathrm{GeV}\boldsymbol{\hat{e}}_x$ (where $\boldsymbol{\hat{e}}_x$ is the unit vector in the $x$-direction) and the transverse coordinate component fixed at $b_x=0.4\,\mathrm{GeV}^{-1}$. The three columns correspond to $x=0.10$, $x=0.25$, and $x=0.40$.}
	\label{6DProtonLLudzby}
\end{figure}

\begin{figure}[htbp]
	\centering
	\subfloat{
		\includegraphics[width=0.31\textwidth]{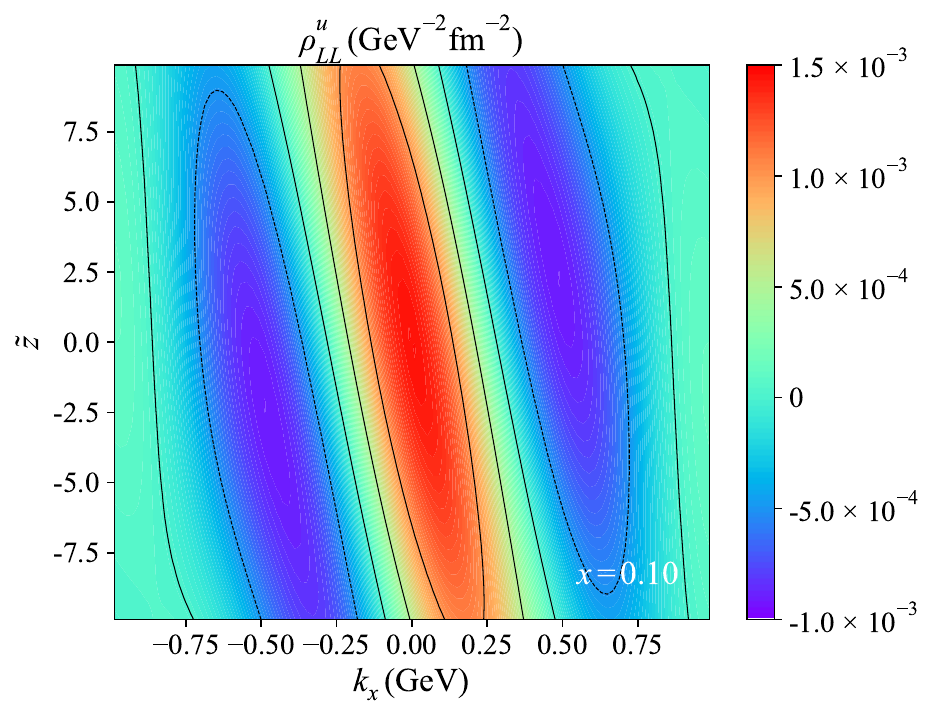}
	}
	\subfloat{
		\includegraphics[width=0.31\textwidth]{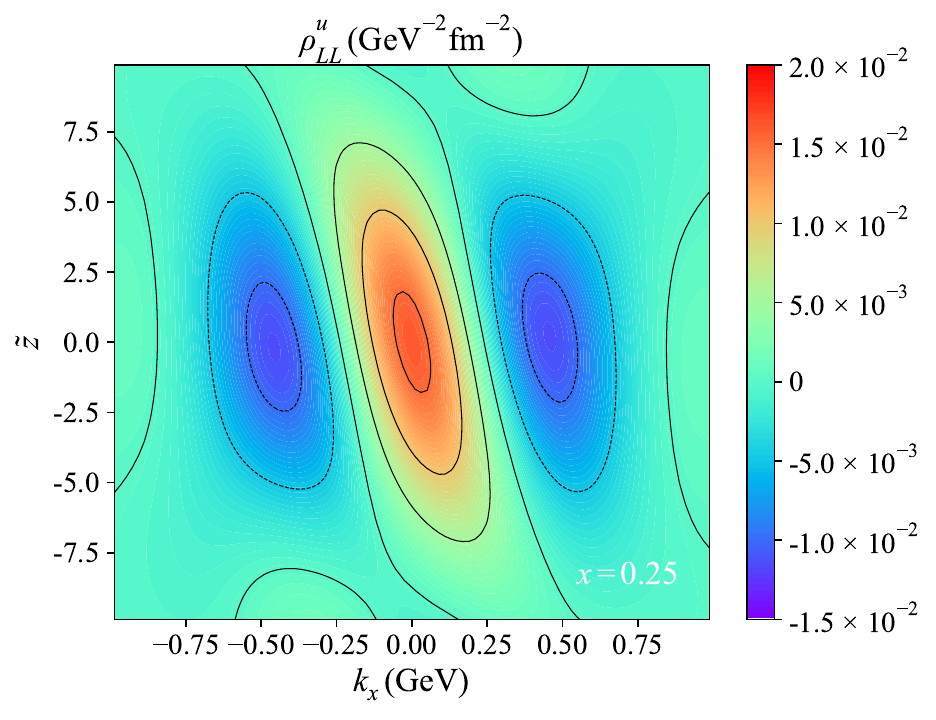}
	}
	\subfloat{
		\includegraphics[width=0.31\textwidth]{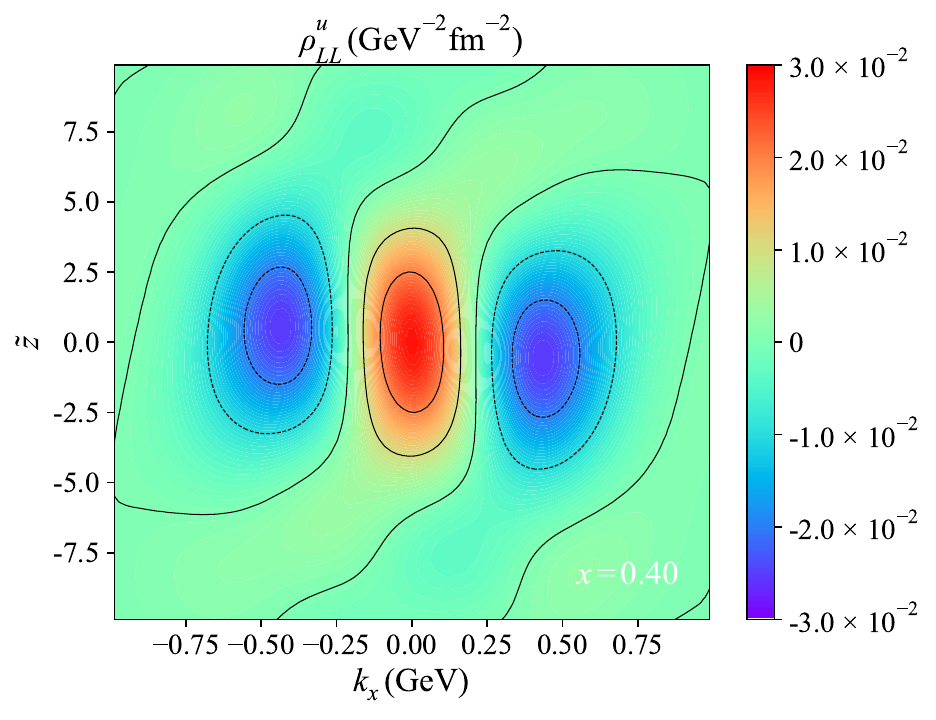}
	}\\
	\subfloat{
		\includegraphics[width=0.31\textwidth]{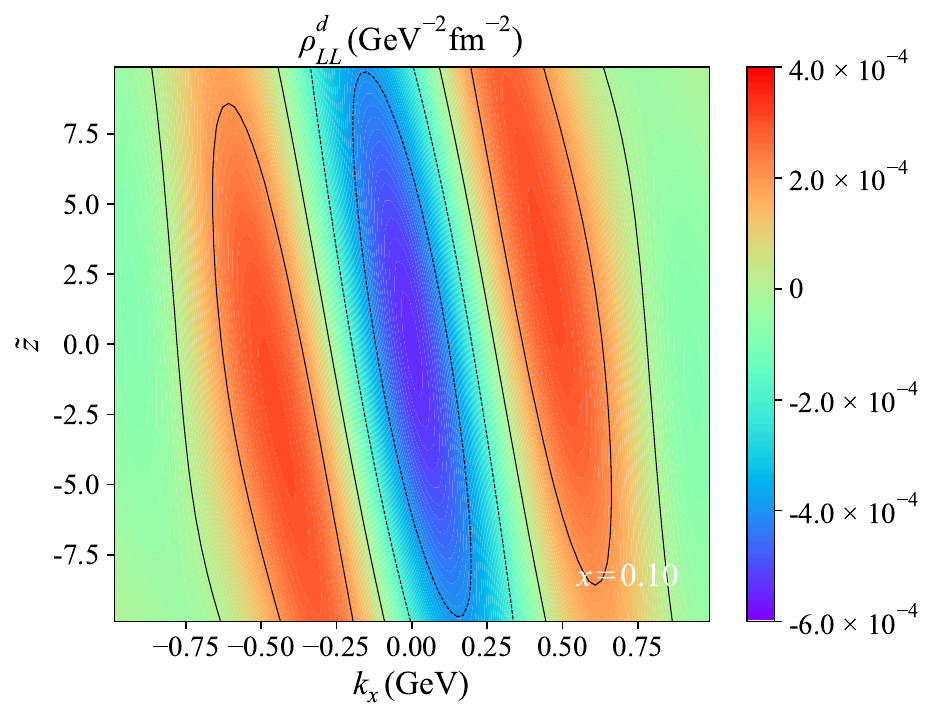}
	}
	\subfloat{
		\includegraphics[width=0.31\textwidth]{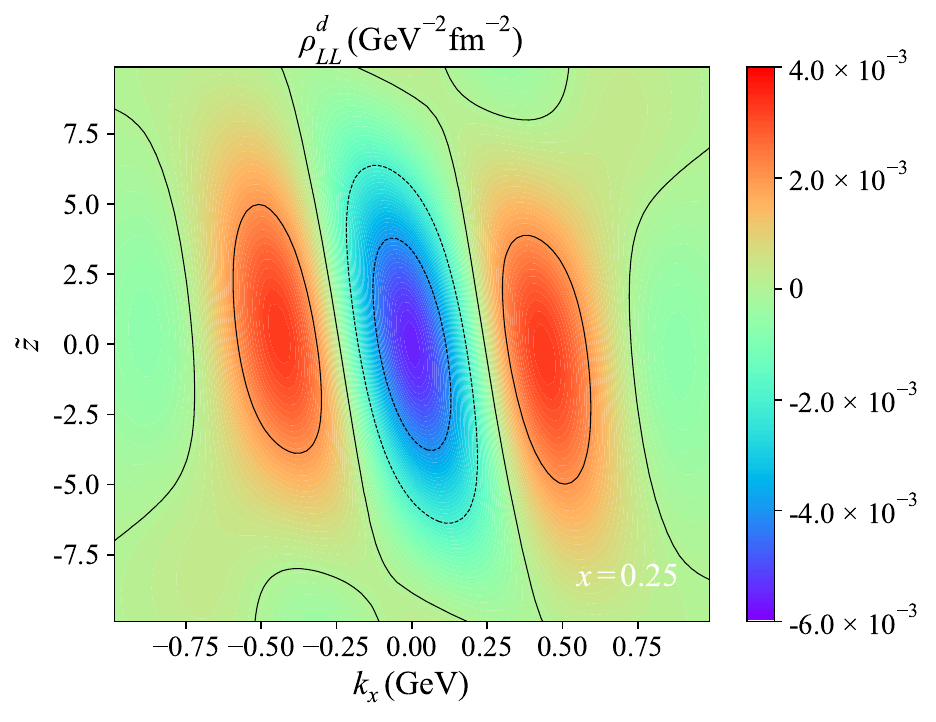}
	}
	\subfloat{
		\includegraphics[width=0.31\textwidth]{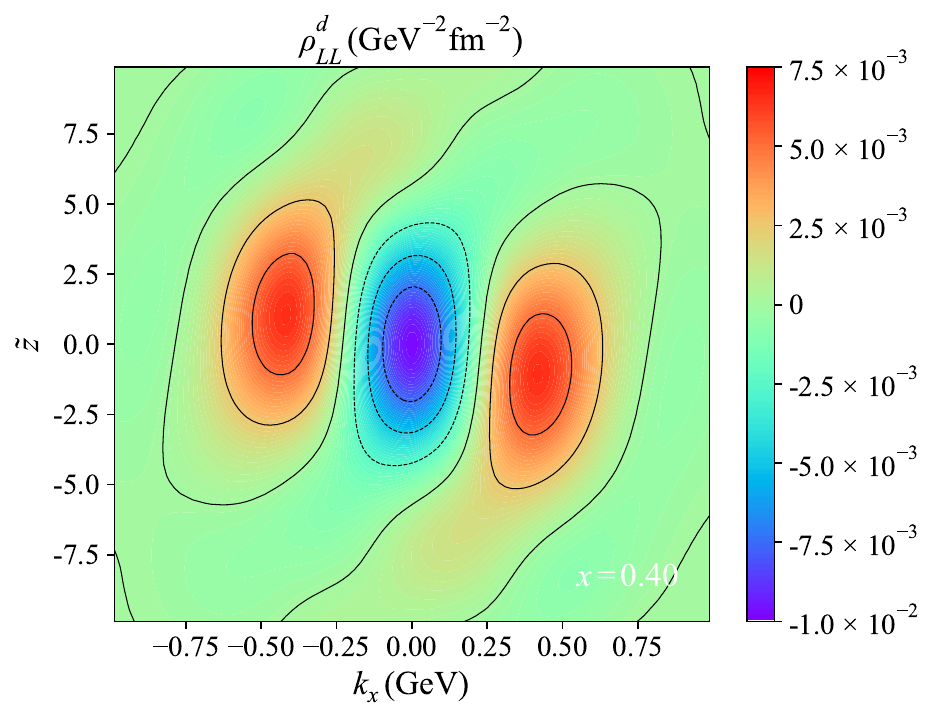}
	}
	\caption{Six-dimensional longitudinal light-front Wigner distribution $\rho_{\mathrm{LL}}\left(\tilde{z},x,\boldsymbol{b}_{\perp}, \boldsymbol{k}_{\perp}\right)$ for $u$ quark (upper panels) and $d$ quark (lower panels). The figure presents the Wigner distributions in the $\tilde{z}-k_x$ plane, with the transverse coordinate fixed at $\boldsymbol{b}_{\perp}=0.4\,\mathrm{GeV}^{-1}\boldsymbol{\hat{e}}_x$ (where $\boldsymbol{\hat{e}}_x$ is the unit vector along the $x$-axis) and the transverse momentum component fixed at $k_y=0.3\,\mathrm{GeV}$. The three columns correspond to $x=0.10$, $x=0.25$, and $x=0.40$.}
	\label{6DProtonLLudzkx}
\end{figure}

\begin{figure}[htbp]
	\centering
	\subfloat{
		\includegraphics[width=0.31\textwidth]{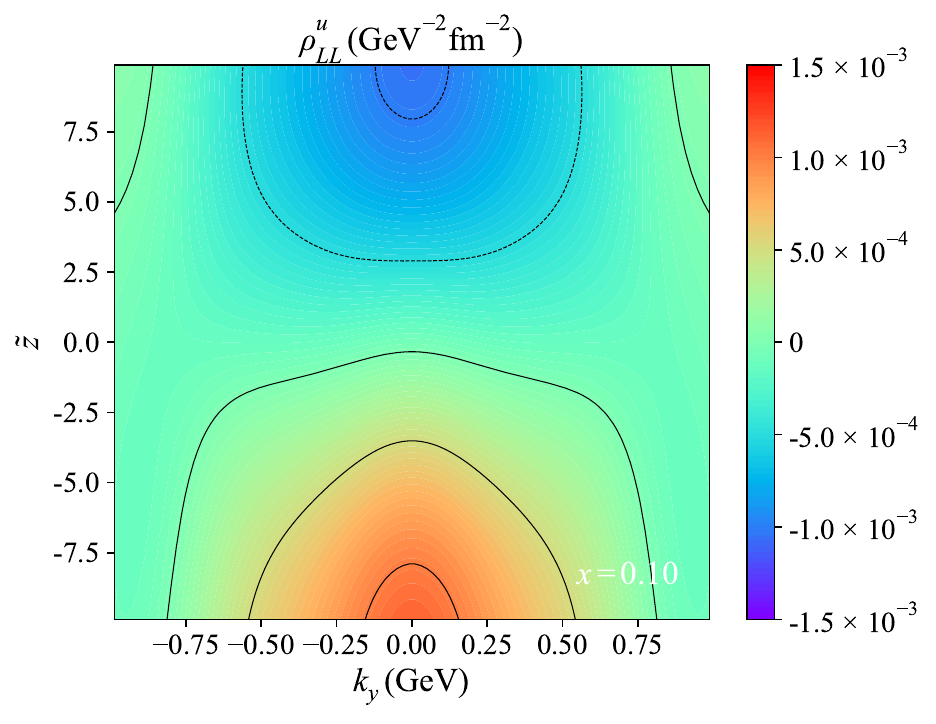}
	}
	\subfloat{
		\includegraphics[width=0.31\textwidth]{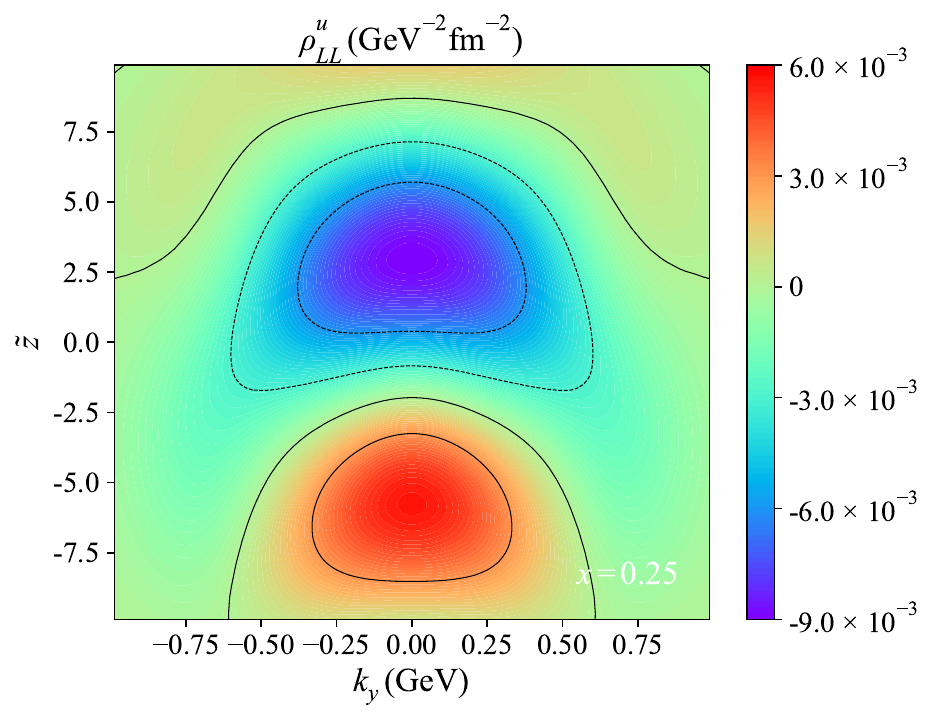}
	}
	\subfloat{
		\includegraphics[width=0.31\textwidth]{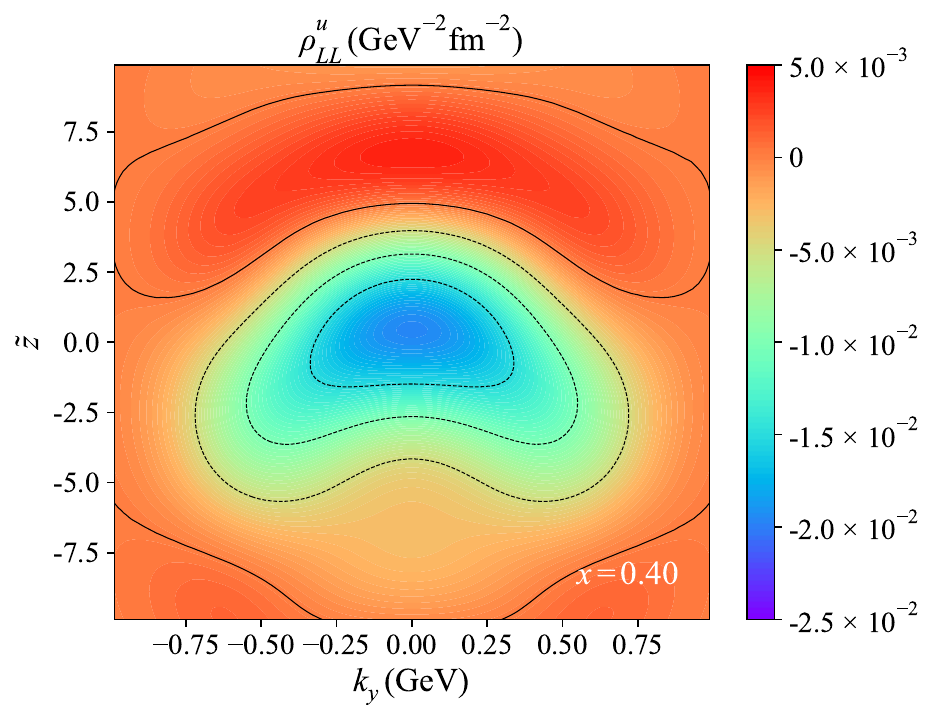}
	}\\
	\subfloat{
		\includegraphics[width=0.31\textwidth]{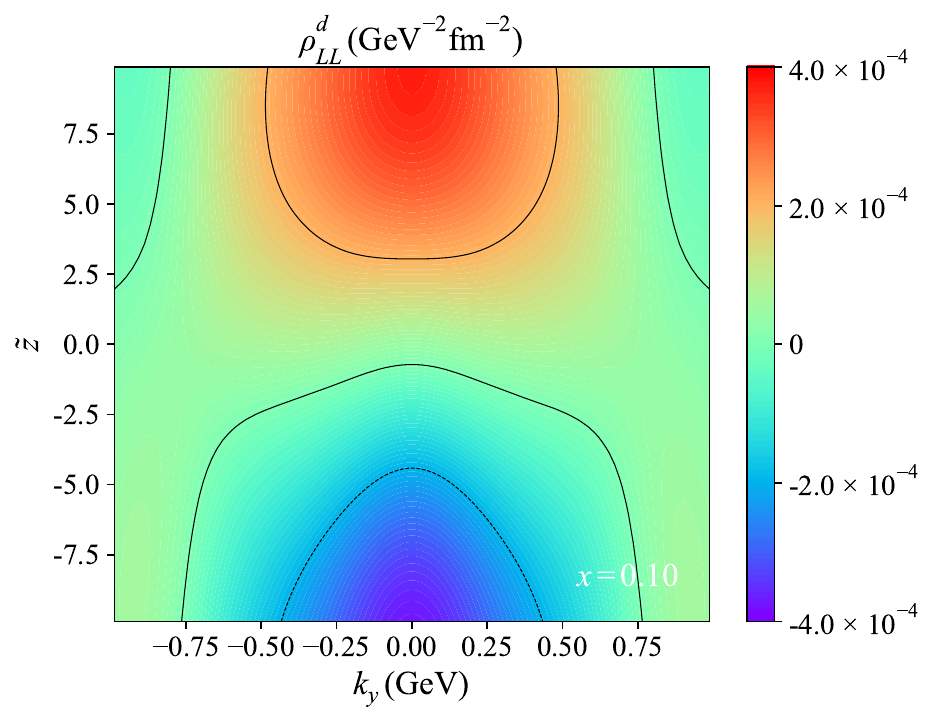}
	}
	\subfloat{
		\includegraphics[width=0.31\textwidth]{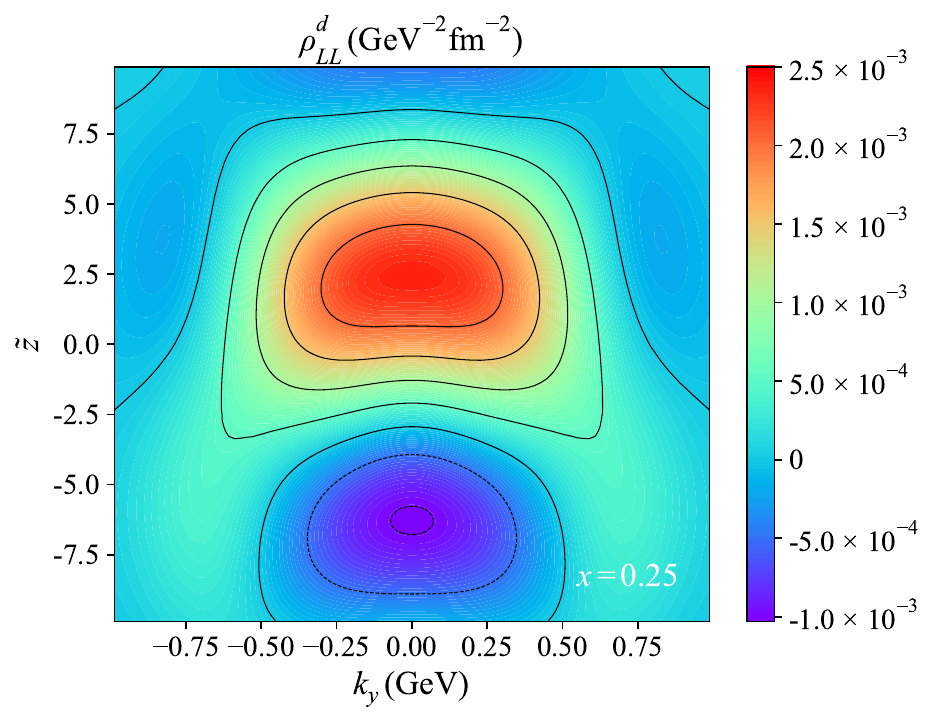}
	}
	\subfloat{
		\includegraphics[width=0.31\textwidth]{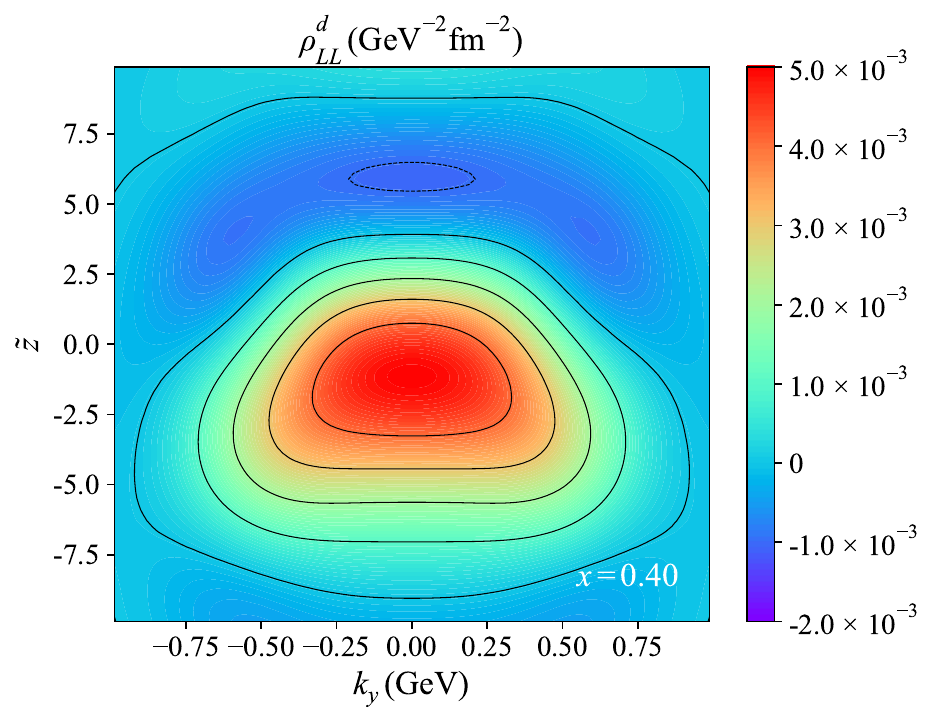}
	}
	\caption{Six-dimensional longitudinal light-front Wigner distribution $\rho_{\mathrm{LL}}\left(\tilde{z},x,\boldsymbol{b}_{\perp}, \boldsymbol{k}_{\perp}\right)$ for $u$ quark (upper panels) and $d$ quark (lower panels). The figure presents the Wigner distributions in the $\tilde{z}-k_y$ plane, with the transverse coordinate fixed at $\boldsymbol{b}_{\perp}=0.4\,\mathrm{GeV}^{-1}\boldsymbol{\hat{e}}_x$ (where $\boldsymbol{\hat{e}}_x$ is the unit vector along the $x$-axis) and the transverse momentum component fixed at $k_x=0.3\,\mathrm{GeV}$. The three columns correspond to $x=0.10$, $x=0.25$, and $x=0.40$.}
	\label{6DProtonLLudzky}
\end{figure}

In Figs.~\ref{6DProtonLLudzbx}--\ref{6DProtonLLudzky}, %In Fig.~\ref{6DProtonLLudzbx}, Fig.~\ref{6DProtonLLudzby}, Fig.~\ref{6DProtonLLudzkx} and Fig.~\ref{6DProtonLLudzky}, 
we plot the six-dimensional longitudinal light-front Wigner distribution $\rho_{\mathrm{LL}}\left(\tilde{z},x,\boldsymbol{b}_{\perp},\boldsymbol{k}_{\perp}\right)$ for the $u$ and $d$ quarks of the proton, displayed in the $\tilde{z}-b_x$, $\tilde{z}-b_y$, $\tilde{z}-k_x$, and $\tilde{z}-k_y$ subspaces, respectively. \new{These six-dimensional longitudinal light-front Wigner distributions describe the spatial and momentum characteristics of longitudinal polarized quarks in a longitudinal polarized proton, which quantifies the helicity correlation.} The numerical results are obtained by fixing the transverse momentum $\boldsymbol{k}_{\perp}$ or the transverse coordinate $\boldsymbol{b}_{\perp}$ at specific values, and the longitudinal momentum fraction $x$ is set at $x = 0.10$, $x = 0.25$, and $x = 0.40$ in the first, second, and third columns, respectively. 

\new{For Fig.~\ref{6DProtonLLudzbx} and Fig.~\ref{6DProtonLLudzkx}, a striking feature emerges in the $\tilde{z}-b_x$ and $\tilde{z}-k_x$ subspaces, where the distributions exhibit exact central symmetry about the origin. The $u$ quark distributions show positive peaks ($\Delta u > 0$), while $d$ quarks display negative peaks ($\Delta d < 0$), consistent with their known helicity contributions. This flavor asymmetry originates from the axial-vector diquark wave function coefficients and agrees well with lattice QCD predictions~\cite{LHPC:2010jcs,Alexandrou:2020sml}. The relative magnitude difference ($\left|\Delta u\right| > \left|\Delta d\right|$) reflects the underlying quark-diquark dynamics in the proton spin structure~\cite{Ma:1997gy,HERMES:2006jyl}. By contrast, for Fig.~\ref{6DProtonLLudzby} and Fig.~\ref{6DProtonLLudzky}, the $\tilde{z}-b_y$ and $\tilde{z}-k_y$ projections reveal more complex behavior, displaying characteristic dipole patterns with respect to $b_x = 0$ that provide direct evidence of spin-orbit coupling. For $u$ quarks, positive lobes at $\tilde{z} > 0$ and negative lobes at $\tilde{z} < 0$ indicate a preference for clockwise and counter-clockwise transverse orbital motion, respectively~\cite{Lorce:2011ni}. These patterns gradually diminish with increasing $x$, showing the strongest effects in the sea quark region ($x=0.1$) and becoming less pronounced for valence-dominated configurations ($x=0.4$).} 

\new{In general, the contrasting behaviors between the $u$ and $d$ quark distributions reflect the opposite signs of the orbital angular momentum, providing a clear indication of their respective motion preferences. It should be noted, however, that this observed behavior is model-dependent, and different choices of the wave function could potentially alter these features.} Despite the presence of multiple peaks in the figures, which are indicative of diffraction patterns along specific directions, the polarization characteristics of the quarks can be inferred from the peak positions. Specifically, the $u$ quark distribution shows a positively polarized peak, while the $d$ quark distribution exhibits a negatively polarized peak, aligning with the known axial charge results. Additionally, the helicity distributions of the quarks are concentrated near the center of the phase space, reflecting the localized nature of the longitudinal polarization.

\new{These Wigner distributions establish important connections to established hadronic structure functions. In the GTMD limit, $\rho_{\mathrm{LL}}$ is related to the twist-two GTMD $G_{1,4}$, probing quark orbital angular momentum. Through $\boldsymbol{k}_\perp$-integration, they relate to the chiral-odd helicity GPD $H(x,\xi,t=-\boldsymbol{\Delta}_{\perp}^2)$ measurable in DVCS experiments, while $\boldsymbol{b}_\perp$ and $\tilde{z}$-integration yields the helicity TMD $g_{1L}(x,\boldsymbol{k}_\perp^2)$ accessible through SIDIS measurements. The full six-dimensional distribution provides a unified framework that simultaneously captures longitudinal momentum, transverse spatial, and spin-orbit correlations.}

\new{The antisymmetry $\rho_{\mathrm{LL}}(\tilde{z},x,\boldsymbol{b}_{\perp}, \boldsymbol{k}_{\perp}) = -\rho_{\mathrm{LL}}(\tilde{z},x,-\boldsymbol{b}_{\perp}, -\boldsymbol{k}_{\perp})$ reflects the parity-odd nature of helicity distributions, while the $\tilde{z}$-dependence encodes longitudinal spin-momentum correlations beyond the collinear helicity PDF. Future high-precision measurements at the EIC will critically test these predictions through flavor-separated helicity PDFs and longitudinal target asymmetries. The quantitative agreement between theory and experiment will offer crucial insights into the proton spin puzzle. While the current light-front spectator model successfully reproduces key observational trends, further refinements including gluon polarization effects at small-$x$ and off-shell corrections at high $\boldsymbol{k}_\perp$ may be necessary for complete theoretical consistency. These improvements will enhance our understanding of quark-gluon dynamics and their role in shaping hadron structure. Additionally, azimuthal asymmetries in DVMP could constrain the $\tilde{z}$-dependence~\cite{Diehl:2013xca,CLAS:2007clm,Kumericki:2016ehc}.}

%\subsubsection{u Quark}

%\subsubsection{d Quark}

\subsection{Longitudinal-transverse Wigner distribution}
\label{LT}

\begin{figure}[htbp]
	\centering
	\subfloat{
		\includegraphics[width=0.31\textwidth]{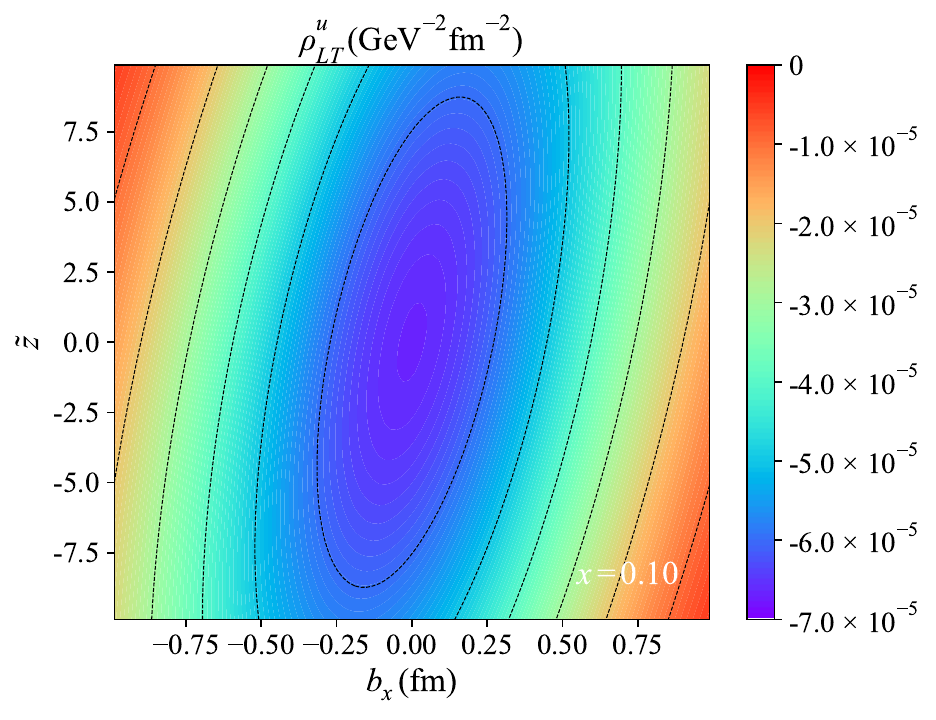}
	}
	\subfloat{
		\includegraphics[width=0.31\textwidth]{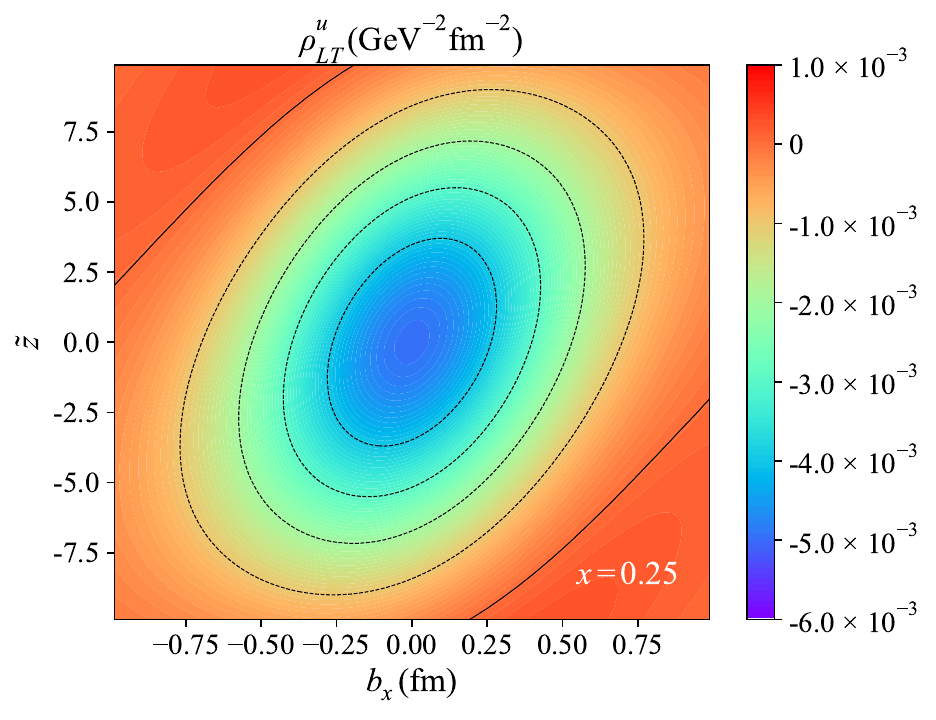}
	}
	\subfloat{
		\includegraphics[width=0.31\textwidth]{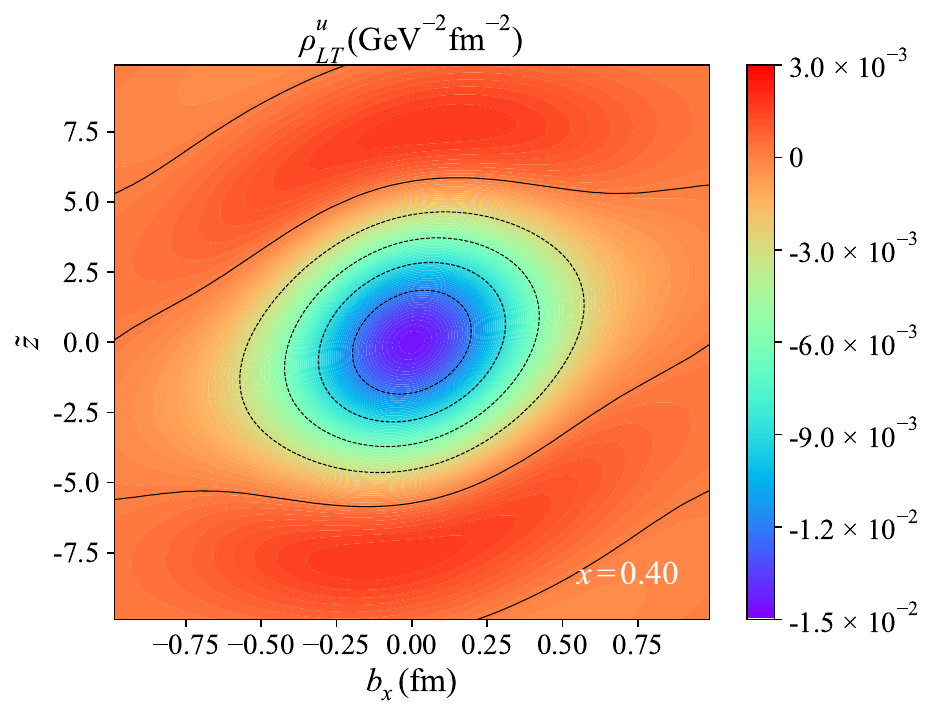}
	}\\
	\subfloat{
		\includegraphics[width=0.31\textwidth]{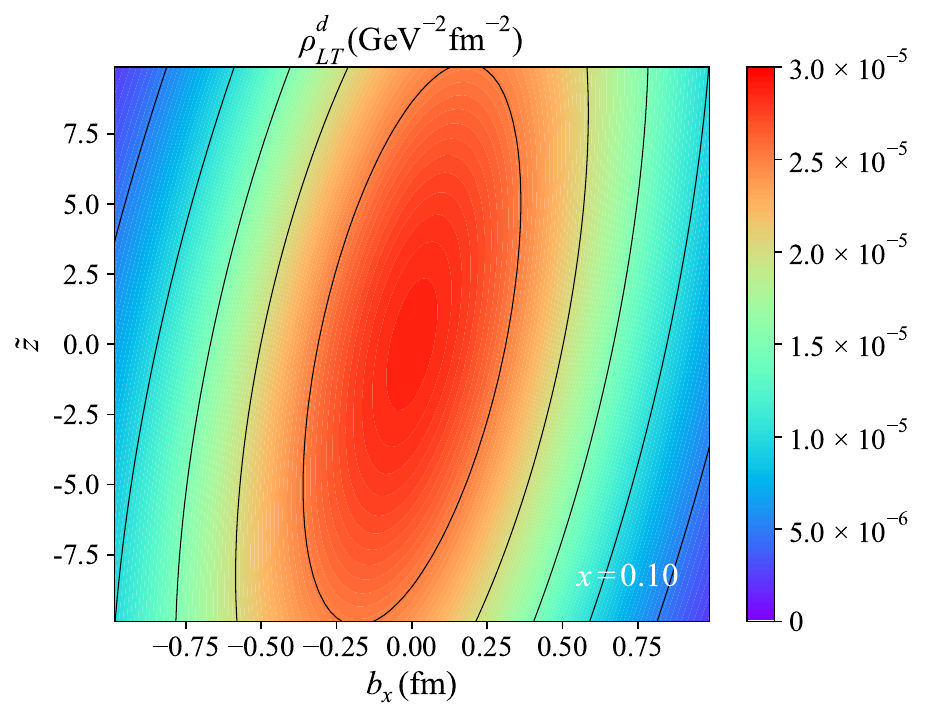}
	}
	\subfloat{
		\includegraphics[width=0.31\textwidth]{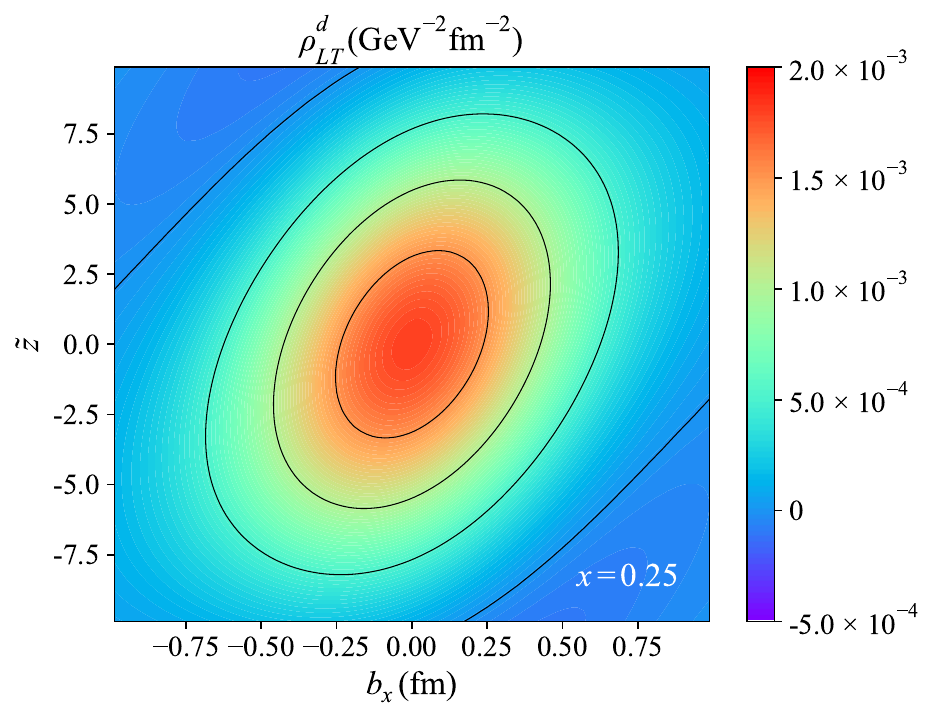}
	}
	\subfloat{
		\includegraphics[width=0.31\textwidth]{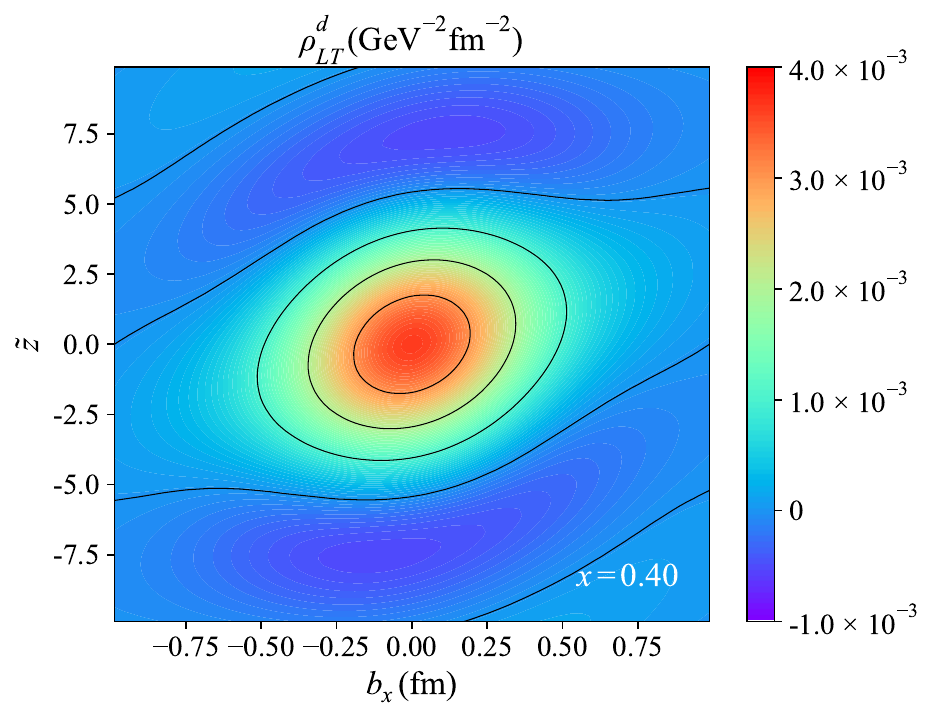}
	}
	\caption{Six-dimensional longitudinal-transverse light-front Wigner distribution $\rho_{\mathrm{LT}}\left(\tilde{z},x,\boldsymbol{b}_{\perp}, \boldsymbol{k}_{\perp}\right)$ for $u$ quark (upper panels) and $d$ quark (lower panels). The figure presents the Wigner distribution in the $\tilde{z}-b_x$ plane, with the transverse momentum fixed at $\boldsymbol{k}_{\perp}=0.3\,\mathrm{GeV}\boldsymbol{\hat{e}}_x$ (where $\boldsymbol{\hat{e}}_x$ is the unit vector in the $x$-direction) and the transverse coordinate component fixed at $b_y=0.4\,\mathrm{GeV}^{-1}$. The three columns correspond to $x=0.10$, $x=0.25$, and $x=0.40$.}
	\label{6DProtonLTudzbx}
\end{figure}

\begin{figure}[htbp]
	\centering
	\subfloat{
		\includegraphics[width=0.31\textwidth]{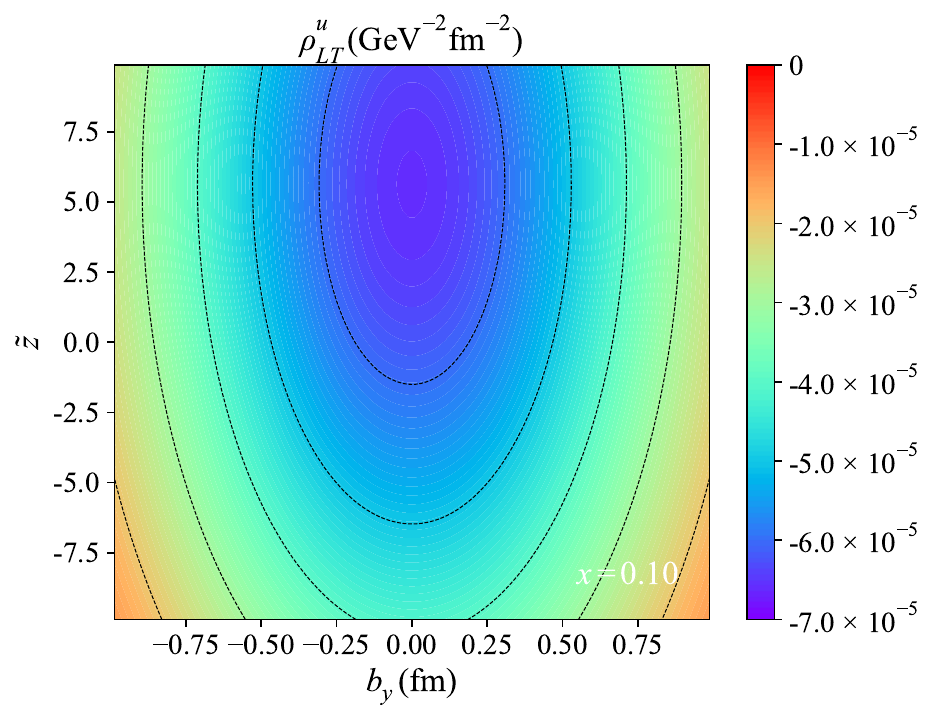}
	}
	\subfloat{
		\includegraphics[width=0.31\textwidth]{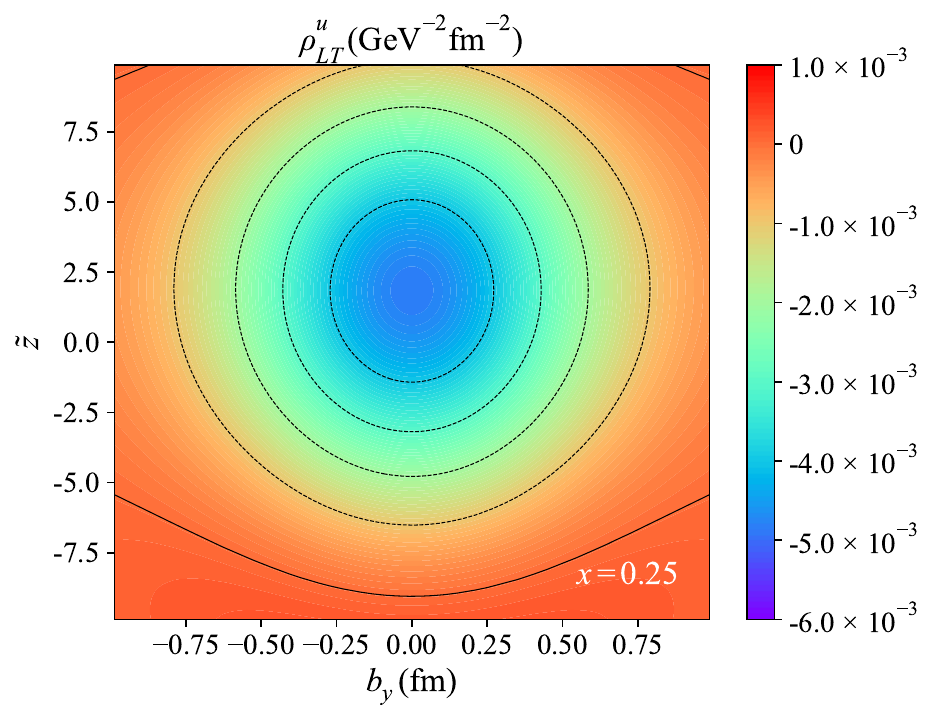}
	}
	\subfloat{
		\includegraphics[width=0.31\textwidth]{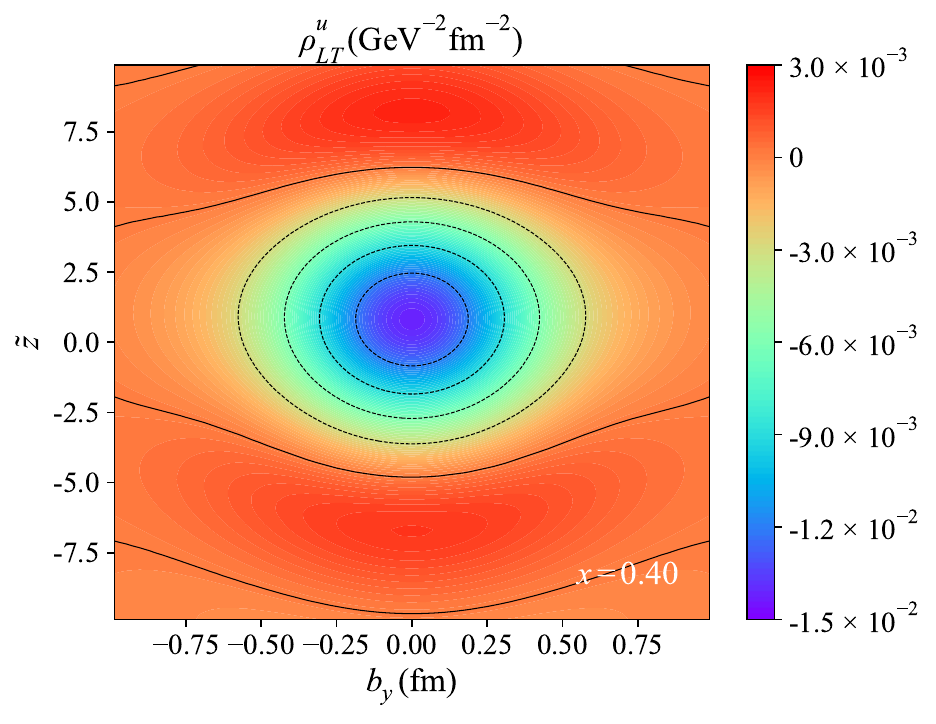}
	}\\
	\subfloat{
		\includegraphics[width=0.31\textwidth]{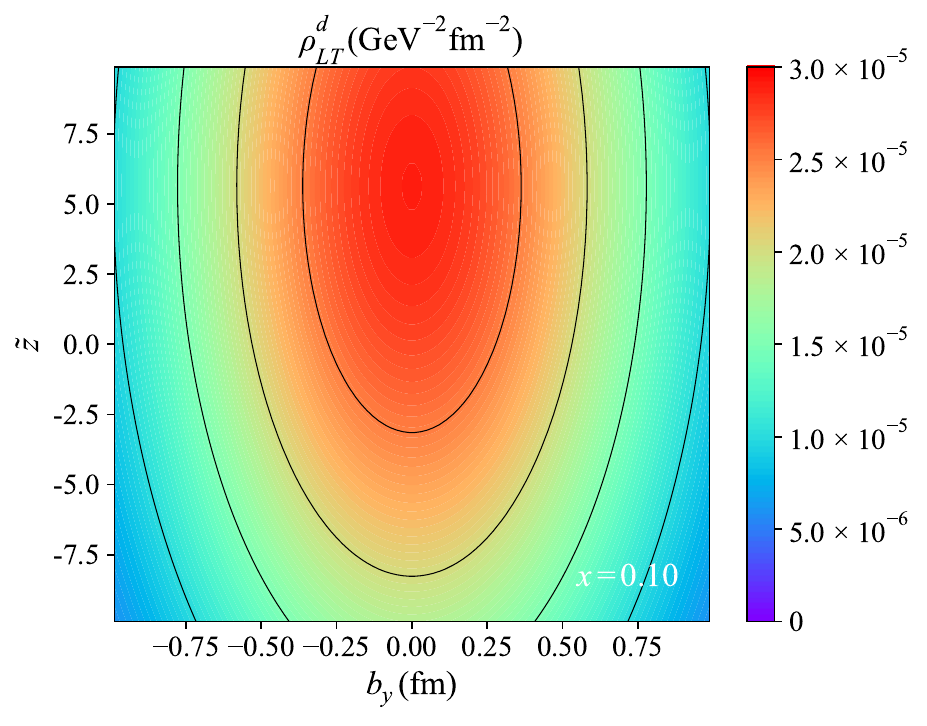}
	}
	\subfloat{
		\includegraphics[width=0.31\textwidth]{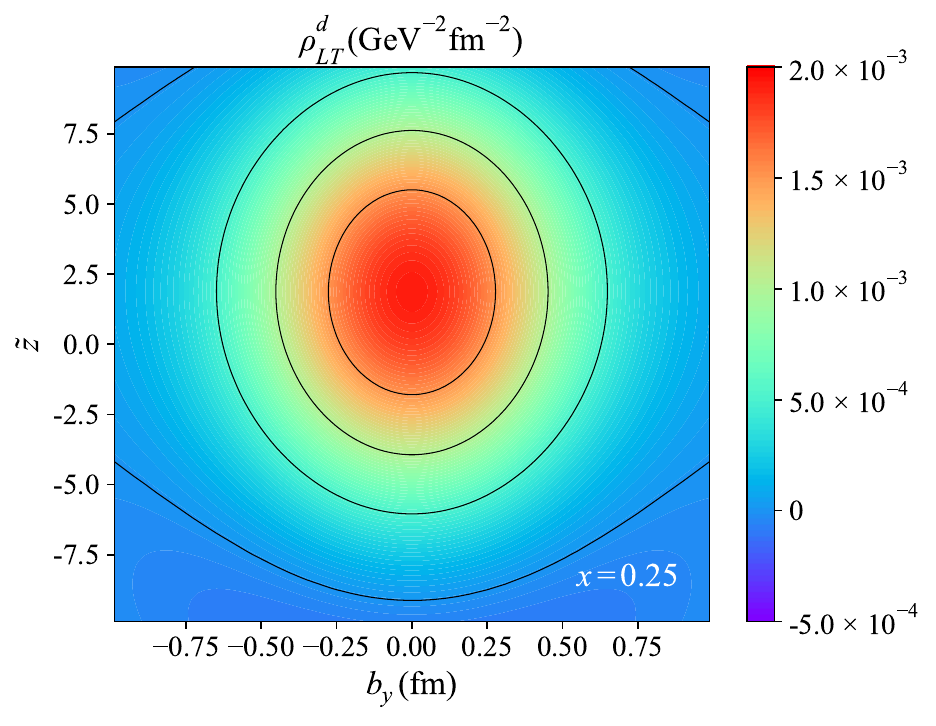}
	}
	\subfloat{
		\includegraphics[width=0.31\textwidth]{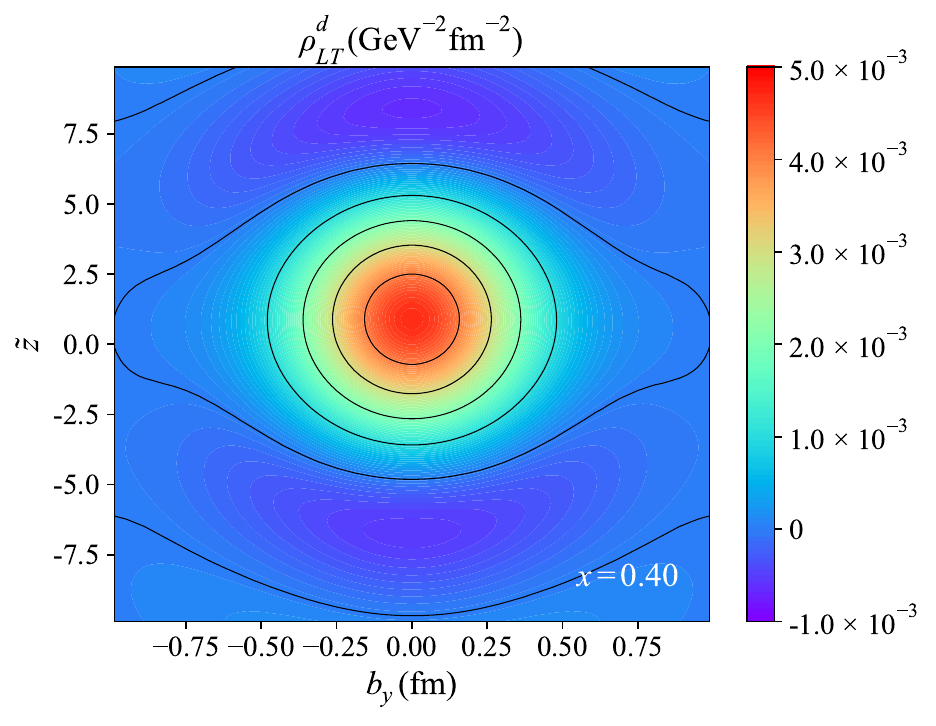}
	}
	\caption{Six-dimensional longitudinal-transverse light-front Wigner distribution $\rho_{\mathrm{LT}}\left(\tilde{z},x,\boldsymbol{b}_{\perp}, \boldsymbol{k}_{\perp}\right)$ for $u$ quark (upper panels) and $d$ quark (lower panels). The figure presents the Wigner distributions in the $\tilde{z}-b_y$ plane, with the transverse momentum fixed at $\boldsymbol{k}_{\perp}=0.3\,\mathrm{GeV}\boldsymbol{\hat{e}}_x$ (where $\boldsymbol{\hat{e}}_x$ is the unit vector in the $x$-direction) and the transverse coordinate component fixed at $b_x=0.4\,\mathrm{GeV}^{-1}$. The three columns correspond to $x=0.10$, $x=0.25$, and $x=0.40$.}
	\label{6DProtonLTudzby}
\end{figure}

\begin{figure}[htbp]
	\centering
	\subfloat{
		\includegraphics[width=0.31\textwidth]{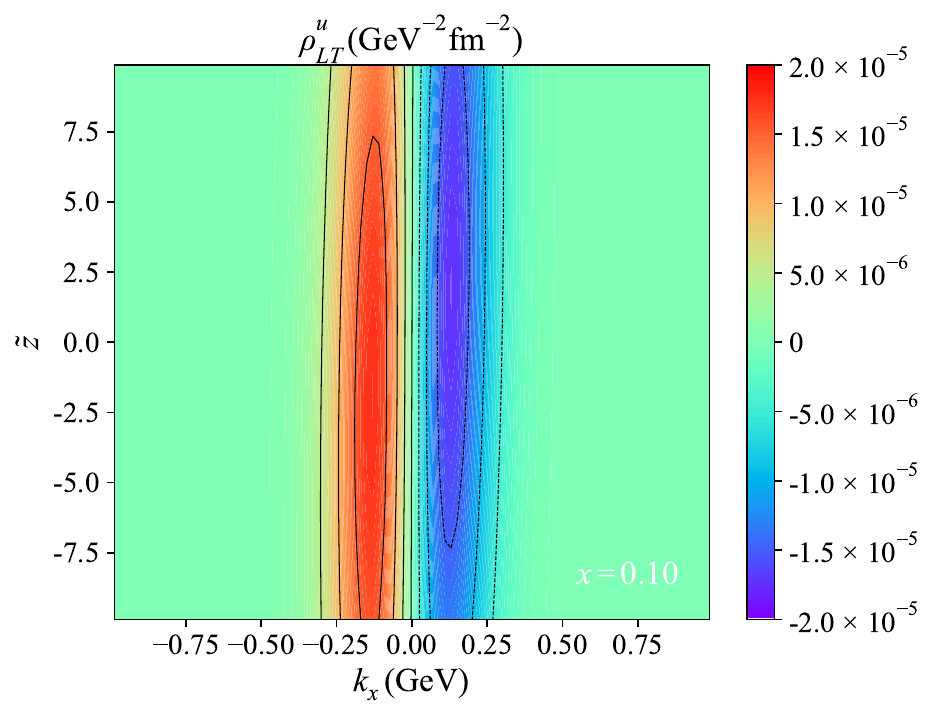}
	}
	\subfloat{
		\includegraphics[width=0.31\textwidth]{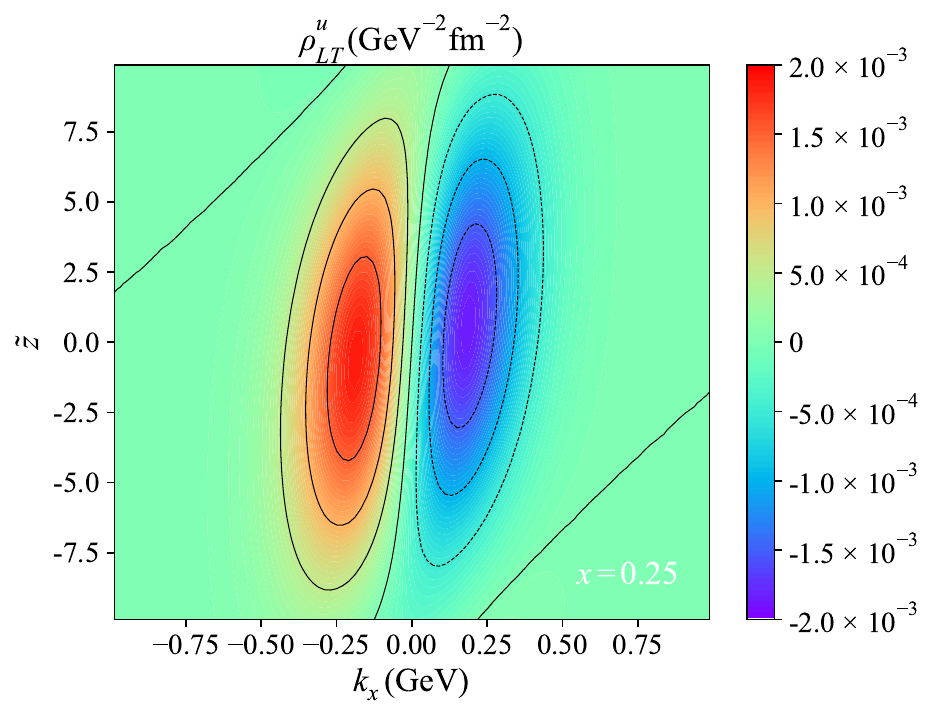}
	}
	\subfloat{
		\includegraphics[width=0.31\textwidth]{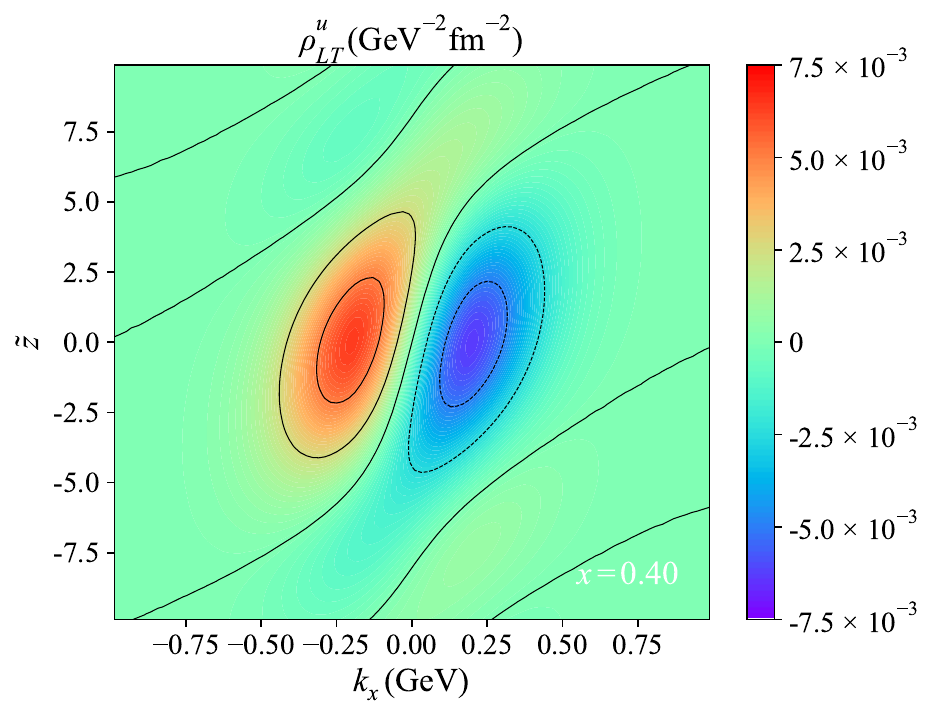}
	}\\
	\subfloat{
		\includegraphics[width=0.31\textwidth]{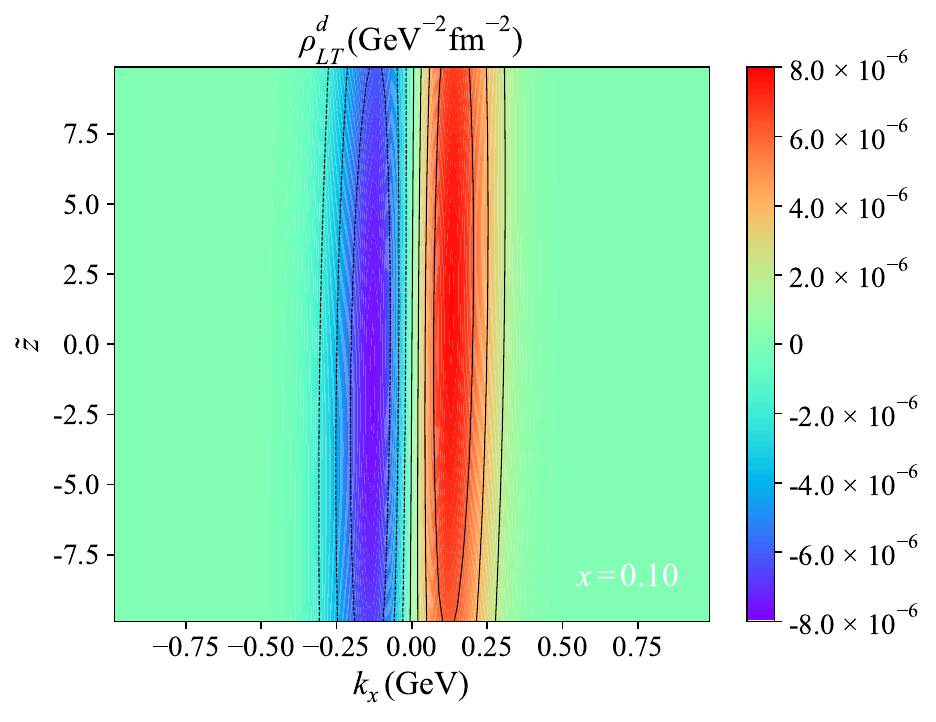}
	}
	\subfloat{
		\includegraphics[width=0.31\textwidth]{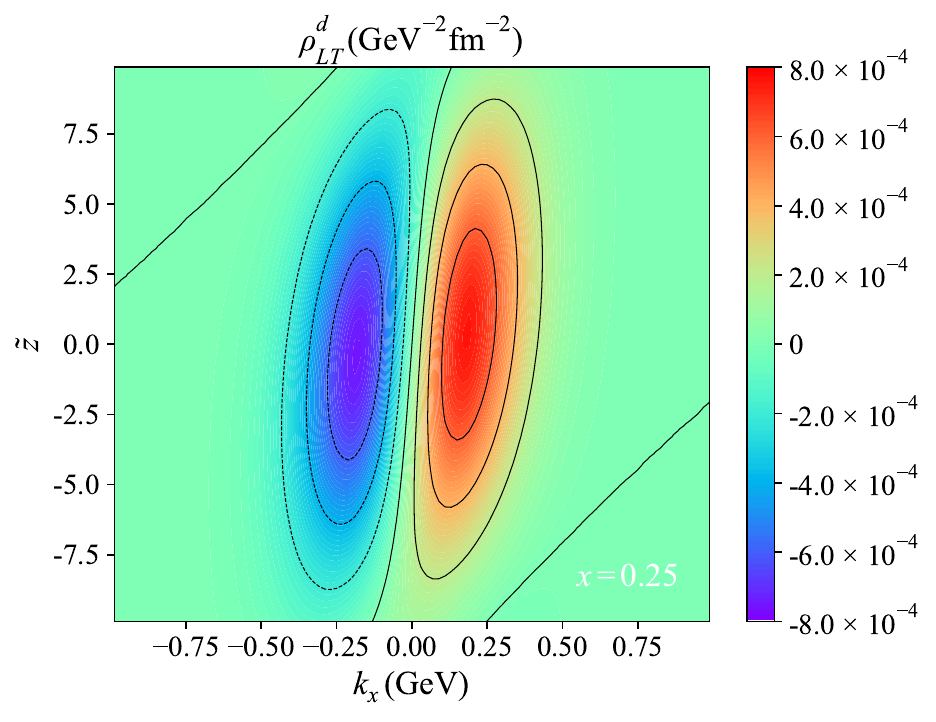}
	}
	\subfloat{
		\includegraphics[width=0.31\textwidth]{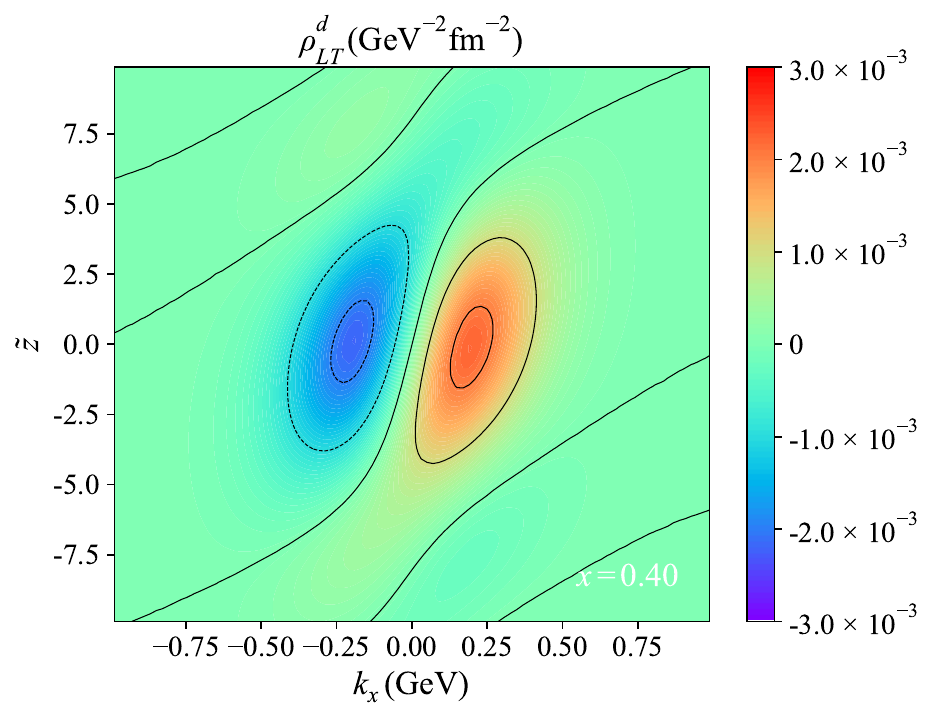}
	}
	\caption{Six-dimensional longitudinal-transverse light-front Wigner distribution $\rho_{\mathrm{LT}}\left(\tilde{z},x,\boldsymbol{b}_{\perp}, \boldsymbol{k}_{\perp}\right)$ for $u$ quark (upper panels) and $d$ quark (lower panels). The figure presents the Wigner distributions in the $\tilde{z}-k_x$ plane, with the transverse coordinate fixed at $\boldsymbol{b}_{\perp}=0.4\,\mathrm{GeV}^{-1}\boldsymbol{\hat{e}}_x$ (where $\boldsymbol{\hat{e}}_x$ is the unit vector along the $x$-axis) and the transverse momentum component fixed at $k_y=0.3\,\mathrm{GeV}$. The three columns correspond to $x=0.10$, $x=0.25$, and $x=0.40$.}
	\label{6DProtonLTudzkx}
\end{figure}

\begin{figure}[htbp]
	\centering
	\subfloat{
		\includegraphics[width=0.31\textwidth]{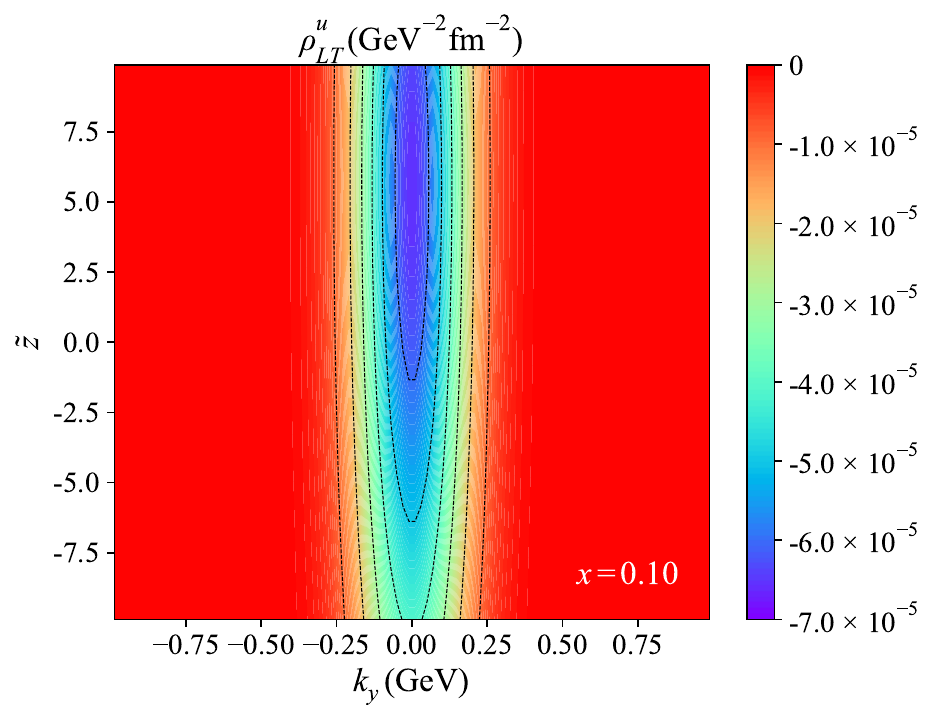}
	}
	\subfloat{
		\includegraphics[width=0.31\textwidth]{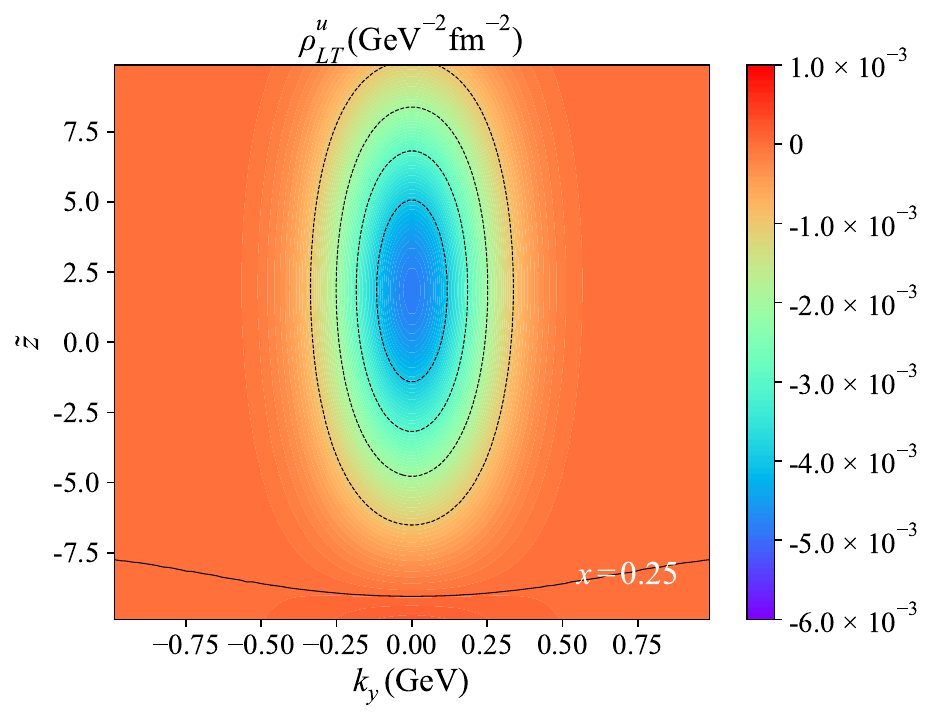}
	}
	\subfloat{
		\includegraphics[width=0.31\textwidth]{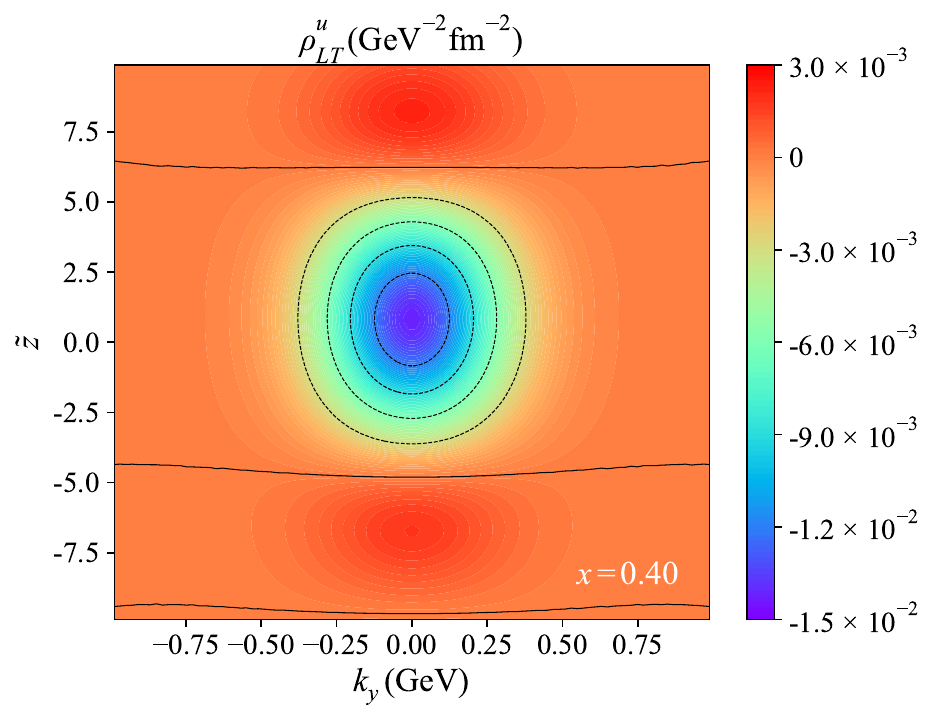}
	}\\
	\subfloat{
		\includegraphics[width=0.31\textwidth]{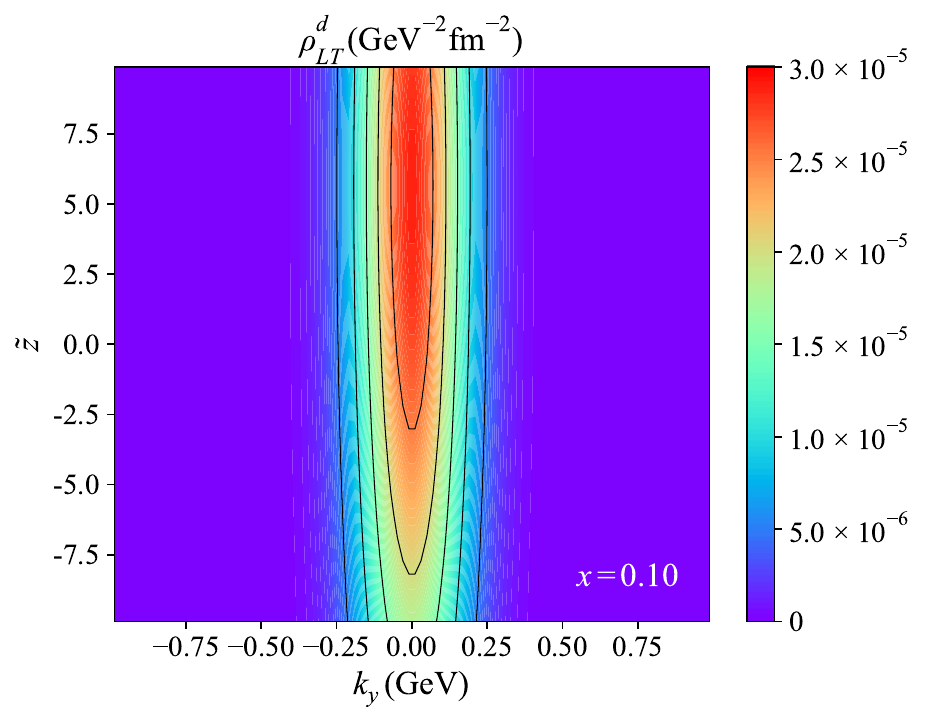}
	}
	\subfloat{
		\includegraphics[width=0.31\textwidth]{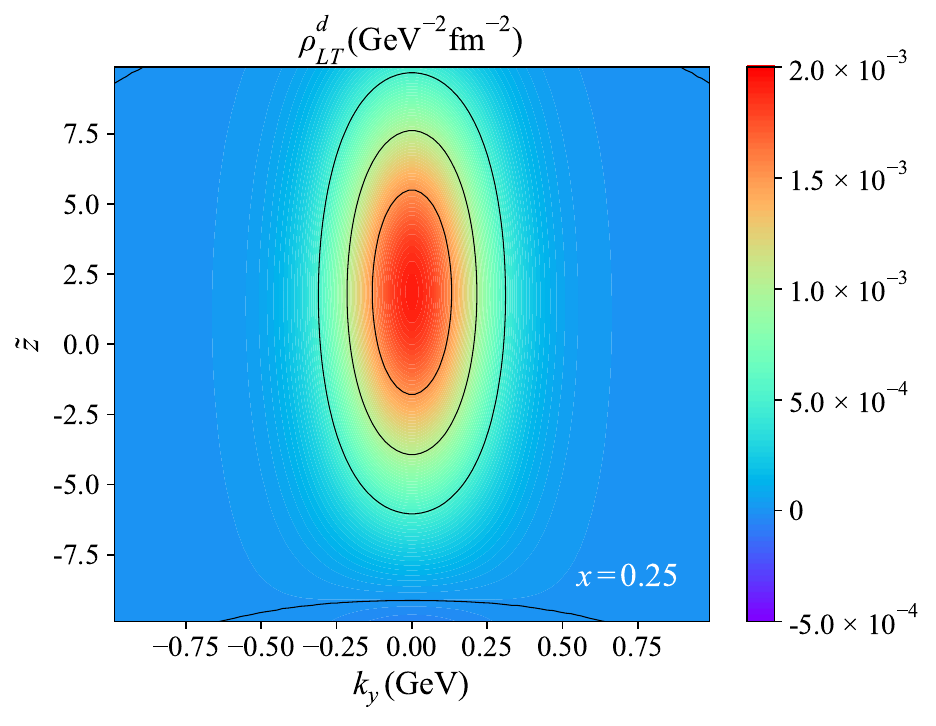}
	}
	\subfloat{
		\includegraphics[width=0.31\textwidth]{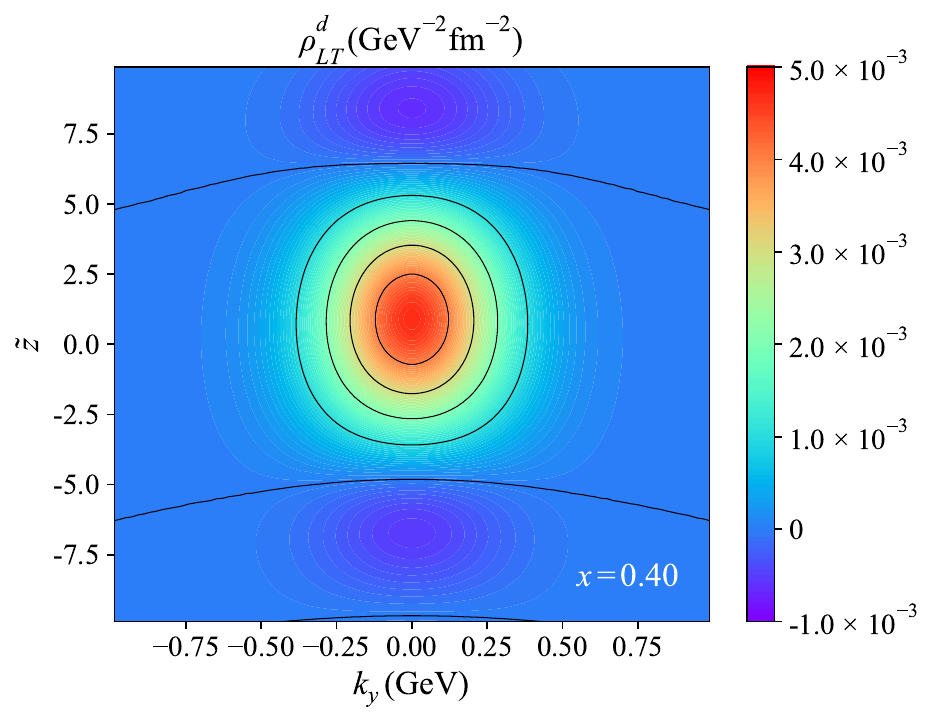}
	}
	\caption{Six-dimensional longitudinal-transverse light-front Wigner distribution $\rho_{\mathrm{LT}}\left(\tilde{z},x,\boldsymbol{b}_{\perp}, \boldsymbol{k}_{\perp}\right)$ for $u$ quark (upper panels) and $d$ quark (lower panels). The figure presents the Wigner distributions in the $\tilde{z}-k_y$ plane, with the transverse coordinate fixed at $\boldsymbol{b}_{\perp}=0.4\,\mathrm{GeV}^{-1}\boldsymbol{\hat{e}}_x$ (where $\boldsymbol{\hat{e}}_x$ is the unit vector along the $x$-axis) and the transverse momentum component fixed at $k_x=0.3\,\mathrm{GeV}$. The three columns correspond to $x=0.10$, $x=0.25$, and $x=0.40$.}
	\label{6DProtonLTudzky}
\end{figure}

In Figs.~\ref{6DProtonLTudzbx}--\ref{6DProtonLTudzky}, %In Fig.~\ref{6DProtonLTudzbx}, Fig.~\ref{6DProtonLTudzby}, Fig.~\ref{6DProtonLTudzkx} and Fig.~\ref{6DProtonLTudzky}, 
we plot the six-dimensional  longitudinal-transverse light-front Wigner distribution $\rho_{\mathrm{LT}}\left(\tilde{z},x,\boldsymbol{b}_{\perp}, \boldsymbol{k}_{\perp}\right)$ for the $u$ and $d$ quarks of the proton, displayed in the $\tilde{z}-b_x$, $\tilde{z}-b_y$, $\tilde{z}-k_x$, and $\tilde{z}-k_y$ subspaces, respectively. The longitudinal-transverse light-front Wigner distributions describe the phase-space correlations between the quark transverse spin and the proton longitudinal spin. The numerical results illustrate the relationship between the longitudinal coordinate $\tilde{z}$ and the transverse coordinates $\boldsymbol{b}_{\perp}$ or transverse momentum components $\boldsymbol{k}_{\perp}$, with specific fixed values for $\boldsymbol{k}_{\perp}$ or $\boldsymbol{b}_{\perp}$. In these figures, the longitudinal momentum fraction $x$ is chosen as $x=0.25$, $x=0.50$, and $x=0.75$ for the first, second, and third columns, respectively.

\new{The antisymmetry $\rho^j_{\mathrm{LT}}\left(\tilde{z},x,\boldsymbol{b}_{\perp}, \boldsymbol{k}_{\perp}\right) = - \rho^j_{\mathrm{LT}}\left(-\tilde{z},x,\boldsymbol{b}_{\perp}, -\boldsymbol{k}_{\perp}\right)$ reflects the parity-odd nature of spin-orbit interactions, while the $\boldsymbol{b}_{\perp} \times \boldsymbol{k}_{\perp}$ dependence implies a connection to quark orbital angular momentum~\eqref{OAM}. For Fig.~\ref{6DProtonLTudzbx} and Fig.~\ref{6DProtonLTudzkx}, the $\tilde{z}-b_x$ and $\tilde{z}-k_x$ subspaces display exact central symmetry about the origin, with peak intensities that directly reflect the quark transverse polarization is negative for $u$ quarks and positive for $d$ quarks, consistent with their known spin contributions. More strikingly, for Fig.~\ref{6DProtonLTudzby} and Fig.~\ref{6DProtonLTudzky}, the $\tilde{z}-b_y$ and $\tilde{z}-k_y$ subspaces projections reveal a distinct dipole structure that provides clear evidence of spin-orbit coupling within the proton; however, the center of symmetry is shifted along the $\tilde{z}$-axis, resulting in a displacement of the peak values. For $u$ quarks, the positive lobes at $\tilde{z}> 0$ and negative lobes at $\tilde{z}< 0$ correspond to right-handed and left-handed orbital motion respectively, while the dipole strength shows a pronounced $x$-dependence, being strongest at $x=0.25$ and gradually weakening at higher momentum fractions.}

\new{These Wigner distributions serve as a crucial bridge connecting to measurable hadronic structure functions: integration over $\boldsymbol{b}_{\perp}$ and $\tilde{z}$ yields the worm-gear function $h^{\perp}_{1L}$ at the TMD limit, while integration over $\boldsymbol{k}_{\perp}$ and $\tilde{z}$ gives access to spin-flip GPDs $H(x,\xi,t=-\boldsymbol{\Delta}_{\perp}^2)$ , with the dipole pattern specifically mapping to the pretzelosity distribution that probes quark orbital angular momentum.} Additionally, the distribution functions can be related to the IPDs $H_{T}$ and $\tilde{H}_{T}$, as well as other relevant distributions in the IPD limit. In previous analyses of the unpolarized-transverse functions, the distribution functions vanish in the TMD limit due to the omission of T-odd contributions. In the present discussion, however, both the TMD and IPDs are connected to the T-even part of the longitudinal-transverse Wigner distributions. These connections provide a framework for further exploring the relationships between these physical observables.

\new{The quantitative predictions from these distributions will be rigorously tested through upcoming precision measurements, particularly in longitudinal-target single spin asymmetries from SIDIS at EIC~\cite{AbdulKhalek:2021gbh} and azimuthal modulations in DVCS at JLab~\cite{Dudek:2012vr}. While the current light-front model successfully captures the essential features of these correlations, several theoretical considerations warrant further investigation, including potential modifications to the dipole structure from T-odd final-state interactions, the validity of the scalar/axial-vector diquark assumption~\cite{Brodsky:2002cx}, and possible contributions from gluon polarization effects at low momentum fractions $(x < 0.25)$~\cite{Ji:2020ena}. The agreement between these predictions and future experimental results will provide crucial tests of our understanding of quark orbital dynamics and their fundamental role in generating the proton spin structure, while also offering opportunities to refine the theoretical framework through comparison with first-principles lattice QCD calculations of GTMDs.}

%\subsubsection{u Quark}

%\subsubsection{d Quark}

\subsection{Transverse-Longitudinal Wigner distribution}
\label{TL}

\begin{figure}[htbp]
	\centering
	\subfloat{
		\includegraphics[width=0.31\textwidth]{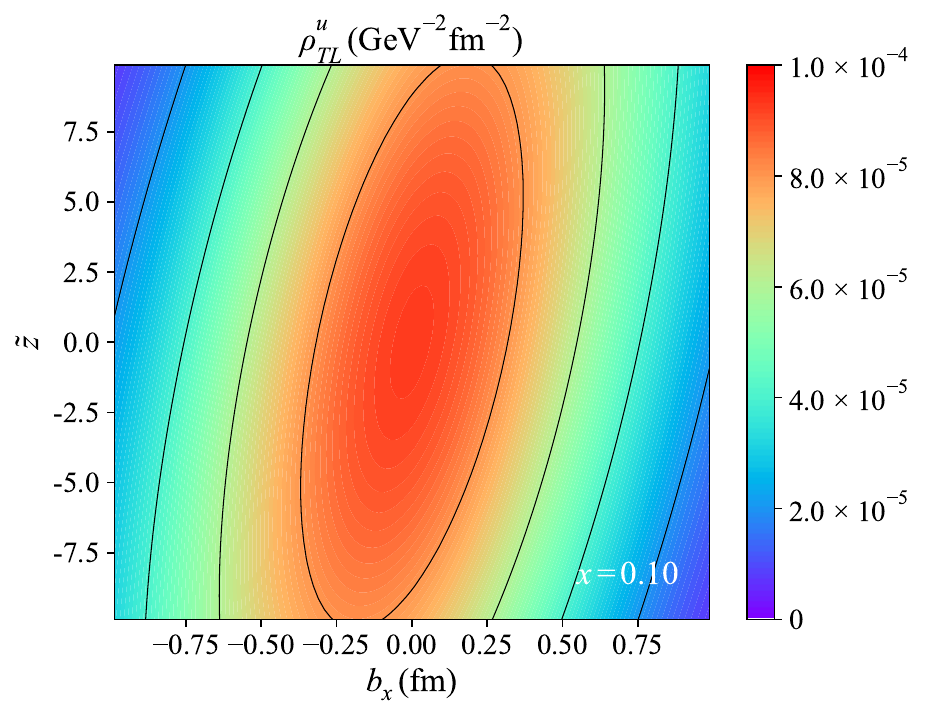}
	}
	\subfloat{
		\includegraphics[width=0.31\textwidth]{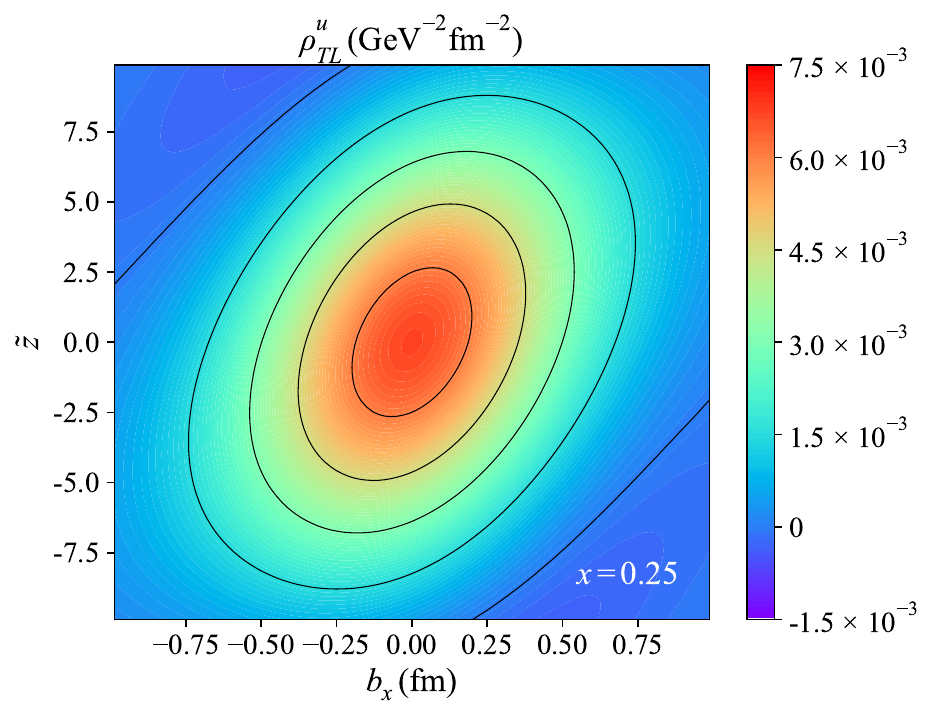}
	}
	\subfloat{
		\includegraphics[width=0.31\textwidth]{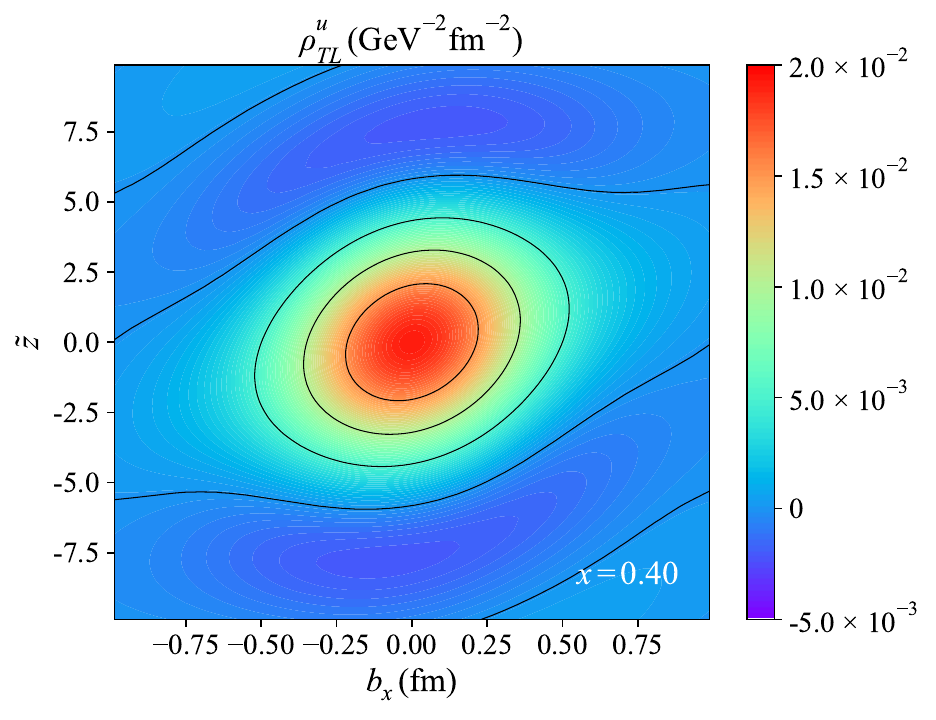}
	}\\
	\subfloat{
		\includegraphics[width=0.31\textwidth]{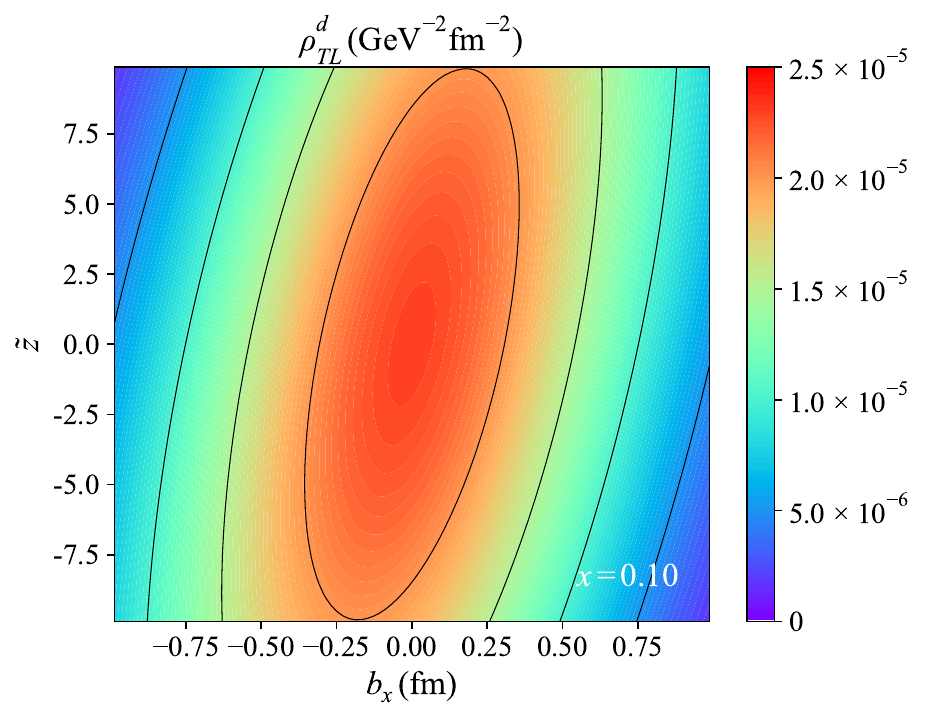}
	}
	\subfloat{
		\includegraphics[width=0.31\textwidth]{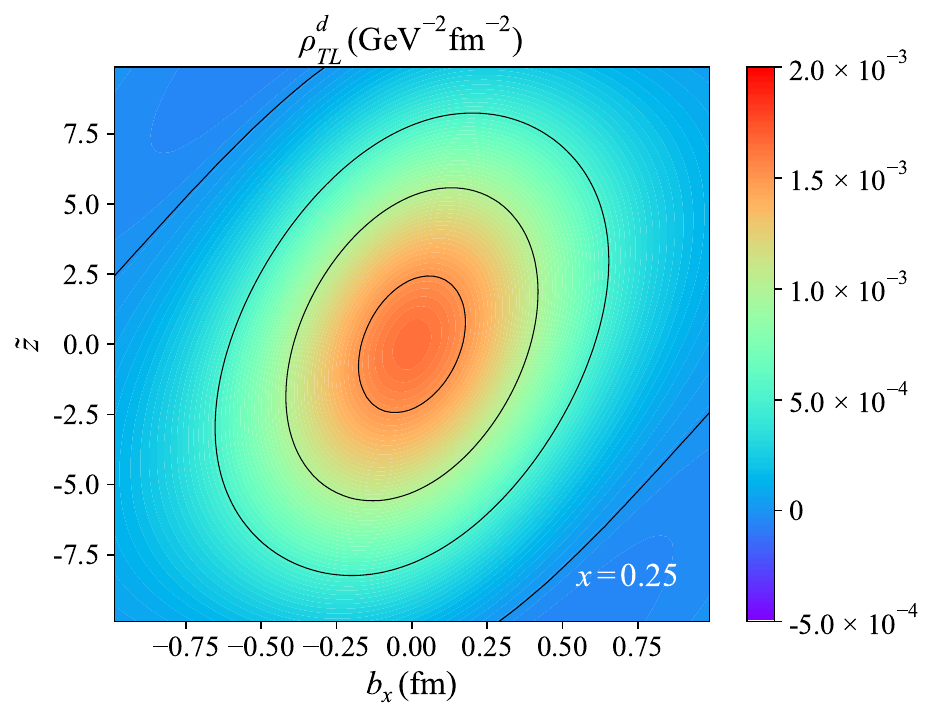}
	}
	\subfloat{
		\includegraphics[width=0.31\textwidth]{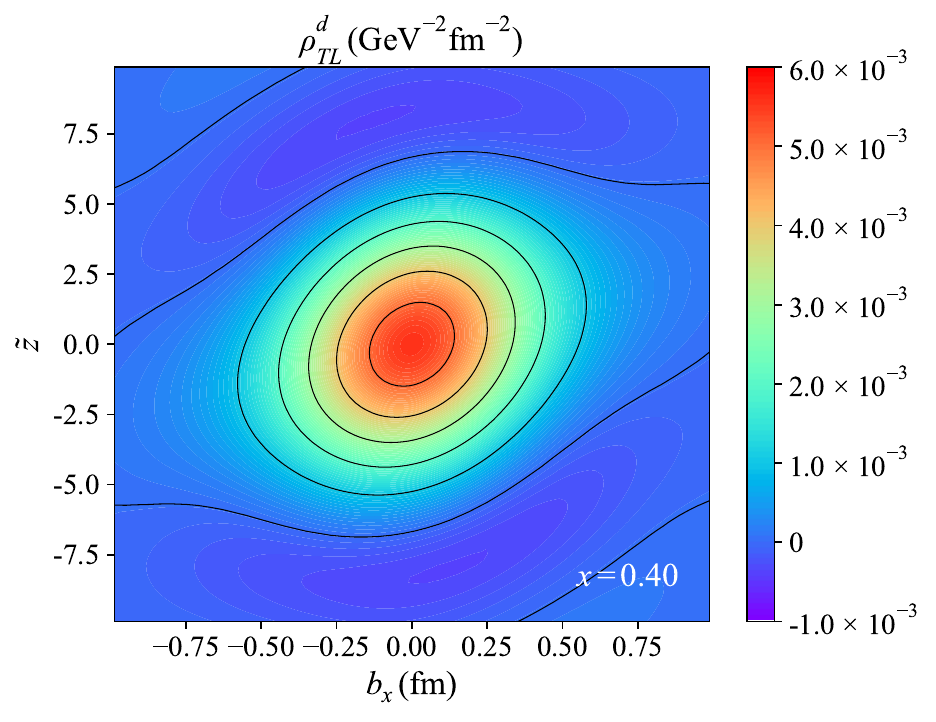}
	}
	\caption{Six-dimensional transverse-longitudinal light-front Wigner distribution $\rho_{\mathrm{TL}}\left(\tilde{z},x,\boldsymbol{b}_{\perp}, \boldsymbol{k}_{\perp}\right)$ for $u$ quark (upper panels) and $d$ quark (lower panels). The figure presents the Wigner distribution in the $\tilde{z}-b_x$ plane, with the transverse momentum fixed at $\boldsymbol{k}_{\perp}=0.3\,\mathrm{GeV}\boldsymbol{\hat{e}}_x$ (where $\boldsymbol{\hat{e}}_x$ is the unit vector in the $x$-direction) and the transverse coordinate component fixed at $b_y=0.4\,\mathrm{GeV}^{-1}$. The three columns correspond to $x=0.10$, $x=0.25$, and $x=0.40$.}
	\label{6DProtonTLudzbx}
\end{figure}

\begin{figure}[htbp]
	\centering
	\subfloat{
		\includegraphics[width=0.31\textwidth]{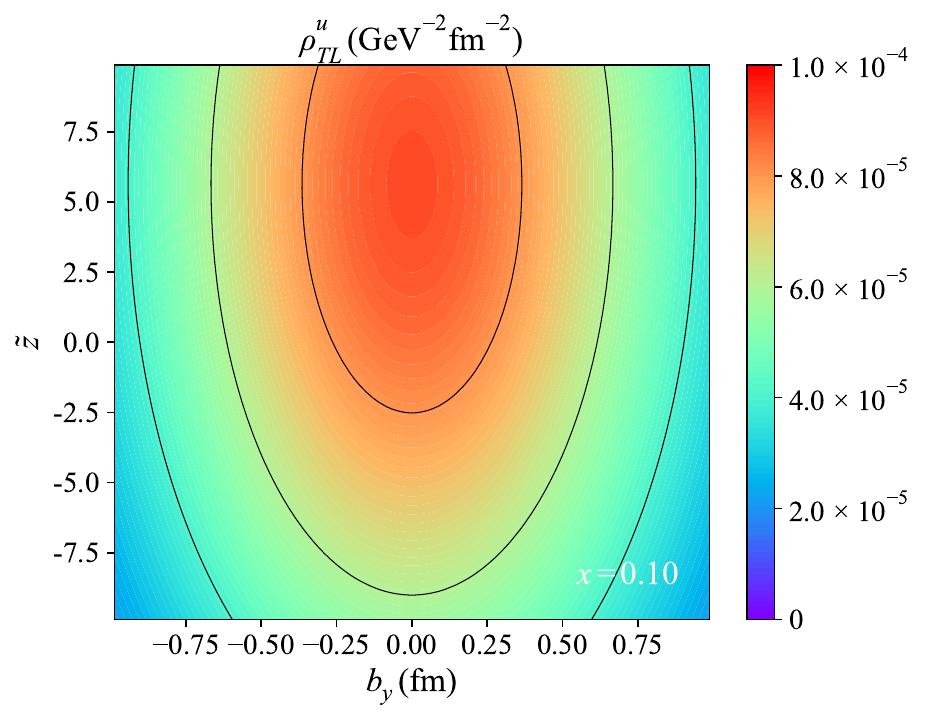}
	}
	\subfloat{
		\includegraphics[width=0.31\textwidth]{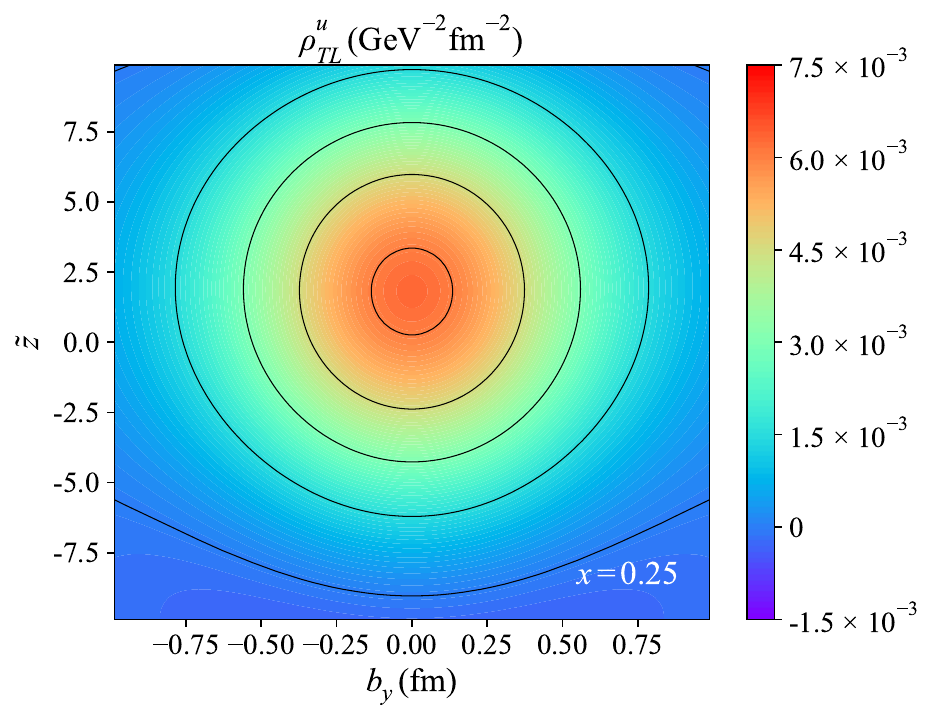}
	}
	\subfloat{
		\includegraphics[width=0.31\textwidth]{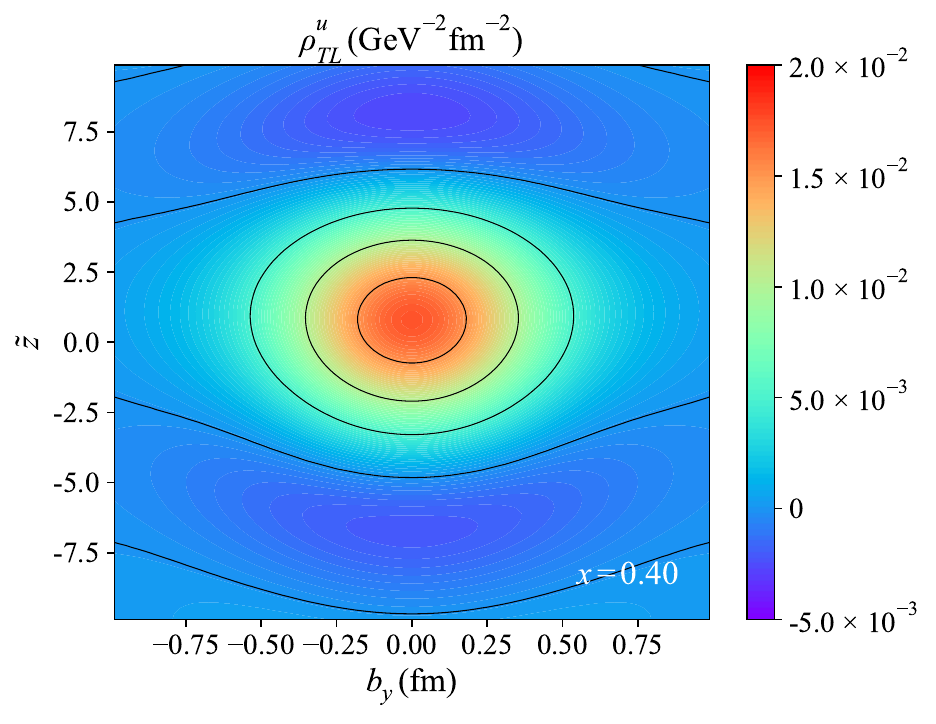}
	}\\
	\subfloat{
		\includegraphics[width=0.31\textwidth]{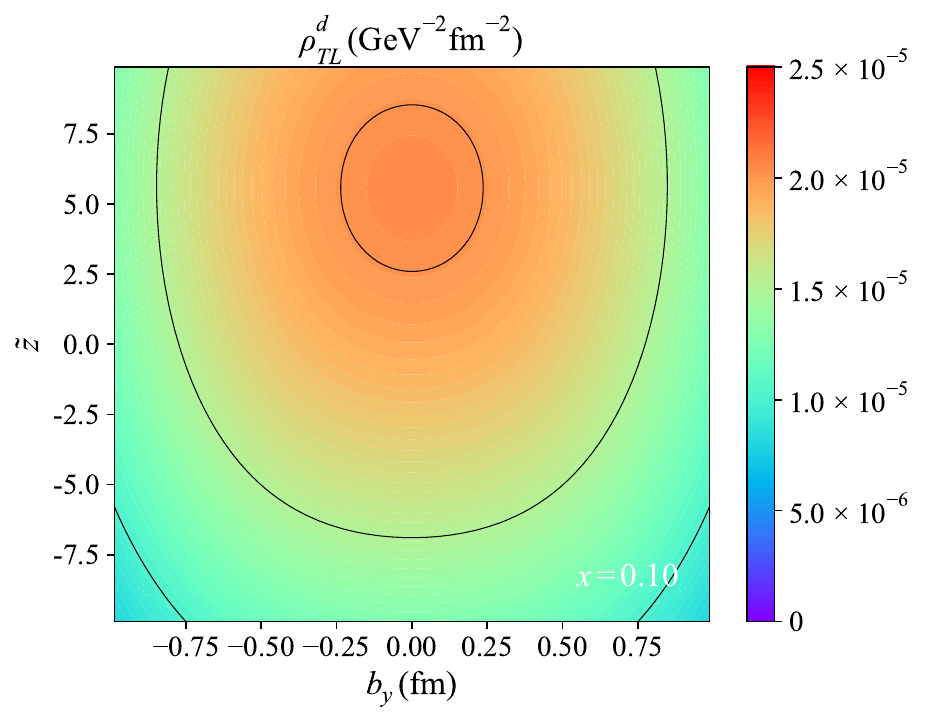}
	}
	\subfloat{
		\includegraphics[width=0.31\textwidth]{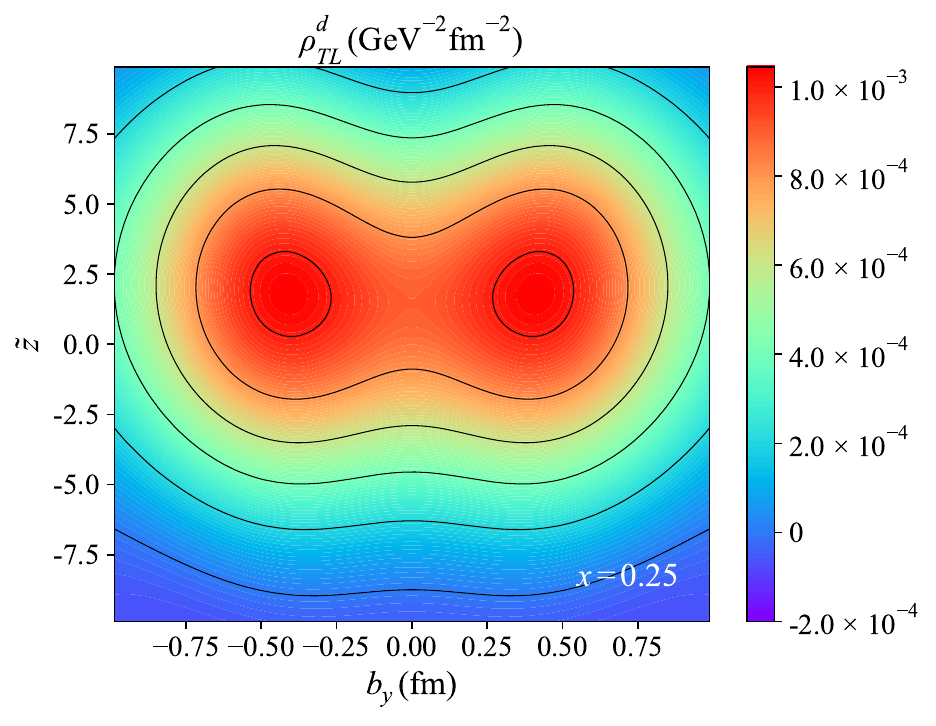}
	}
	\subfloat{
		\includegraphics[width=0.31\textwidth]{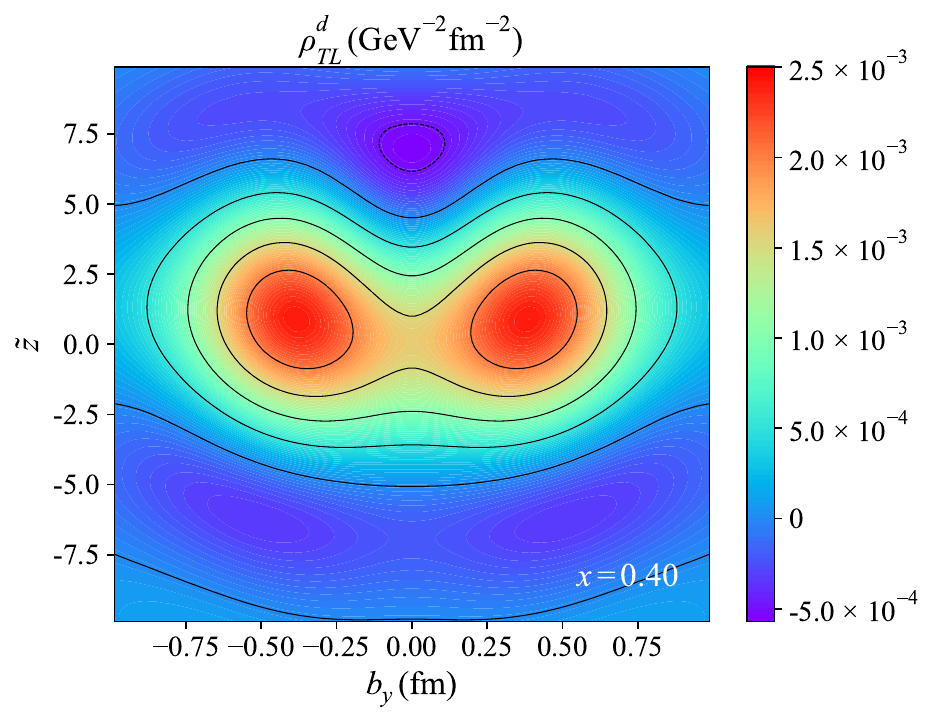}
	}
	\caption{Six-dimensional transverse-longitudinal light-front Wigner distribution $\rho_{\mathrm{TL}}\left(\tilde{z},x,\boldsymbol{b}_{\perp}, \boldsymbol{k}_{\perp}\right)$ for $u$ quark (upper panels) and $d$ quark (lower panels). The figure presents the Wigner distributions in the $\tilde{z}-b_y$ plane, with the transverse momentum fixed at $\boldsymbol{k}_{\perp}=0.3\,\mathrm{GeV}\boldsymbol{\hat{e}}_x$ (where $\boldsymbol{\hat{e}}_x$ is the unit vector in the $x$-direction) and the transverse coordinate component fixed at $b_x=0.4\,\mathrm{GeV}^{-1}$. The three columns correspond to $x=0.10$, $x=0.25$, and $x=0.40$.}
	\label{6DProtonTLudzby}
\end{figure}

\begin{figure}[htbp]
	\centering
	\subfloat{
		\includegraphics[width=0.31\textwidth]{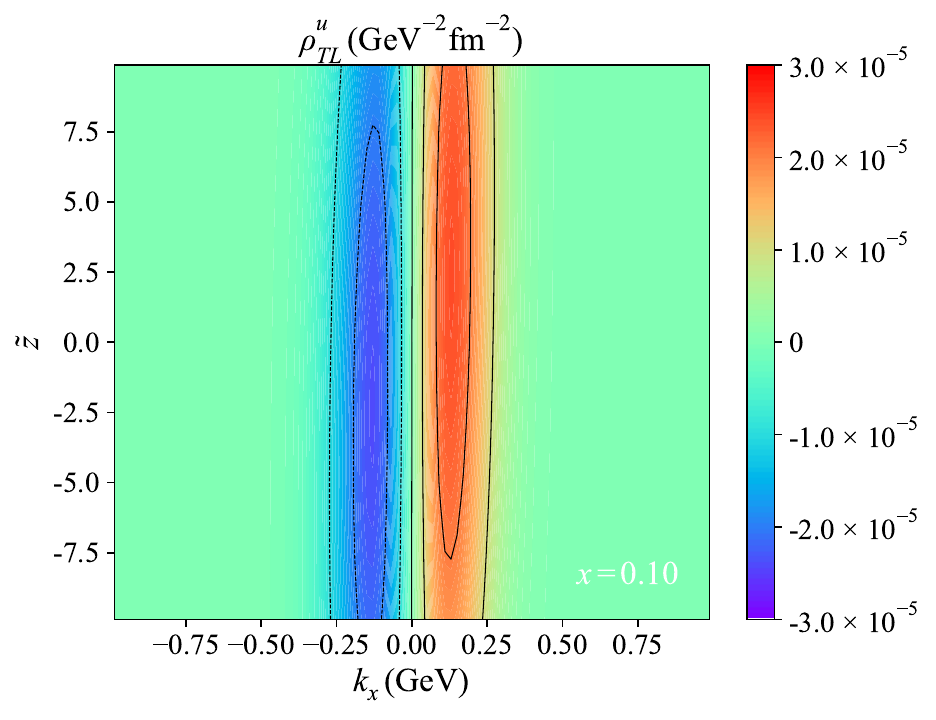}
	}
	\subfloat{
		\includegraphics[width=0.31\textwidth]{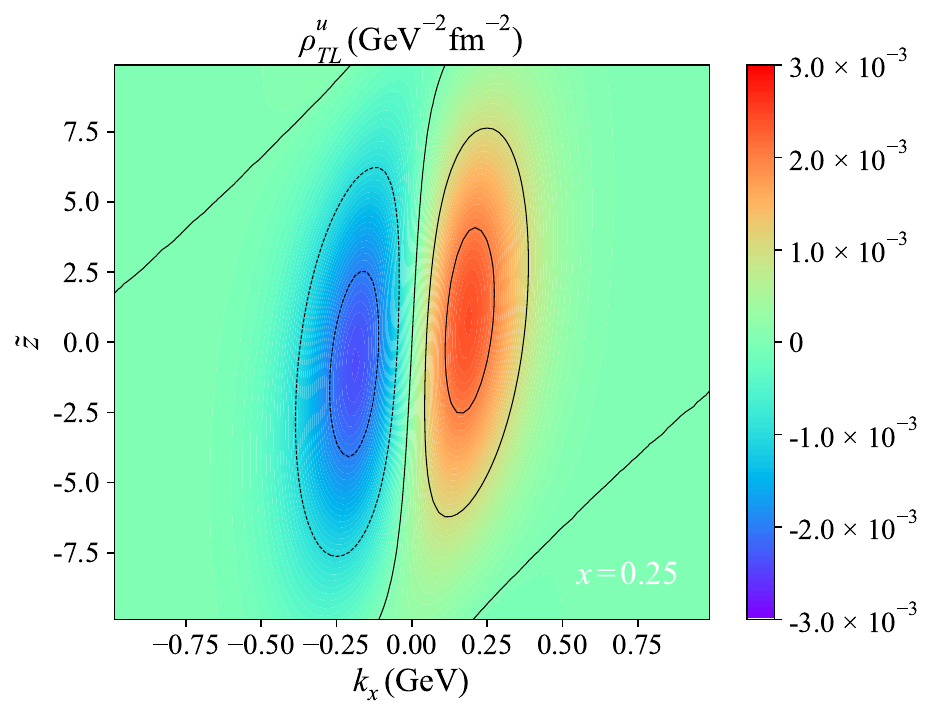}
	}
	\subfloat{
		\includegraphics[width=0.31\textwidth]{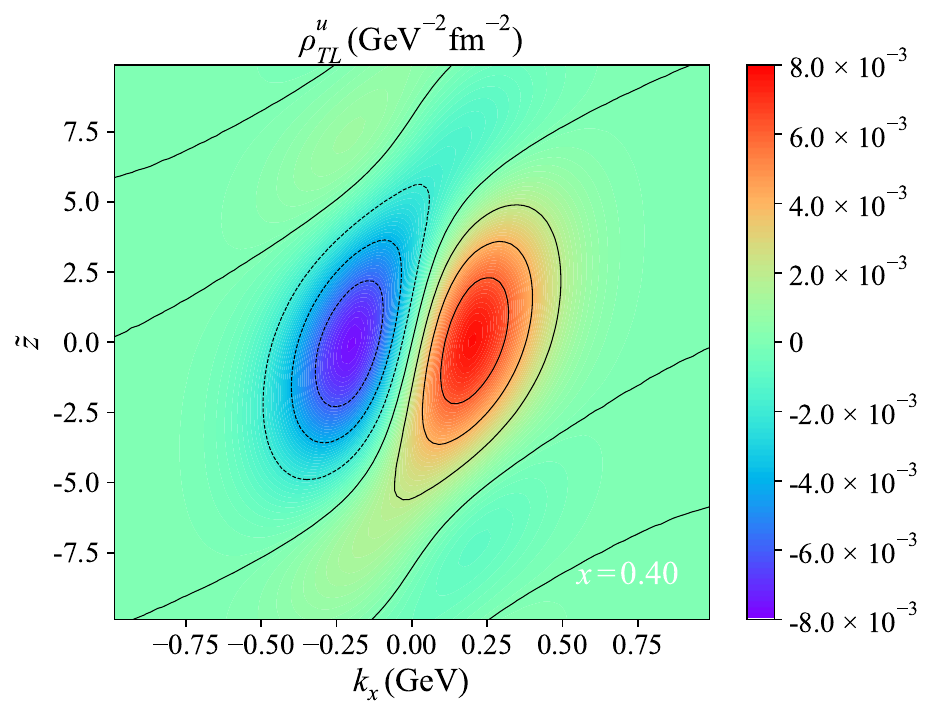}
	}\\
	\subfloat{
		\includegraphics[width=0.31\textwidth]{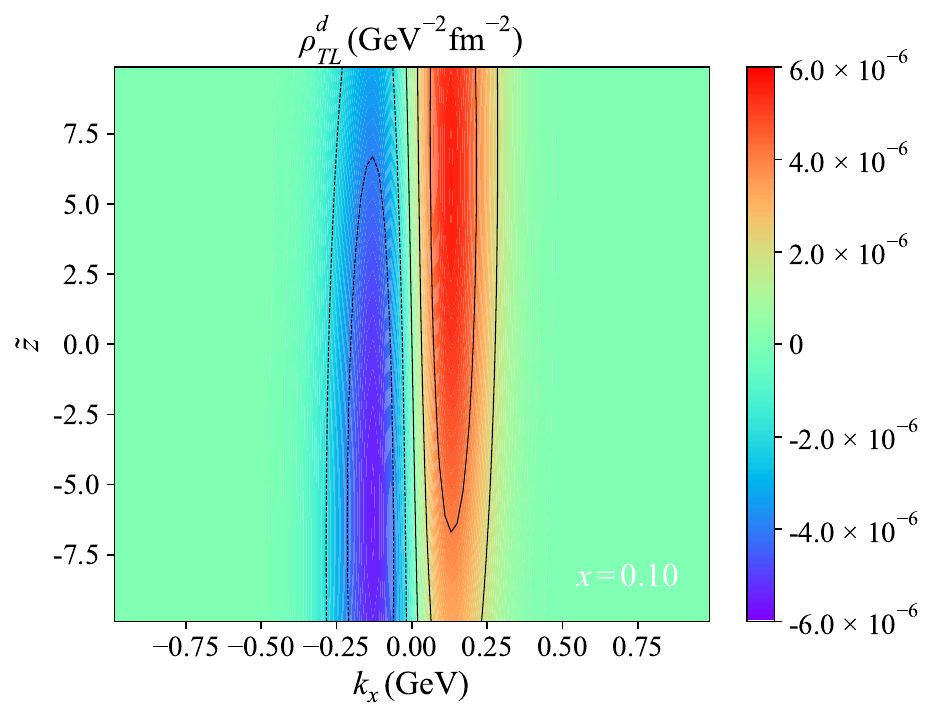}
	}
	\subfloat{
		\includegraphics[width=0.31\textwidth]{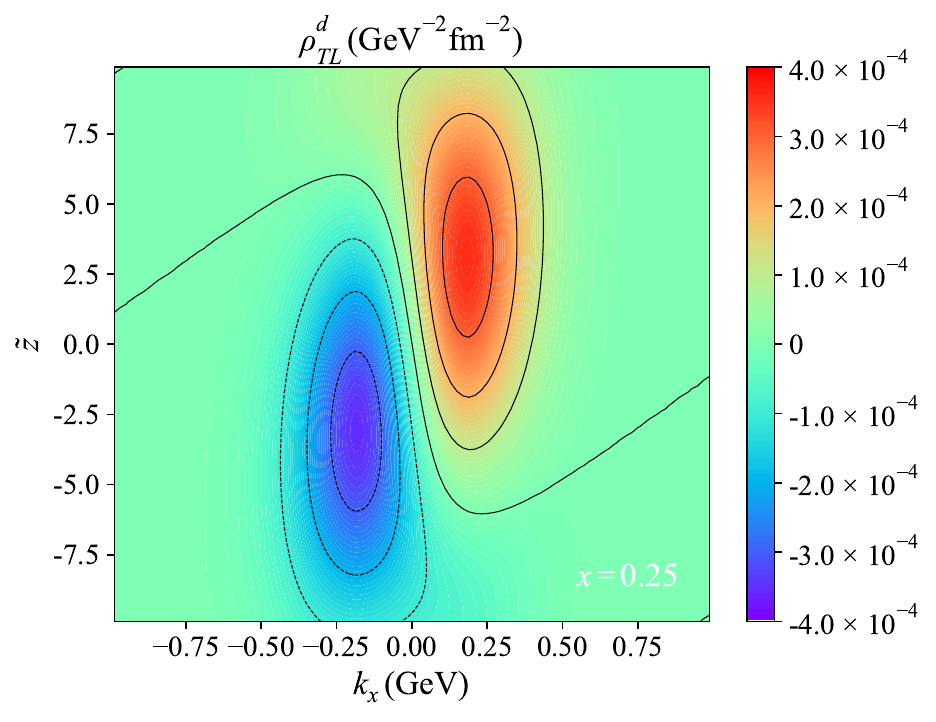}
	}
	\subfloat{
		\includegraphics[width=0.31\textwidth]{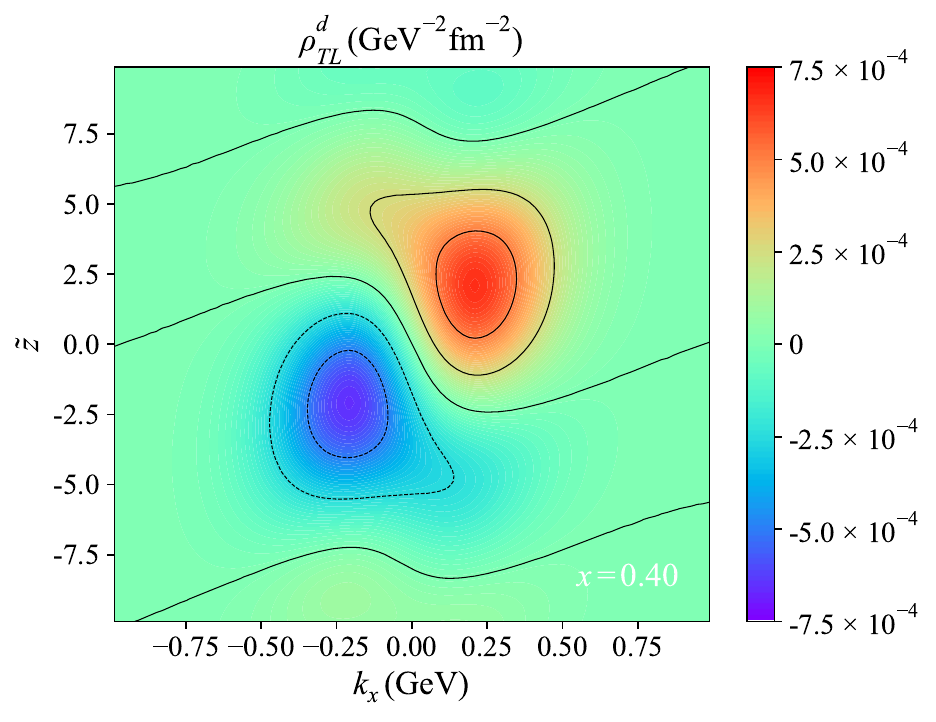}
	}
	\caption{Six-dimensional transverse-longitudinal light-front Wigner distribution $\rho_{\mathrm{TL}}\left(\tilde{z},x,\boldsymbol{b}_{\perp}, \boldsymbol{k}_{\perp}\right)$ for $u$ quark (upper panels) and $d$ quark (lower panels). The figure presents the Wigner distributions in the $\tilde{z}-k_x$ plane, with the transverse coordinate fixed at $\boldsymbol{b}_{\perp}=0.4\,\mathrm{GeV}^{-1}\boldsymbol{\hat{e}}_x$ (where $\boldsymbol{\hat{e}}_x$ is the unit vector along the $x$-axis) and the transverse momentum component fixed at $k_y=0.3\,\mathrm{GeV}$. The three columns correspond to $x=0.10$, $x=0.25$, and $x=0.40$.}
	\label{6DProtonTLudzkx}
\end{figure}

\begin{figure}[htbp]
	\centering
	\subfloat{
		\includegraphics[width=0.31\textwidth]{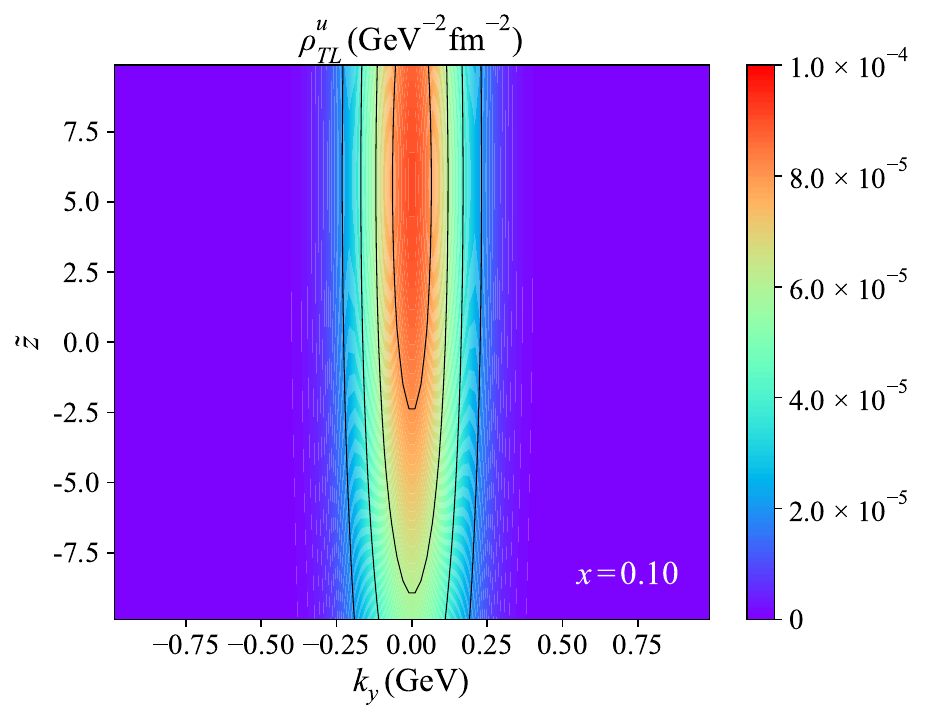}
	}
	\subfloat{
		\includegraphics[width=0.31\textwidth]{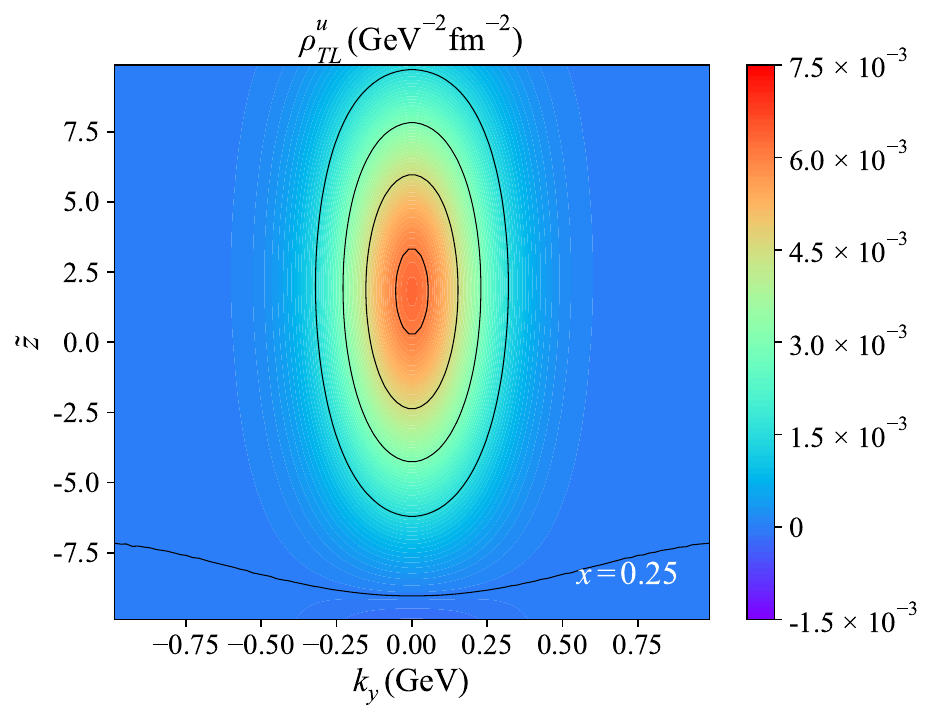}
	}
	\subfloat{
		\includegraphics[width=0.31\textwidth]{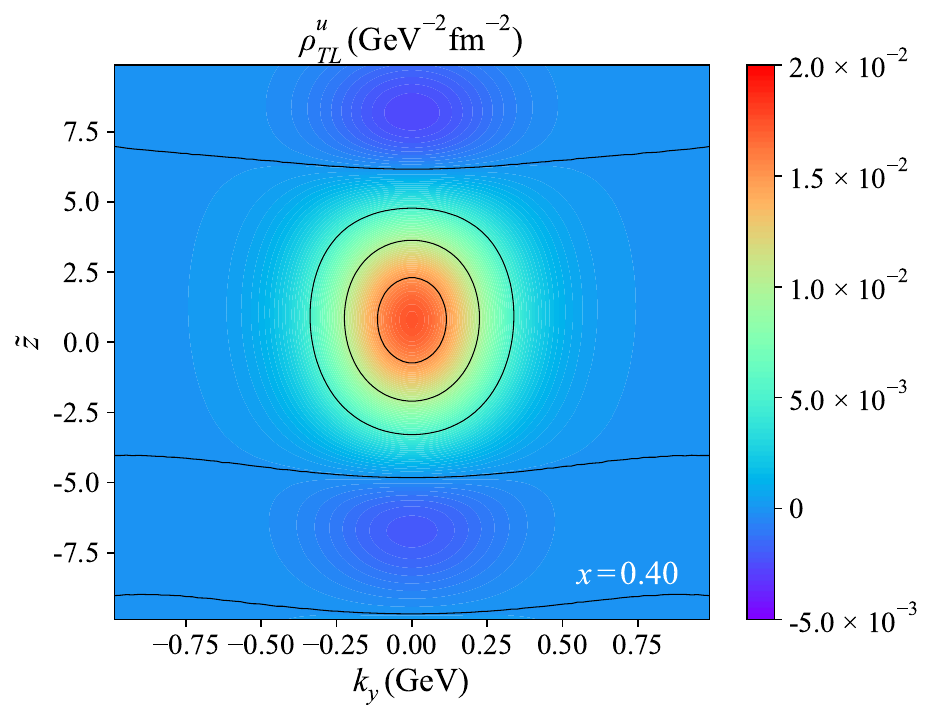}
	}\\
	\subfloat{
		\includegraphics[width=0.31\textwidth]{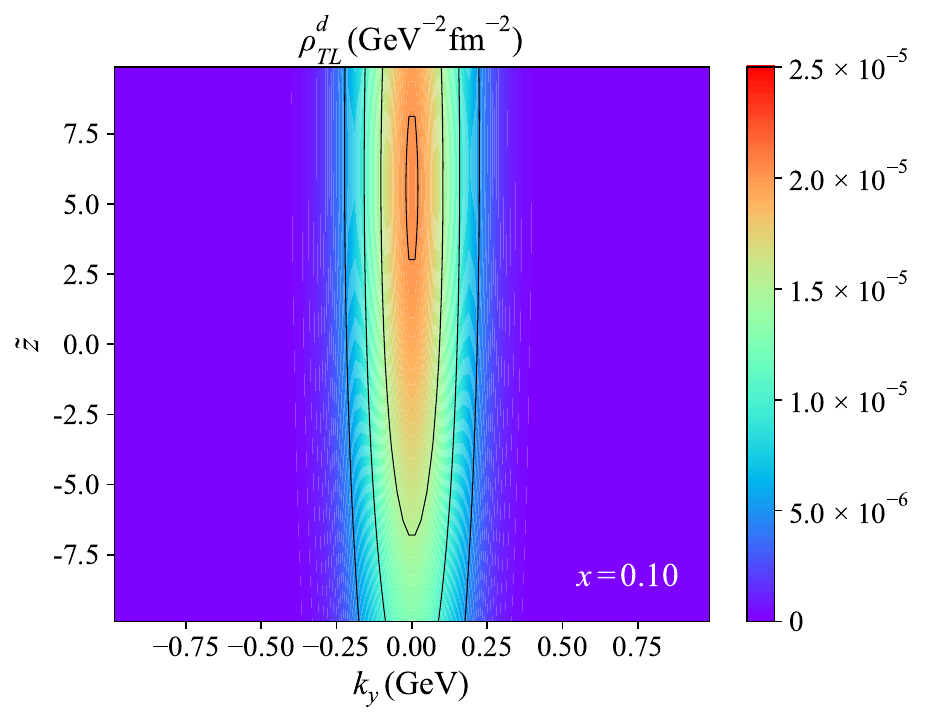}
	}
	\subfloat{
		\includegraphics[width=0.31\textwidth]{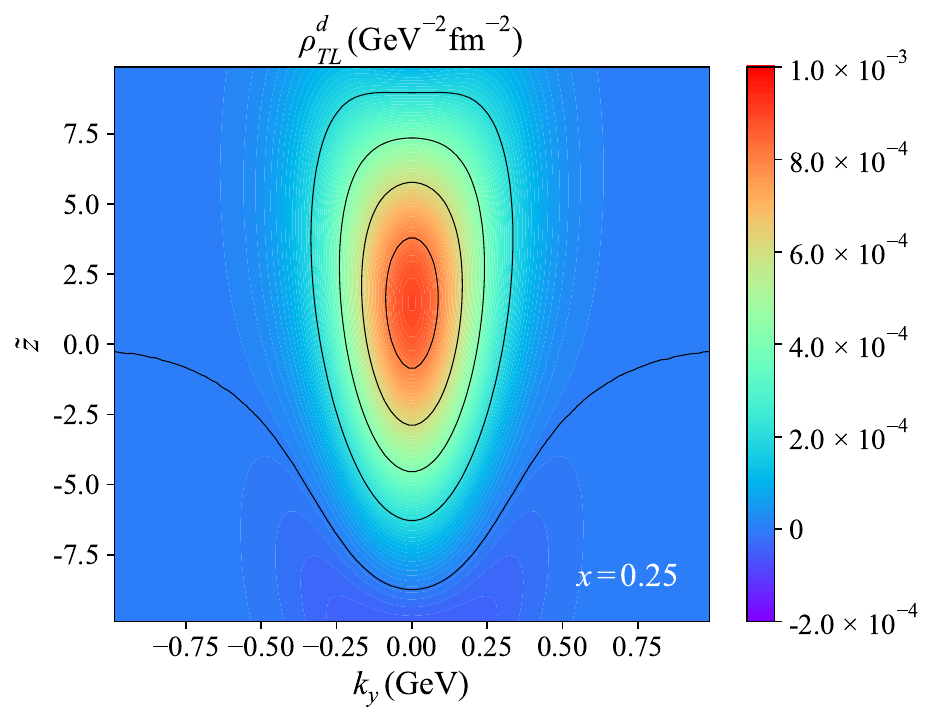}
	}
	\subfloat{
		\includegraphics[width=0.31\textwidth]{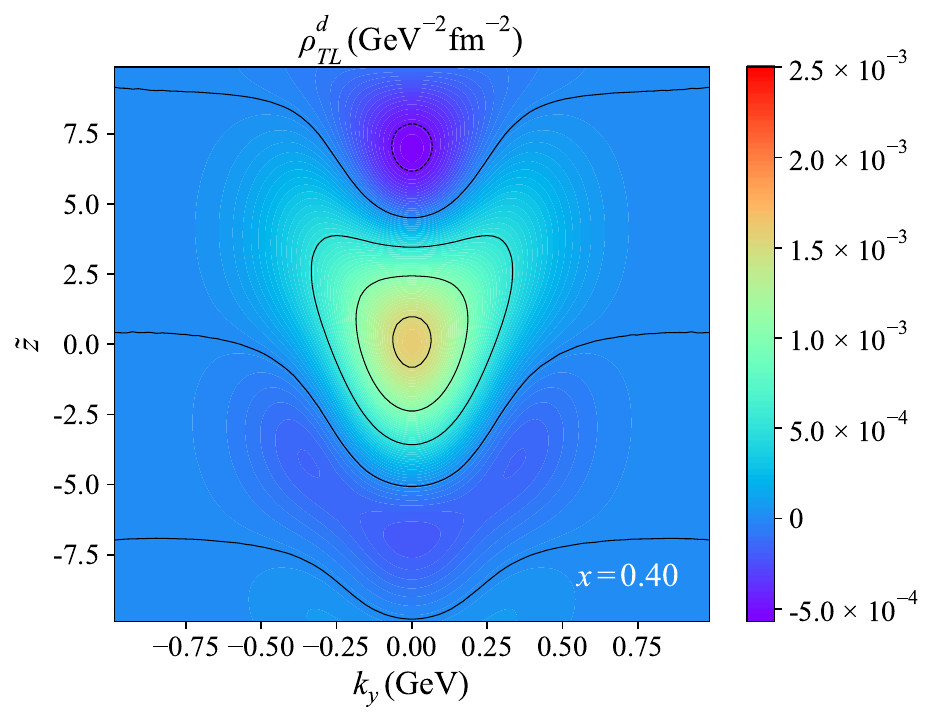}
	}
	\caption{Six-dimensional transverse-longitudinal light-front Wigner distribution $\rho_{\mathrm{TL}}\left(\tilde{z},x,\boldsymbol{b}_{\perp}, \boldsymbol{k}_{\perp}\right)$ for $u$ quark (upper panels) and $d$ quark (lower panels). The figure presents the Wigner distributions in the $\tilde{z}-k_y$ plane, with the transverse coordinate fixed at $\boldsymbol{b}_{\perp}=0.4\,\mathrm{GeV}^{-1}\boldsymbol{\hat{e}}_x$ (where $\boldsymbol{\hat{e}}_x$ is the unit vector along the $x$-axis) and the transverse momentum component fixed at $k_x=0.3\,\mathrm{GeV}$. The three columns correspond to $x=0.10$, $x=0.25$, and $x=0.40$.}
	\label{6DProtonTLudzky}
\end{figure}

In Figs.~\ref{6DProtonTLudzbx}--\ref{6DProtonTLudzky}, %In Fig.~\ref{6DProtonTLudzbx}, Fig.~\ref{6DProtonTLudzby}, Fig.~\ref{6DProtonTLudzkx} and Fig.~\ref{6DProtonTLudzky}, 
we plot the six-dimensional transverse-longitudinal light-front Wigner distribution $\rho_{\mathrm{TL}}\left(\tilde{z},x,\boldsymbol{b}_{\perp}, \boldsymbol{k}_{\perp}\right)$ for the $u$ and $d$ quarks of the proton, displayed in the $\tilde{z}-b_x$, $\tilde{z}-b_y$, $\tilde{z}-k_x$, and $\tilde{z}-k_y$ subspaces, respectively. The six-dimensional transverse-longitudinal light-front Wigner distributions characterize the phase-space correlations for a longitudinal-polarized quark in a transverse-polarized proton. The numerical results presented are obtained by fixing the transverse momentum $\boldsymbol{k}_{\perp}$ or the transverse coordinate $\boldsymbol{b}_{\perp}$ at specific values, and the longitudinal momentum fraction $x$ is set at $x = 0.10$, $x = 0.25$, and $x = 0.40$ in the first, second, and third columns, respectively.

In the $\tilde{z}-b_x$ and $\tilde{z}-k_x$ subspaces (Fig.~\ref{6DProtonTLudzbx} and Fig.~\ref{6DProtonTLudzkx}), the distributions exhibit centrosymmetry about the origin, with the peak values consistently located at the coordinate origin for each fixed $x$. In contrast, the distributions in the $\tilde{z}-b_y$ and $\tilde{z}-k_y$ subspaces (Fig.~\ref{6DProtonTLudzby} and Fig.~\ref{6DProtonTLudzky}) display a distinct dipole-symmetric pattern about $b_x = 0$ that provides unambiguous evidence of spin-orbit coupling within the proton.

By integrating over the three-dimensional coordinate space, the six-dimensional transverse-longitudinal Wigner distribution can be reduced to the other worm-gear function, i.e., the transverse-helicity $g_{1L}$, \new{whereas the distribution function is related to the IPDs $\tilde{H}$ and $E$ as well as other distributions at the IPD limit~\cite{Burkardt:2002hr,Diehl:2013xca}, which is known to be especially sensitive to quark orbital angular momentum and has been constrained by recent COMPASS measurements at intermediate $x$ values~\cite{COMPASS:2017jbv,COMPASS:2012ozz}.} Since it follows that the transverse-helicity TMD $g_{1L}$ and the IPDs $\tilde{H}$ and $E$ only depend on the T-even part of the transverse-longitudinal distribution, these connections offer a framework to explore potential relationships between these physical observables. However, the current model does not include the T-odd components, which might contain additional information and merit further investigation.

%\subsubsection{u Quark}

%\subsection{Transverse-polarized proton}

%\subsubsection{d Quark}

\section{Summary and Conclusion}
\label{Sec:SC}

In this paper, we investigate the six-dimensional light-front Wigner distribution functions of protons (spin-1/2 hadrons) within the framework of the light-front quark spectator-diquark model. This study represents a specific calculation of the recently proposed six-dimensional light-front Wigner functions for all twist-two quark Wigner distributions in this theoretical context. The light-front quark spectator-diquark model is a widely utilized approach for exploring the structure of hadrons, conceptualizing them as two-body systems composed of quarks and a spectator diquark. Our analysis provides the Wigner distribution function as a function of the longitudinal coordinate $\tilde{z}$, the longitudinal momentum $x$, the transverse coordinate 
$\boldsymbol{b}_{\perp}$, and the transverse momentum 
$\boldsymbol{k}_{\perp}$ for all polarization cases. This calculation marks the first application of this new physical quantity to the study of proton structure, carrying profound physical implications.

As a phase space distribution, the Wigner distribution encompasses more comprehensive physical information compared to traditional momentum space distributions, such as TMDs and GPDs. The Wigner distribution function presented in this paper allows us to accurately describe the joint distribution of three-dimensional coordinate space and three-dimensional momentum space of the partons within the hadron. Furthermore, by integrating the six-dimensional light-front Wigner distribution functions, we can recover other known distribution functions. Thus, the six-dimensional light-front Wigner distribution functions defined herein retain all relevant information and serve as the most comprehensive distribution functions currently known, including TMDs and GPDs. Its integration can yield new multidimensional distribution functions for partons that incorporate longitudinal spatial coordinate. Consequently, we can view the Wigner function as a bridge to other functions that describe hadron structures, and our findings significantly contribute to a more complete understanding of the internal structure of hadrons.

It is important to note that the six-dimensional light-front Wigner distribution function explored in this study is not a probability distribution in the traditional sense; rather, it is a quasi-probability distribution. In addition to the Wigner distribution, other forms of phase space distribution functions exist, such as the Husimi distribution function. The Husimi distribution can be interpreted as a Weierstrass transformation of the Wigner distribution, ensuring that it remains non-negative and thus qualifies as a true probability distribution function in phase space. Further investigation into its definition may provide valuable insights into the nature of non-probability distributions, as discussed in this paper.

Unlike traditional momentum distribution functions, directly observing spatial distributions poses significant experimental challenges. Typically, the approach involves converting observations of spatial distributions into momentum distributions, such as interpreting the charge distribution of the proton through form factor measurements. Since the six-dimensional light-front Wigner distribution functions defined in this paper are the three-dimensional Fourier transform of the GTMDs, measurements of GTMDs at non-zero skewness can be considered. Examples include diffractive di-jet production in DIS~\cite{Hatta:2016dxp}, the exclusive double Drell-Yan process~\cite{Bhattacharya:2017bvs}, and virtual photon-nucleus quasi-elastic scattering~\cite{Zhou:2016rnt}. These experimental efforts will be instrumental in observing new boost-invariant physical quantities. Additionally, as highlighted in previous analyses, there exists a special relationship between the Wigner distribution and OAM. Therefore, it may be experimentally feasible to indirectly measure the Wigner function by assessing the OAM, as discussed in Ref.~\cite{Bhattacharya:2023hbq}.

\acknowledgments{This work is supported by National Natural Science Foundation of China under Grants No.~12335006, No.~12175117, No.~12321005 and No.~12075003 and by Shandong Province Natural Science Foundation under Grant No. ZFJH202303.}

\section*{Appendix}
\label{Appendix}

\subsection{Unpolarized-Longitudinal distribution}

\begin{figure}[htbp]
	\centering
	\subfloat{
		\includegraphics[width=0.31\textwidth]{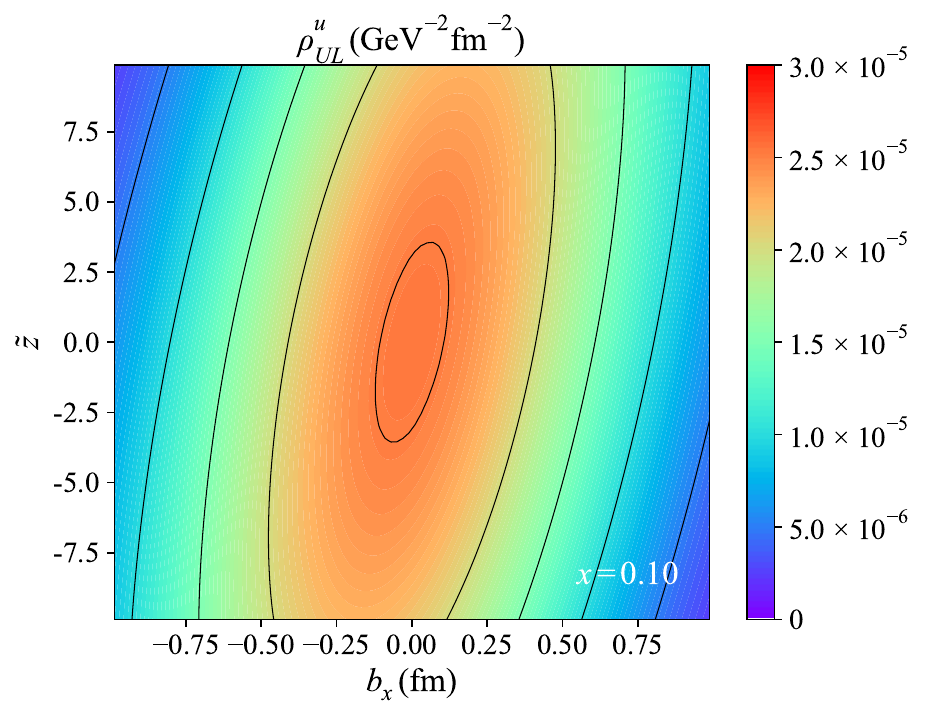}
	}
	\subfloat{
		\includegraphics[width=0.31\textwidth]{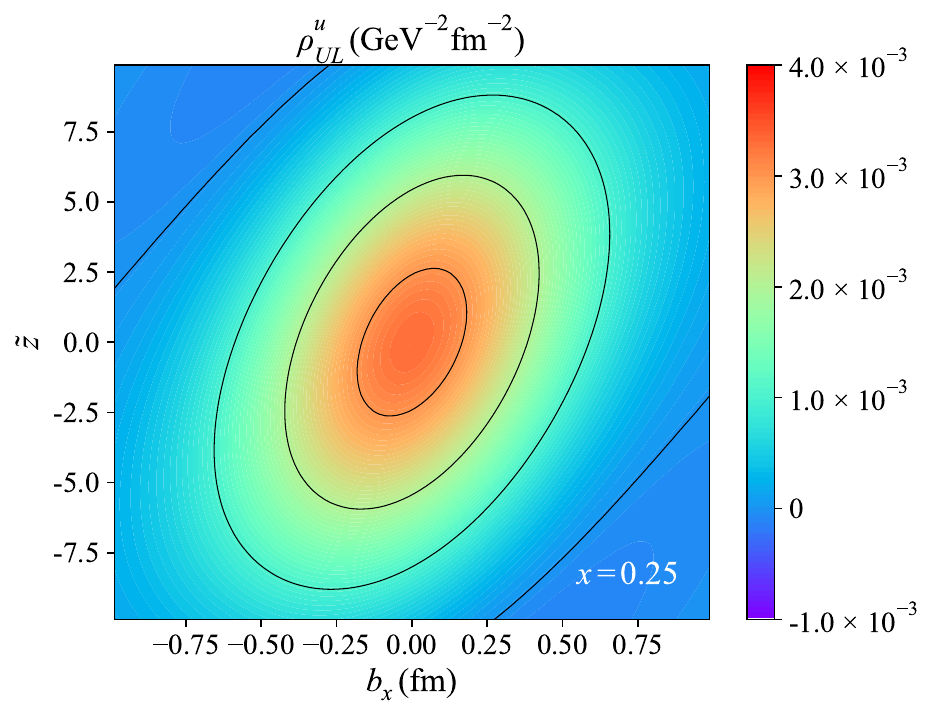}
	}
	\subfloat{
		\includegraphics[width=0.31\textwidth]{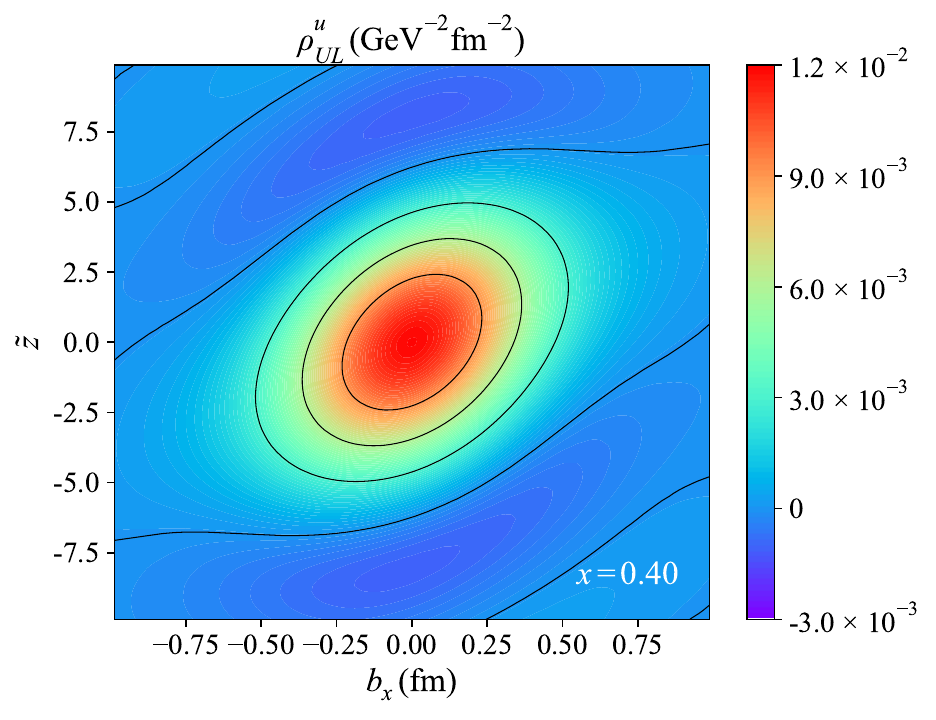}
	}\\
	\subfloat{
		\includegraphics[width=0.31\textwidth]{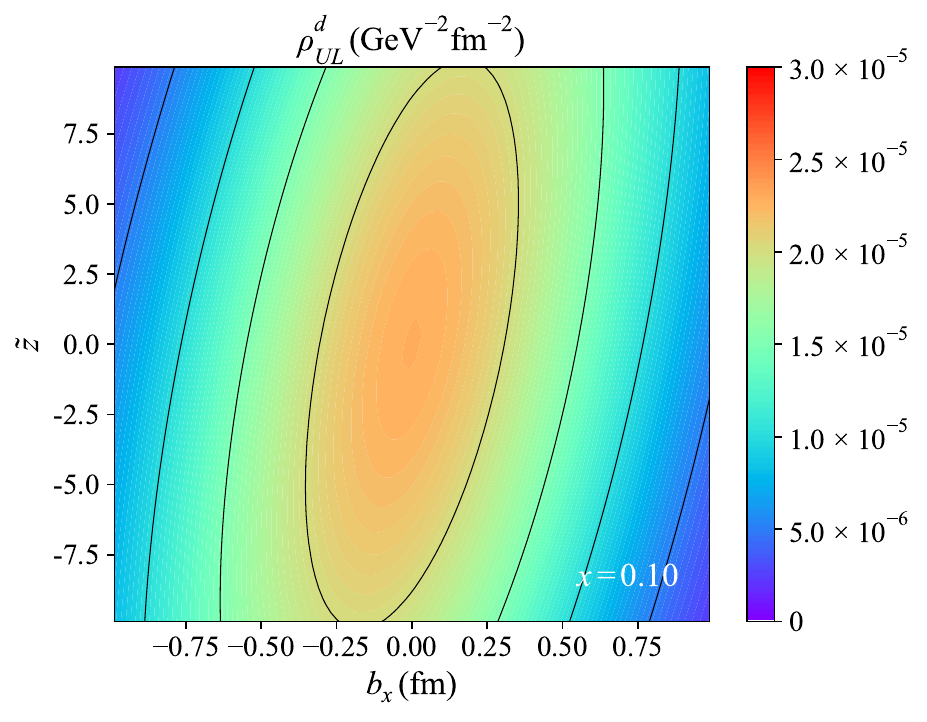}
	}
	\subfloat{
		\includegraphics[width=0.31\textwidth]{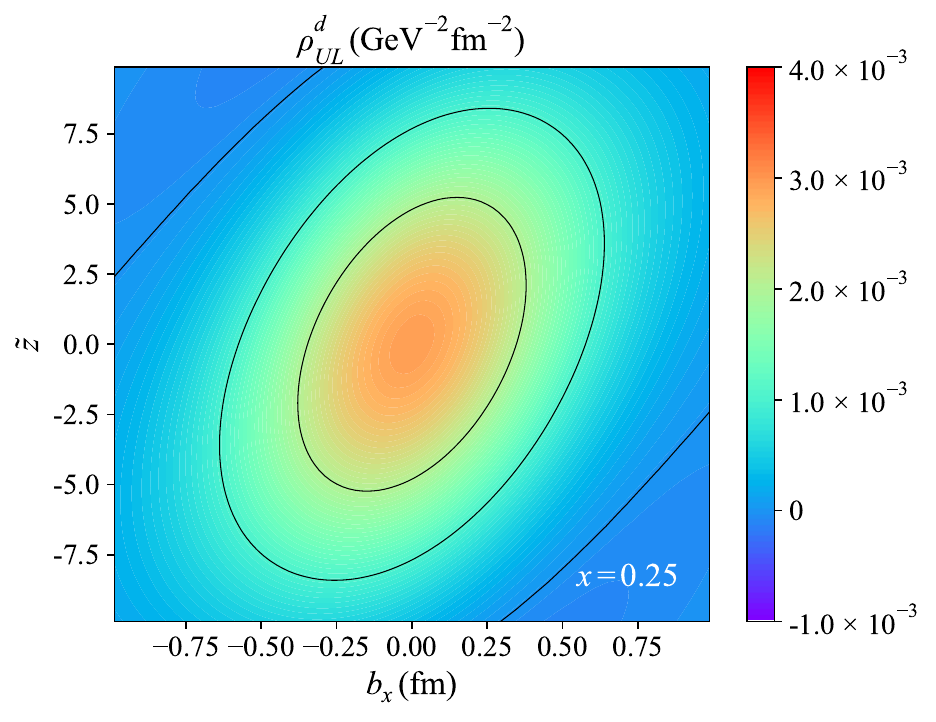}
	}
	\subfloat{
		\includegraphics[width=0.31\textwidth]{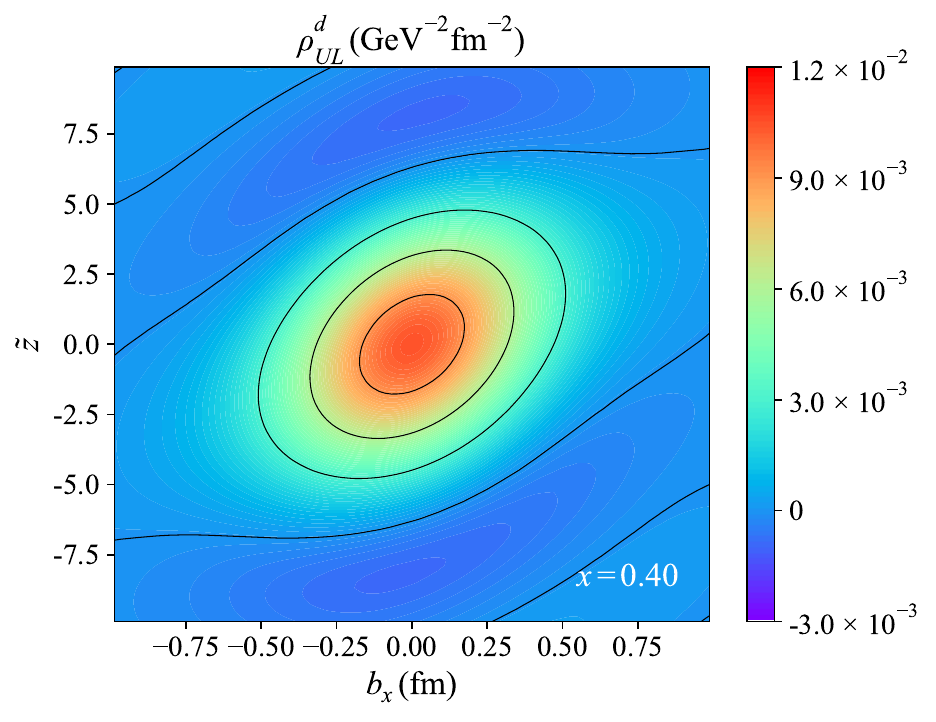}
	}
	\caption{Six-dimensional unpolarized-longitudinal light-front Wigner distribution $\rho_{\mathrm{UL}}\left(\tilde{z},x,\boldsymbol{b}_{\perp}, \boldsymbol{k}_{\perp}\right)$ for $u$ quark (upper panels) and $d$ quark (lower panels). The figure presents the Wigner distribution in the $\tilde{z}-b_x$ plane, with the transverse momentum fixed at $\boldsymbol{k}_{\perp}=0.3\,\mathrm{GeV}\boldsymbol{\hat{e}}_x$ (where $\boldsymbol{\hat{e}}_x$ is the unit vector in the $x$-direction) and the transverse coordinate component fixed at $b_y=0.4\,\mathrm{GeV}^{-1}$. The three columns correspond to $x=0.10$, $x=0.25$, and $x=0.40$.}
	\label{6DProtonULudzbx}
\end{figure}

\begin{figure}[htbp]
	\centering
	\subfloat{
		\includegraphics[width=0.31\textwidth]{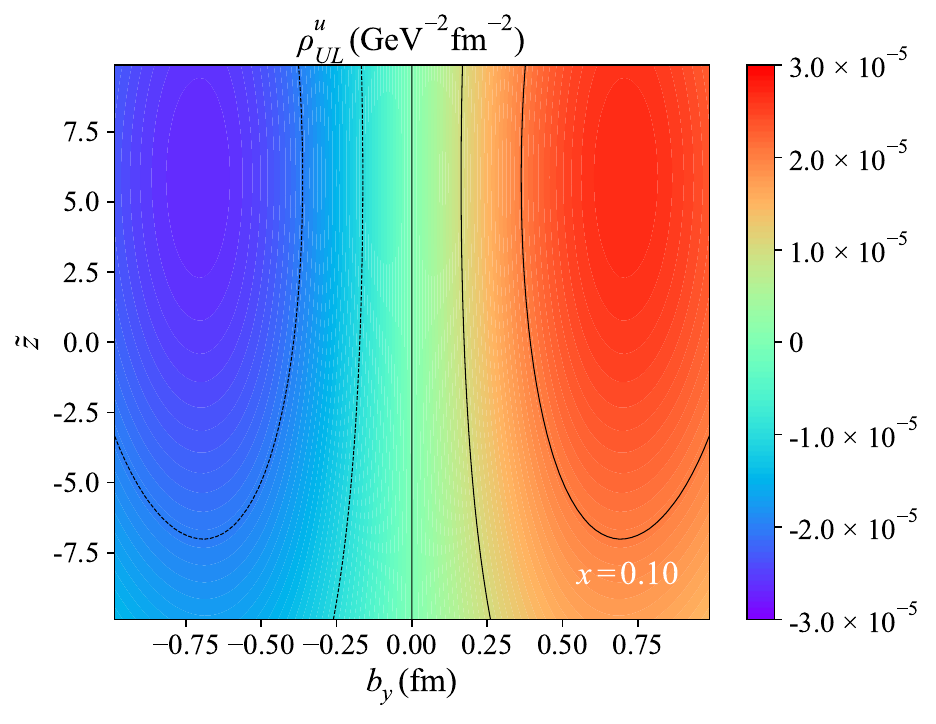}
	}
	\subfloat{
		\includegraphics[width=0.31\textwidth]{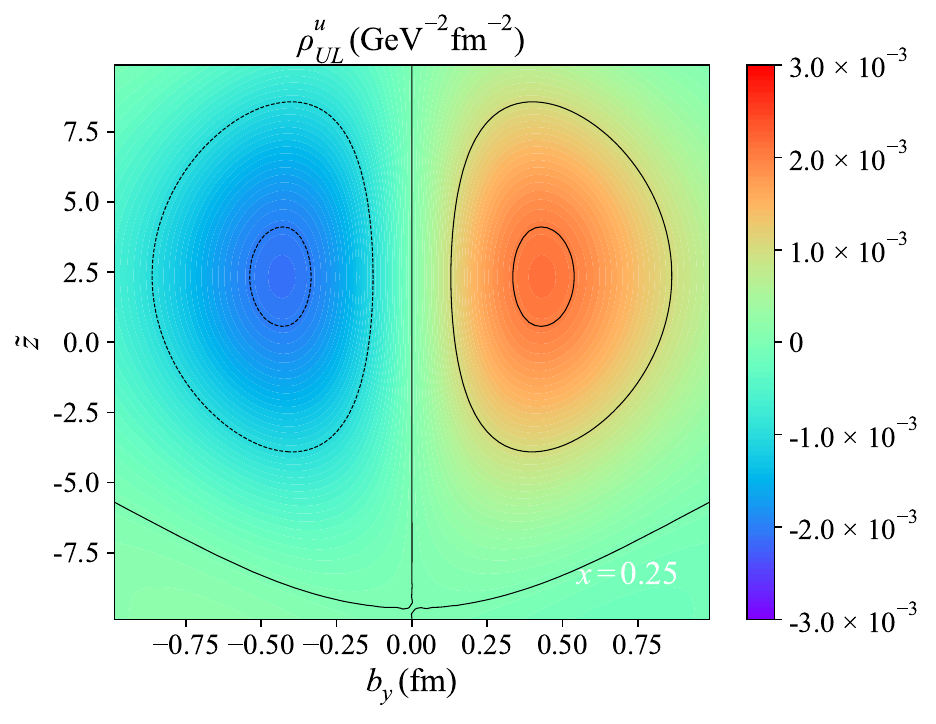}
	}
	\subfloat{
		\includegraphics[width=0.31\textwidth]{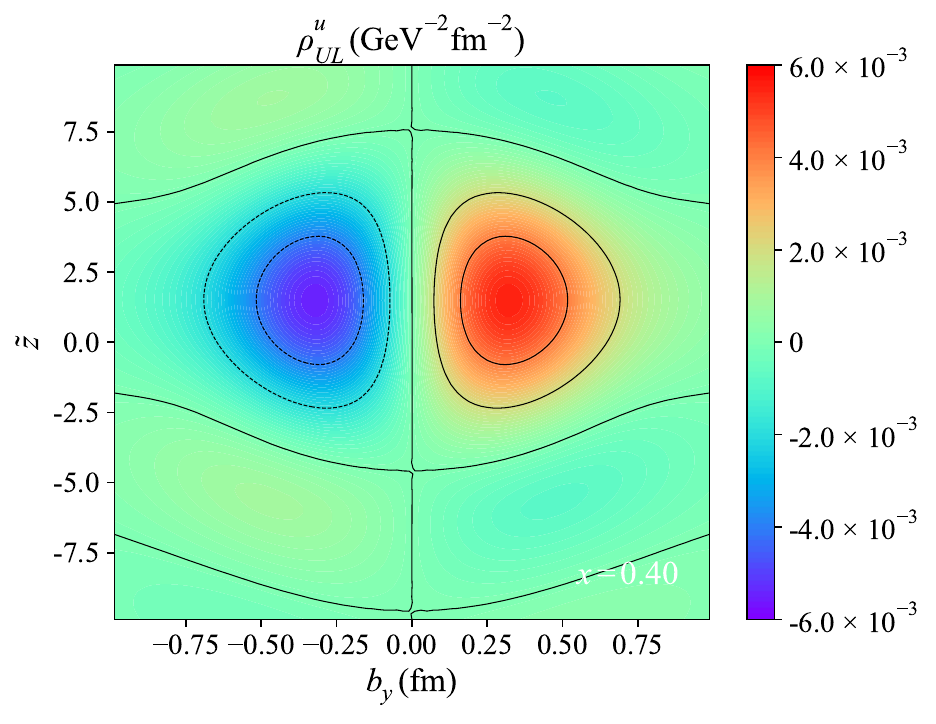}
	}\\
	\subfloat{
		\includegraphics[width=0.31\textwidth]{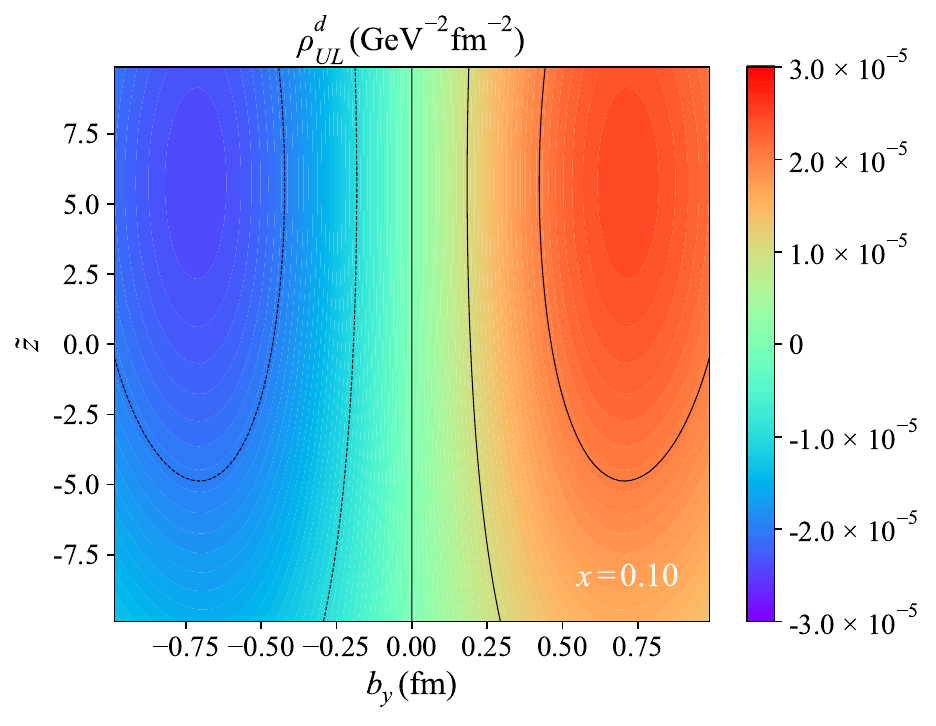}
	}
	\subfloat{
		\includegraphics[width=0.31\textwidth]{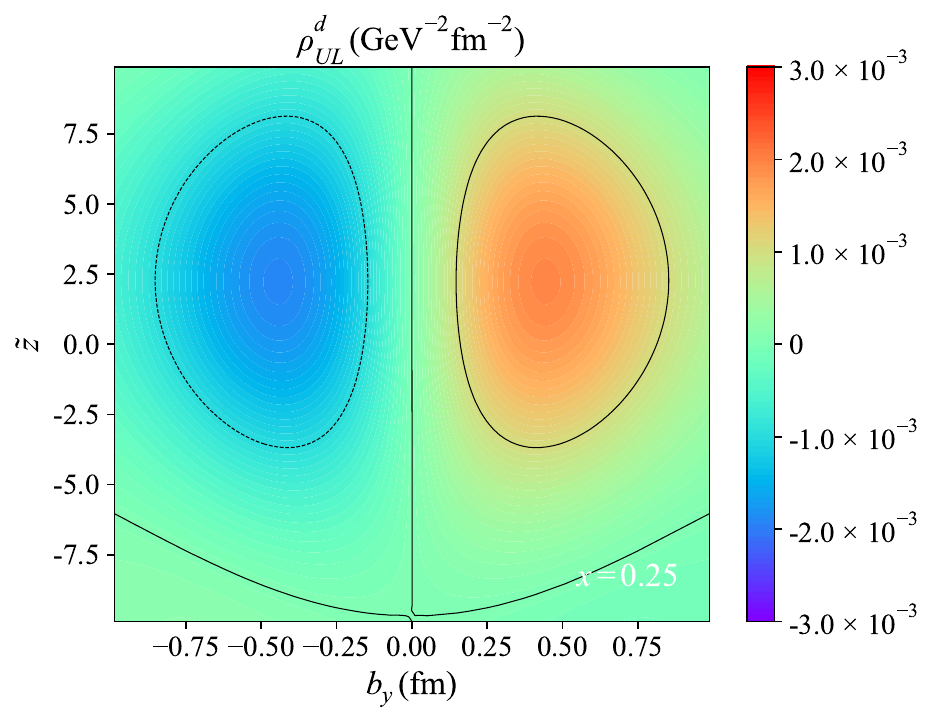}
	}
	\subfloat{
		\includegraphics[width=0.31\textwidth]{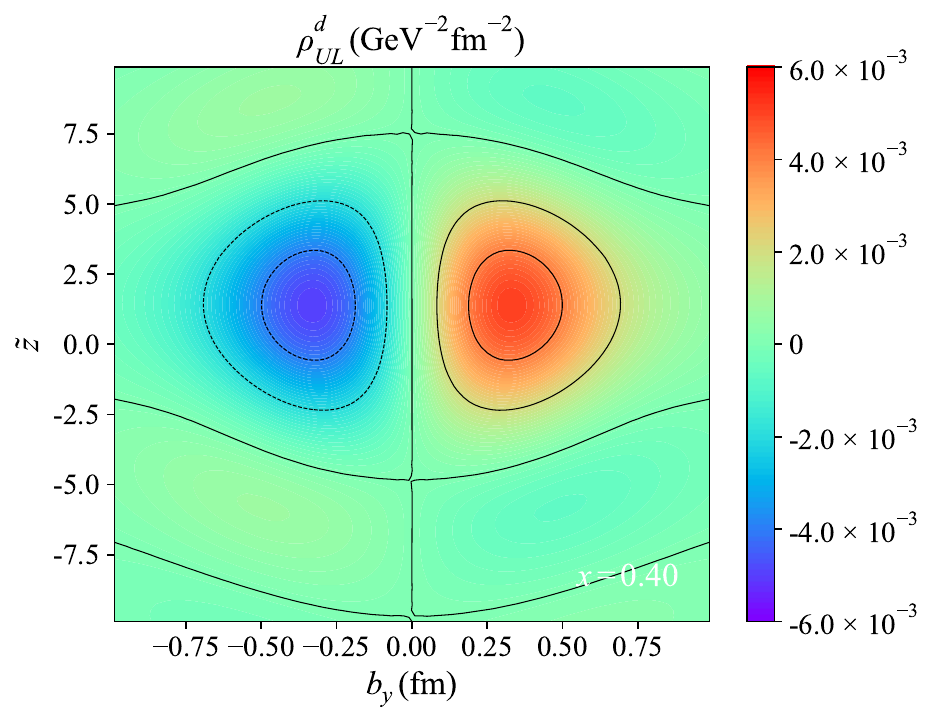}
	}
	\caption{Six-dimensional unpolarized-longitudinal light-front Wigner distribution $\rho_{\mathrm{UL}}\left(\tilde{z},x,\boldsymbol{b}_{\perp}, \boldsymbol{k}_{\perp}\right)$ for $u$ quark (upper panels) and $d$ quark (lower panels). The figure presents the Wigner distributions in the $\tilde{z}-b_y$ plane, with the transverse momentum fixed at $\boldsymbol{k}_{\perp}=0.3\,\mathrm{GeV}\boldsymbol{\hat{e}}_x$ (where $\boldsymbol{\hat{e}}_x$ is the unit vector in the $x$-direction) and the transverse coordinate component fixed at $b_x=0.4\,\mathrm{GeV}^{-1}$. The three columns correspond to $x=0.10$, $x=0.25$, and $x=0.40$.}
	\label{6DProtonULudzby}
\end{figure}

\begin{figure}[htbp]
	\centering
	\subfloat{
		\includegraphics[width=0.31\textwidth]{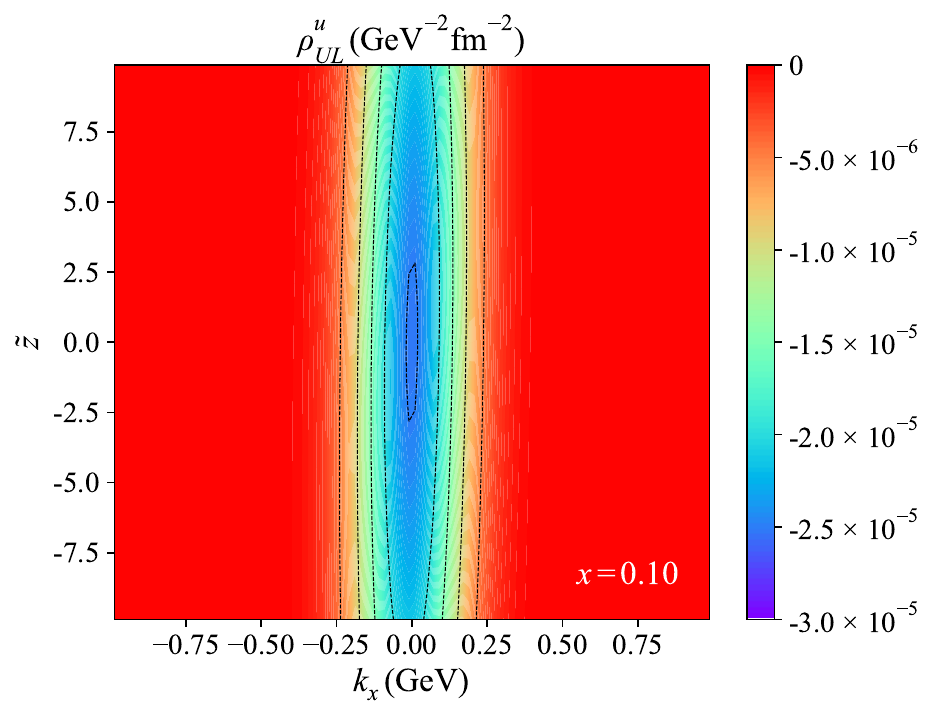}
	}
	\subfloat{
		\includegraphics[width=0.31\textwidth]{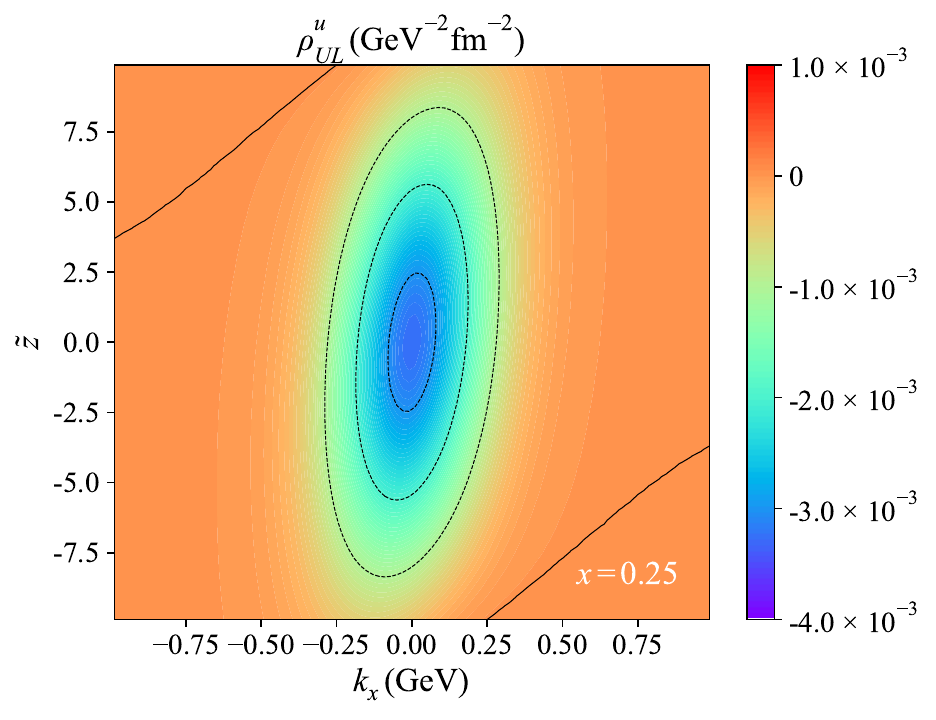}
	}
	\subfloat{
		\includegraphics[width=0.31\textwidth]{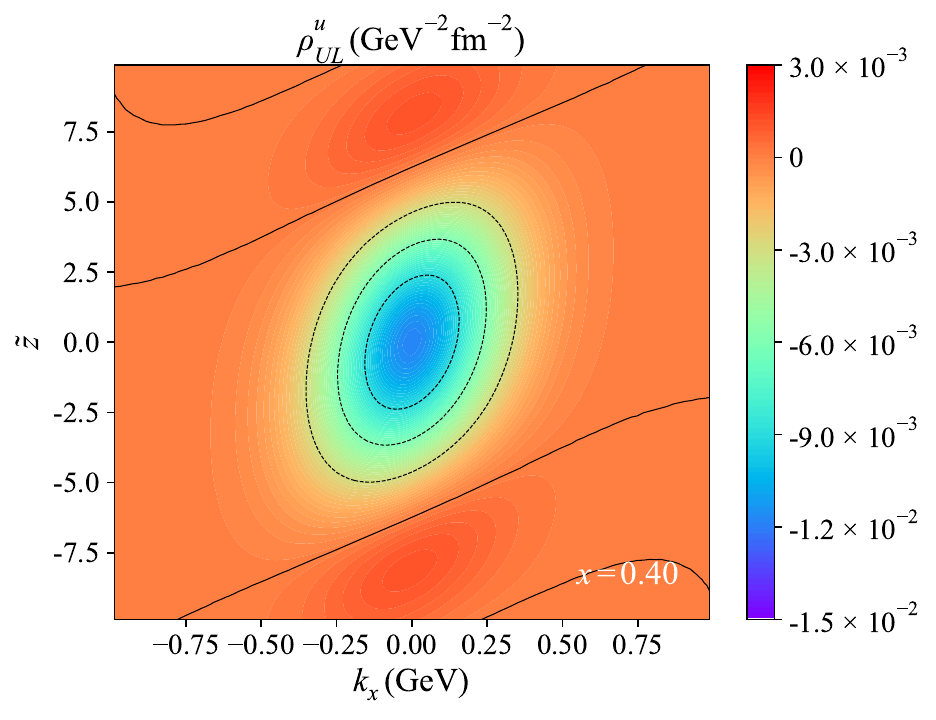}
	}\\
	\subfloat{
		\includegraphics[width=0.31\textwidth]{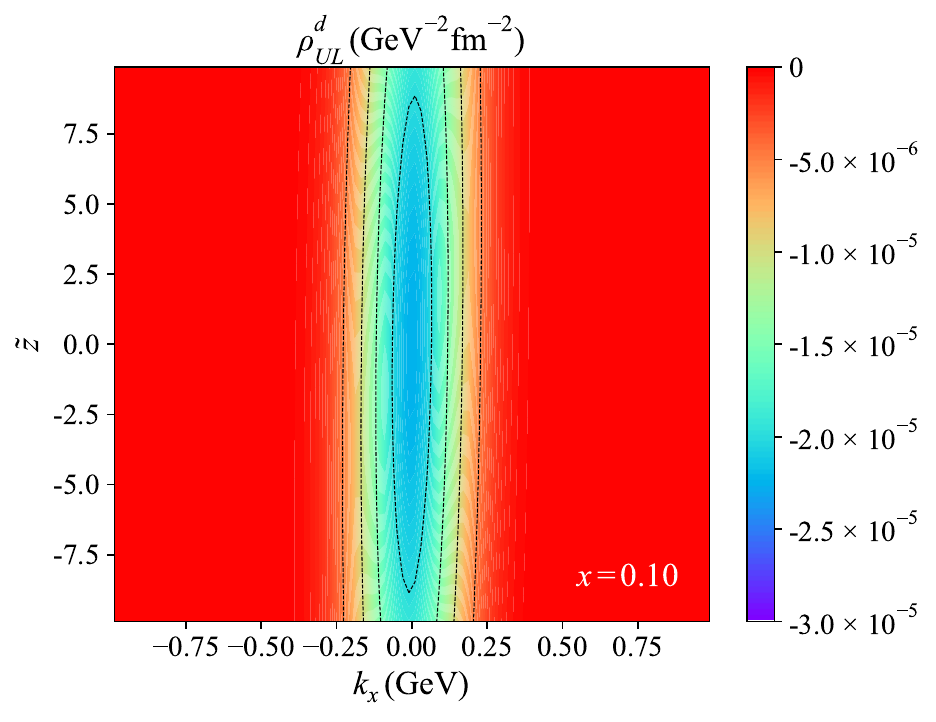}
	}
	\subfloat{
		\includegraphics[width=0.31\textwidth]{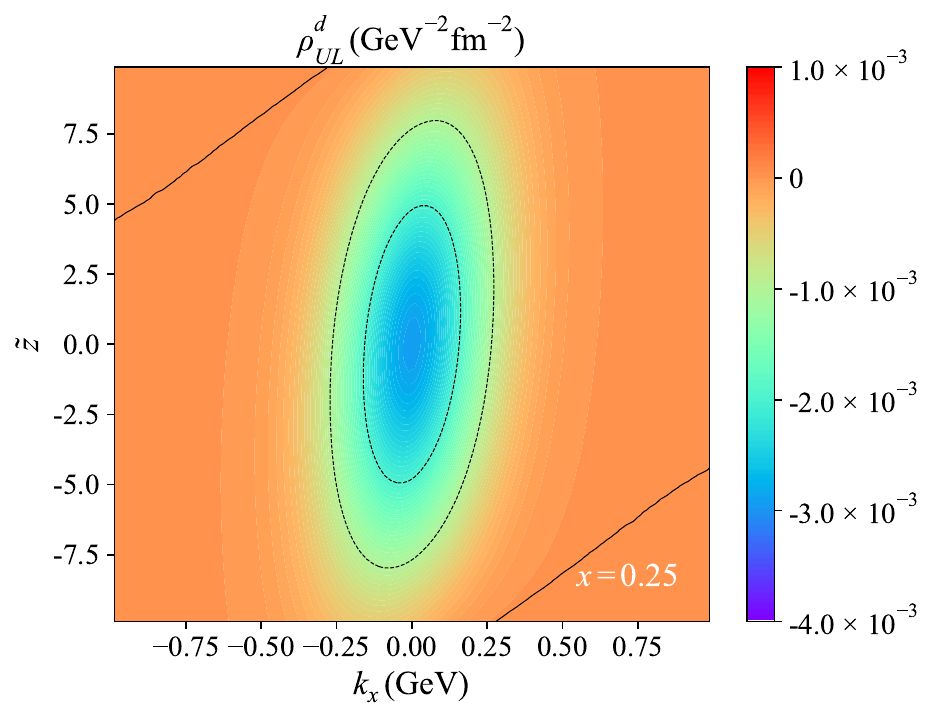}
	}
	\subfloat{
		\includegraphics[width=0.31\textwidth]{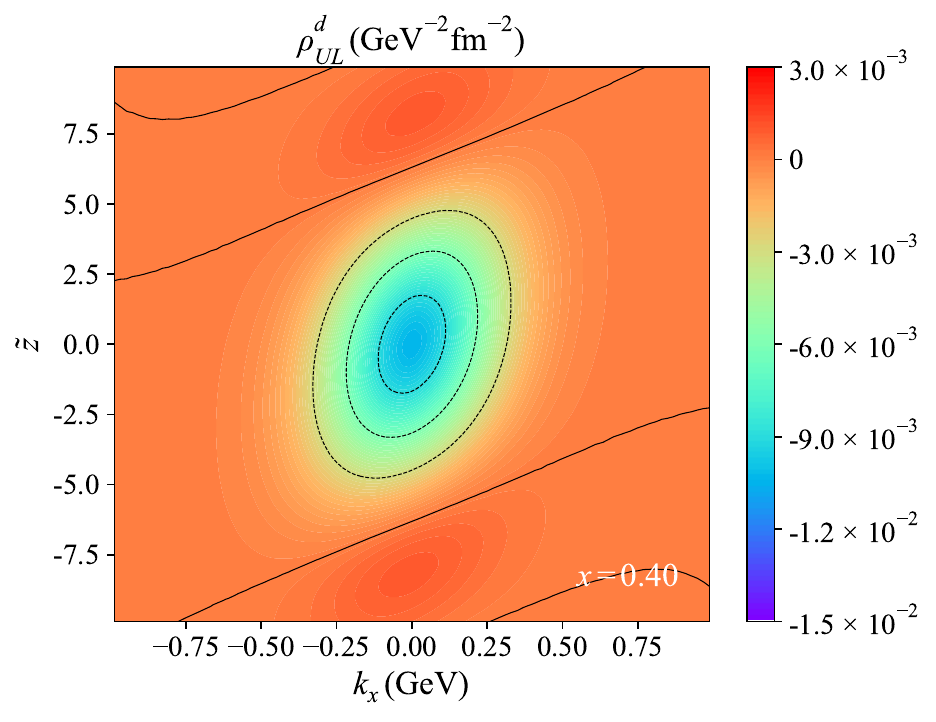}
	}
	\caption{Six-dimensional unpolarized-longitudinal light-front Wigner distribution $\rho_{\mathrm{UL}}\left(\tilde{z},x,\boldsymbol{b}_{\perp}, \boldsymbol{k}_{\perp}\right)$ for $u$ quark (upper panels) and $d$ quark (lower panels). The figure presents the Wigner distributions in the $\tilde{z}-k_x$ plane, with the transverse coordinate fixed at $\boldsymbol{b}_{\perp}=0.4\,\mathrm{GeV}^{-1}\boldsymbol{\hat{e}}_x$ (where $\boldsymbol{\hat{e}}_x$ is the unit vector along the $x$-axis) and the transverse momentum component fixed at $k_y=0.3\,\mathrm{GeV}$. The three columns correspond to $x=0.10$, $x=0.25$, and $x=0.40$.}
	\label{6DProtonULudzkx}
\end{figure}

\begin{figure}[htbp]
	\centering
	\subfloat{
		\includegraphics[width=0.31\textwidth]{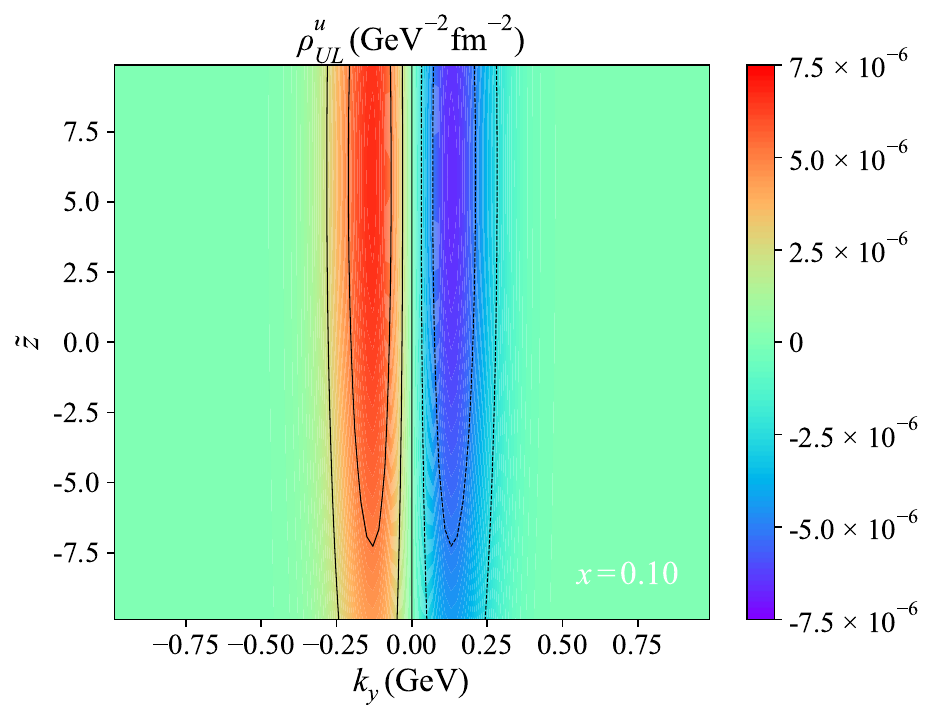}
	}
	\subfloat{
		\includegraphics[width=0.31\textwidth]{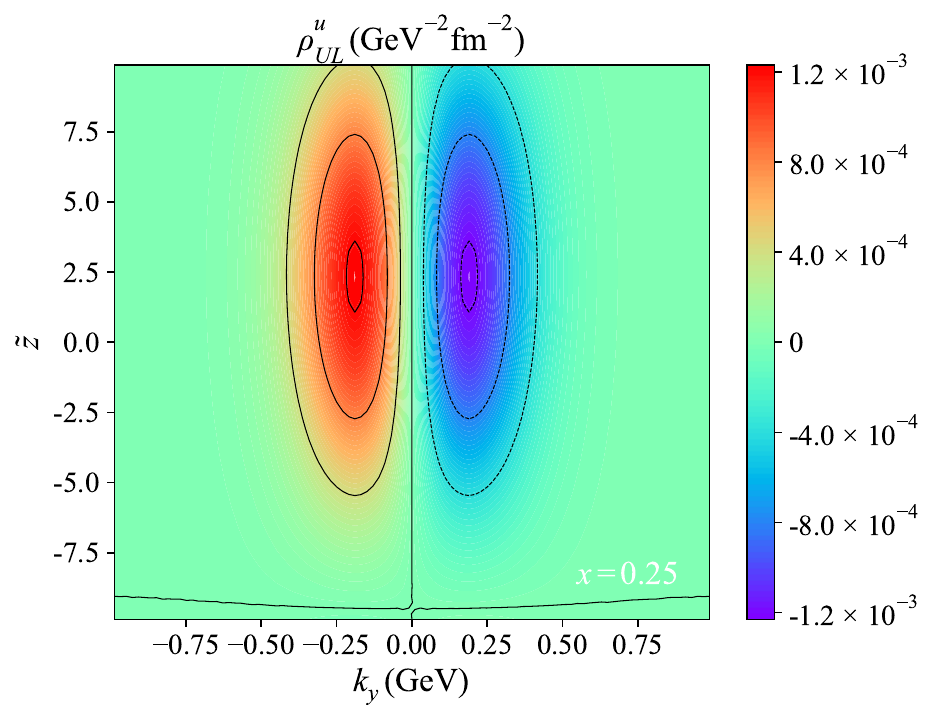}
	}
	\subfloat{
		\includegraphics[width=0.31\textwidth]{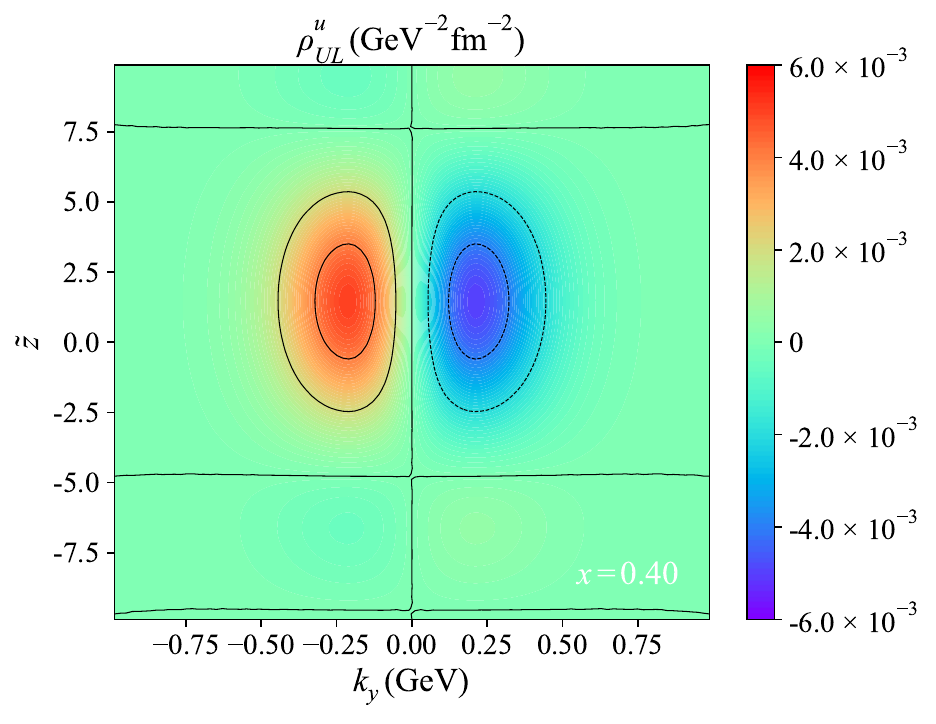}
	}\\
	\subfloat{
		\includegraphics[width=0.31\textwidth]{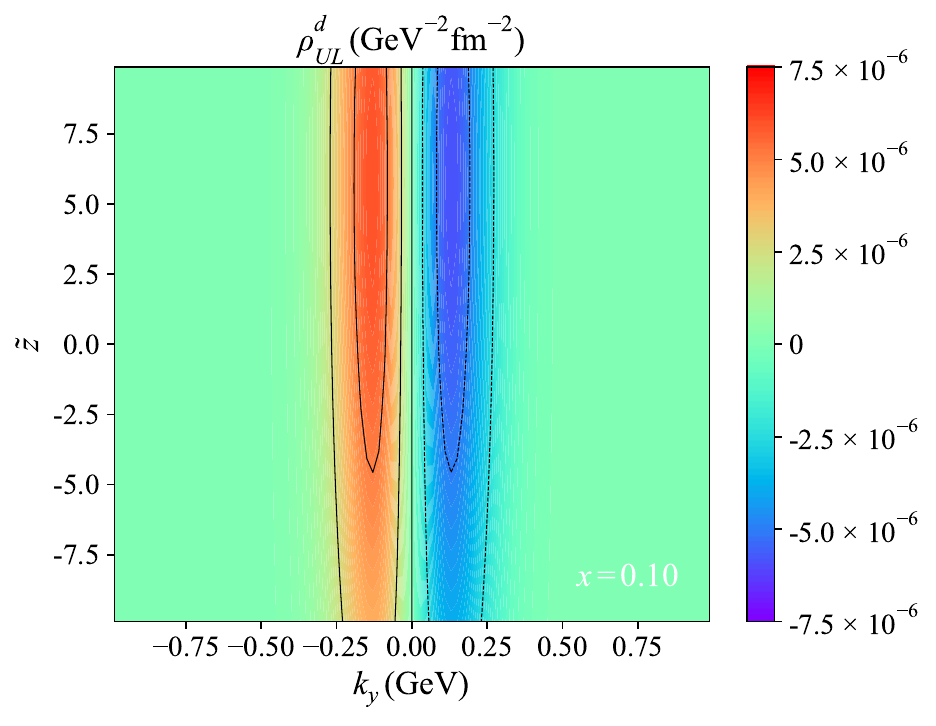}
	}
	\subfloat{
		\includegraphics[width=0.31\textwidth]{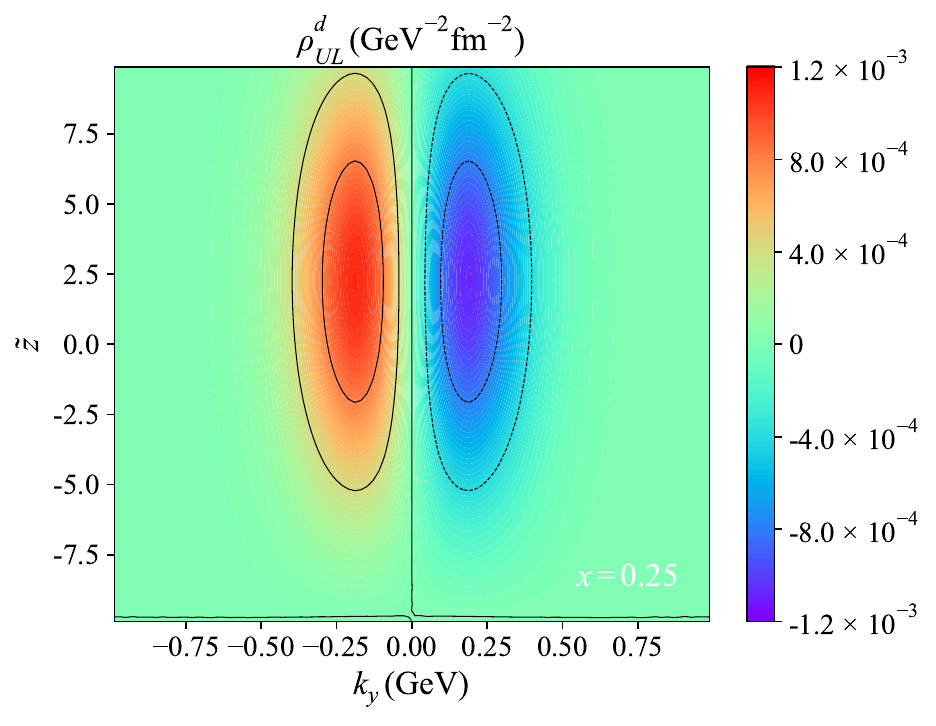}
	}
	\subfloat{
		\includegraphics[width=0.31\textwidth]{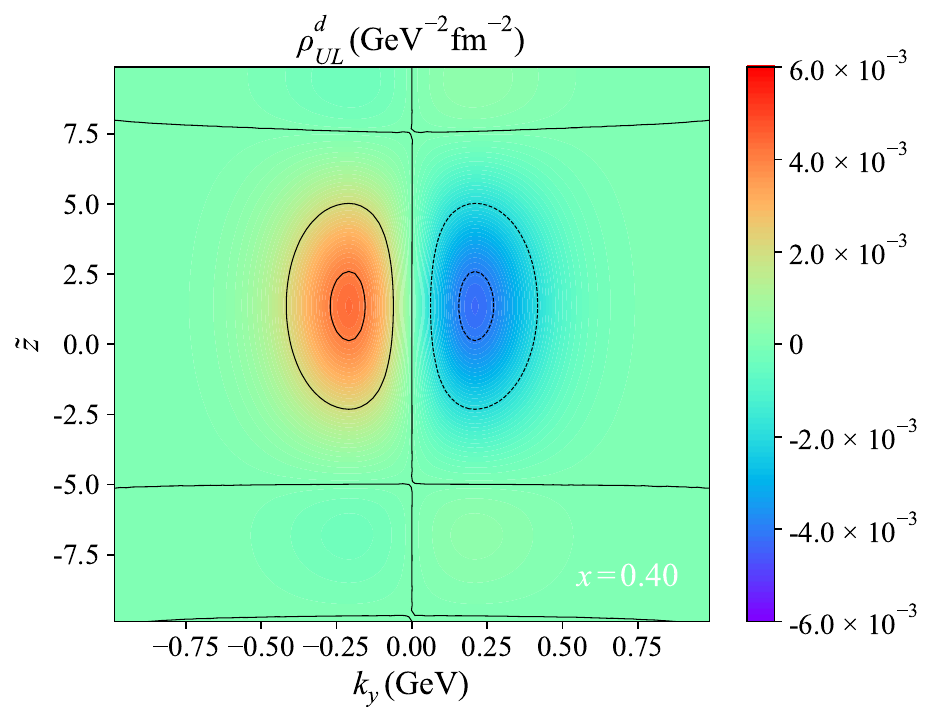}
	}
	\caption{Six-dimensional unpolarized-longitudinal light-front Wigner distribution $\rho_{\mathrm{UL}}\left(\tilde{z},x,\boldsymbol{b}_{\perp}, \boldsymbol{k}_{\perp}\right)$ for $u$ quark (upper panels) and $d$ quark (lower panels). The figure presents the Wigner distributions in the $\tilde{z}-k_y$ plane, with the transverse coordinate fixed at $\boldsymbol{b}_{\perp}=0.4\,\mathrm{GeV}^{-1}\boldsymbol{\hat{e}}_x$ (where $\boldsymbol{\hat{e}}_x$ is the unit vector along the $x$-axis) and the transverse momentum component fixed at $k_x=0.3\,\mathrm{GeV}$. The three columns correspond to $x=0.10$, $x=0.25$, and $x=0.40$.}
	\label{6DProtonULudzky}
\end{figure}

In Figs.~\ref{6DProtonULudzbx}--\ref{6DProtonULudzky}, %, Fig.~\ref{6DProtonULudzby}, Fig.~\ref{6DProtonULudzkx}
we plot the six-dimensional unpolarized-longitudinal light-front Wigner distribution $\rho_{\mathrm{UL}}\left(\tilde{z},x,\boldsymbol{b}_{\perp}, \boldsymbol{k}_{\perp}\right)$ for the $u$ and $d$ quarks of the proton, displayed in the $\tilde{z}-b_x$, $\tilde{z}-b_y$, $\tilde{z}-k_x$, and $\tilde{z}-k_y$ subspaces, respectively. The six-dimensional unpolarized-longitudinal light-front Wigner distributions represent the correlation of the longitudinal-polarized quark in an unpolarized proton. The numerical results for the relationship between longitudinal coordinates and transverse coordinates or transverse momentum are shown for fixed values of the transverse momentum $\boldsymbol{k}_{\perp}$ or the transverse coordinate $\boldsymbol{b}_{\perp}$, and the longitudinal momentum fraction $x$ is set at $x = 0.10$, $x = 0.25$, and $x = 0.40$ in the first, second, and third columns, respectively.

\new{In Fig.~\ref{6DProtonULudzbx} and Fig.~\ref{6DProtonULudzkx}, the six-dimensional unpolarized-longitudinal light-front Wigner distributions exhibit a characteristic extremum near $\tilde{z}=0$ about the origin in the $\tilde{z}-b_x$ and $\tilde{z}-k_x$ subspaces, reflecting the localization of quarks in longitudinal position space. This non-vanishing behavior at $\tilde{z}=0$ originates from the Melosh rotation effects and the specific form of the BHL wave function used in our model, consistent with general expectations in light-front quantization. The extremum shows distinct $x$-dependence, appearing as a broad peak in the sea quark dominated region $(x=0.10)$ and evolving into a sharper feature at higher $x$ values where valence quarks dominate $(x=0.40)$. By contrast, in Fig.~\ref{6DProtonULudzby} and Fig.~\ref{6DProtonULudzky}, the distributions take on a dipole-symmetric shape about $b_x = 0$ that becomes increasingly pronounced at lower momentum fractions. This pattern provides direct evidence of spin-orbit coupling, where the negative sign of the correlation indicates a counterintuitive orbital motion that opposes classical expectations - a phenomenon attributed to relativistic effects in the light-front formalism. The dipole strength diminishes gradually as $x$ increases from $0.10$ to $0.40$, reflecting the transition from sea quark to valence quark dominated regimes.}

Upon integrating over the longitudinal coordinate $\tilde{z}$, the six-dimensional unpolarized light-front Wigner distributions manifest a dipole structure in both the transverse coordinate space $\boldsymbol{b}_{\perp}$ and transverse momentum space $\boldsymbol{k}_{\perp}$ for each fixed $x$. Since the distribution functions integrate to zero in both the $\boldsymbol{b}_{\perp}$ and $\boldsymbol{k}_{\perp}$ spaces, they vanish at the TMD and IPD limit. As a result, this distribution cannot be directly associated with TMDs or IPDs at leading twist. Consequently, the information contained in this distribution is not observable at the leading twist level, although sub-leading twist effects may still provide useful insights. Therefore, extracting relevant information from this distribution experimentally at the leading twist is not feasible, but such information might be accessible through sub-leading twist effects.

The results depicted in these figures suggest a negative spin-orbit correlation within the light-front quark spectator-diquark model. This phenomenon arises from the spin structure of the quark and spectator, a feature commonly observed in such models. This behavior appears to be relatively independent of the specific choice of the momentum-space wave function or model parameters. However, it should be noted that this conclusion is model-dependent, and alternative models might yield different results. Therefore, further experimental verifications are required to confirm these findings. \new{While the current light-front model captures the essential qualitative features, several theoretical aspects warrant further investigation. In particular, the treatment of higher-twist effects and quark-gluon correlations at low $x$, along with a more rigorous incorporation of Melosh rotation effects, could refine the quantitative predictions.}

%\subsubsection{u Quark}

%\subsubsection{d Quark}

\subsection{Unpolarized-transverse Wigner distribution}

\begin{figure}[htbp]
	\centering
	\subfloat{
		\includegraphics[width=0.31\textwidth]{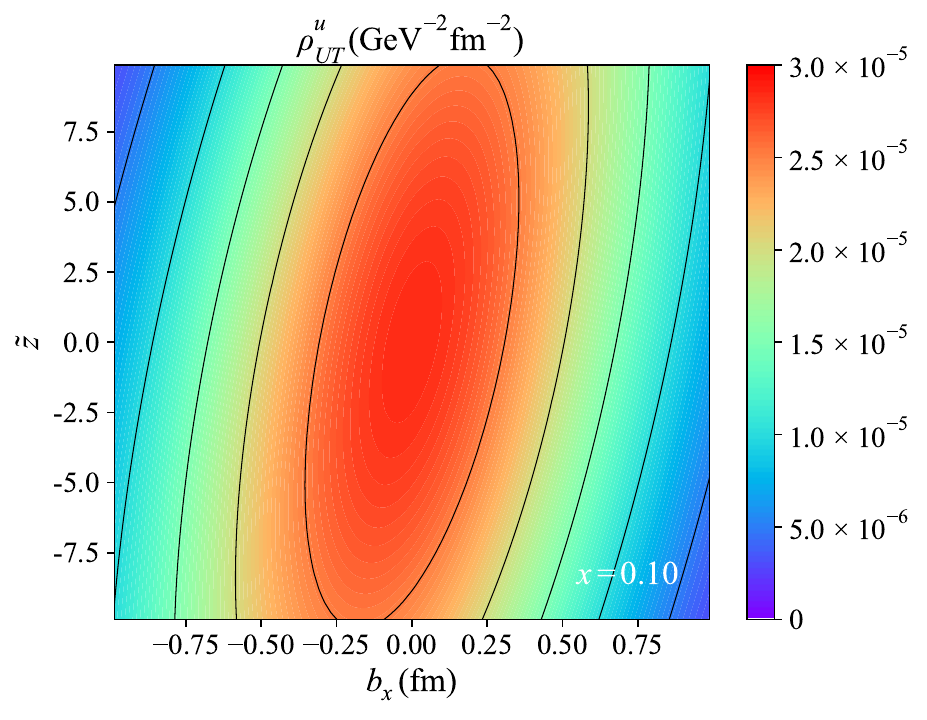}
	}
	\subfloat{
		\includegraphics[width=0.31\textwidth]{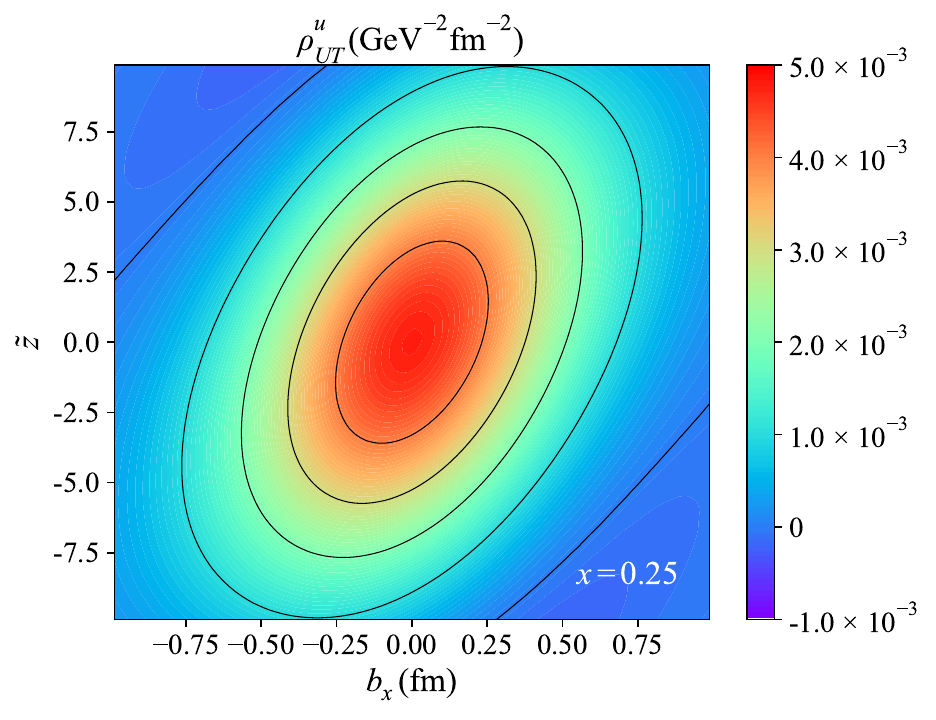}
	}
	\subfloat{
		\includegraphics[width=0.31\textwidth]{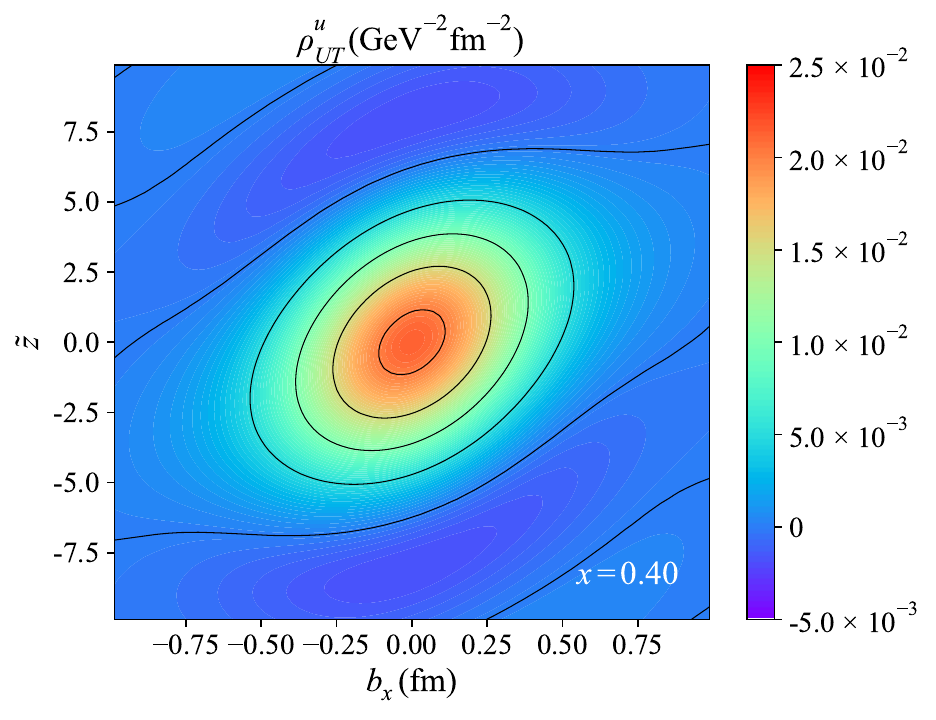}
	}\\
	\subfloat{
		\includegraphics[width=0.31\textwidth]{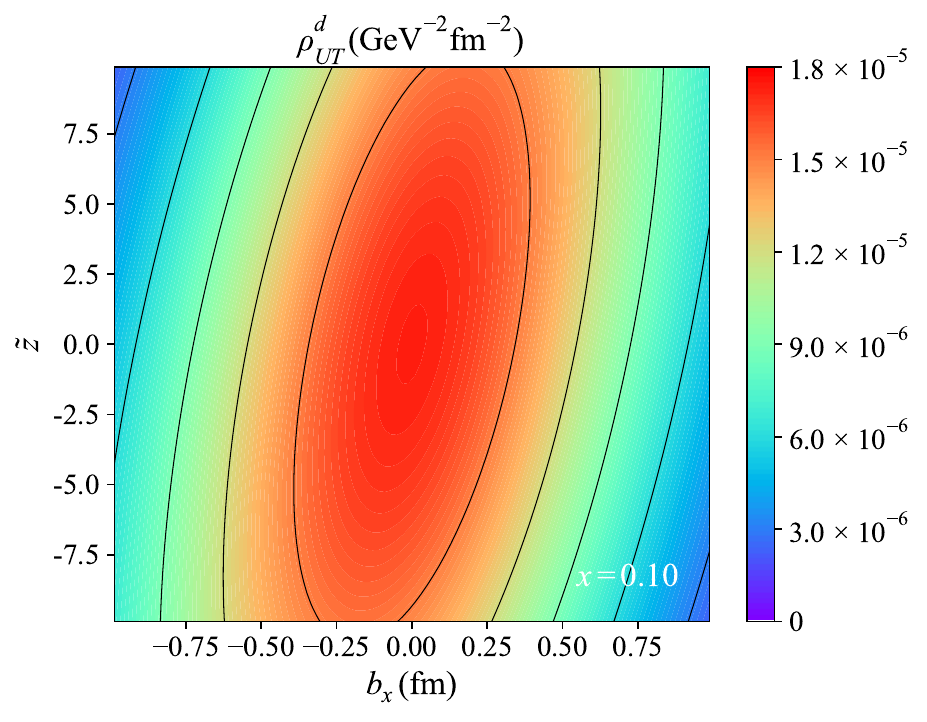}
	}
	\subfloat{
		\includegraphics[width=0.31\textwidth]{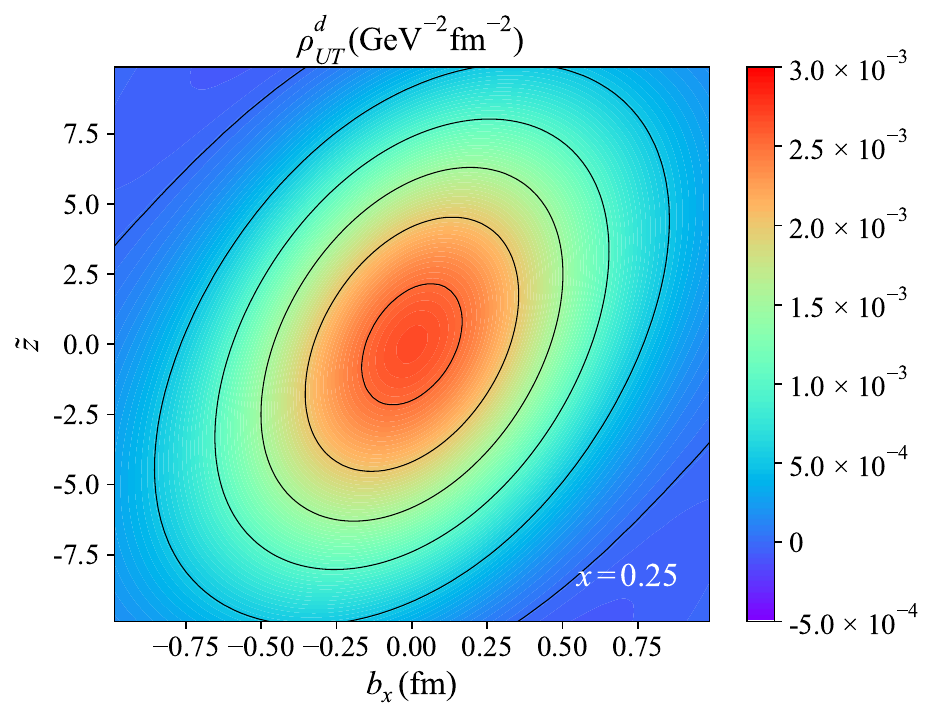}
	}
	\subfloat{
		\includegraphics[width=0.31\textwidth]{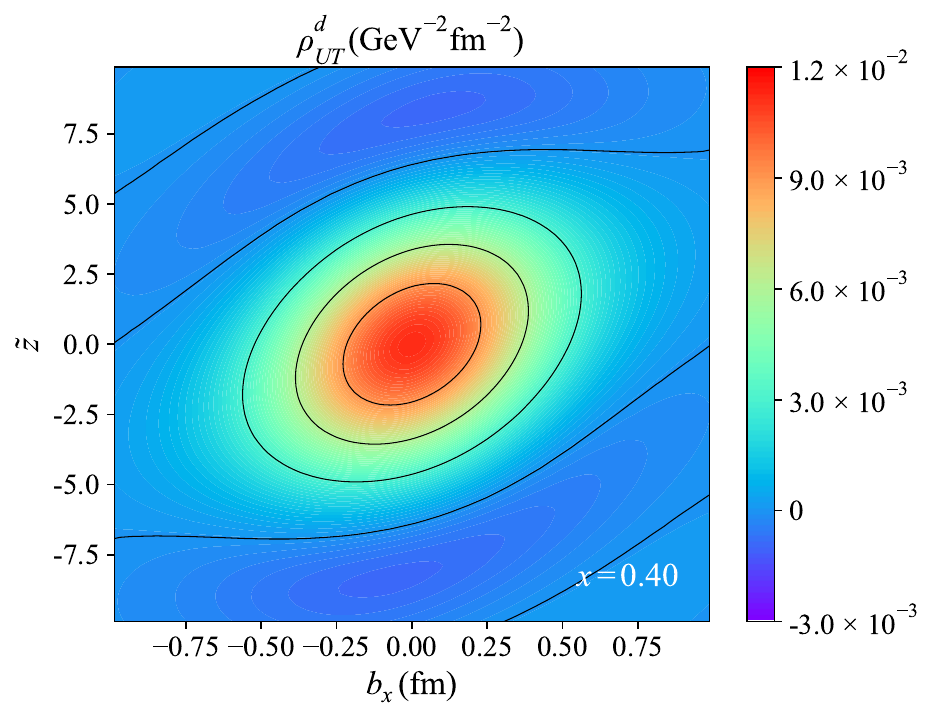}
	}
	\caption{Six-dimensional unpolarized-transverse light-front Wigner distribution $\rho_{\mathrm{UT}}\left(\tilde{z},x,\boldsymbol{b}_{\perp}, \boldsymbol{k}_{\perp}\right)$ for $u$ quark (upper panels) and $d$ quark (lower panels). The figure presents the Wigner distribution in the $\tilde{z}-b_x$ plane, with the transverse momentum fixed at $\boldsymbol{k}_{\perp}=0.3\,\mathrm{GeV}\boldsymbol{\hat{e}}_x$ (where $\boldsymbol{\hat{e}}_x$ is the unit vector in the $x$-direction) and the transverse coordinate component fixed at $b_y=0.4\,\mathrm{GeV}^{-1}$. The three columns correspond to $x=0.10$, $x=0.25$, and $x=0.40$.}
	\label{6DProtonUTudzbx}
\end{figure}

\begin{figure}[htbp]
	\centering
	\subfloat{
		\includegraphics[width=0.31\textwidth]{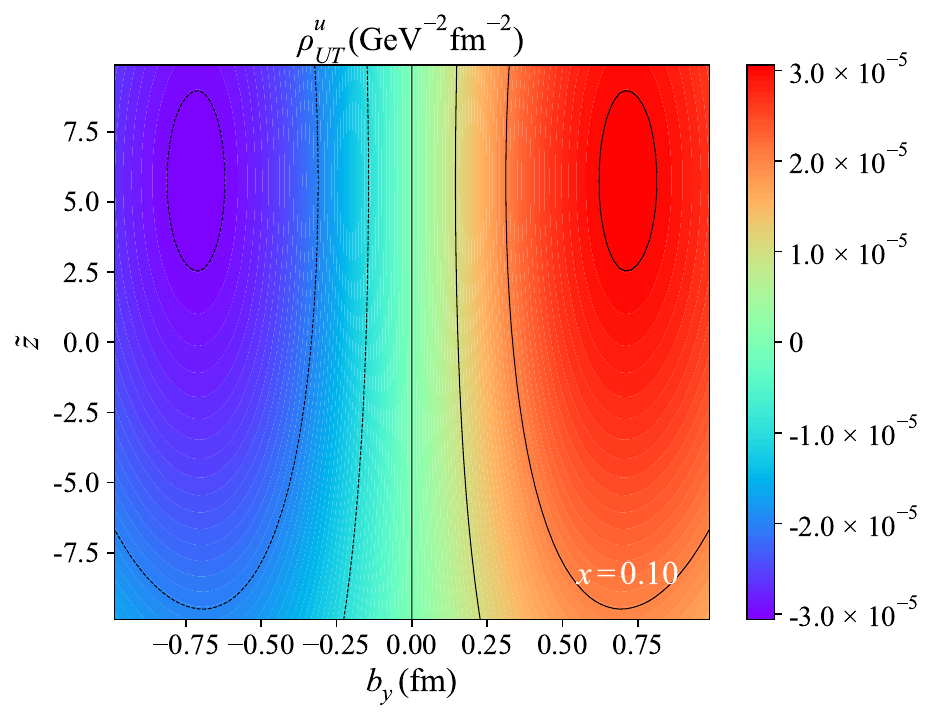}
	}
	\subfloat{
		\includegraphics[width=0.31\textwidth]{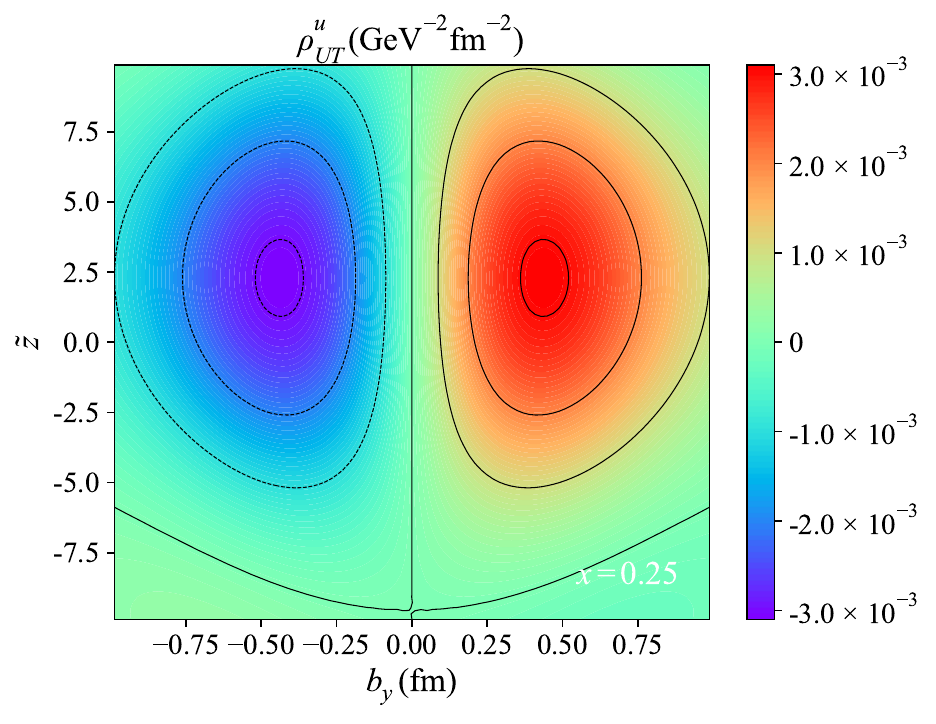}
	}
	\subfloat{
		\includegraphics[width=0.31\textwidth]{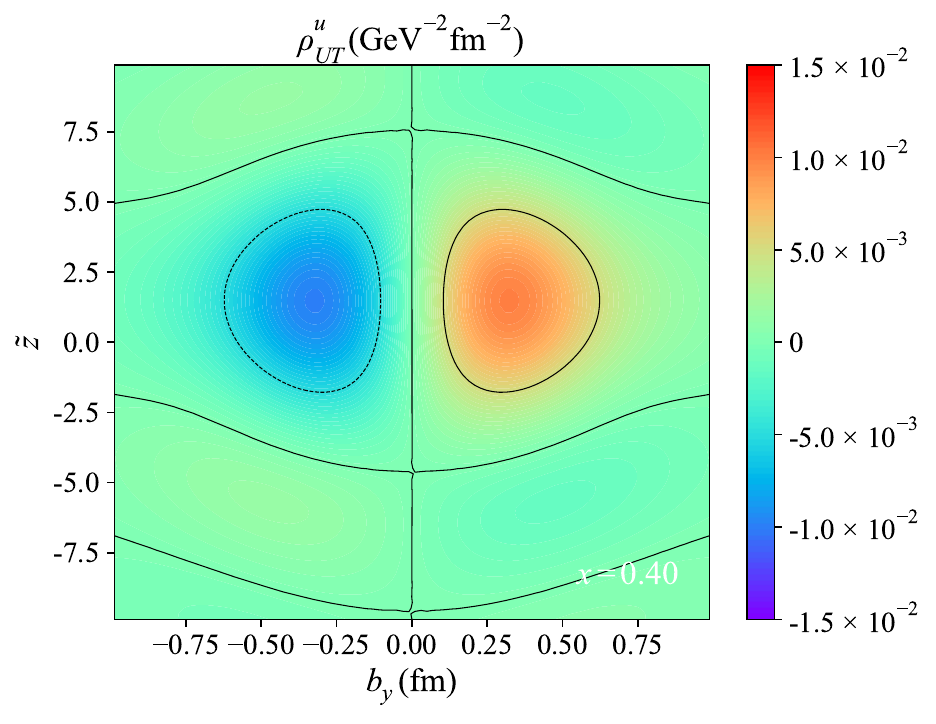}
	}\\
	\subfloat{
		\includegraphics[width=0.31\textwidth]{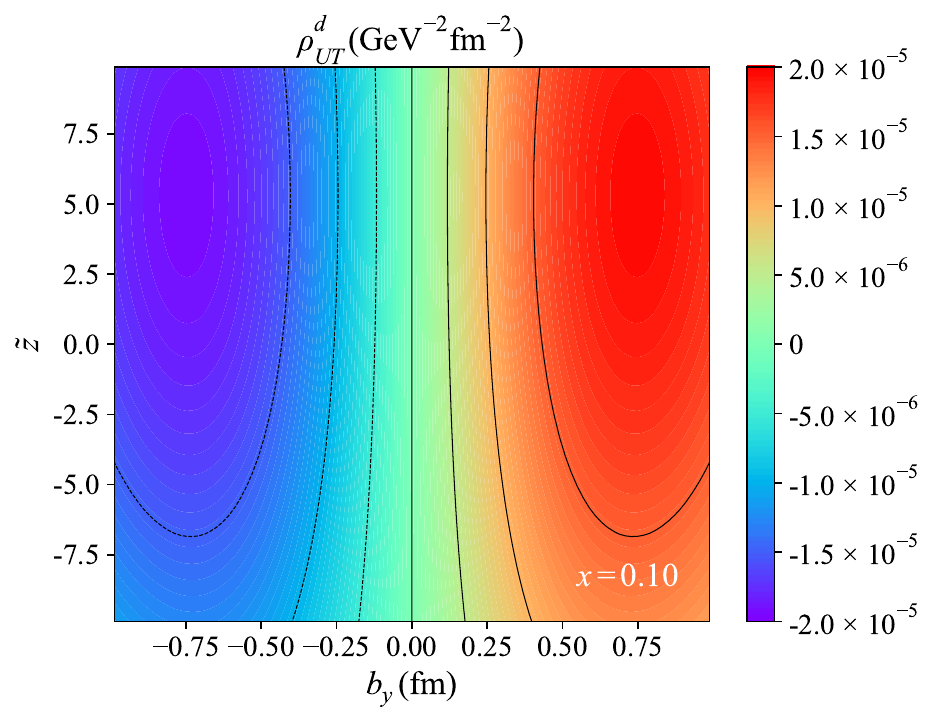}
	}
	\subfloat{
		\includegraphics[width=0.31\textwidth]{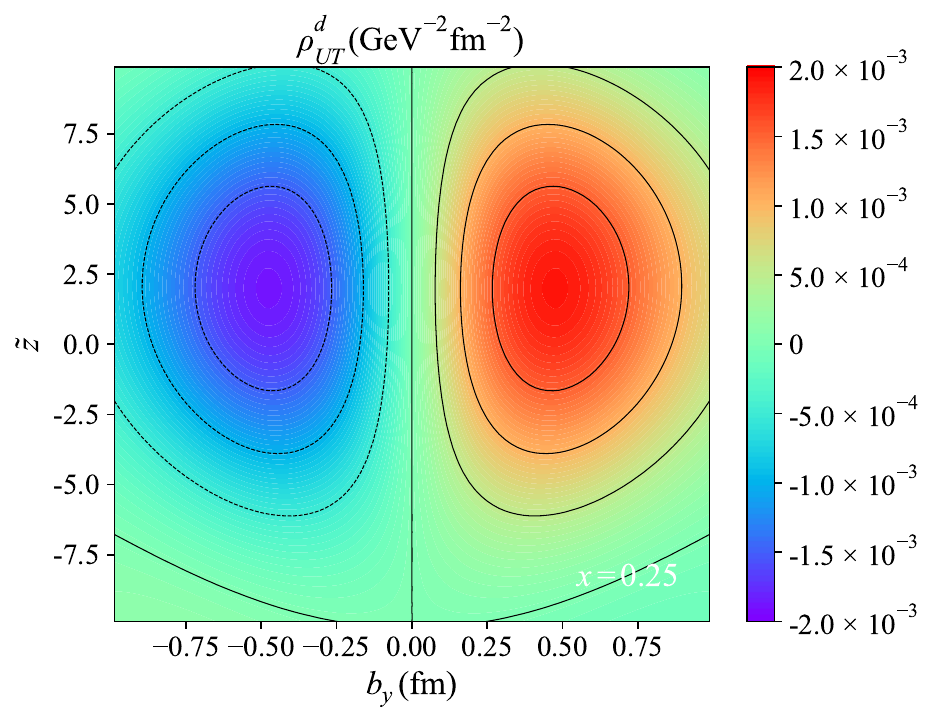}
	}
	\subfloat{
		\includegraphics[width=0.31\textwidth]{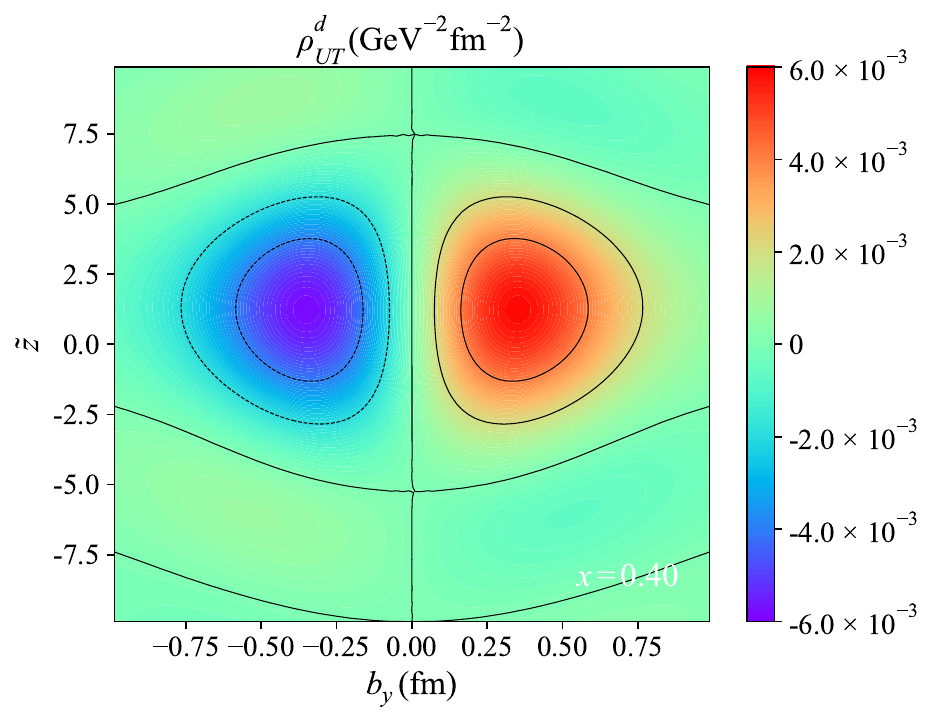}
	}
	\caption{Six-dimensional unpolarized-transverse light-front Wigner distribution $\rho_{\mathrm{UT}}\left(\tilde{z},x,\boldsymbol{b}_{\perp}, \boldsymbol{k}_{\perp}\right)$ for $u$ quark (upper panels) and $d$ quark (lower panels). The figure presents the Wigner distributions in the $\tilde{z}-b_y$ plane, with the transverse momentum fixed at $\boldsymbol{k}_{\perp}=0.3\,\mathrm{GeV}\boldsymbol{\hat{e}}_x$ (where $\boldsymbol{\hat{e}}_x$ is the unit vector in the $x$-direction) and the transverse coordinate component fixed at $b_x=0.4\,\mathrm{GeV}^{-1}$. The three columns correspond to $x=0.10$, $x=0.25$, and $x=0.40$.}
	\label{6DProtonUTudzby}
\end{figure}

\begin{figure}[htbp]
	\centering
	\subfloat{
		\includegraphics[width=0.31\textwidth]{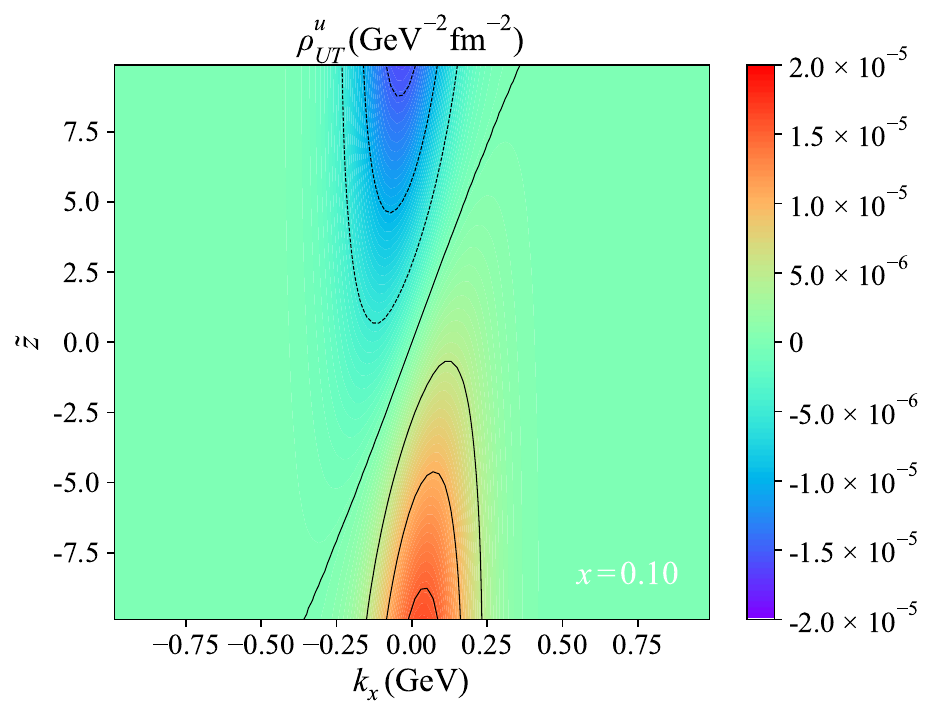}
	}
	\subfloat{
		\includegraphics[width=0.31\textwidth]{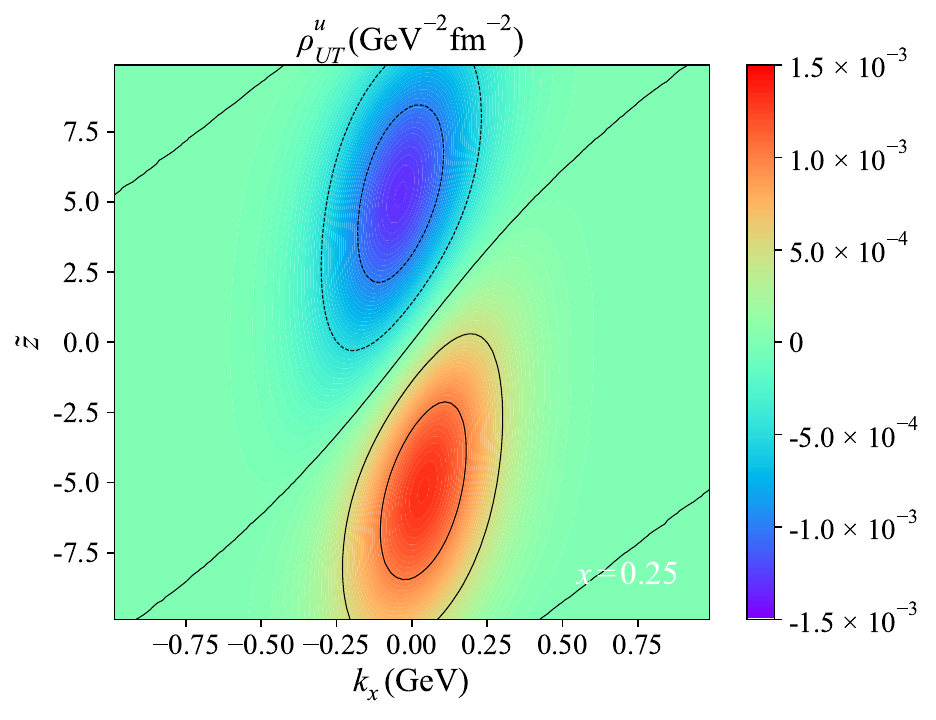}
	}
	\subfloat{
		\includegraphics[width=0.31\textwidth]{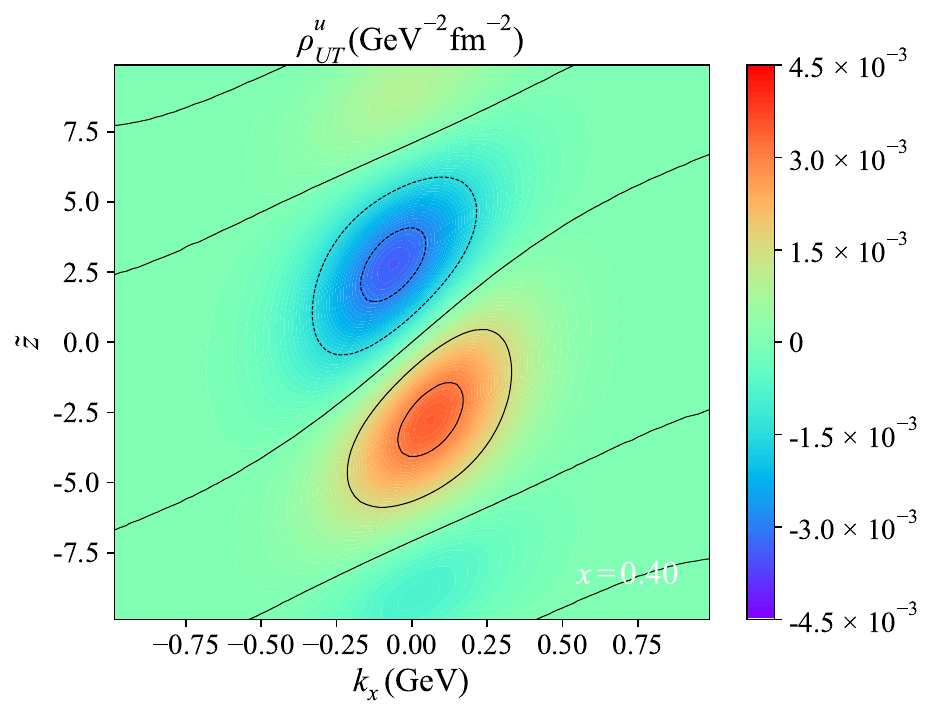}
	}\\
	\subfloat{
		\includegraphics[width=0.31\textwidth]{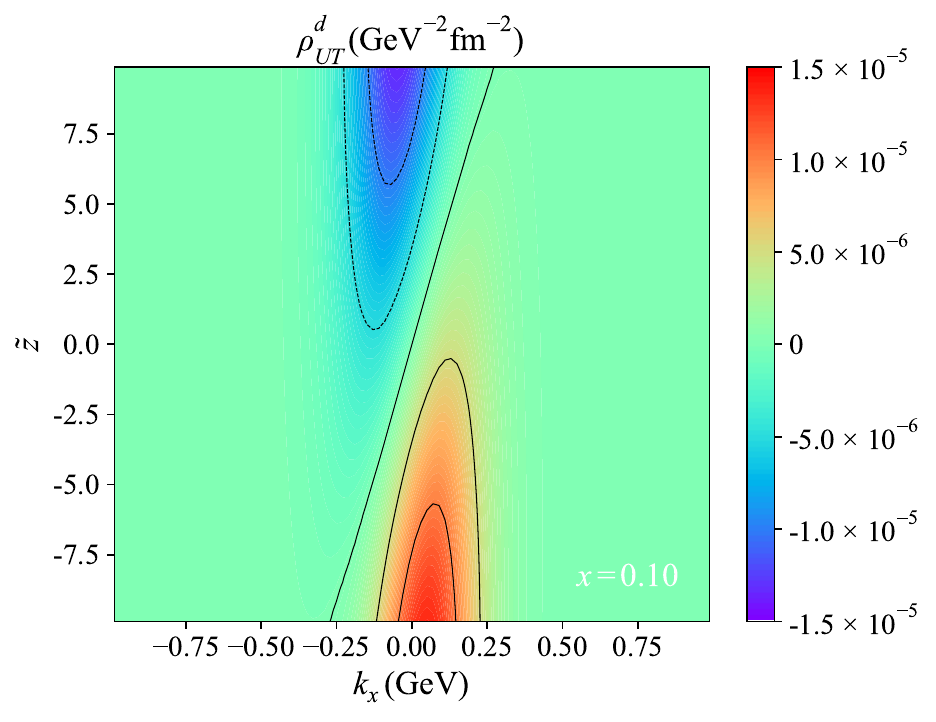}
	}
	\subfloat{
		\includegraphics[width=0.31\textwidth]{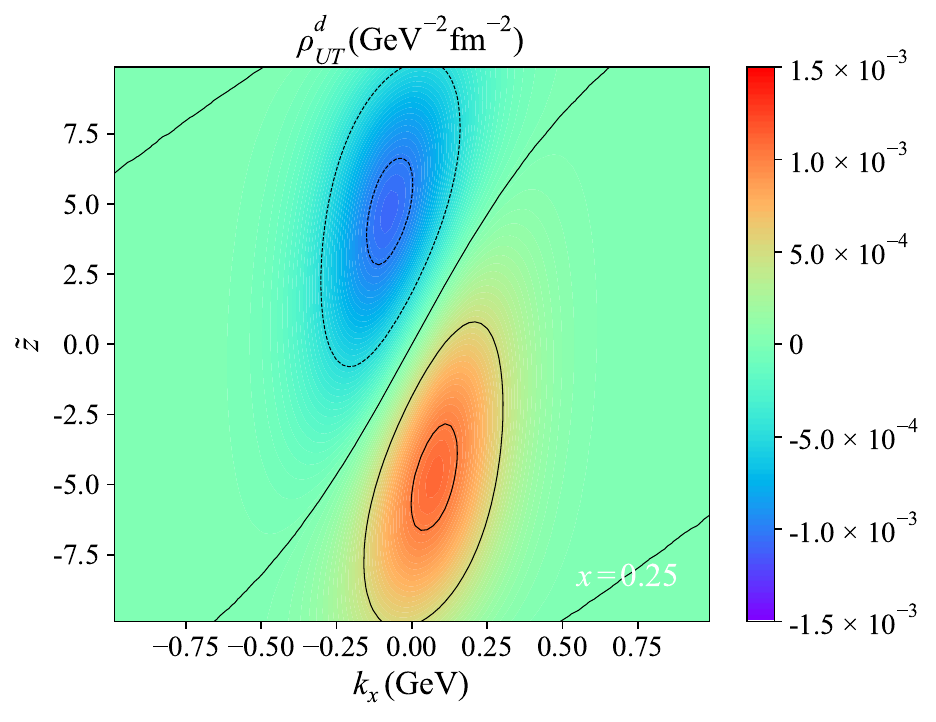}
	}
	\subfloat{
		\includegraphics[width=0.31\textwidth]{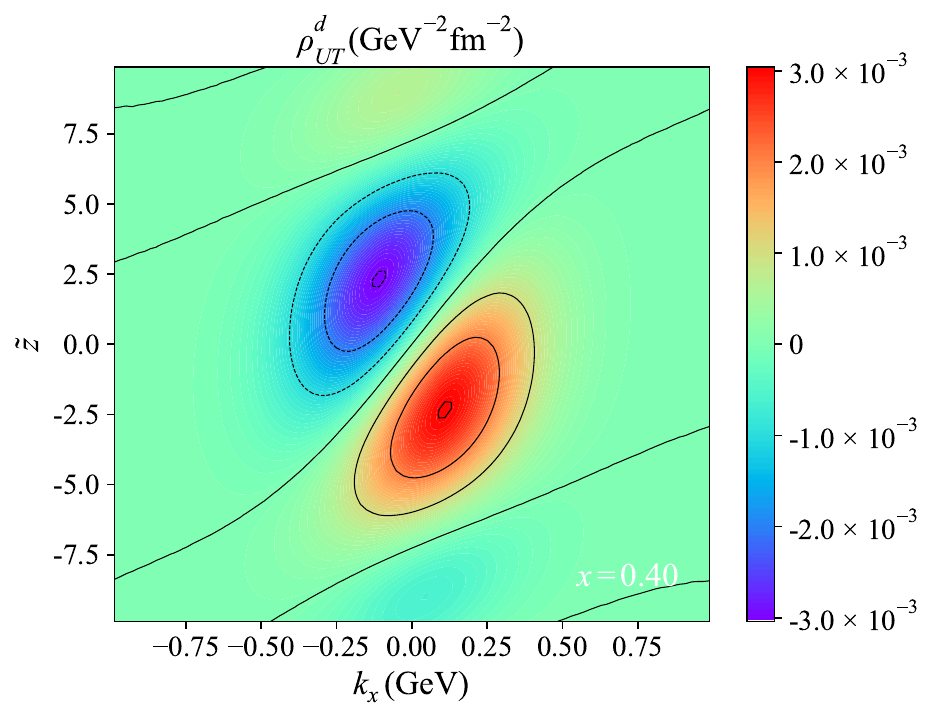}
	}
	\caption{Six-dimensional unpolarized-transverse light-front Wigner distribution $\rho_{\mathrm{UT}}\left(\tilde{z},x,\boldsymbol{b}_{\perp}, \boldsymbol{k}_{\perp}\right)$ for $u$ quark (upper panels) and $d$ quark (lower panels). The figure presents the Wigner distributions in the $\tilde{z}-k_x$ plane, with the transverse coordinate fixed at $\boldsymbol{b}_{\perp}=0.4\,\mathrm{GeV}^{-1}\boldsymbol{\hat{e}}_x$ (where $\boldsymbol{\hat{e}}_x$ is the unit vector along the $x$-axis) and the transverse momentum component fixed at $k_y=0.3\,\mathrm{GeV}$. The three columns correspond to $x=0.10$, $x=0.25$, and $x=0.40$.}
	\label{6DProtonUTudzkx}
\end{figure}

\begin{figure}[htbp]
	\centering
	\subfloat{
		\includegraphics[width=0.31\textwidth]{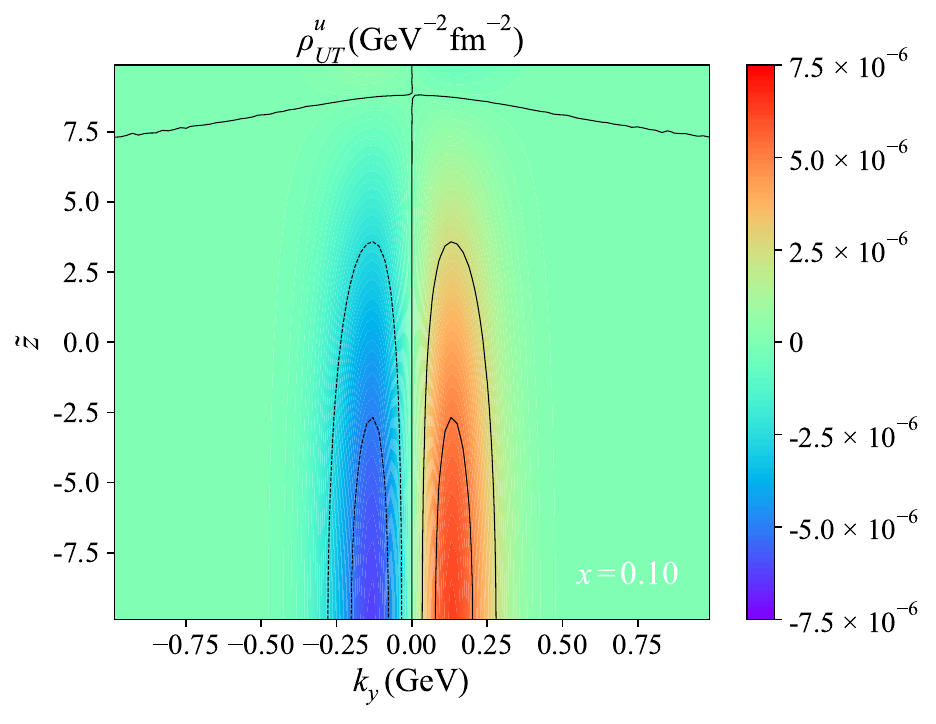}
	}
	\subfloat{
		\includegraphics[width=0.31\textwidth]{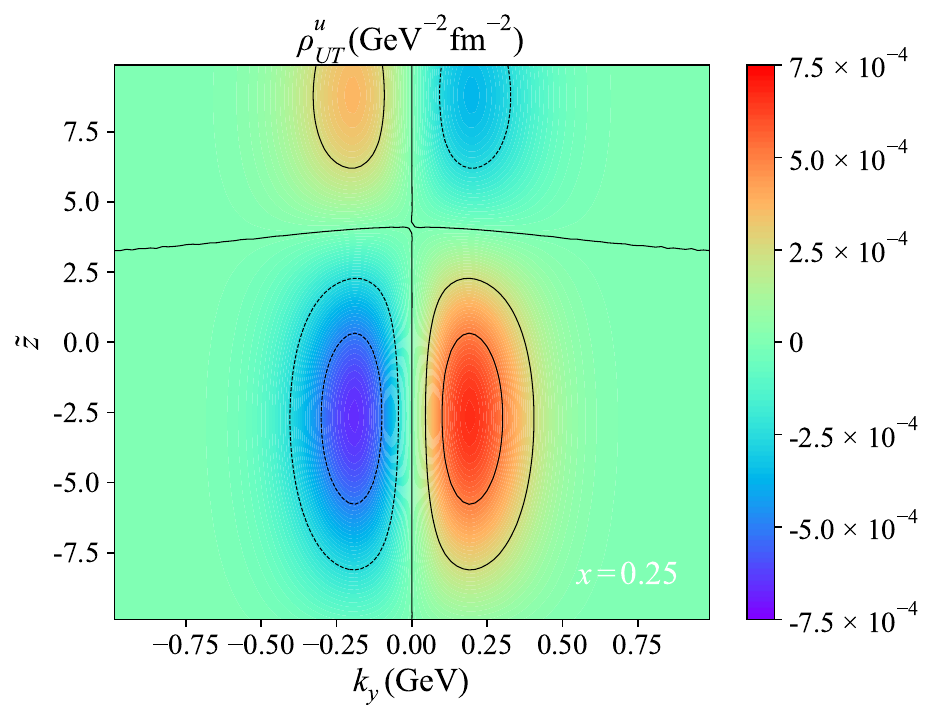}
	}
	\subfloat{
		\includegraphics[width=0.31\textwidth]{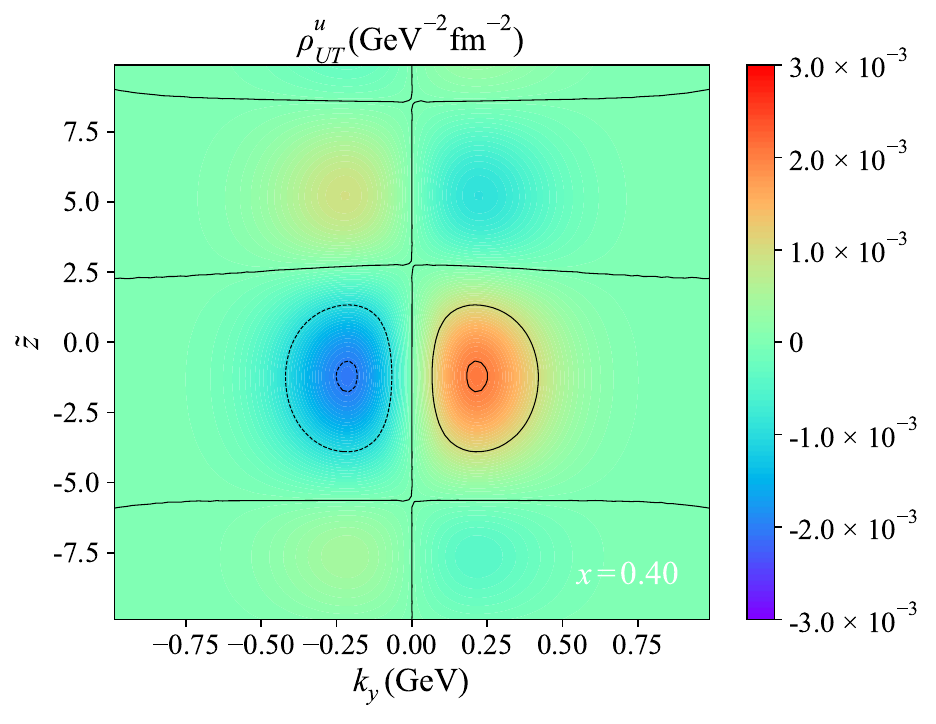}
	}\\
	\subfloat{
		\includegraphics[width=0.31\textwidth]{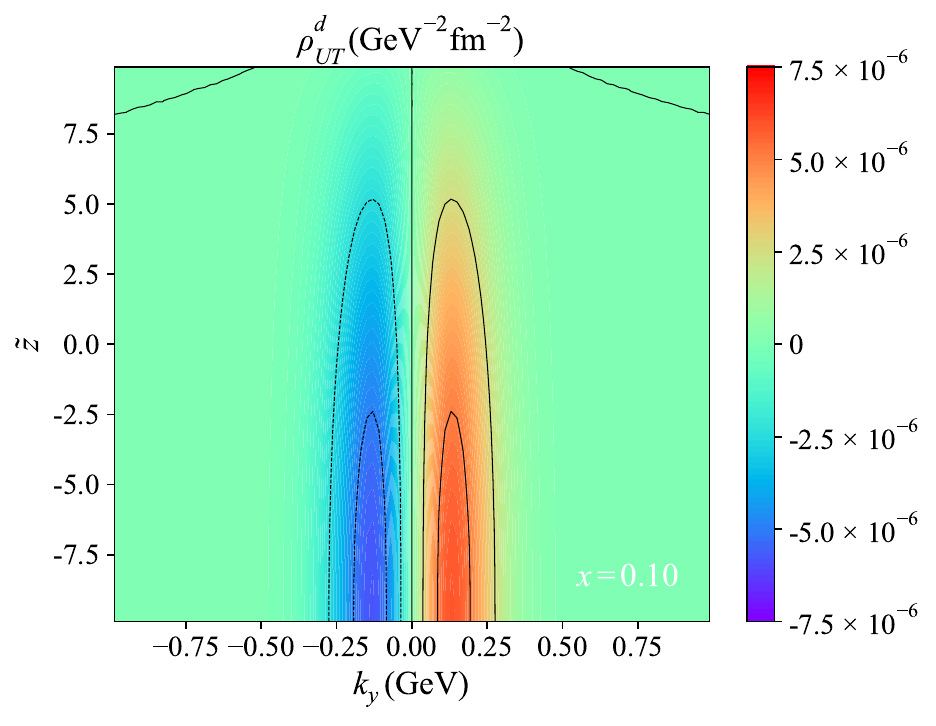}
	}
	\subfloat{
		\includegraphics[width=0.31\textwidth]{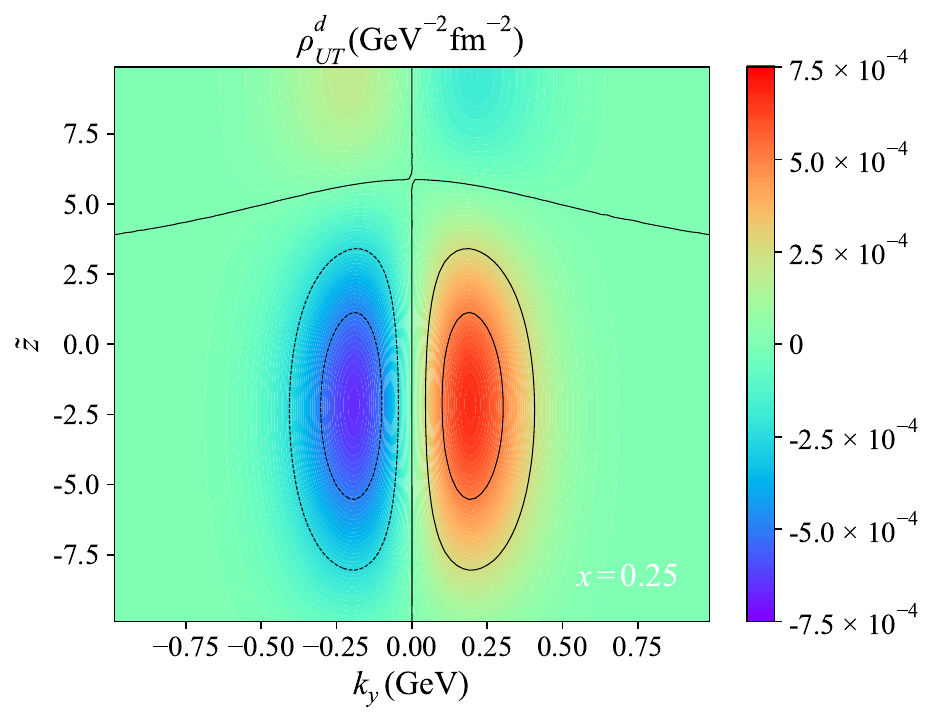}
	}
	\subfloat{
		\includegraphics[width=0.31\textwidth]{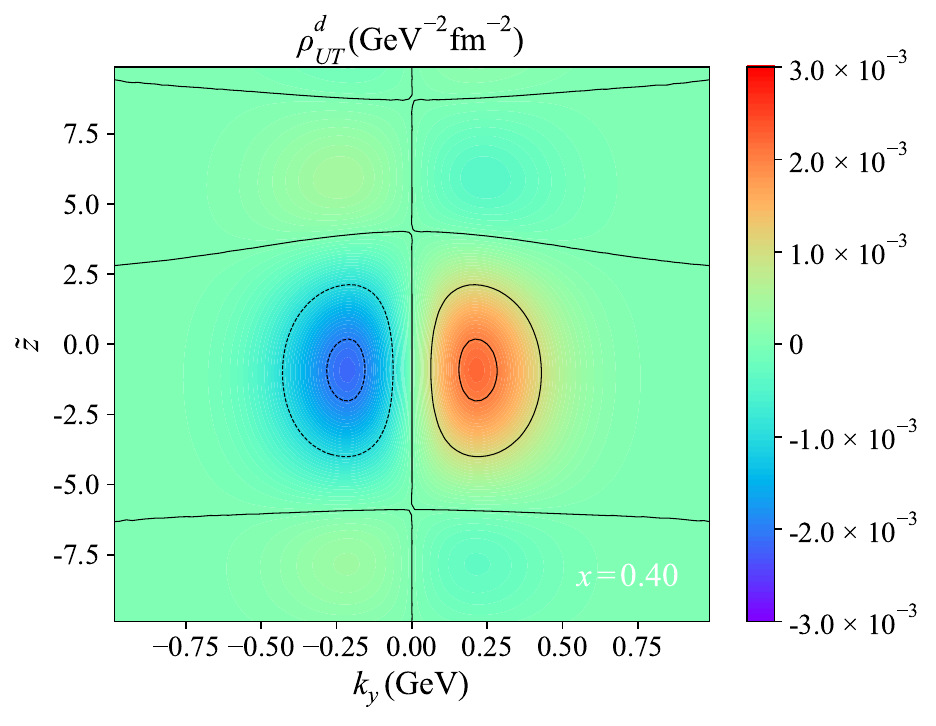}
	}
	\caption{Six-dimensional unpolarized-transverse light-front Wigner distribution $\rho_{\mathrm{UT}}\left(\tilde{z},x,\boldsymbol{b}_{\perp}, \boldsymbol{k}_{\perp}\right)$ for $u$ quark (upper panels) and $d$ quark (lower panels). The figure presents the Wigner distributions in the $\tilde{z}-k_y$ plane, with the transverse coordinate fixed at $\boldsymbol{b}_{\perp}=0.4\,\mathrm{GeV}^{-1}\boldsymbol{\hat{e}}_x$ (where $\boldsymbol{\hat{e}}_x$ is the unit vector along the $x$-axis) and the transverse momentum component fixed at $k_x=0.3\,\mathrm{GeV}$. The three columns correspond to $x=0.10$, $x=0.25$, and $x=0.40$.}
	\label{6DProtonUTudzky}
\end{figure}

In Figs.~\ref{6DProtonUTudzbx}--\ref{6DProtonUTudzky}, % In Fig.~\ref{6DProtonUTudzbx}, Fig.~\ref{6DProtonUTudzby}, Fig.~\ref{6DProtonUTudzkx} and Fig.~\ref{6DProtonUTudzky}, 
we plot the six-dimensional unpolarized-transverse light-front Wigner distribution $\rho_{\mathrm{UT}}\left(\tilde{z},x,\boldsymbol{b}_{\perp}, \boldsymbol{k}_{\perp}\right)$ for the $u$ and $d$ quarks of the proton, displayed in the $\tilde{z}-b_x$, $\tilde{z}-b_y$, $\tilde{z}-k_x$, and $\tilde{z}-k_y$ subspaces, respectively. The six-dimensional unpolarized-transverse light-front Wigner distributions represent the transverse-polarized quark in an unpolarized proton. The numerical results illustrating the relationship between longitudinal coordinates and transverse coordinates or transverse momentum are shown for fixed values of the transverse momentum $\boldsymbol{k}_{\perp}$ or the transverse coordinate $\boldsymbol{b}_{\perp}$, and the longitudinal momentum fraction $x$ is set at $x = 0.10$, $x = 0.25$, and $x = 0.40$ in the first, second, and third columns, respectively.

In Fig.~\ref{6DProtonUTudzbx} and Fig.~\ref{6DProtonUTudzkx}, the six-dimensional unpolarized-transverse light-front Wigner distributions exhibit centrosymmetry about the coordinate origin in the $\tilde{z}-b_x$ and $\tilde{z}-k_x$ subspaces, with the maximum values located at the center (coordinate origin) for each fixed $x$, \new{reflecting the isotropic transverse distribution when viewed along the polarization axis.} In Fig.~\ref{6DProtonUTudzby} and Fig.~\ref{6DProtonUTudzky}, after integrating over the longitudinal coordinates $\tilde{z}$ for each fixed $x$, the distributions reveal a dipole-symmetric shape about $b_x=0$ \new{that becomes increasingly pronounced at lower momentum fractions, providing direct evidence of spin-orbit coupling in the proton quark structure. The observed dipole pattern, particularly its $x$-dependence, carries significant physical meaning. At $x=0.10$ in the sea quark dominated region, the dipole moment is most pronounced, decreasing substantially for valence quarks at $x=0.40$. This evolution suggests stronger spin-orbit correlations in the sea quark sector compared to valence quarks, consistent with predictions from light-front relativistic models incorporating Melosh rotations. In addition, Fig.~\ref{6DProtonUTudzky} exhibits particularly intriguing behavior in the $\tilde{z}-k_y$ projection, where the distribution shows a distinct quadrupole-like structure. This higher-order multipole pattern suggests the existence of more complex spin-orbit correlations beyond the simple dipole approximation, potentially revealing new aspects of quark orbital dynamics in the valence-dominated regime. This evolution may reflect the changing balance between single-quark relativistic effects and multi-quark correlation effects across different $x$ regions.}

\new{These distributions connect to physically observable quantities through several important relationships.} On the one hand,  the six-dimensional unpolarized-transverse light-front Wigner distribution can be reduced to the Boer–Mulders function $h^{\perp}_{1}$ at the TMD limit. On the other hand, this distribution function is related to the $\tilde{H}_{T}$ together with some other distributions. The function $h^{\perp}_{1}$ corresponds to the T-odd part, while $\tilde{H}_{T}$ corresponds to the T-even part. However, the present analysis considers only the leading-order contribution, and due to time-reversal symmetry, the T-odd part vanishes, resulting in a value of zero after integrating over the three-dimensional coordinate space at the TMD limit.

In general, the transverse spin of a quarks has no correlation with its parallel transverse coordinates. However, by analyzing the spin structure of the quarks, we observe that the intrinsic transverse coordinates of the quarks are aligned with their polarization in the distribution. This alignment is reflected in the symmetry of the distribution, where the behavior is not correlated with the direction of the quark transverse momentum. Notably, this behavior may change if a nontrivial Wilson line is introduced into the theory, potentially leading to different results.

%\subsubsection{u Quark}

\subsection{Longitudinal-Unpolarized Wigner distribution}

\begin{figure}[htbp]
	\centering
	\subfloat{
		\includegraphics[width=0.31\textwidth]{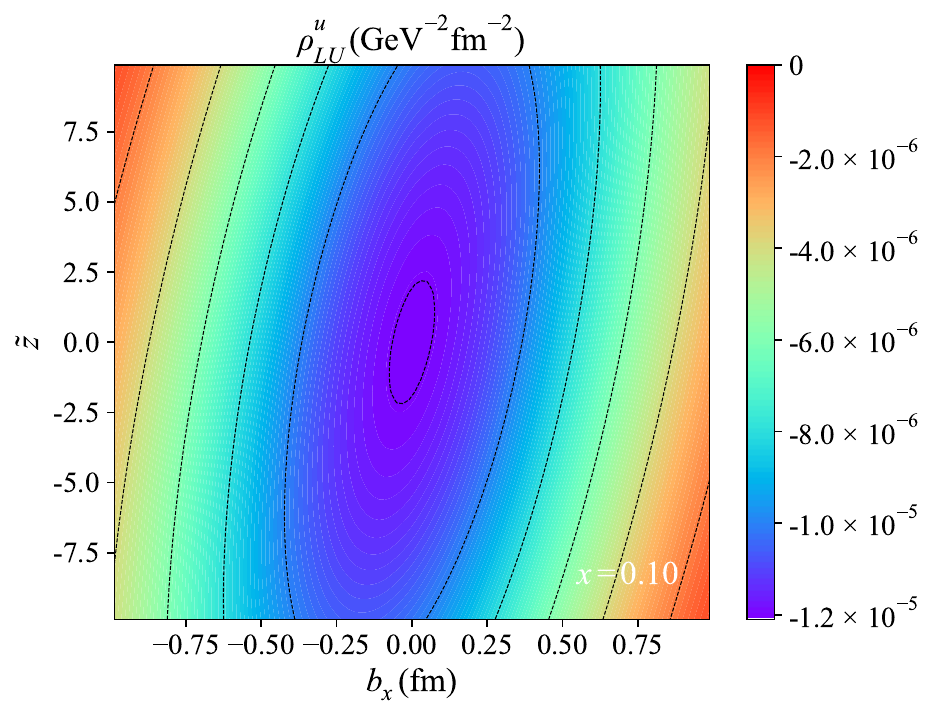}
	}
	\subfloat{
		\includegraphics[width=0.31\textwidth]{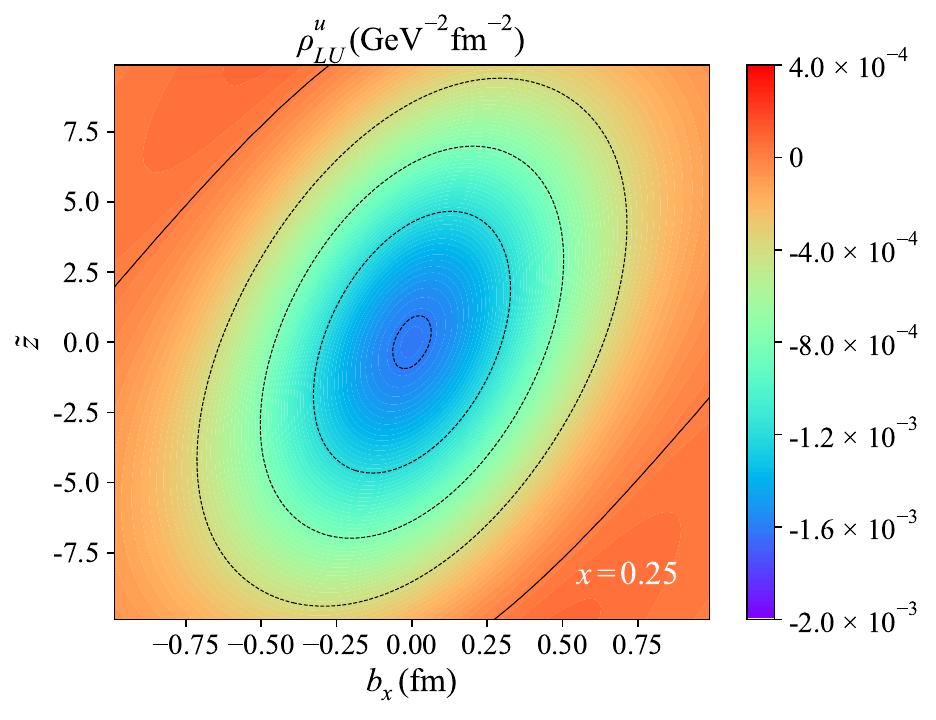}
	}
	\subfloat{
		\includegraphics[width=0.31\textwidth]{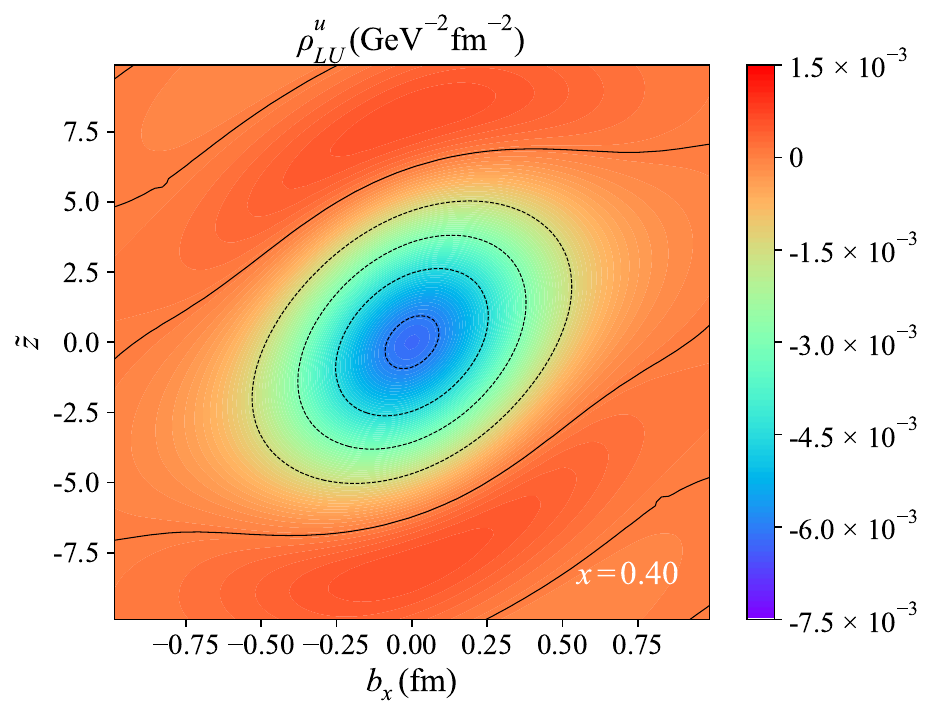}
	}\\
	\subfloat{
		\includegraphics[width=0.31\textwidth]{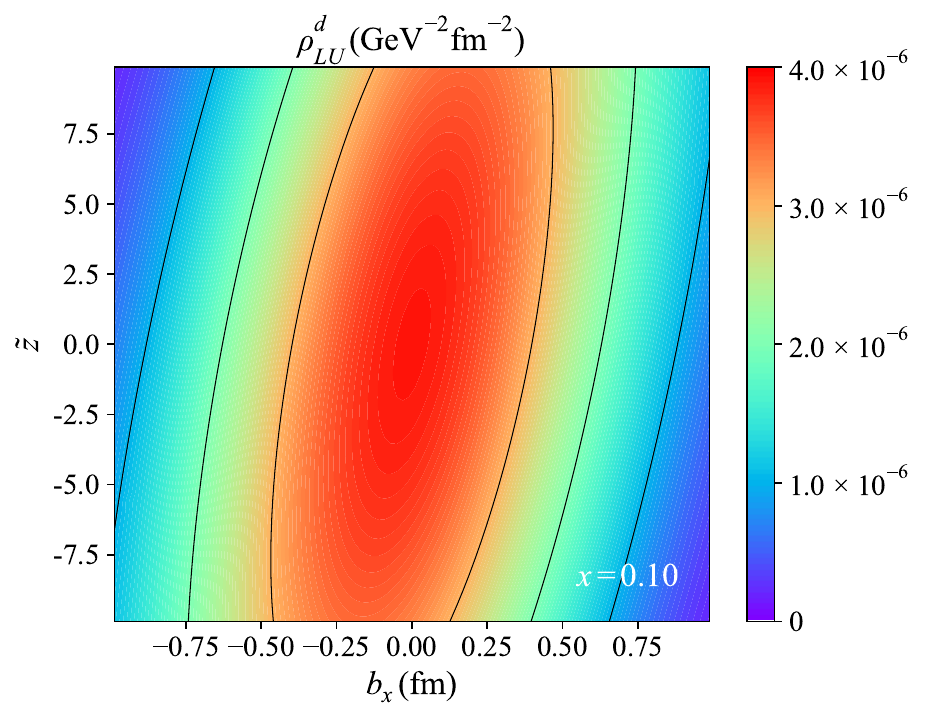}
	} 
	\subfloat{
		\includegraphics[width=0.31\textwidth]{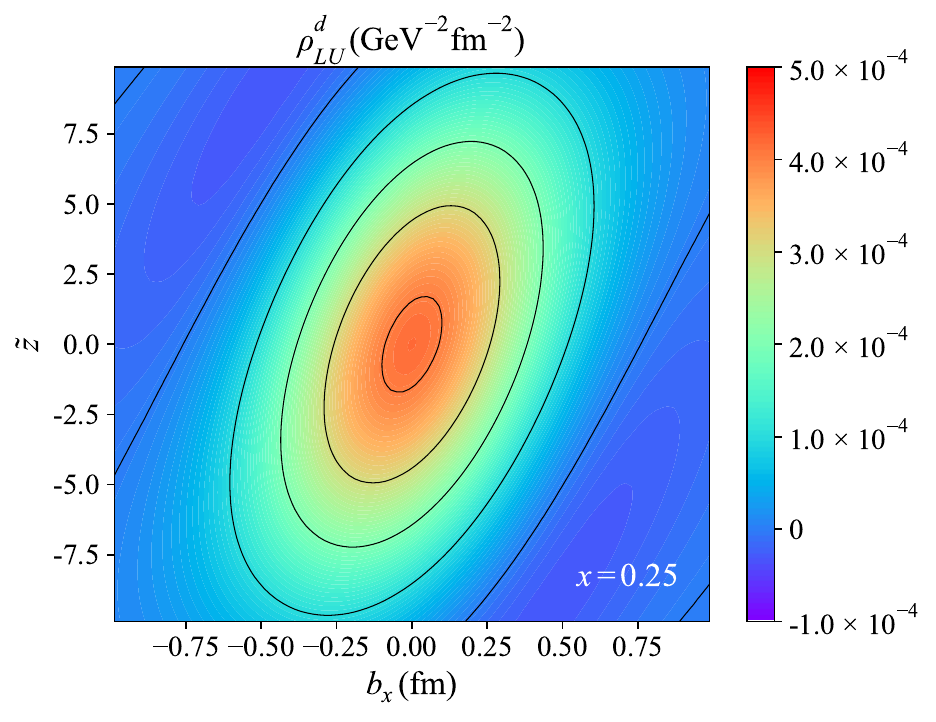}
	}
	\subfloat{
		\includegraphics[width=0.31\textwidth]{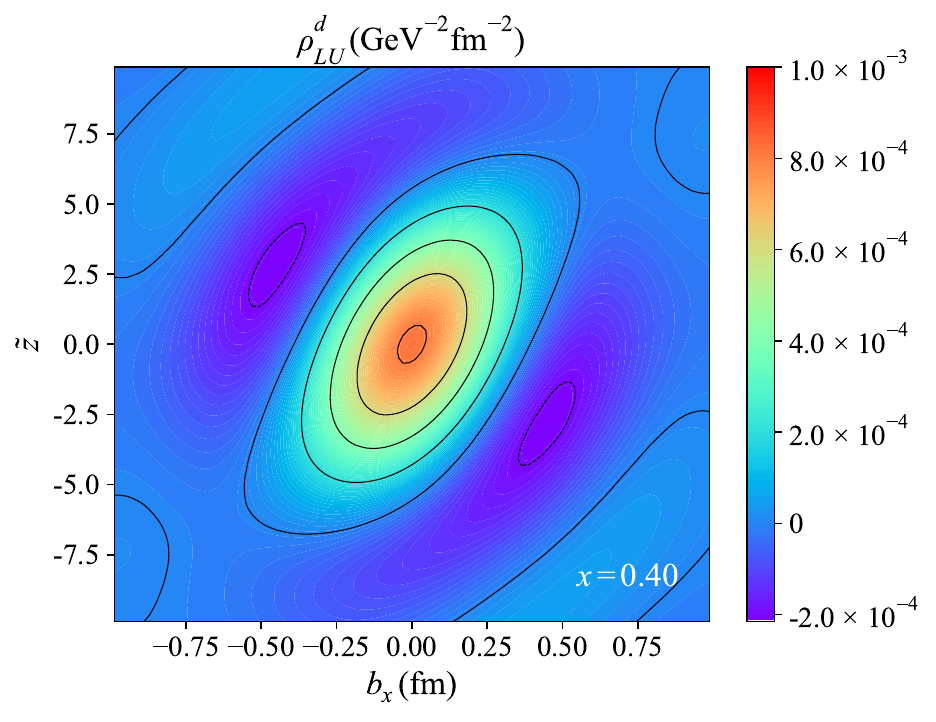}
	}
	\caption{Six-dimensional longitudinal-unpolarized light-front Wigner distribution $\rho_{\mathrm{LU}}\left(\tilde{z},x,\boldsymbol{b}_{\perp}, \boldsymbol{k}_{\perp}\right)$ for $u$ quark (upper panels) and $d$ quark (lower panels). The figure presents the Wigner distribution in the $\tilde{z}-b_x$ plane, with the transverse momentum fixed at $\boldsymbol{k}_{\perp}=0.3\,\mathrm{GeV}\boldsymbol{\hat{e}}_x$ (where $\boldsymbol{\hat{e}}_x$ is the unit vector in the $x$-direction) and the transverse coordinate component fixed at $b_y=0.4\,\mathrm{GeV}^{-1}$. The three columns correspond to $x=0.10$, $x=0.25$, and $x=0.40$.}
	\label{6DProtonLUudzbx}
\end{figure}

\begin{figure}[htbp]
	\centering
	\subfloat{
		\includegraphics[width=0.31\textwidth]{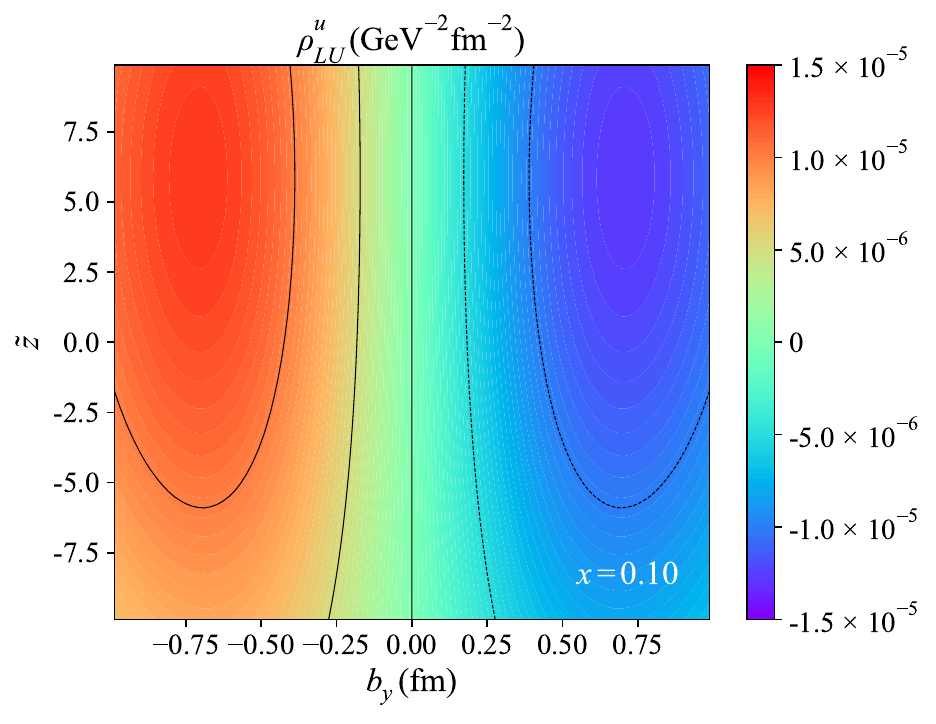}
	}
	\subfloat{
		\includegraphics[width=0.31\textwidth]{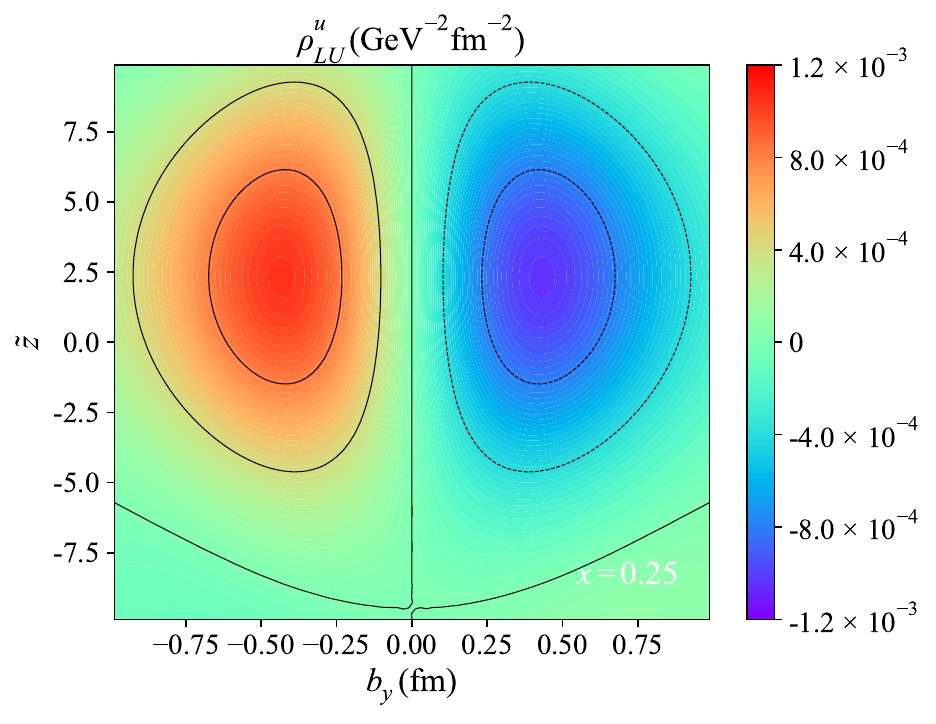}
	}
	\subfloat{
		\includegraphics[width=0.31\textwidth]{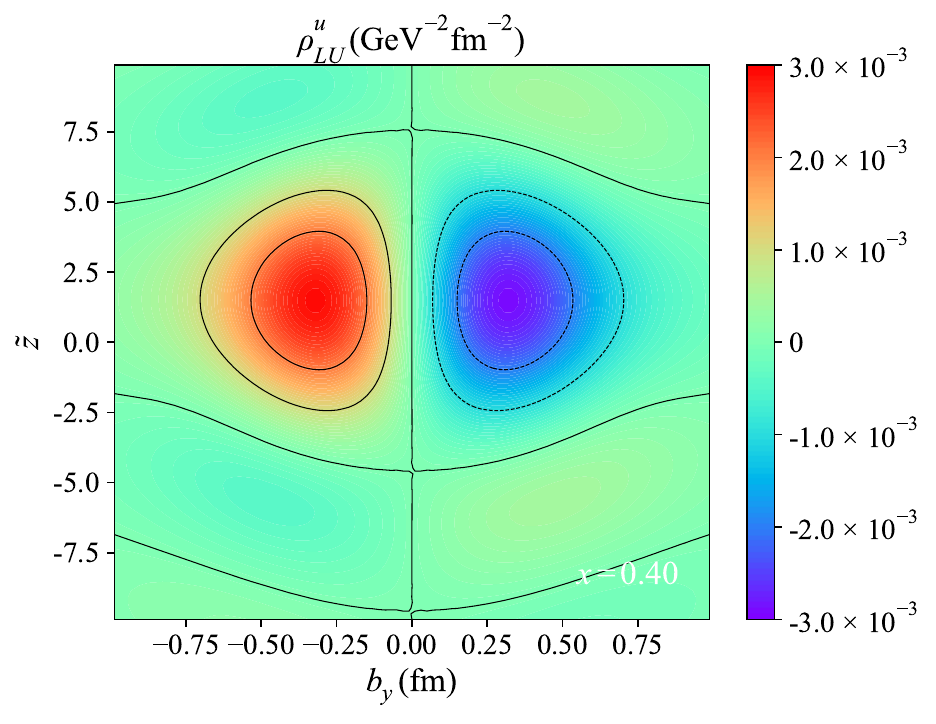}
	}\\
	\subfloat{
		\includegraphics[width=0.31\textwidth]{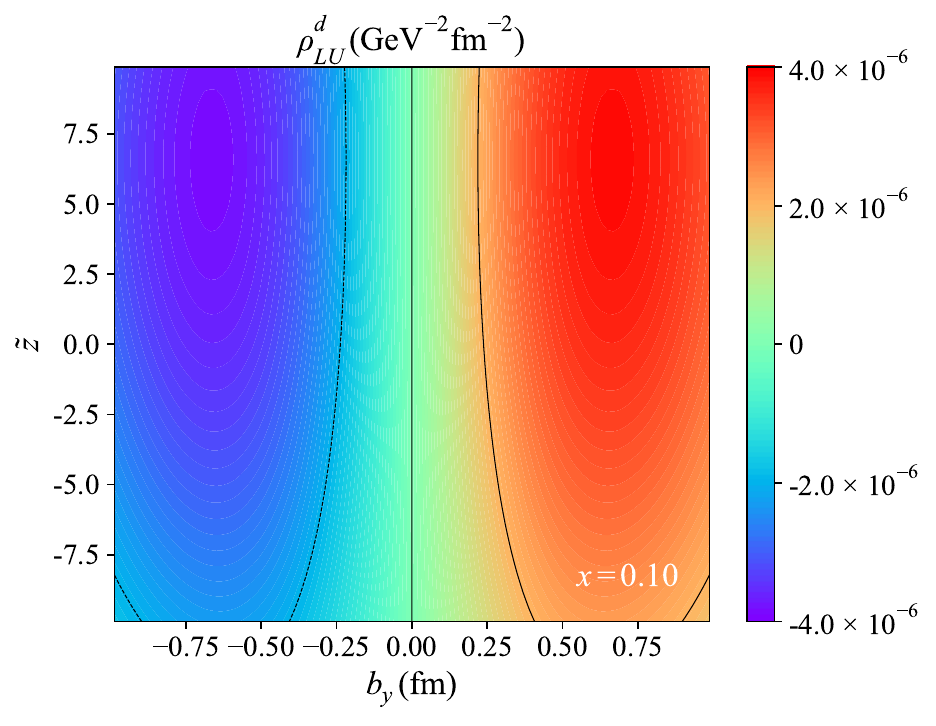}
	}
	\subfloat{
		\includegraphics[width=0.31\textwidth]{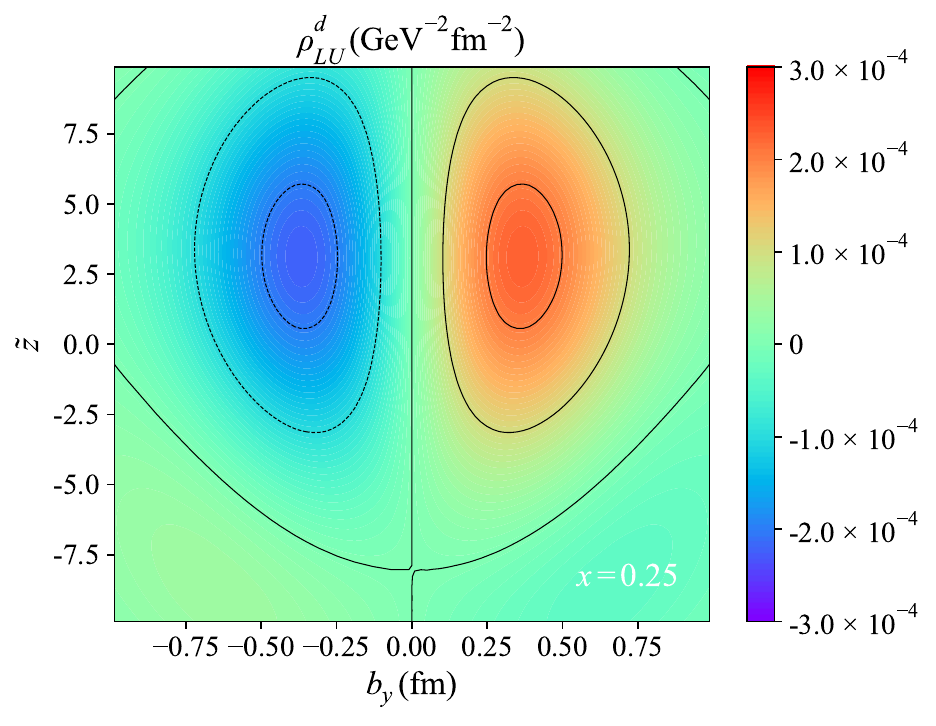}
	}
	\subfloat{
		\includegraphics[width=0.31\textwidth]{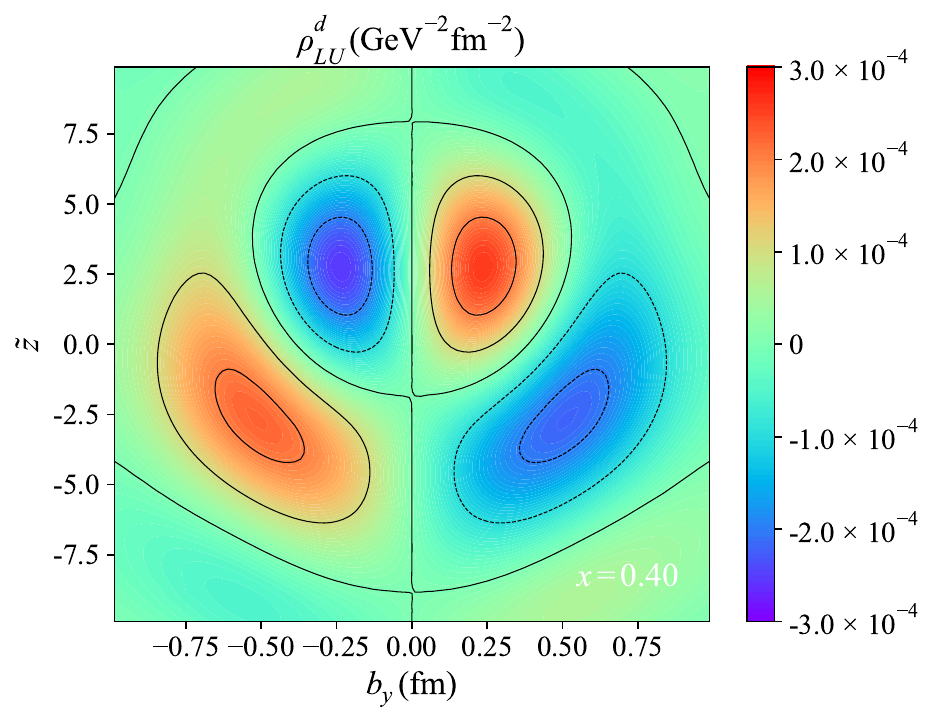}
	}
	\caption{Six-dimensional longitudinal-unpolarized light-front Wigner distribution $\rho_{\mathrm{LU}}\left(\tilde{z},x,\boldsymbol{b}_{\perp}, \boldsymbol{k}_{\perp}\right)$ for $u$ quark (upper panels) and $d$ quark (lower panels). The figure presents the Wigner distributions in the $\tilde{z}-b_y$ plane, with the transverse momentum fixed at $\boldsymbol{k}_{\perp}=0.3\,\mathrm{GeV}\boldsymbol{\hat{e}}_x$ (where $\boldsymbol{\hat{e}}_x$ is the unit vector in the $x$-direction) and the transverse coordinate component fixed at $b_x=0.4\,\mathrm{GeV}^{-1}$. The three columns correspond to $x=0.10$, $x=0.25$, and $x=0.40$.}
	\label{6DProtonLUudzby}
\end{figure}

\begin{figure}[htbp]
	\centering
	\subfloat{
		\includegraphics[width=0.31\textwidth]{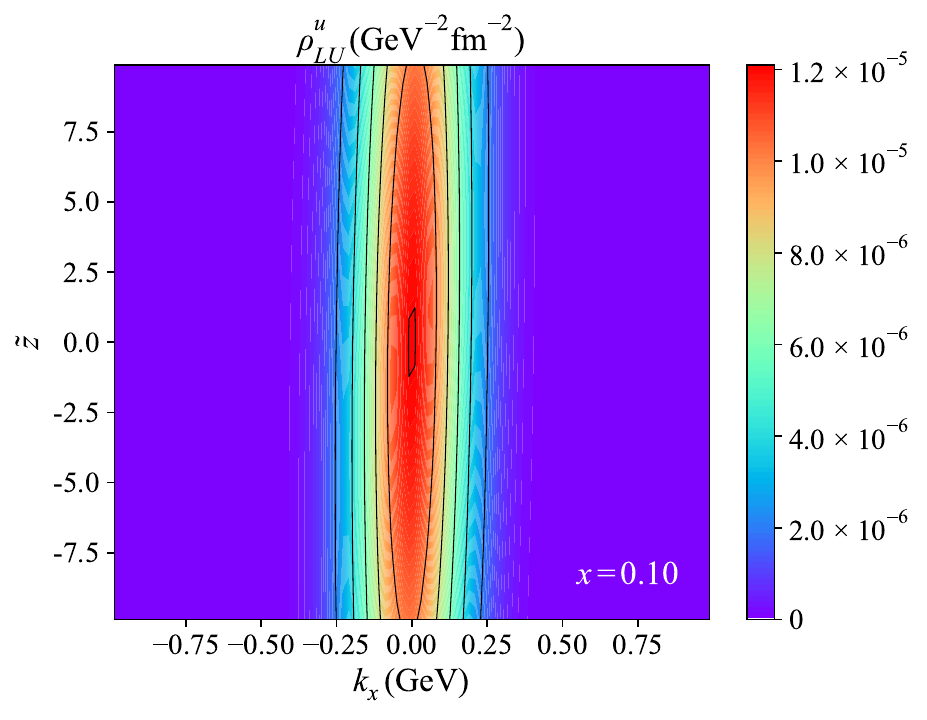}
	}
	\subfloat{
		\includegraphics[width=0.31\textwidth]{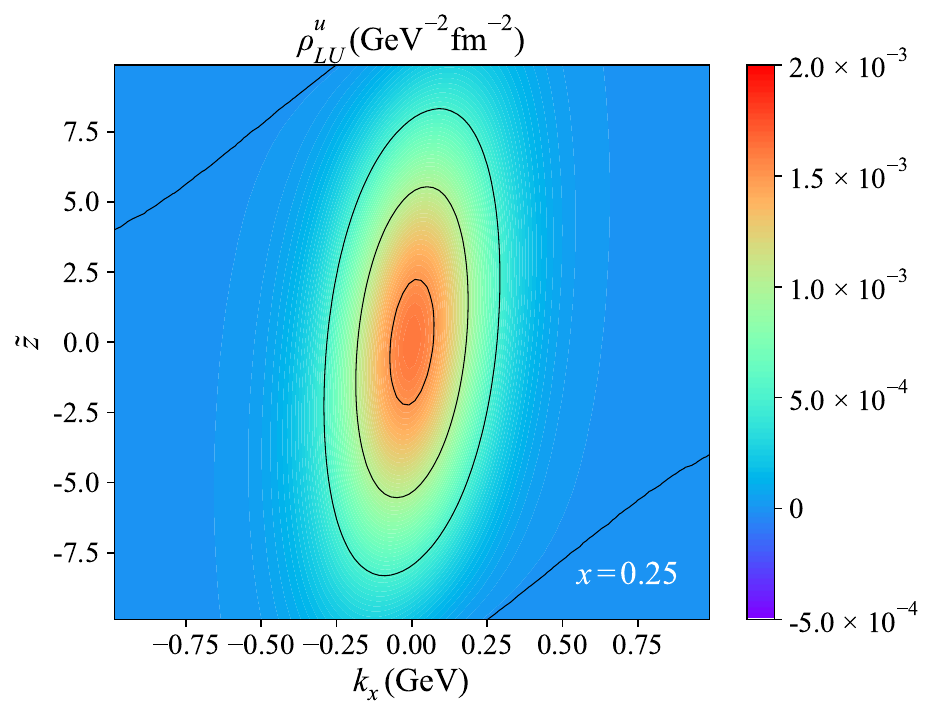}
	}
	\subfloat{
		\includegraphics[width=0.31\textwidth]{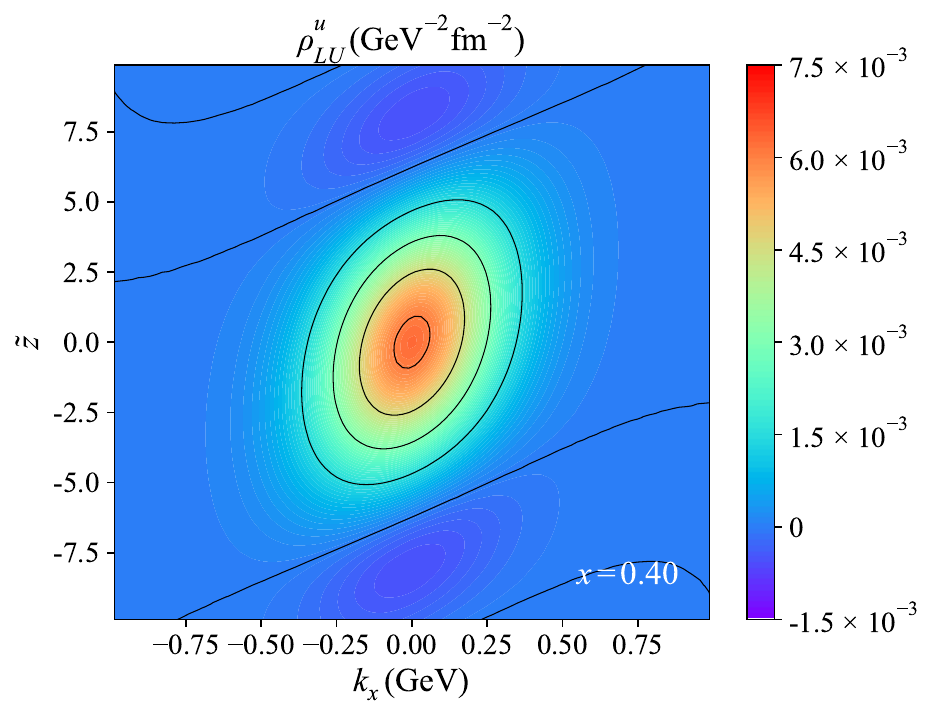}
	}\\
	\subfloat{
		\includegraphics[width=0.31\textwidth]{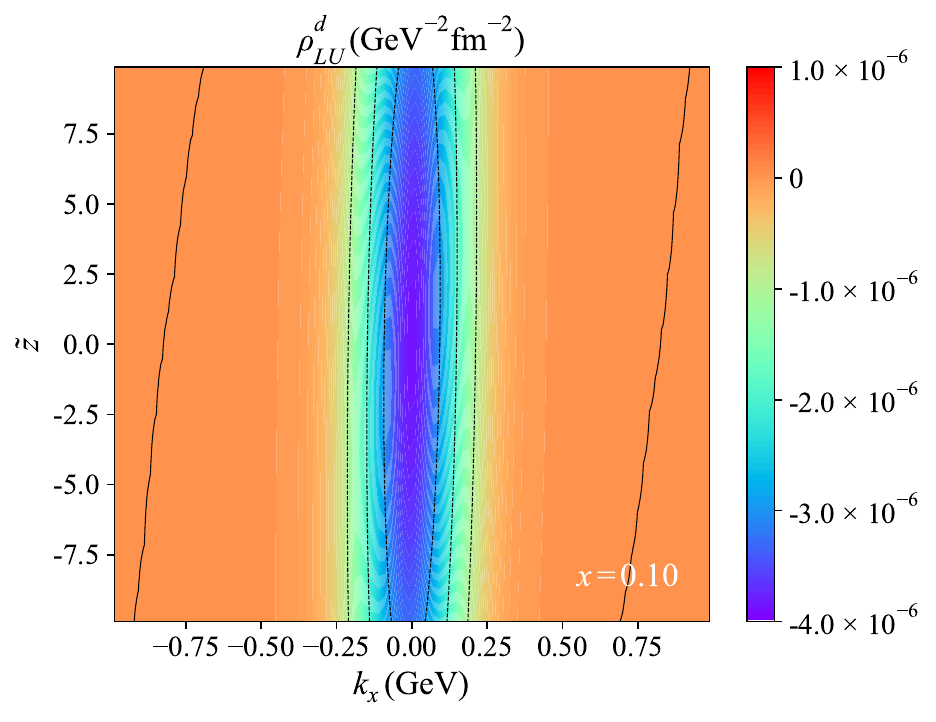}
	}
	\subfloat{
		\includegraphics[width=0.31\textwidth]{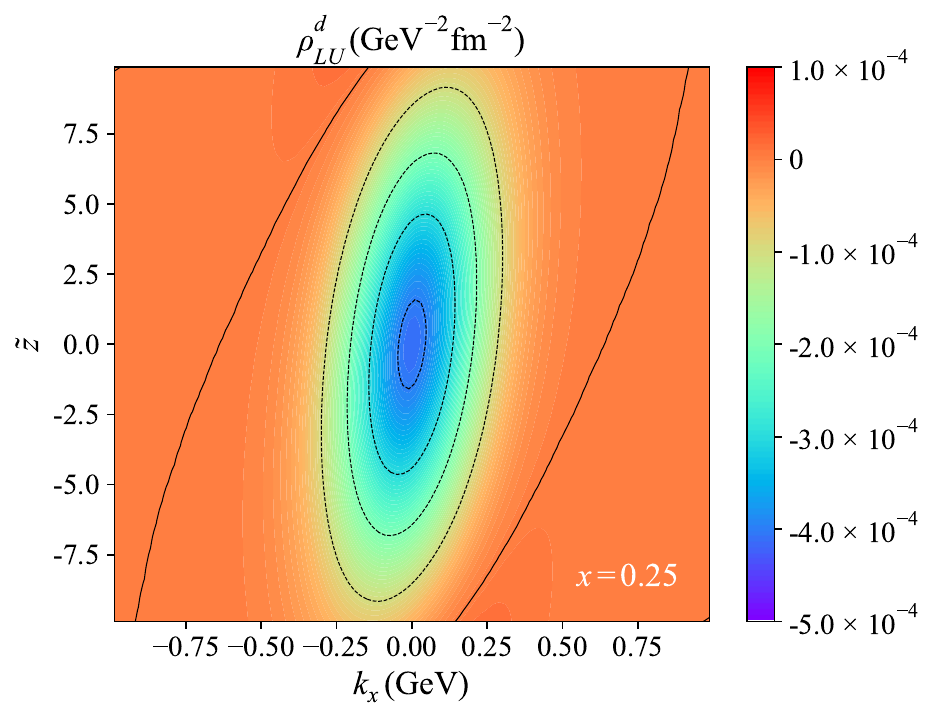}
	}
	\subfloat{
		\includegraphics[width=0.31\textwidth]{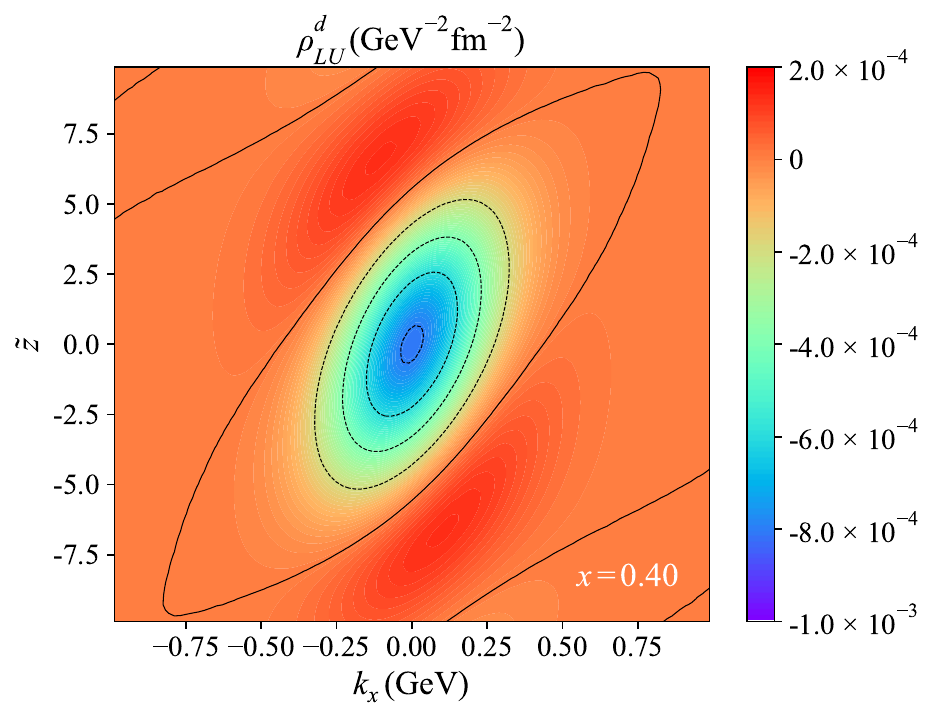}
	}
	\caption{Six-dimensional longitudinal-unpolarized light-front Wigner distribution $\rho_{\mathrm{LU}}\left(\tilde{z},x,\boldsymbol{b}_{\perp}, \boldsymbol{k}_{\perp}\right)$ for $u$ quark (upper panels) and $d$ quark (lower panels). The figure presents the Wigner distributions in the $\tilde{z}-k_x$ plane, with the transverse coordinate fixed at $\boldsymbol{b}_{\perp}=0.4\,\mathrm{GeV}^{-1}\boldsymbol{\hat{e}}_x$ (where $\boldsymbol{\hat{e}}_x$ is the unit vector along the $x$-axis) and the transverse momentum component fixed at $k_y=0.3\,\mathrm{GeV}$. The three columns correspond to $x=0.10$, $x=0.25$, and $x=0.40$.}
	\label{6DProtonLUudzkx}
\end{figure}

\begin{figure}[htbp]
	\centering
	\subfloat{
		\includegraphics[width=0.31\textwidth]{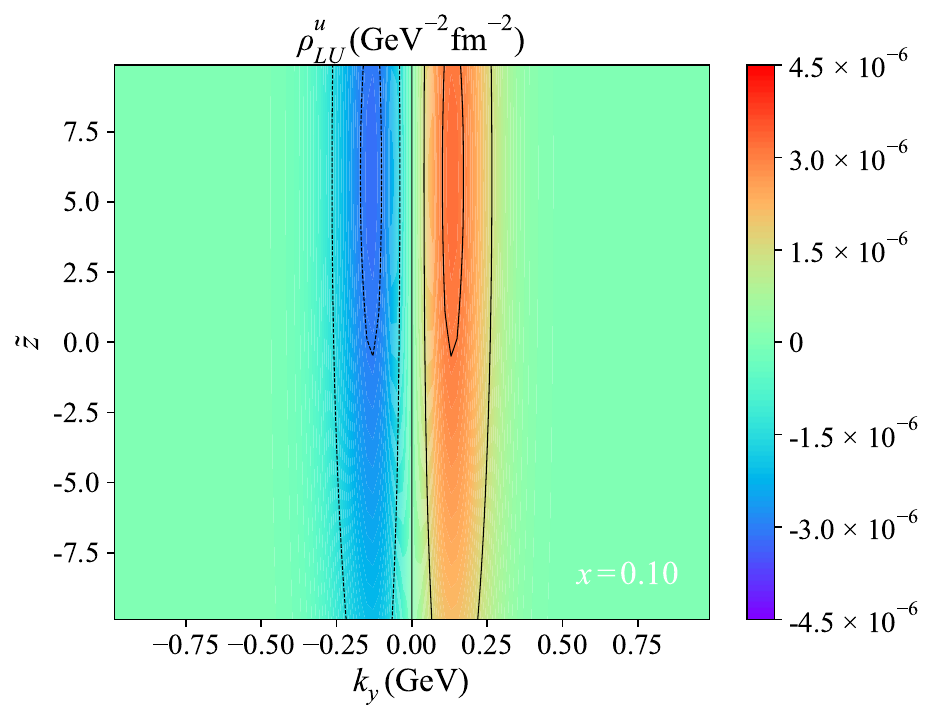}
	}
	\subfloat{
		\includegraphics[width=0.31\textwidth]{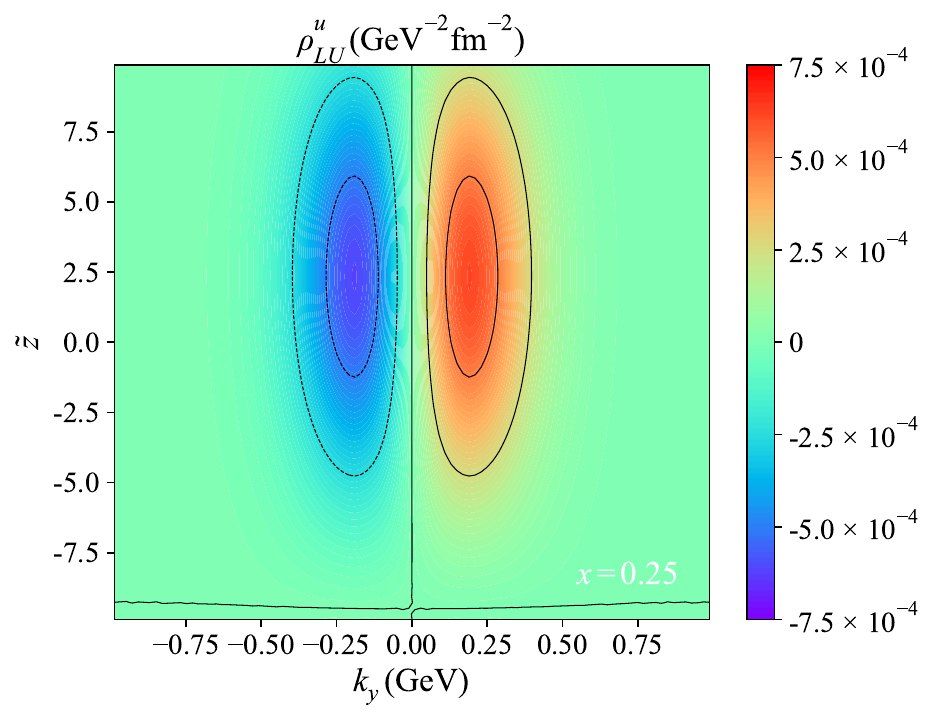}
	}
	\subfloat{
		\includegraphics[width=0.31\textwidth]{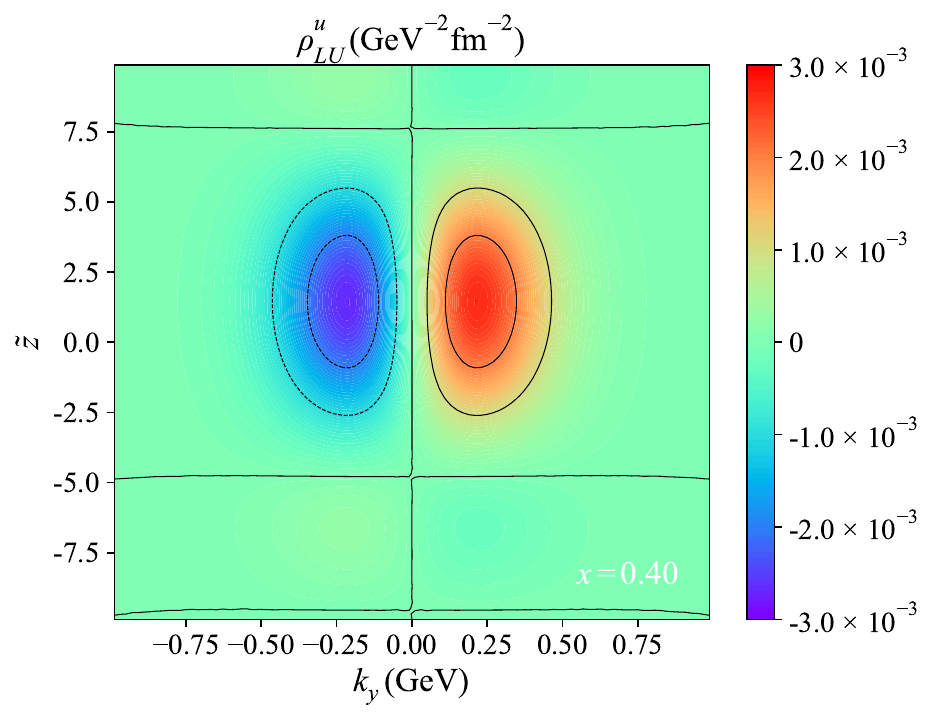}
	}\\
	\subfloat{
		\includegraphics[width=0.31\textwidth]{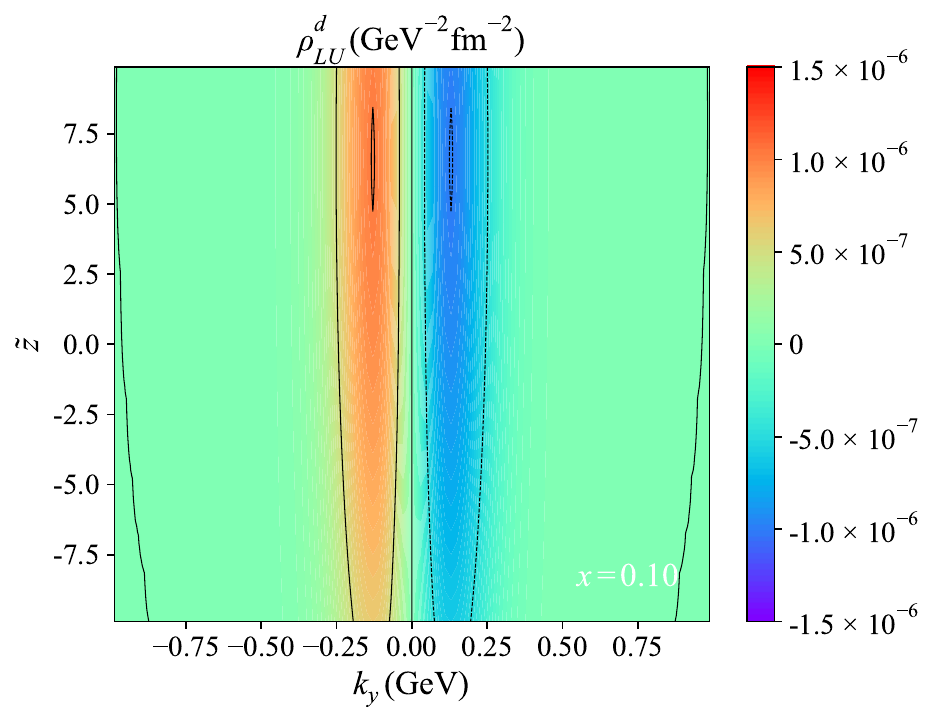}
	}
	\subfloat{
		\includegraphics[width=0.31\textwidth]{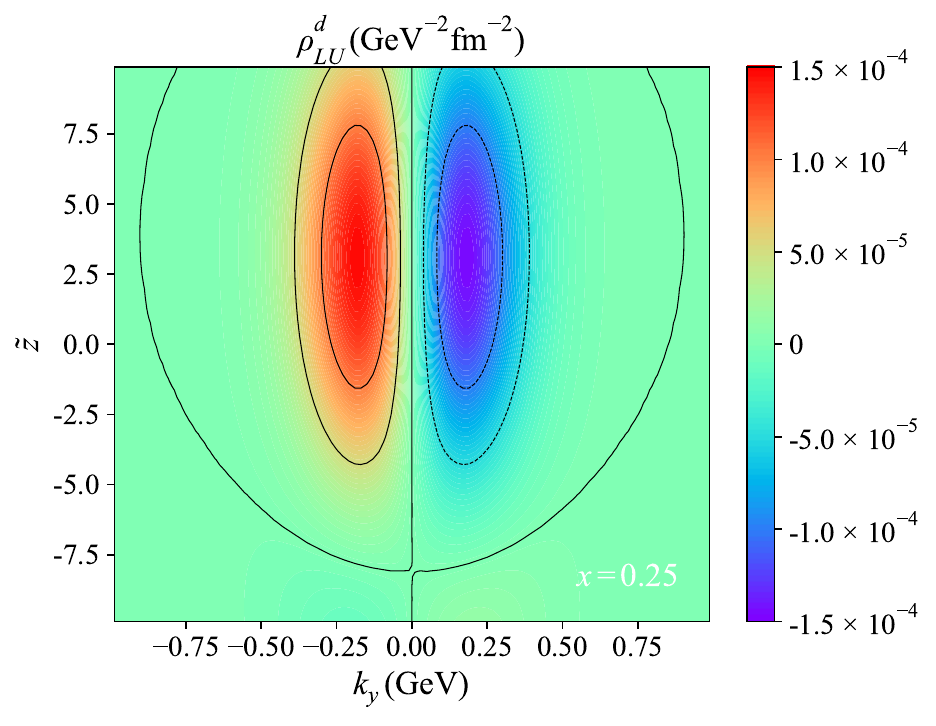}
		.
	}
	\subfloat{
		\includegraphics[width=0.31\textwidth]{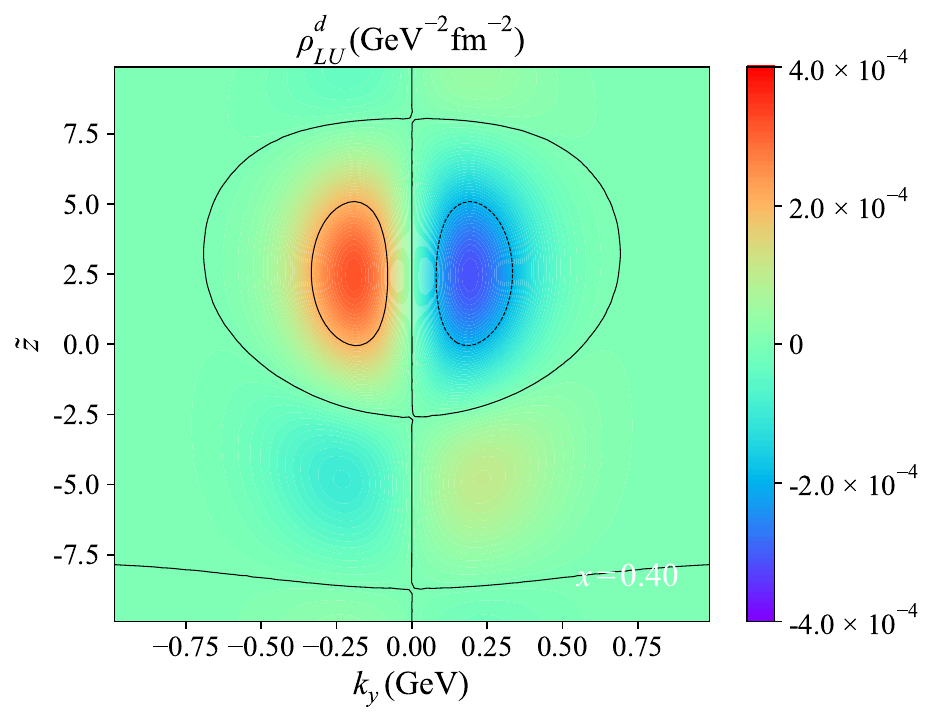}
	}
	\caption{Six-dimensional longitudinal-unpolarized light-front Wigner distribution $\rho_{\mathrm{LU}}\left(\tilde{z},x,\boldsymbol{b}_{\perp}, \boldsymbol{k}_{\perp}\right)$ for $u$ quark (upper panels) and $d$ quark (lower panels). The figure presents the Wigner distributions in the $\tilde{z}-k_y$ plane, with the transverse coordinate fixed at $\boldsymbol{b}_{\perp}=0.4\,\mathrm{GeV}^{-1}\boldsymbol{\hat{e}}_x$ (where $\boldsymbol{\hat{e}}_x$ is the unit vector along the $x$-axis) and the transverse momentum component fixed at $k_x=0.3\,\mathrm{GeV}$. The three columns correspond to $x=0.10$, $x=0.25$, and $x=0.40$.}
	\label{6DProtonLUudzky}
\end{figure}

In Figs.~\ref{6DProtonLUudzbx}--\ref{6DProtonLUudzky}, % In Fig.~\ref{6DProtonLUudzbx}, Fig.~\ref{6DProtonLUudzby}, Fig.~\ref{6DProtonLUudzkx} and Fig.~\ref{6DProtonLUudzky}, 
we plot the six-dimensional longitudinal-unpolarized light-front Wigner distribution $\rho_{\mathrm{LU}}\left(\tilde{z},x,\boldsymbol{b}_{\perp}, \boldsymbol{k}_{\perp}\right)$ for the $u$ and $d$ quarks of the proton, displayed in the $\tilde{z}-b_x$, $\tilde{z}-b_y$, $\tilde{z}-k_x$, and $\tilde{z}-k_y$ subspaces, respectively. The six-dimensional longitudinal-unpolarized light-front Wigner distributions describe quark phase-space distributions in a longitudinal-polarized proton without any information of quark spins. The numerical results are shown for fixed values of the transverse momentum $\boldsymbol{k}_{\perp}$ or the transverse coordinate $\boldsymbol{b}_{\perp}$, and the longitudinal momentum fraction $x$ is set at $x = 0.10$, $x = 0.25$, and $x = 0.40$ in the first, second, and third columns, respectively. As the longitudinal-unpolarized distribution is independent of quark spin, it offers a unique perspective in the study of orbital angular momentum~\cite{Lorce:2011ni}. This distribution, which captures the quark spatial and momentum distribution in a longitudinal-polarized proton, does not provide any information about the spin alignment of quarks.

In Fig.~\ref{6DProtonLUudzbx} and Fig.~\ref{6DProtonLUudzkx}, the six-dimensional longitudinal-unpolarized light-front Wigner distribution exhibits centrosymmetry with respect to the origin in the $\tilde{z}-b_x$ and $\tilde{z}-k_x$ subspaces, with the maximum values occurring at the center of the coordinate system for each fixed value of $x$. In contrast, Fig.~\ref{6DProtonLUudzby} and Fig.~\ref{6DProtonLUudzky} display a dipole-symmetric distribution around $b_x = 0$, which shows strong $x$-dependence. \new{This dipole pattern directly manifests the correlation between quark orbital motion and the proton spin direction, with its $x$-evolution suggesting different mechanisms of orbital angular momentum generation in sea versus valence quark regimes.}

\new{The physical significance of these distributions becomes particularly clear when examining their connection to quark orbital angular momentum through the relation in Eq.~\eqref{OAM}, where the cross product $(\boldsymbol{b}_{\perp} \times \boldsymbol{k}_{\perp})_{z}$ quantifies the quark orbital motion. Our calculations reveal an important flavor dependence, with $u$ and $d$ quarks contributing orbital angular momentum of opposite sign, a finding consistent with both Jaffe-Manohar and Ji decompositions of proton spin~\cite{Jaffe:1989jz,Ji:1996ek}. This flavor asymmetry likely originates from the proton axial-vector diquark component, which affects $u$ and $d$ quarks differently in our light-front spectator model framework.} Specifically, the distribution vanishes at the TMD or IPD limit, indicating that the phase-space behavior captured by this distribution is not observable in leading-twist TMDs or IPDs. Therefore, this distribution is expected to be most relevant for studies involving higher-order twist contributions~\cite{Jaffe:1996zw}.

%\subsubsection{u Quark}

%\subsubsection{d Quark}

\subsection{Transverse-unpolarized Wigner distribution}

\begin{figure}[htbp]
	\centering
	\subfloat{
		\includegraphics[width=0.31\textwidth]{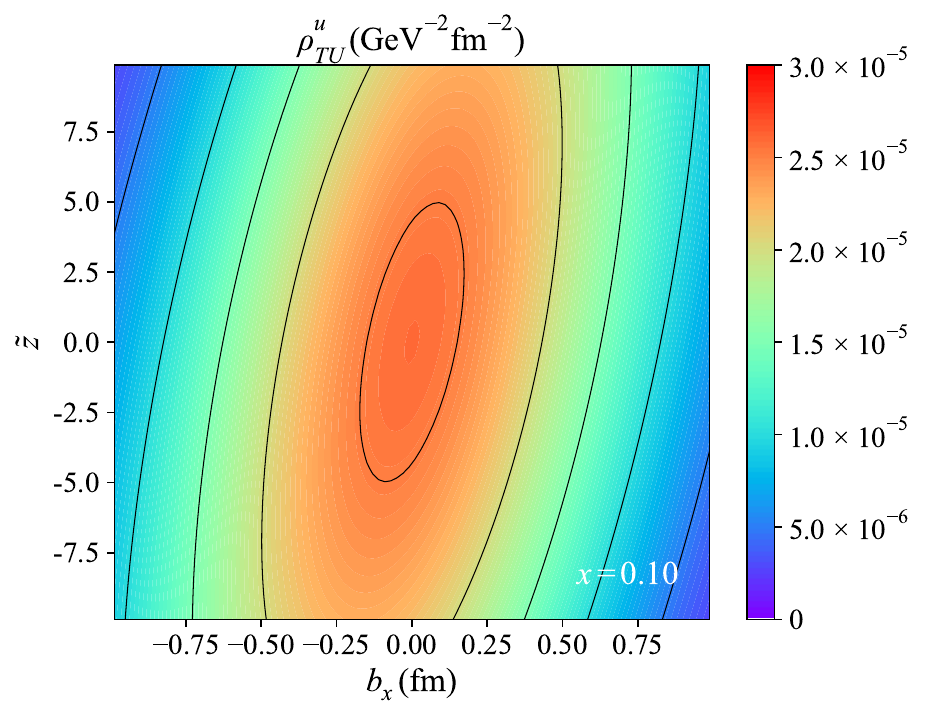}
	}
	\subfloat{
		\includegraphics[width=0.31\textwidth]{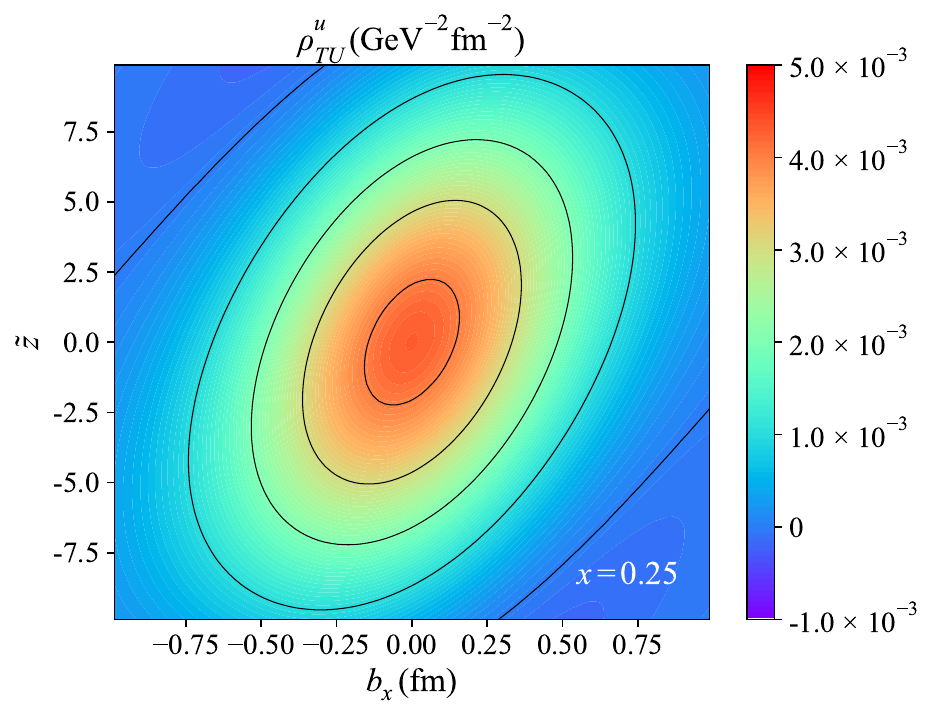}
	}
	\subfloat{
		\includegraphics[width=0.31\textwidth]{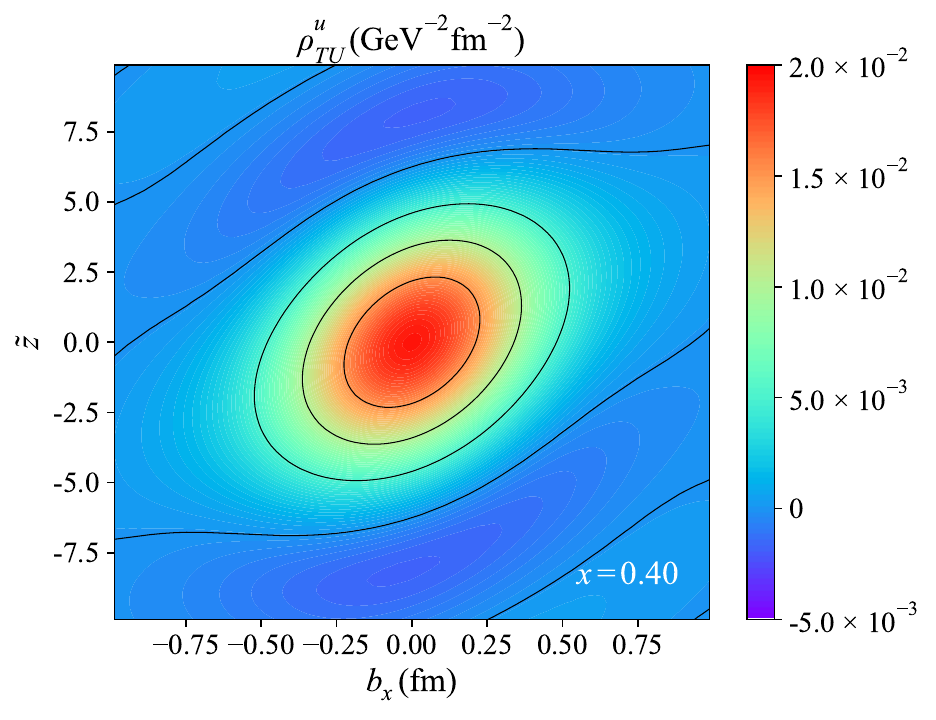}
	}\\
	\subfloat{
		\includegraphics[width=0.31\textwidth]{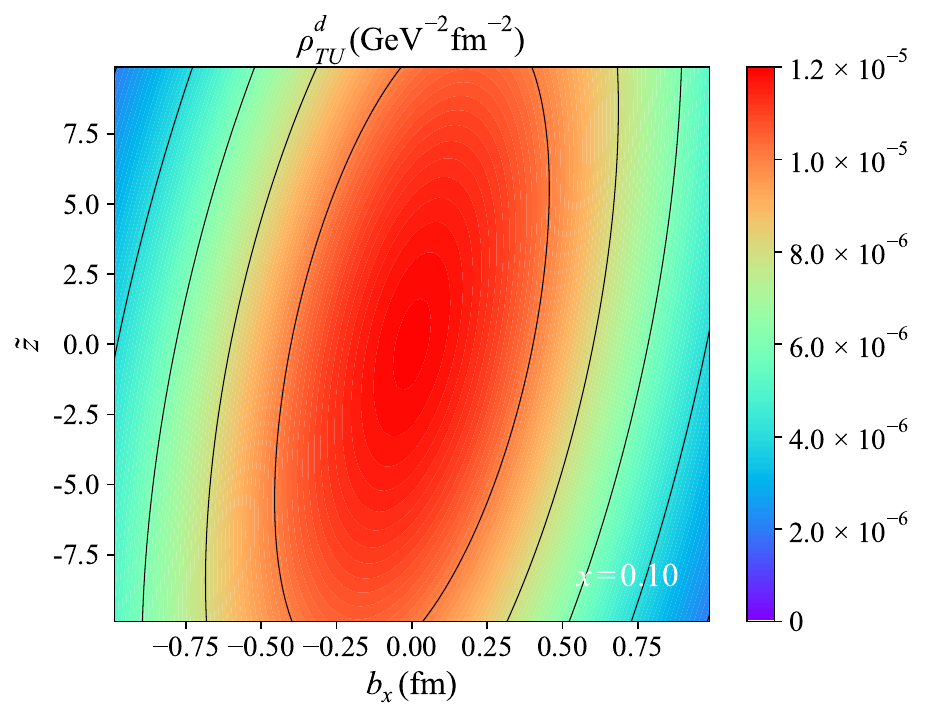}
	}
	\subfloat{
		\includegraphics[width=0.31\textwidth]{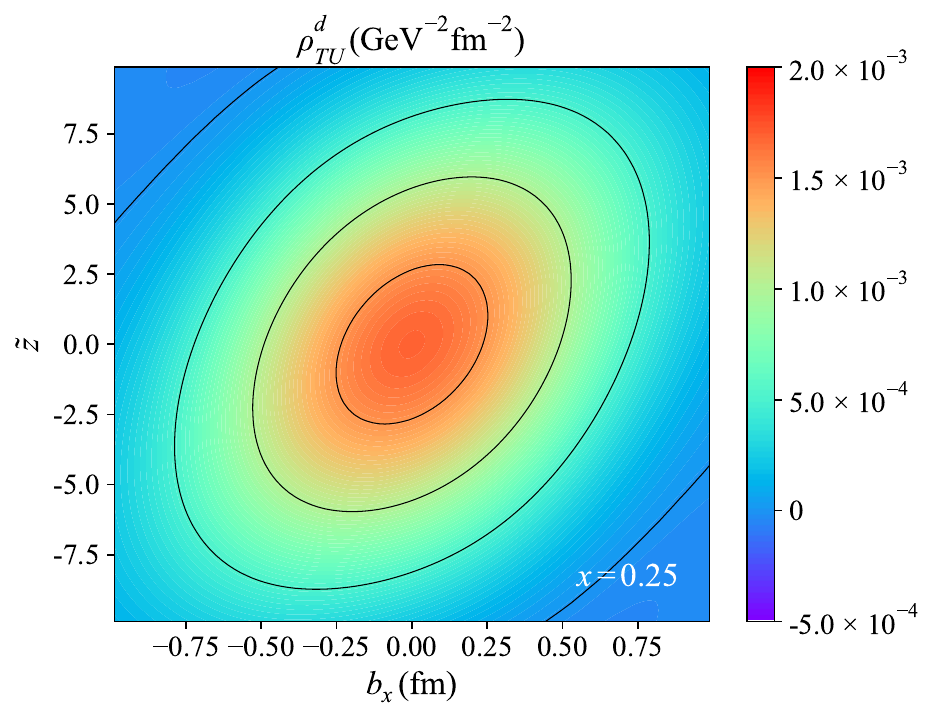}
	}
	\subfloat{
		\includegraphics[width=0.31\textwidth]{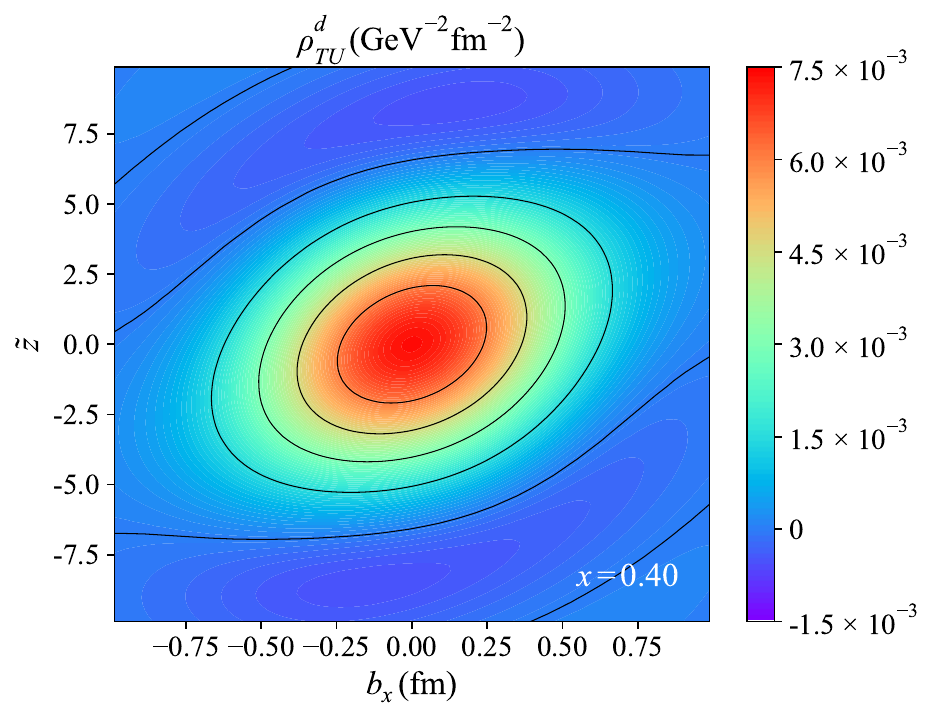}
	}
	\caption{Six-dimensional transverse-unpolarized light-front Wigner distribution $\rho_{\mathrm{TU}}\left(\tilde{z},x,\boldsymbol{b}_{\perp}, \boldsymbol{k}_{\perp}\right)$ for $u$ quark (upper panels) and $d$ quark (lower panels). The figure presents the Wigner distribution in the $\tilde{z}-b_x$ plane, with the transverse momentum fixed at $\boldsymbol{k}_{\perp}=0.3\,\mathrm{GeV}\boldsymbol{\hat{e}}_x$ (where $\boldsymbol{\hat{e}}_x$ is the unit vector in the $x$-direction) and the transverse coordinate component fixed at $b_y=0.4\,\mathrm{GeV}^{-1}$. The three columns correspond to $x=0.10$, $x=0.25$, and $x=0.40$.}
	\label{6DProtonTUudzbx}
\end{figure}

\begin{figure}[htbp]
	\centering
	\subfloat{
		\includegraphics[width=0.31\textwidth]{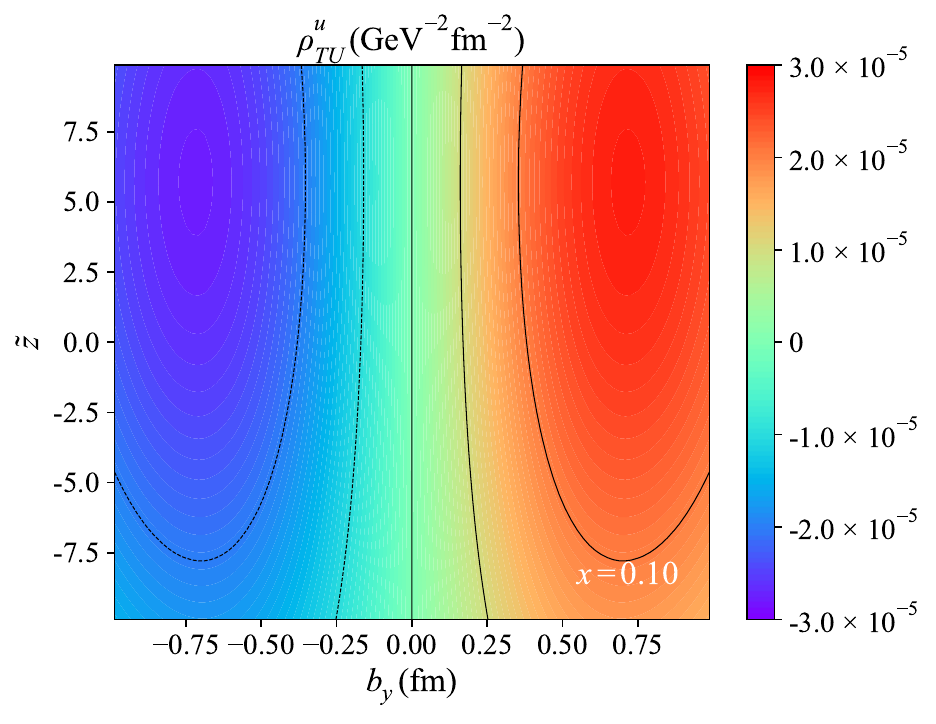}
	}
	\subfloat{
		\includegraphics[width=0.31\textwidth]{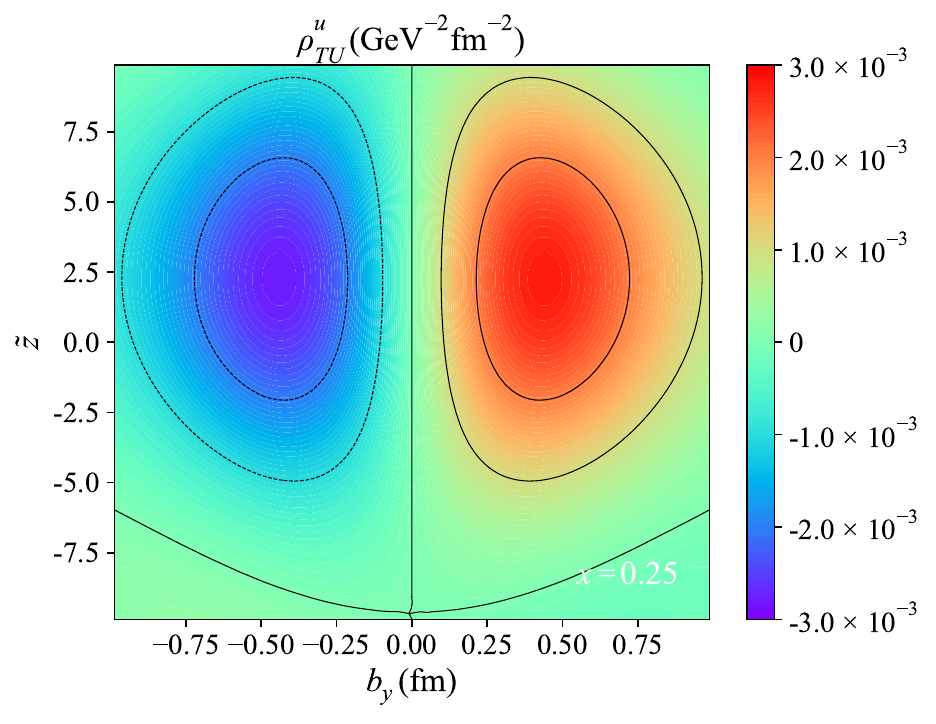}
	}
	\subfloat{
		\includegraphics[width=0.31\textwidth]{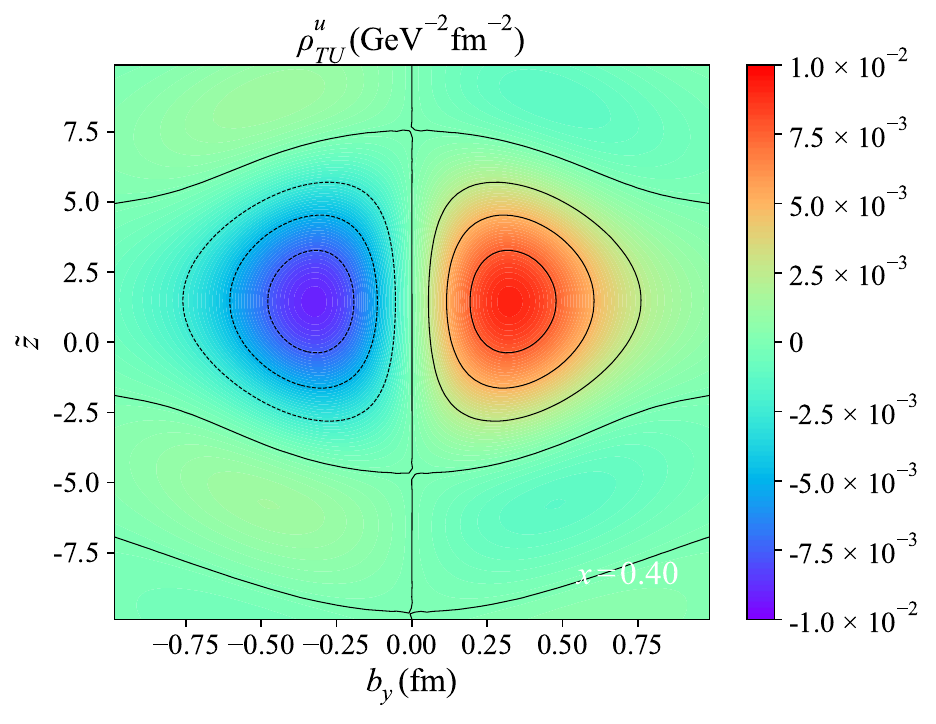}
	}\\
	\subfloat{
		\includegraphics[width=0.31\textwidth]{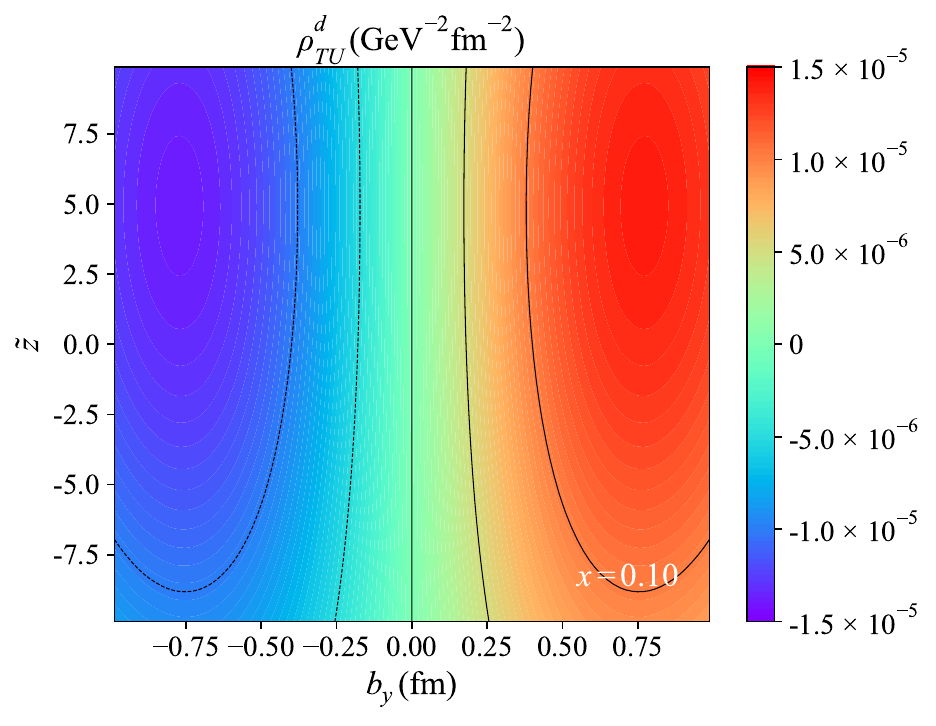}
	}
	\subfloat{
		\includegraphics[width=0.31\textwidth]{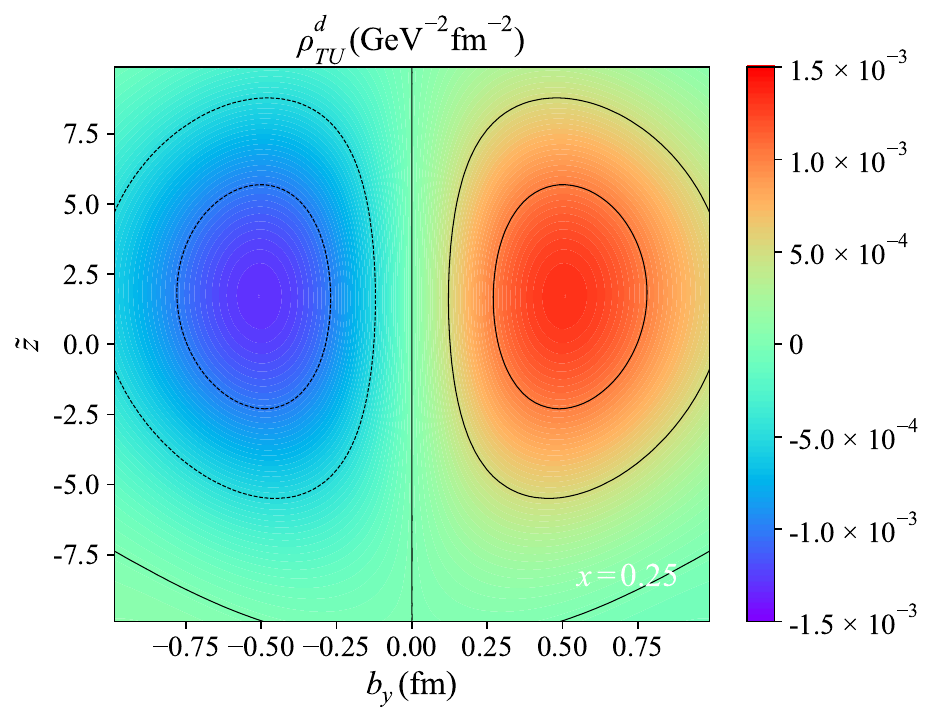}
	}
	\subfloat{
		\includegraphics[width=0.31\textwidth]{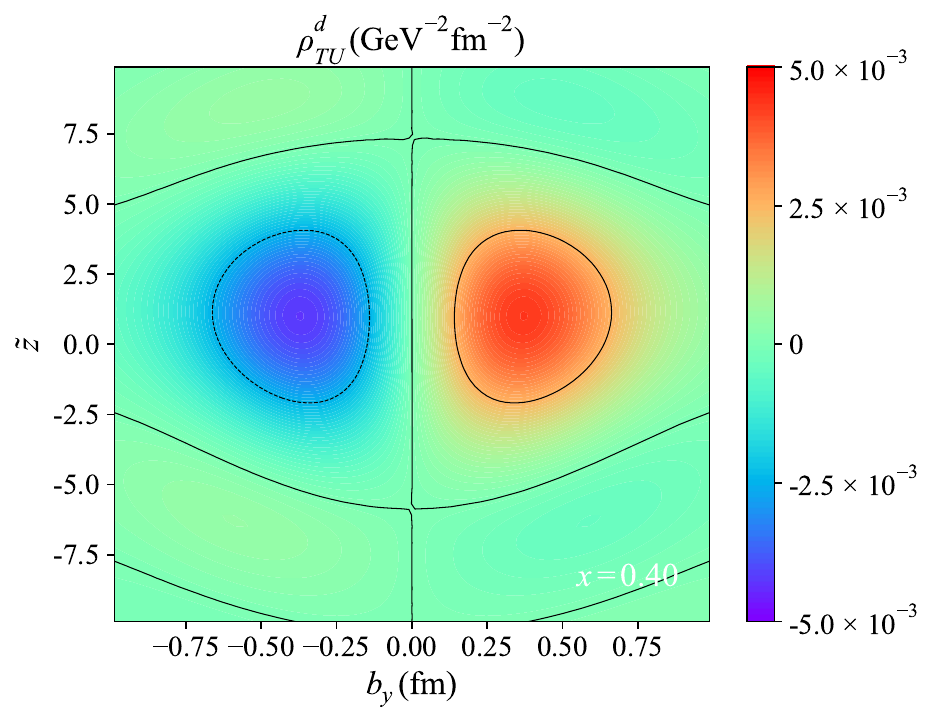}
	}
	\caption{Six-dimensional transverse-unpolarized light-front Wigner distribution $\rho_{\mathrm{TU}}\left(\tilde{z},x,\boldsymbol{b}_{\perp}, \boldsymbol{k}_{\perp}\right)$ for $u$ quark (upper panels) and $d$ quark (lower panels). The figure presents the Wigner distributions in the $\tilde{z}-b_y$ plane, with the transverse momentum fixed at $\boldsymbol{k}_{\perp}=0.3\,\mathrm{GeV}\boldsymbol{\hat{e}}_x$ (where $\boldsymbol{\hat{e}}_x$ is the unit vector in the $x$-direction) and the transverse coordinate component fixed at $b_x=0.4\,\mathrm{GeV}^{-1}$. The three columns correspond to $x=0.10$, $x=0.25$, and $x=0.40$.}
	\label{6DProtonTUudzby}
\end{figure}

\begin{figure}[htbp]
	\centering
	\subfloat{
		\includegraphics[width=0.31\textwidth]{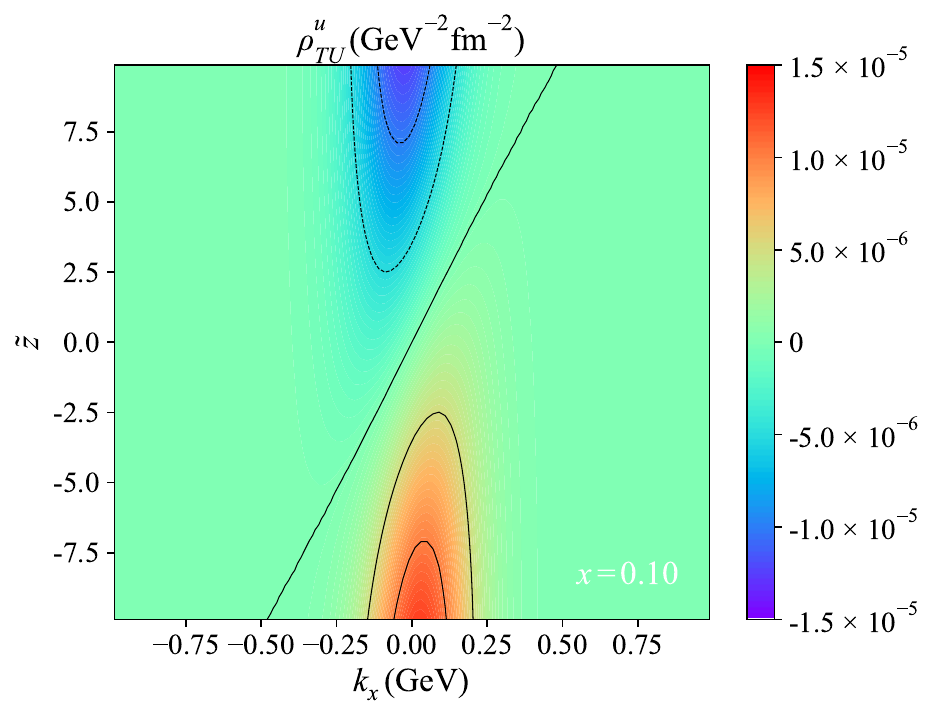}
	}
	\subfloat{
		\includegraphics[width=0.31\textwidth]{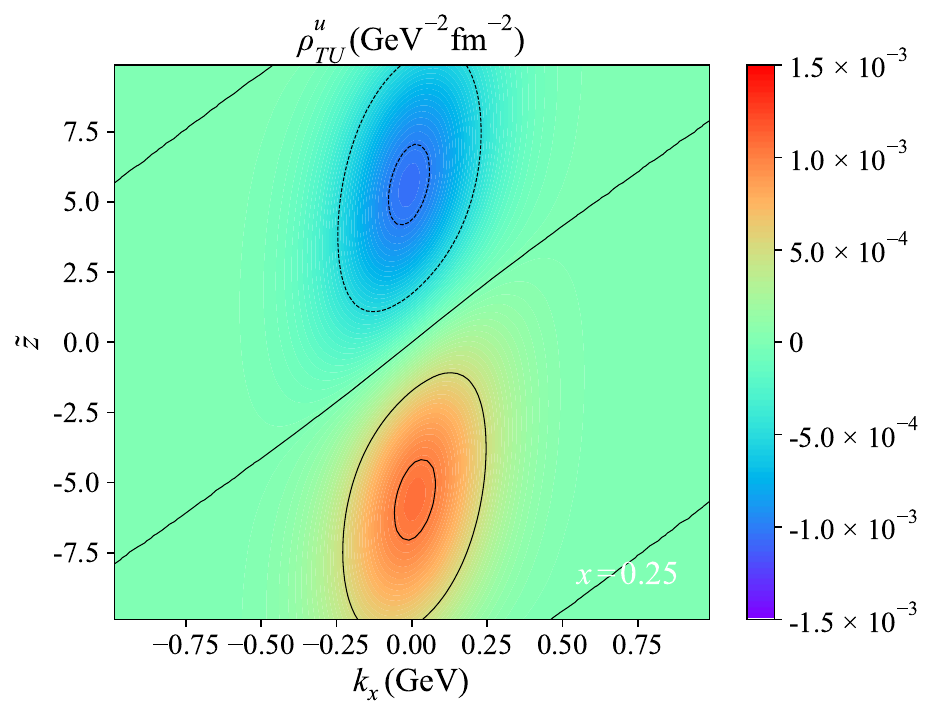}
	}
	\subfloat{
		\includegraphics[width=0.31\textwidth]{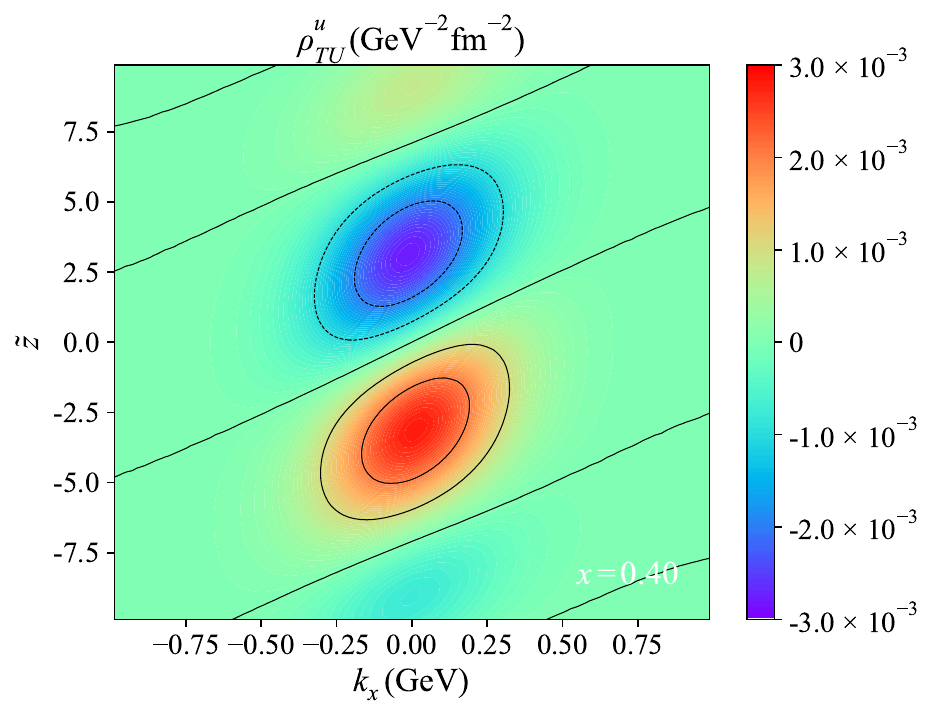}
	}\\
	\subfloat{
		\includegraphics[width=0.31\textwidth]{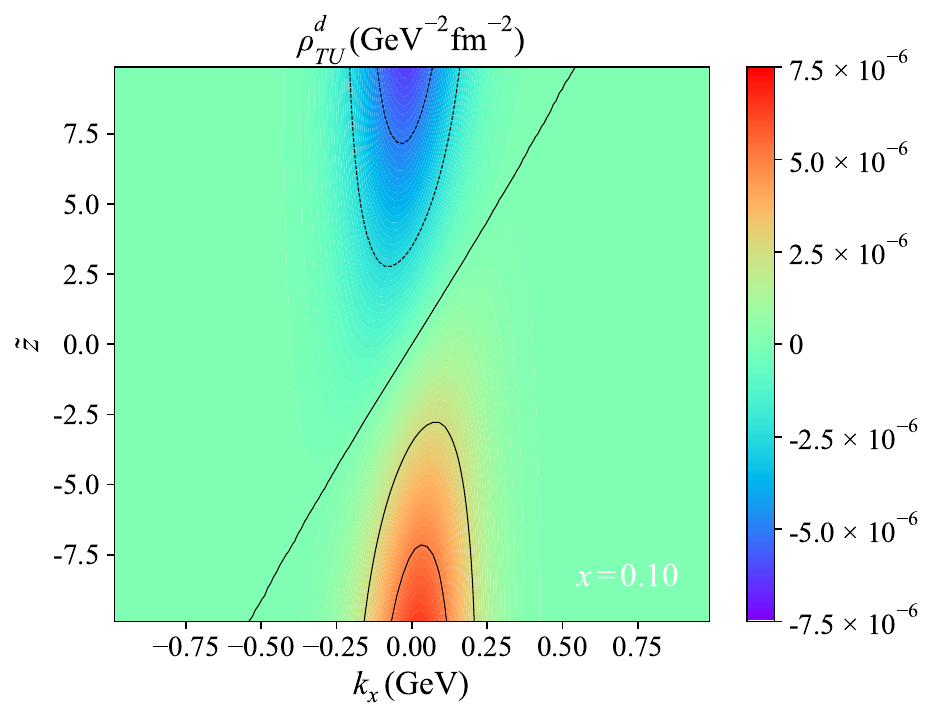}
	}
	\subfloat{
		\includegraphics[width=0.31\textwidth]{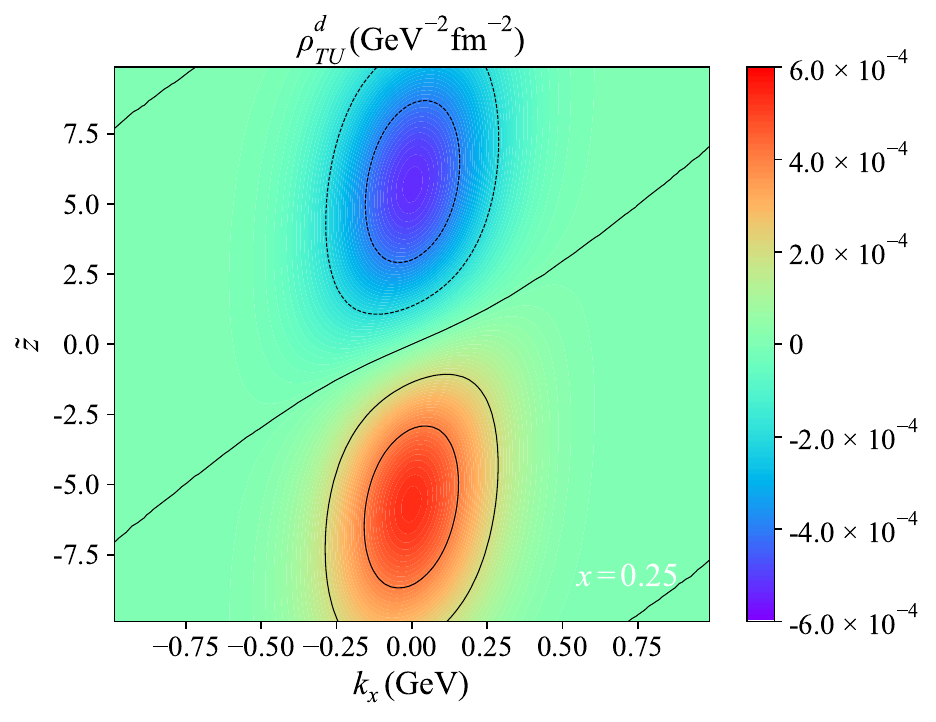}
	}
	\subfloat{
		\includegraphics[width=0.31\textwidth]{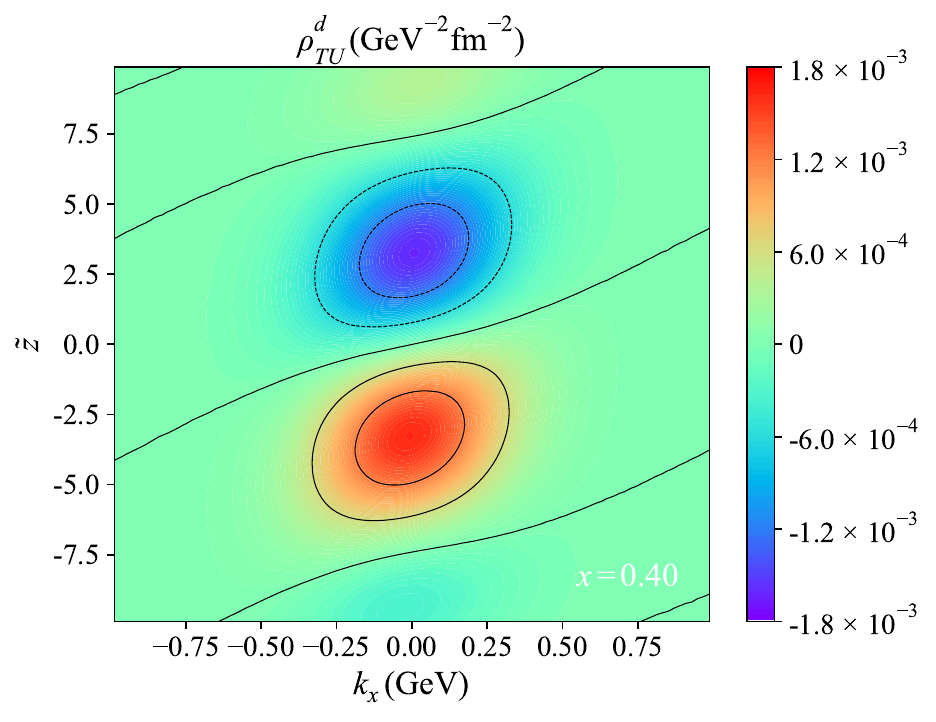}
	}
	\caption{Six-dimensional transverse-unpolarized light-front Wigner distribution $\rho_{\mathrm{TU}}\left(\tilde{z},x,\boldsymbol{b}_{\perp}, \boldsymbol{k}_{\perp}\right)$ for $u$ quark (upper panels) and $d$ quark (lower panels). The figure presents the Wigner distributions in the $\tilde{z}-k_x$ plane, with the transverse coordinate fixed at $\boldsymbol{b}_{\perp}=0.4\,\mathrm{GeV}^{-1}\boldsymbol{\hat{e}}_x$ (where $\boldsymbol{\hat{e}}_x$ is the unit vector along the $x$-axis) and the transverse momentum component fixed at $k_y=0.3\,\mathrm{GeV}$. The three columns correspond to $x=0.10$, $x=0.25$, and $x=0.40$.}
	\label{6DProtonTUudzkx}
\end{figure}

\begin{figure}[htbp]
	\centering
	\subfloat{
		\includegraphics[width=0.31\textwidth]{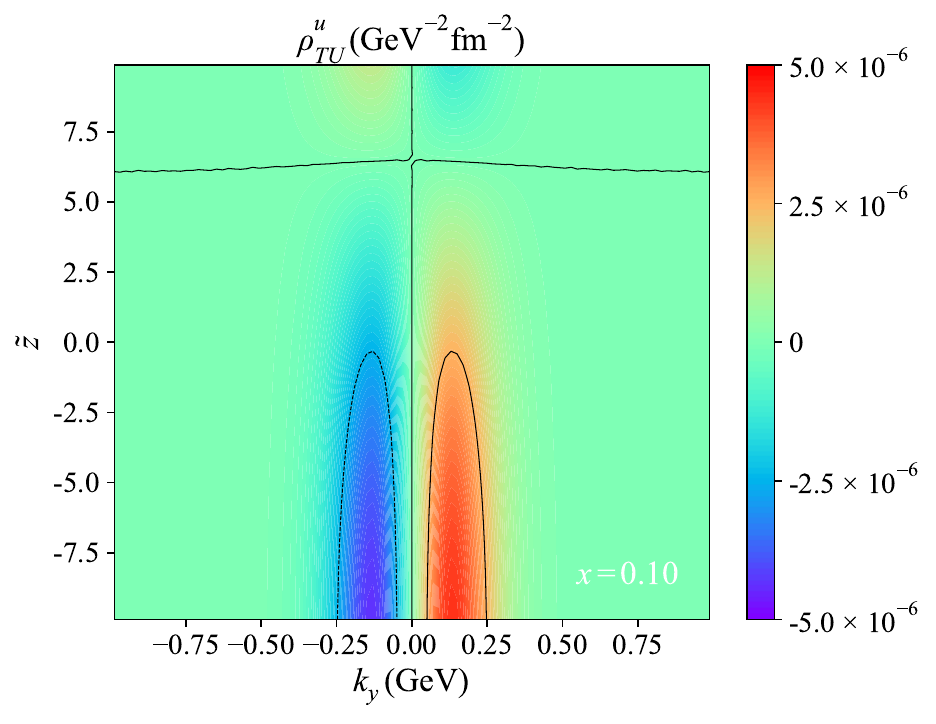}
	}
	\subfloat{
		\includegraphics[width=0.31\textwidth]{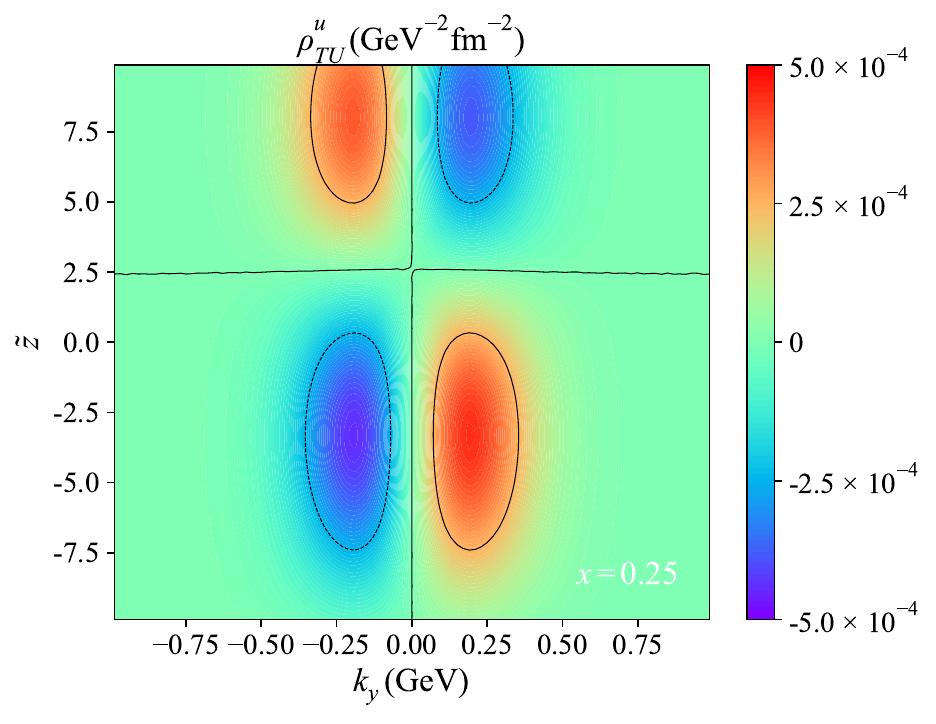}
	}
	\subfloat{
		\includegraphics[width=0.31\textwidth]{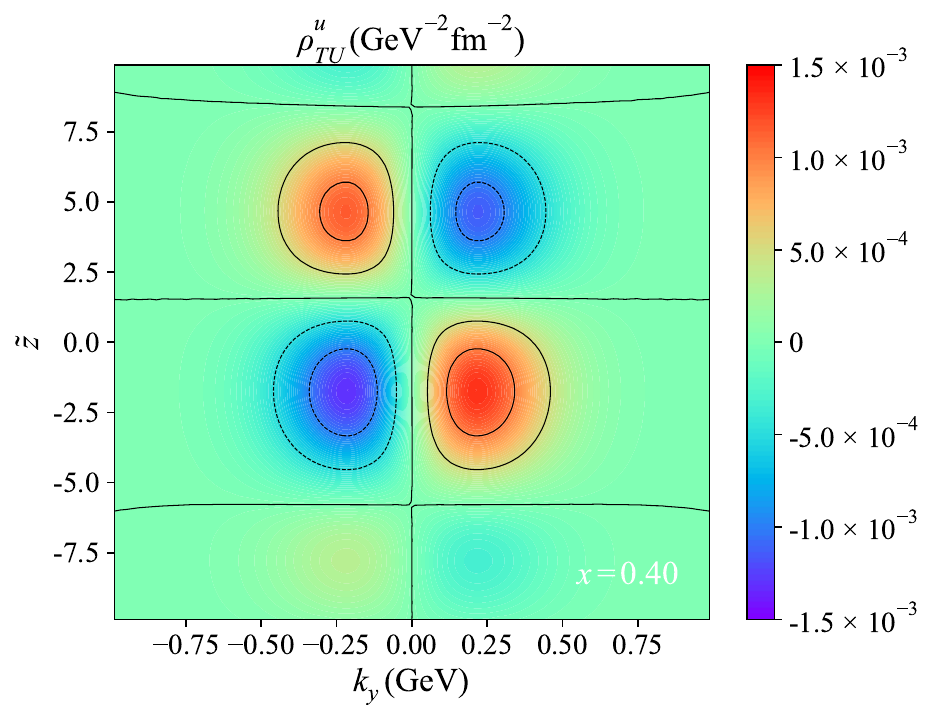}
	}\\
	\subfloat{
		\includegraphics[width=0.31\textwidth]{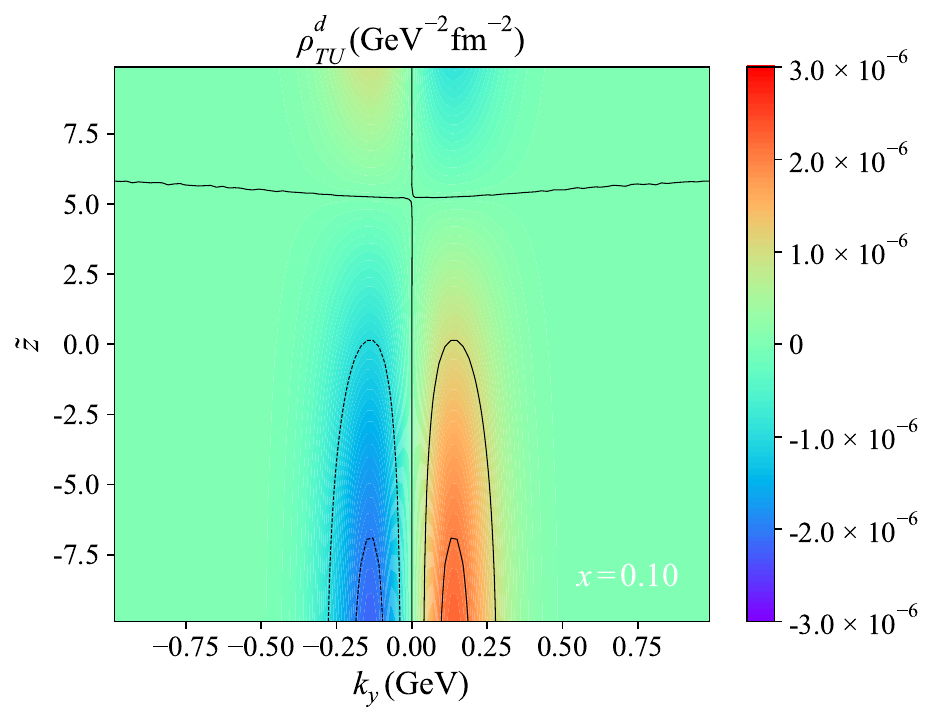}
	}
	\subfloat{
		\includegraphics[width=0.31\textwidth]{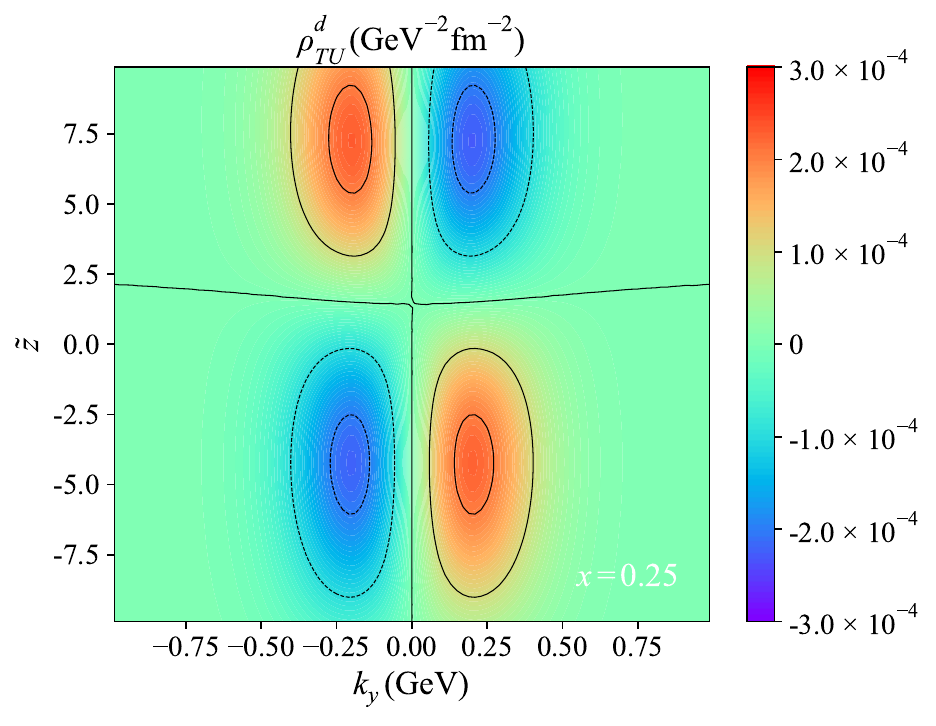}
	}
	\subfloat{
		\includegraphics[width=0.31\textwidth]{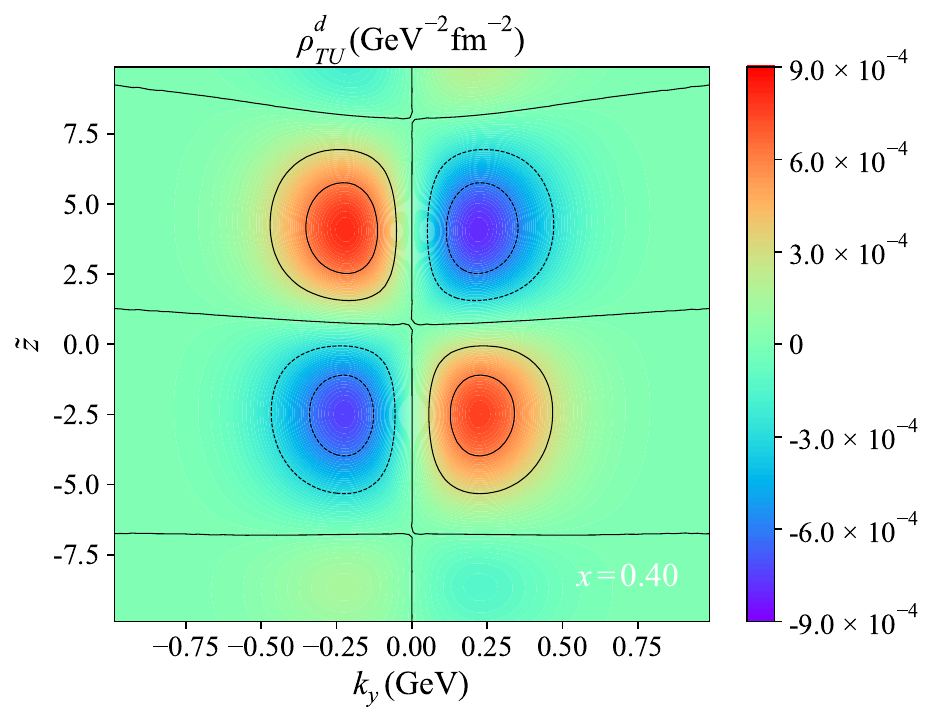}
	}
	\caption{Six-dimensional transverse-unpolarized light-front Wigner distribution $\rho_{\mathrm{TU}}\left(\tilde{z},x,\boldsymbol{b}_{\perp}, \boldsymbol{k}_{\perp}\right)$ for $u$ quark (upper panels) and $d$ quark (lower panels). The figure presents the Wigner distributions in the $\tilde{z}-k_y$ plane, with the transverse coordinate fixed at $\boldsymbol{b}_{\perp}=0.4\,\mathrm{GeV}^{-1}\boldsymbol{\hat{e}}_x$ (where $\boldsymbol{\hat{e}}_x$ is the unit vector along the $x$-axis) and the transverse momentum component fixed at $k_x=0.3\,\mathrm{GeV}$. The three columns correspond to $x=0.10$, $x=0.25$, and $x=0.40$.}
	\label{6DProtonTUudzky}
\end{figure}

In Figs.~\ref{6DProtonTUudzbx}--\ref{6DProtonTUudzky}, % In Fig.~\ref{6DProtonTUudzbx}, Fig.~\ref{6DProtonTUudzby}, Fig.~\ref{6DProtonTUudzkx} and Fig.~\ref{6DProtonTUudzky}, 
we plot the six-dimensional transverse-unpolarized light-front Wigner distribution $\rho_{\mathrm{TU}}\left(\tilde{z},x,\boldsymbol{b}_{\perp}, \boldsymbol{k}_{\perp}\right)$ for the $u$ and $d$ quarks of the proton, displayed in the $\tilde{z}-b_x$, $\tilde{z}-b_y$, $\tilde{z}-k_x$, and $\tilde{z}-k_y$ subspaces, respectively. The six-dimensional transverse-unpolarized light-front Wigner distributions characterize an unpolarized quark in a transverse-polarized proton, providing insight into the correlation between the quark transverse distribution and the proton transverse spin. The numerical results are shown for fixed values of the transverse momentum $\boldsymbol{k}_{\perp}$ or the transverse coordinate $\boldsymbol{b}_{\perp}$, and the longitudinal momentum fraction $x$ is set at $x = 0.10$, $x = 0.25$, and $x = 0.40$ in the first, second, and third columns, respectively. The proton polarization introduces a privileged direction in the transverse plane, which is along the $x$-direction.

For Fig.~\ref{6DProtonTUudzbx} and Fig.~\ref{6DProtonTUudzkx}, the six-dimensional transverse-unpolarized light-front Wigner distribution exhibits centrosymmetry about the origin in the $\tilde{z}-b_x$ and $\tilde{z}-k_x$ subspaces, with the maximum values located at the center of the coordinate system for each fixed value of $x$. \new{This symmetry reflects the underlying rotational invariance when viewed along the polarization axis. In contrast, Fig.~\ref{6DProtonTUudzby} and Fig.~\ref{6DProtonTUudzky} display a distinct dipole-symmetric pattern about $b_x=0$ in the $\tilde{z}-b_y$ and $\tilde{z}-k_y$ subspaces, with the dipole strength showing a pronounced $x$-dependence. Notably, the structure in Fig.~\ref{6DProtonTUudzky} displays a more pronounced dipole symmetry for fixed values of $k_x$ around $\tilde{z} = 5$. Additionally, the non-positive properties of the distribution are preserved, as expected for the Wigner distribution.}

\new{The flavor dependence of these distributions reveals important physical differences between $u$ and $d$ quarks. The $u$ quark distributions show stronger dipole magnitudes compared to $d$ quarks, consistent with their dominant coupling to scalar diquarks in the proton spin-flavor wavefunction. Notably, the $d$ quark distributions exhibit negative dipole lobes in certain projections, indicating opposite spin-orbit correlations compared to $u$ quarks. These features are particularly visible in the $\tilde{z}-k_y$ subspace, where the distribution shows a quadrupole-like structure at higher $x$ values, suggesting more complex orbital dynamics in the valence quark regime. The non-positive definite nature of these distributions is preserved throughout, as expected for genuine Wigner functions that encode quantum interference effects.}

Through integration over three-dimensional coordinate space, the six-dimensional transverse-unpolarized light-front Wigner distribution can be related to a naive T-odd distribution at the TMD limit, i.e., the Sivers function $f^{\perp}_{1T}$, whereas the distribution function can also be associated with the IPDs $H$ and $E$ as well as other distributions at the IPD limit. It follows that the Sivers TMD corresponds to the T-odd part, while the IPDs $H$ and $E$ are linked to the T-even part. Consequently, the TMD and IPD limits correspond to different components of the six-dimensional transverse-unpolarized Wigner distribution. Recent theoretical and experimental studies~\cite{JeffersonLabHallA:2013mjr} have indicated a potential correlation between the Sivers function $f^{\perp}_{1T}$ and the GPD $E$, as suggested by some model calculations~\cite{Burkardt:2003je,Bacchetta:2011gx}.

\new{In our previous theoretical analysis, we found that the transverse-unpolarized light-front Wigner distribution vanishes at the TMD limit. This limitation could be attributed to the omission of the Wilson line effect, which may be crucial for capturing the T-odd dynamics~\cite{Lu:2012gu}.} Further exploration of non-trivial gauge links may reveal a deeper connection between the Sivers TMD and GPDs. Based on the previous discussion, this distribution becomes zero when the quark transverse coordinate aligns with the proton polarization, indicating no correlation between the quark transverse spin and the proton transverse spin. This result is independent of the quark transverse momentum direction. New discoveries may emerge if additional non-trivial Wilson lines are incorporated into the theoretical framework.

%\subsubsection{u Quark}

%\subsubsection{d Quark}

\subsection{Transverse Wigner distribution}
\label{sub:TT}

\new{In the following we discuss the six-dimensional transverse light-front Wigner distribution, focusing on the two degrees of freedom in the transverse polarization of both quarks and protons. These distributions provide critical insights into the spin-spin correlations between transversely polarized quarks and protons, which the relative orientation of spins reveals fundamentally different aspects of hadronic structure. We systematically examine two distinct polarization configurations: (1) parallel alignment ($\rho_{\mathrm{TT}}$) where both quark and proton spins are oriented along the $x$-direction (Sec.~\ref{sub:TT}), and (2) orthogonal alignment ($\rho_{\mathrm{TT}}^{\perp}$) where they are polarized along perpendicular axes (Sec.~\ref{sub:TTt}). This decomposition captures the complete tensor structure of transverse spin correlations in the light-front formalism.}

\begin{figure}[htbp]
	\centering
	\subfloat{
		\includegraphics[width=0.31\textwidth]{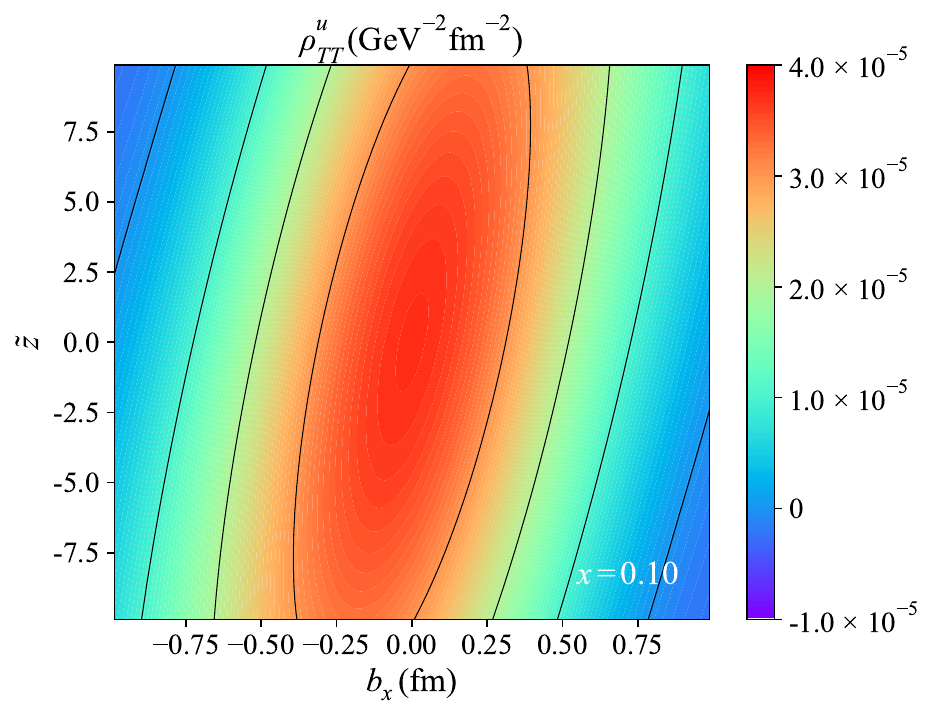}
	}
	\subfloat{
		\includegraphics[width=0.31\textwidth]{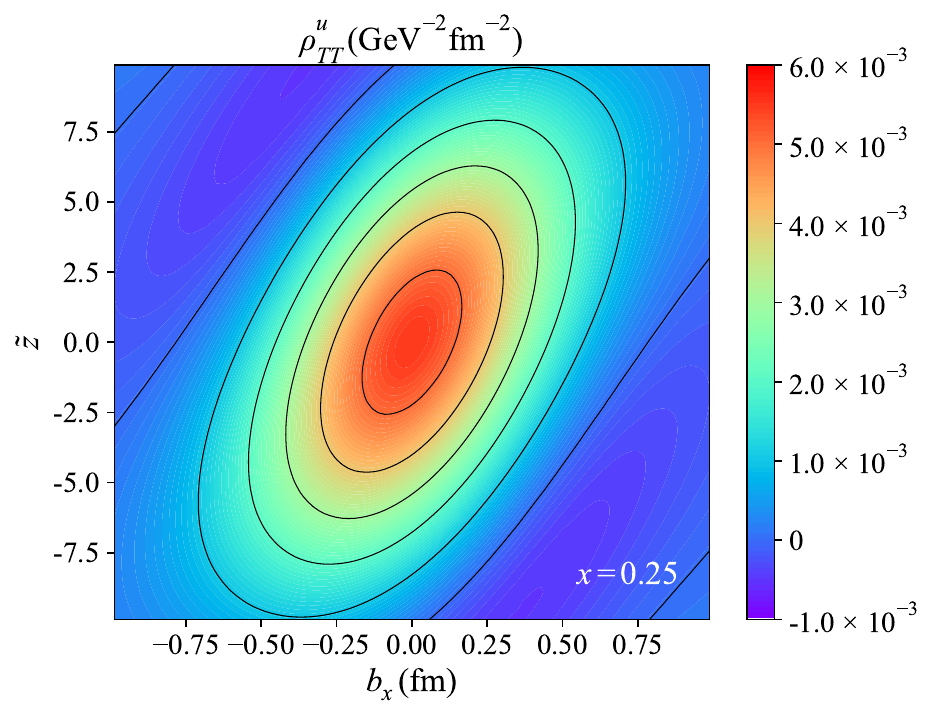}
	}
	\subfloat{
		\includegraphics[width=0.31\textwidth]{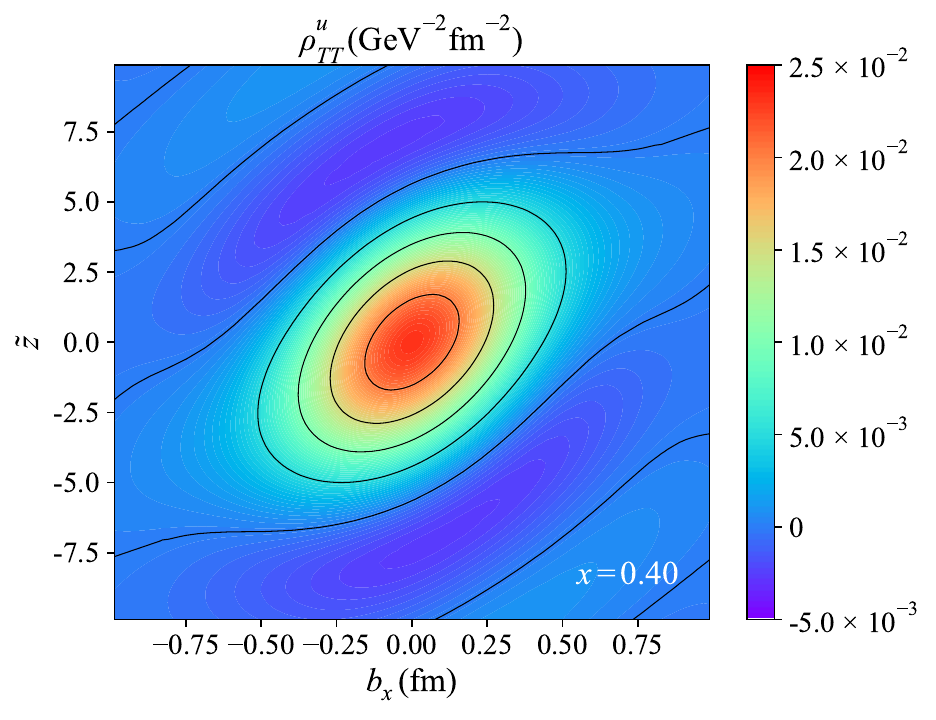}
	}\\
	\subfloat{
		\includegraphics[width=0.31\textwidth]{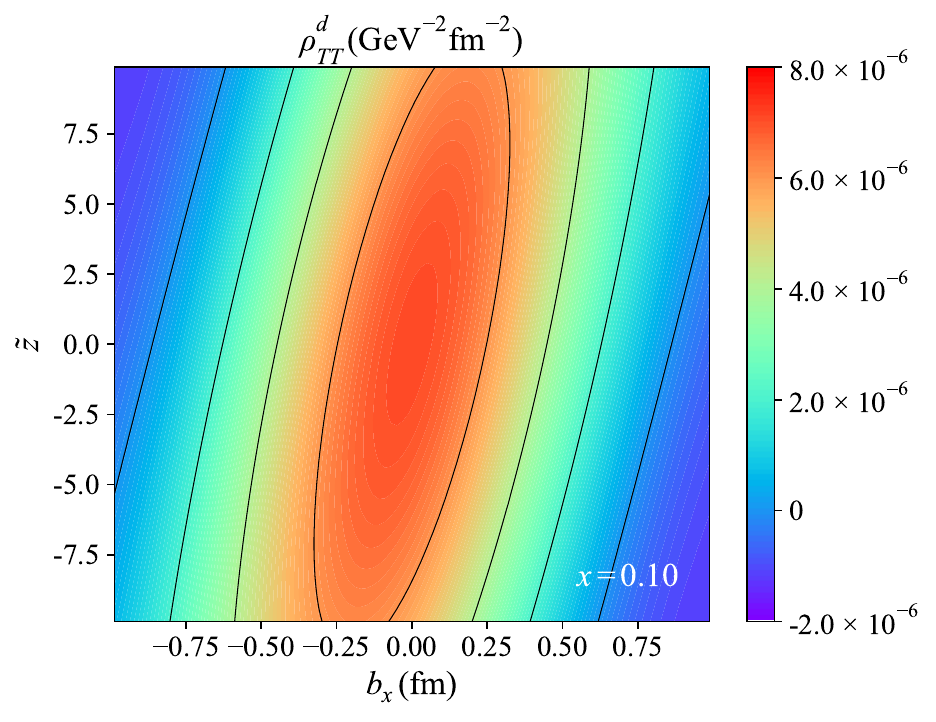}
	}
	\subfloat{
		\includegraphics[width=0.31\textwidth]{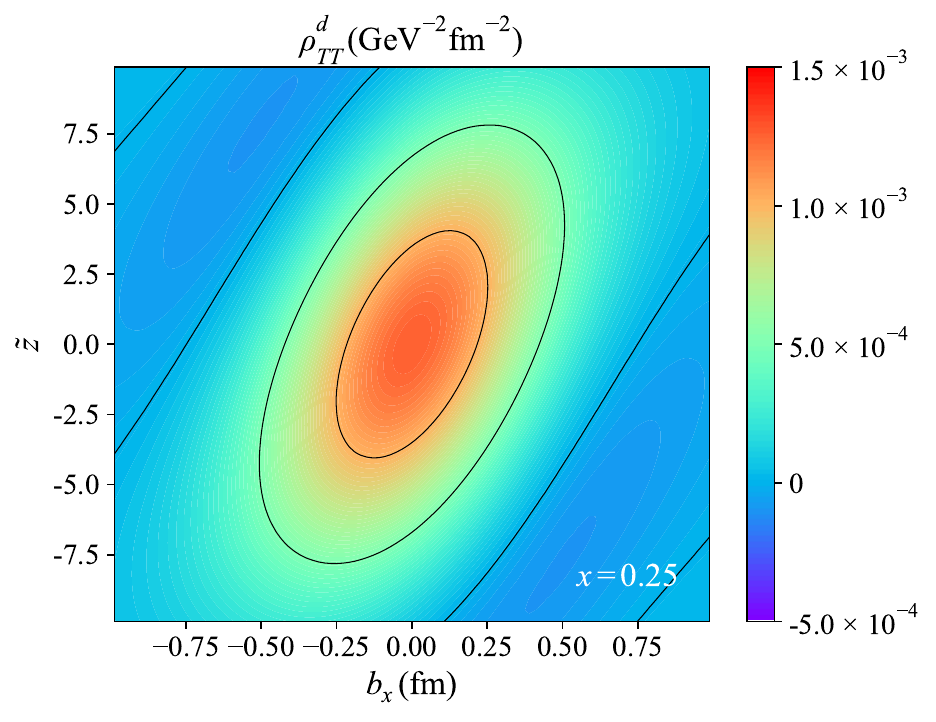}
	}
	\subfloat{
		\includegraphics[width=0.31\textwidth]{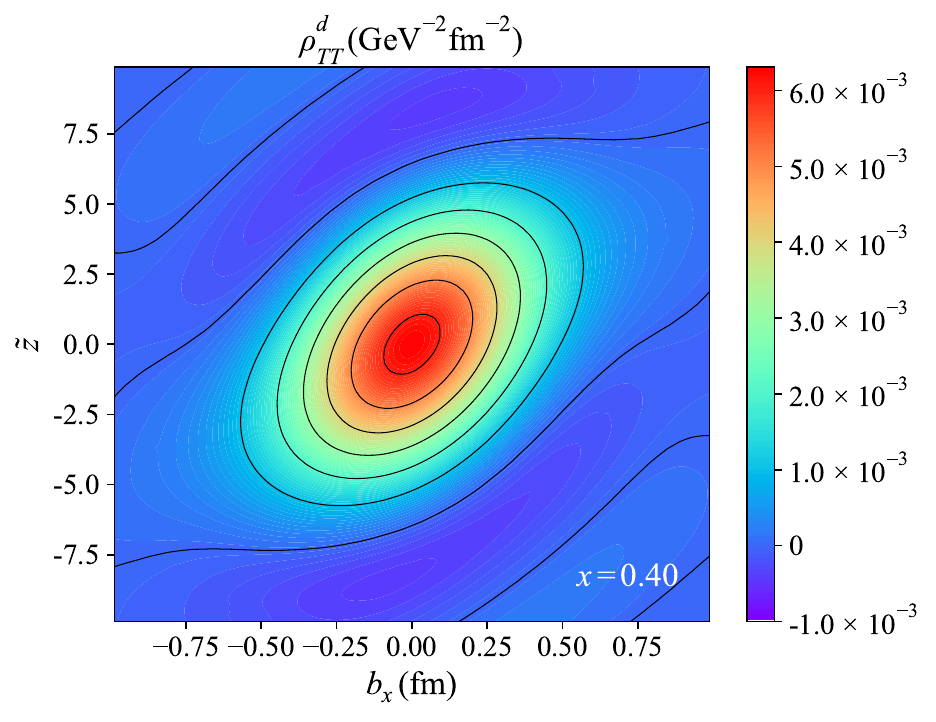}
	}
	\caption{Six-dimensional transverse light-front Wigner distribution $\rho_{\mathrm{TT}}\left(\tilde{z},x,\boldsymbol{b}_{\perp}, \boldsymbol{k}_{\perp}\right)$ for $u$ quark (upper panels) and $d$ quark (lower panels). The figure presents the Wigner distribution in the $\tilde{z}-b_x$ plane, with the transverse momentum fixed at $\boldsymbol{k}_{\perp}=0.3\,\mathrm{GeV}\boldsymbol{\hat{e}}_x$ (where $\boldsymbol{\hat{e}}_x$ is the unit vector in the $x$-direction) and the transverse coordinate component fixed at $b_y=0.4\,\mathrm{GeV}^{-1}$. The three columns correspond to $x=0.10$, $x=0.25$, and $x=0.40$.}
	\label{6DProtonTTudzbx}
\end{figure}

\begin{figure}[htbp]
	\centering
	\subfloat{
		\includegraphics[width=0.31\textwidth]{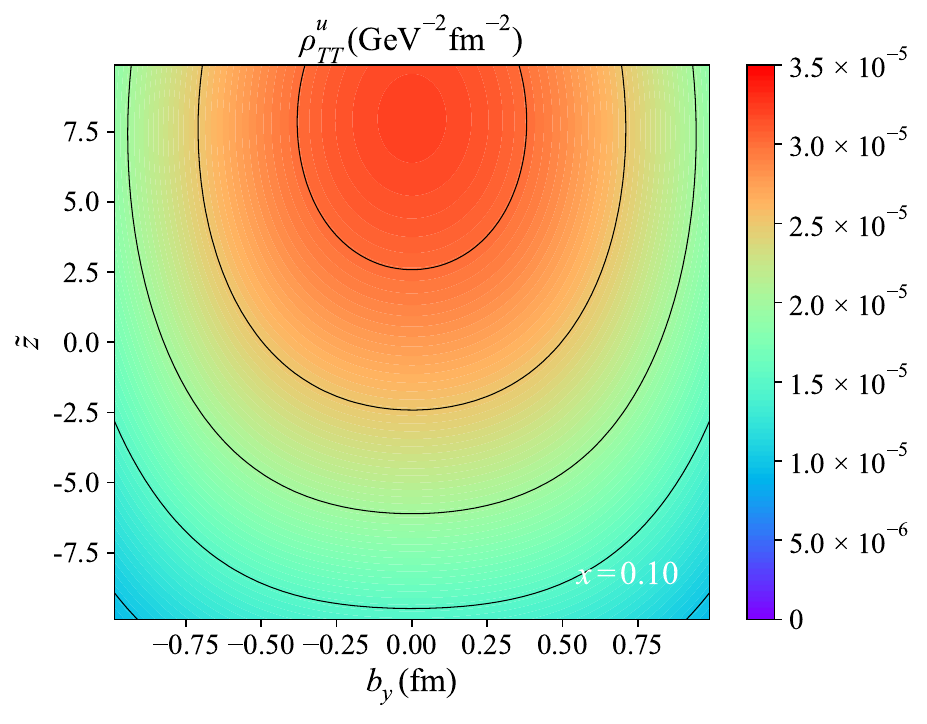}
	}
	\subfloat{
		\includegraphics[width=0.31\textwidth]{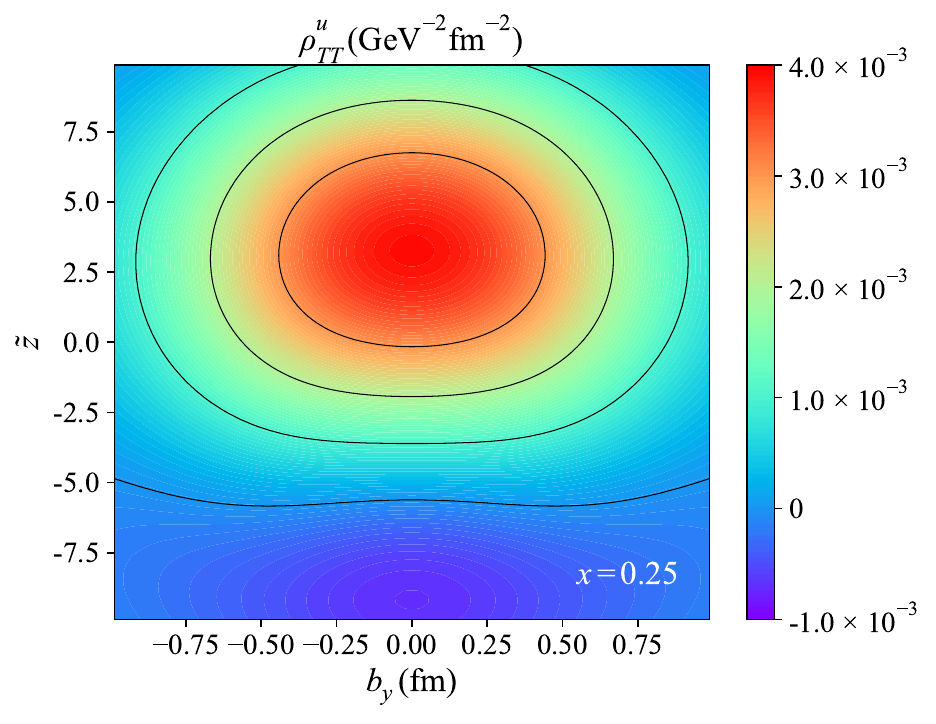}
	}
	\subfloat{
		\includegraphics[width=0.31\textwidth]{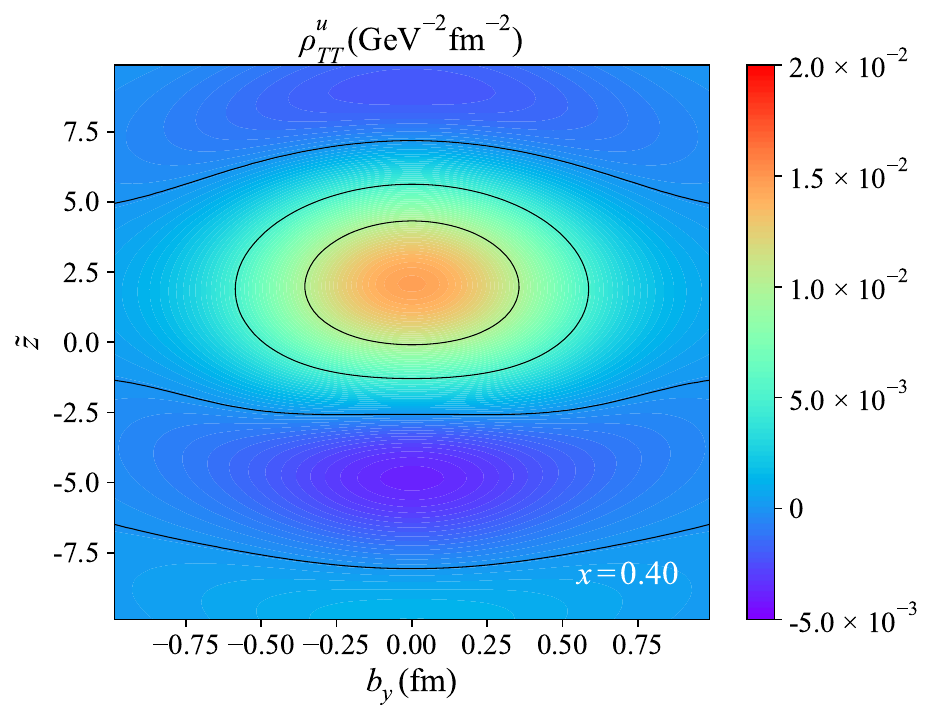}
	}\\
	\subfloat{
		\includegraphics[width=0.31\textwidth]{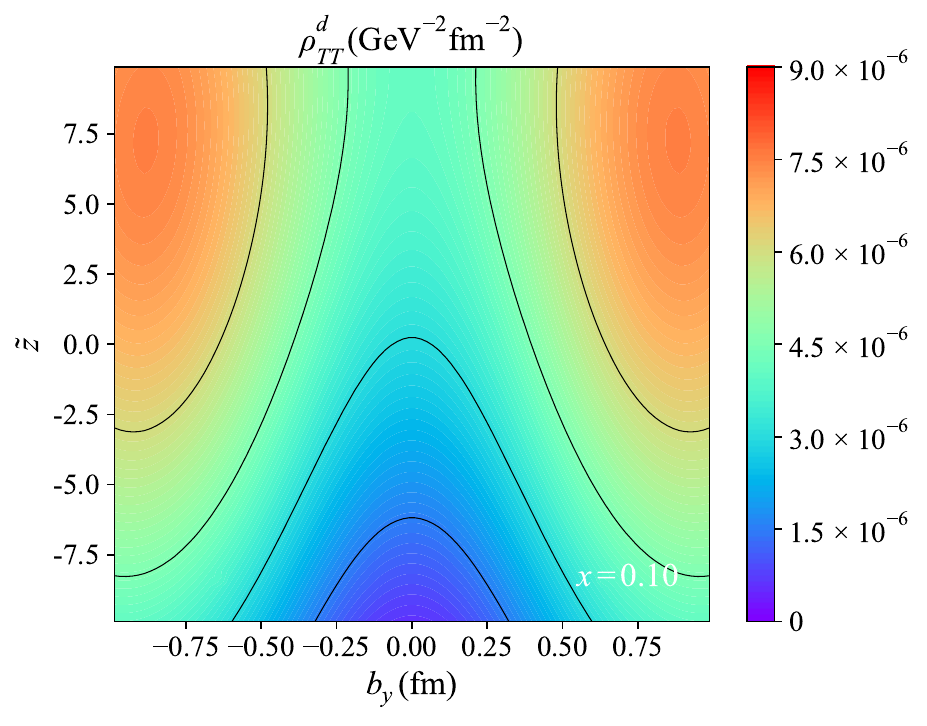}
	}
	\subfloat{
		\includegraphics[width=0.31\textwidth]{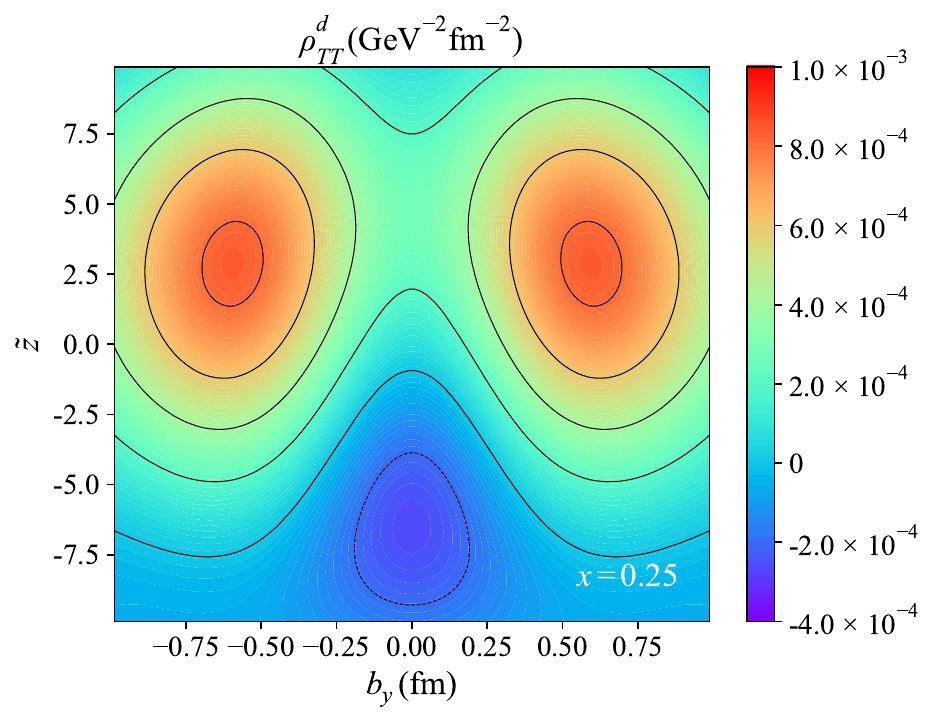}
	}
	\subfloat{
		\includegraphics[width=0.31\textwidth]{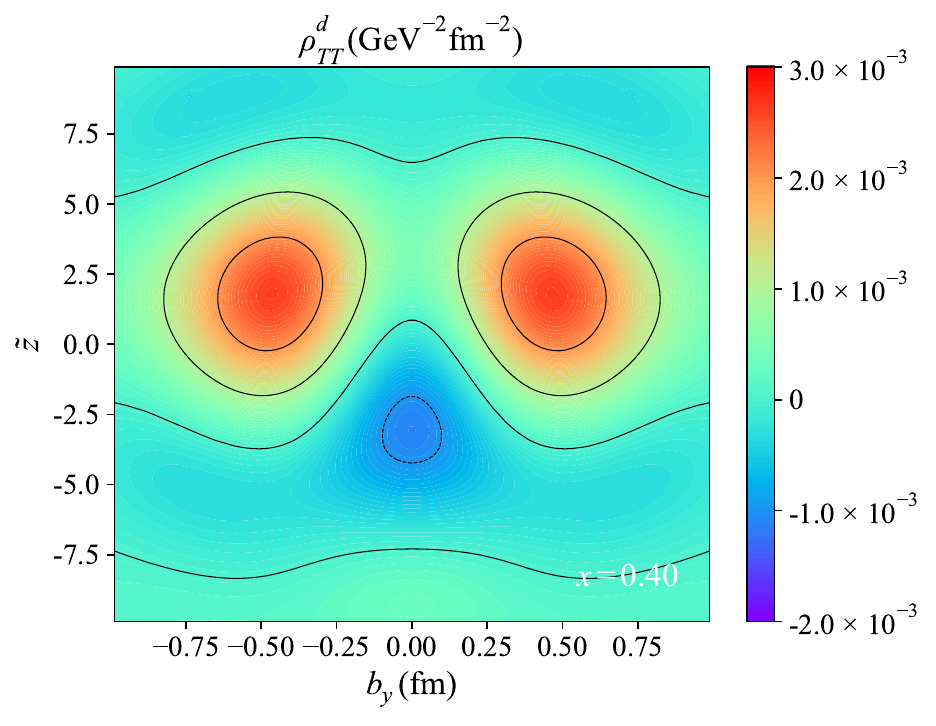}
	}
	\caption{Six-dimensional transverse light-front Wigner distribution $\rho_{\mathrm{TT}}\left(\tilde{z},x,\boldsymbol{b}_{\perp}, \boldsymbol{k}_{\perp}\right)$ for $u$ quark (upper panels) and $d$ quark (lower panels). The figure presents the Wigner distributions in the $\tilde{z}-b_y$ plane, with the transverse momentum fixed at $\boldsymbol{k}_{\perp}=0.3\,\mathrm{GeV}\boldsymbol{\hat{e}}_x$ (where $\boldsymbol{\hat{e}}_x$ is the unit vector in the $x$-direction) and the transverse coordinate component fixed at $b_x=0.4\,\mathrm{GeV}^{-1}$. The three columns correspond to $x=0.10$, $x=0.25$, and $x=0.40$.}
	\label{6DProtonTTudzby}
\end{figure}

\begin{figure}[htbp]
	\centering
	\subfloat{
		\includegraphics[width=0.31\textwidth]{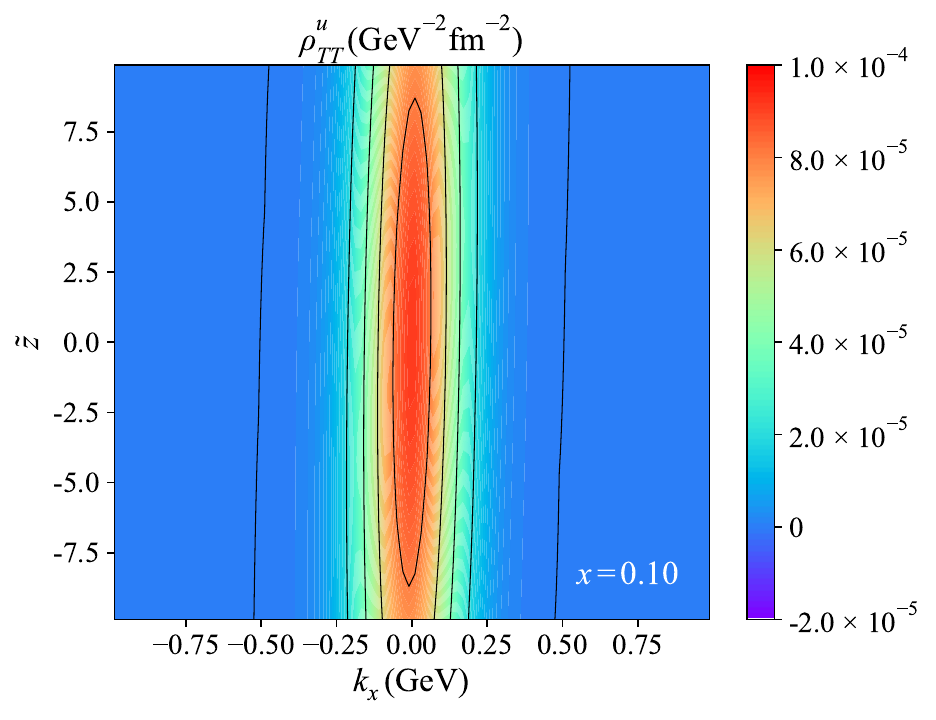}
	}
	\subfloat{
		\includegraphics[width=0.31\textwidth]{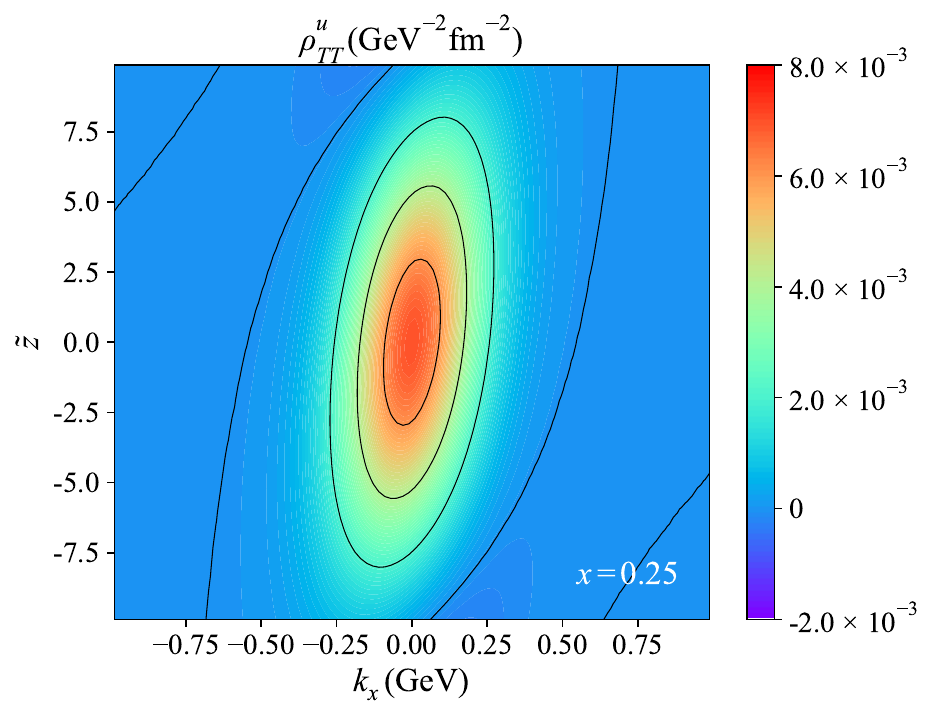}
	}
	\subfloat{
		\includegraphics[width=0.31\textwidth]{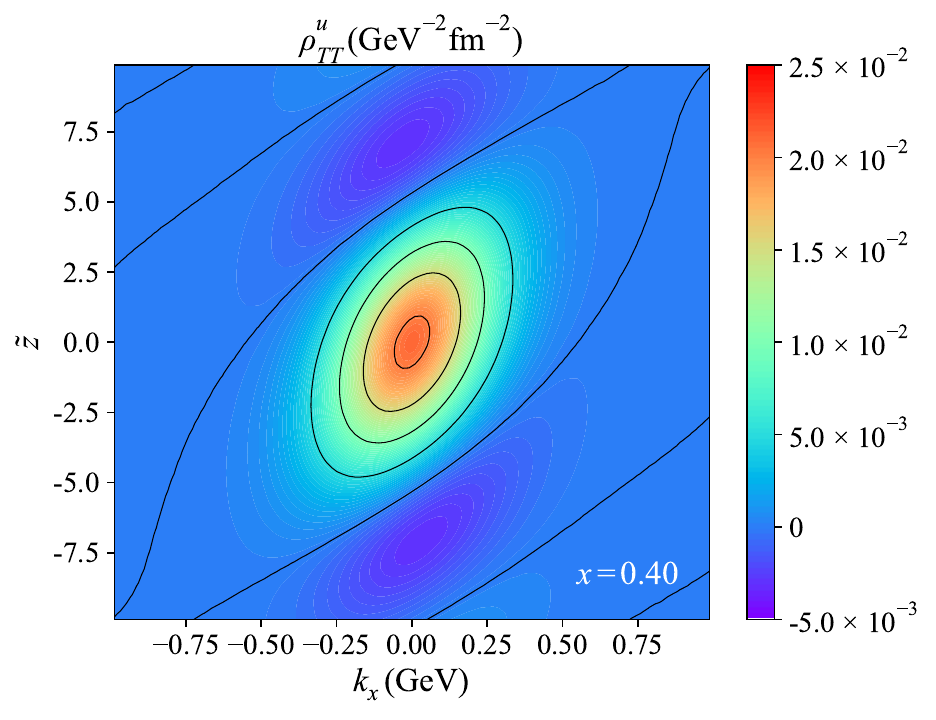}
	}\\
	\subfloat{
		\includegraphics[width=0.31\textwidth]{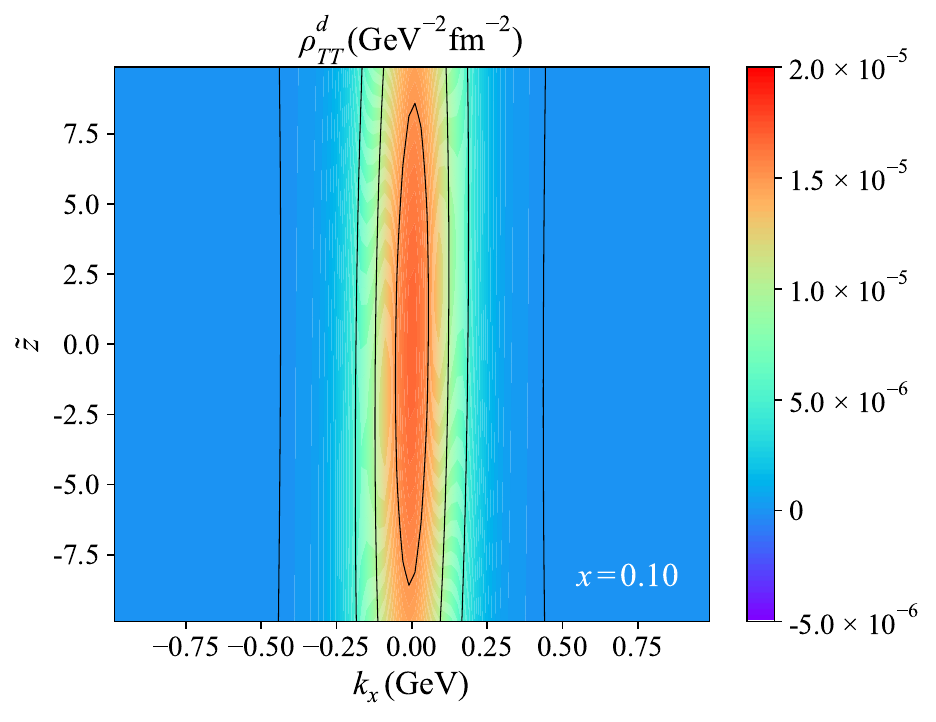}
	}
	\subfloat{
		\includegraphics[width=0.31\textwidth]{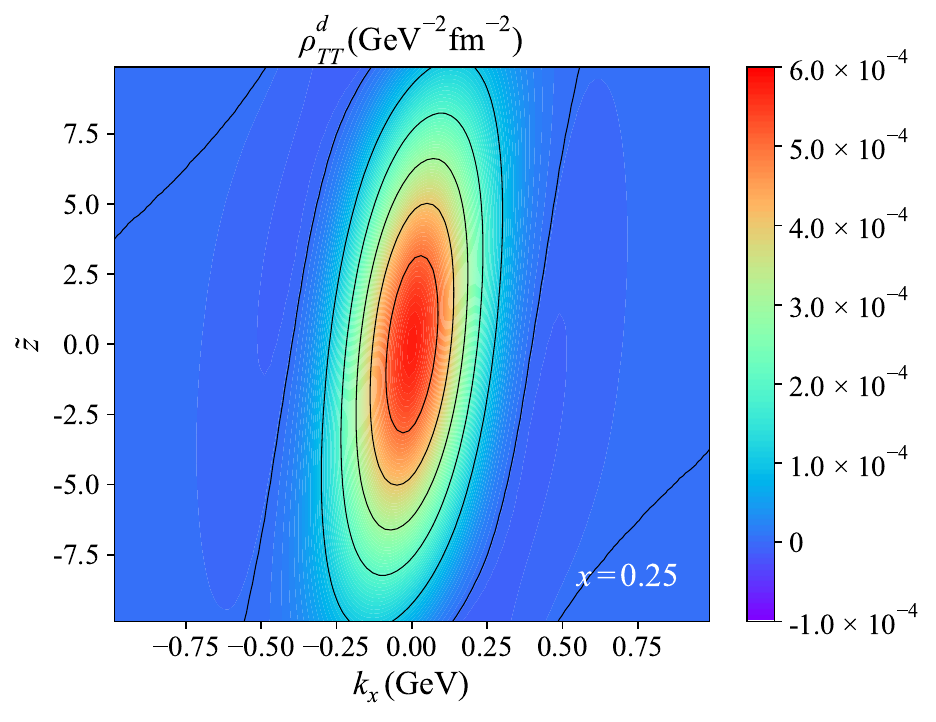}
	}
	\subfloat{
		\includegraphics[width=0.31\textwidth]{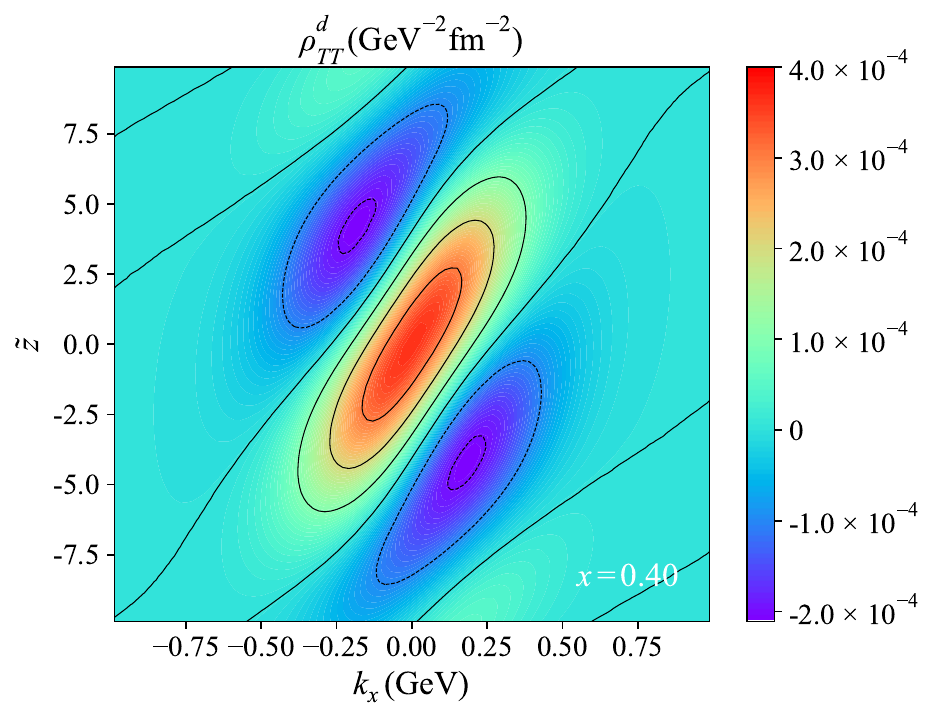}
	}
	\caption{Six-dimensional transverse light-front Wigner distribution $\rho_{\mathrm{TT}}\left(\tilde{z},x,\boldsymbol{b}_{\perp}, \boldsymbol{k}_{\perp}\right)$ for $u$ quark (upper panels) and $d$ quark (lower panels). The figure presents the Wigner distributions in the $\tilde{z}-k_x$ plane, with the transverse coordinate fixed at $\boldsymbol{b}_{\perp}=0.4\,\mathrm{GeV}^{-1}\boldsymbol{\hat{e}}_x$ (where $\boldsymbol{\hat{e}}_x$ is the unit vector along the $x$-axis) and the transverse momentum component fixed at $k_y=0.3\,\mathrm{GeV}$. The three columns correspond to $x=0.10$, $x=0.25$, and $x=0.40$.}
	\label{6DProtonTTudzkx}
\end{figure}

\begin{figure}[htbp]
	\centering
	\subfloat{
		\includegraphics[width=0.31\textwidth]{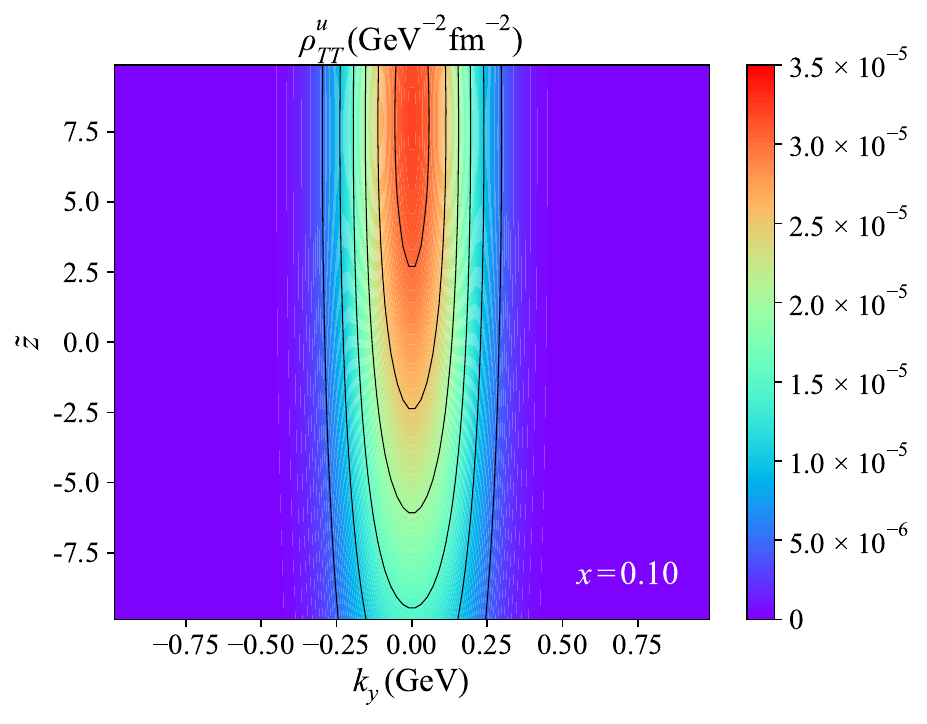}
	}
	\subfloat{
		\includegraphics[width=0.31\textwidth]{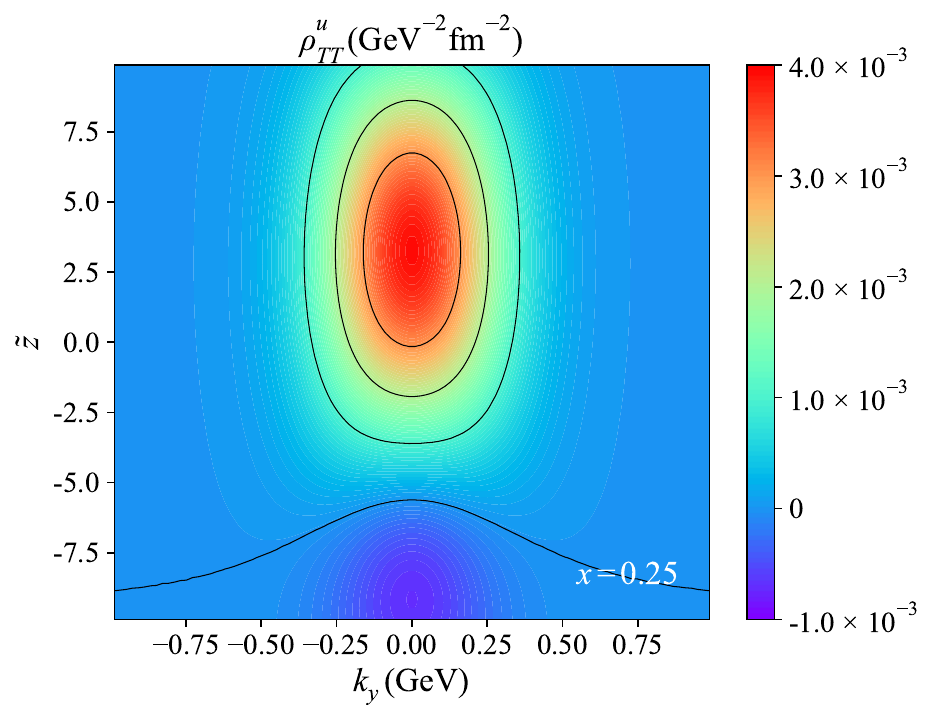}
	}
	\subfloat{
		\includegraphics[width=0.31\textwidth]{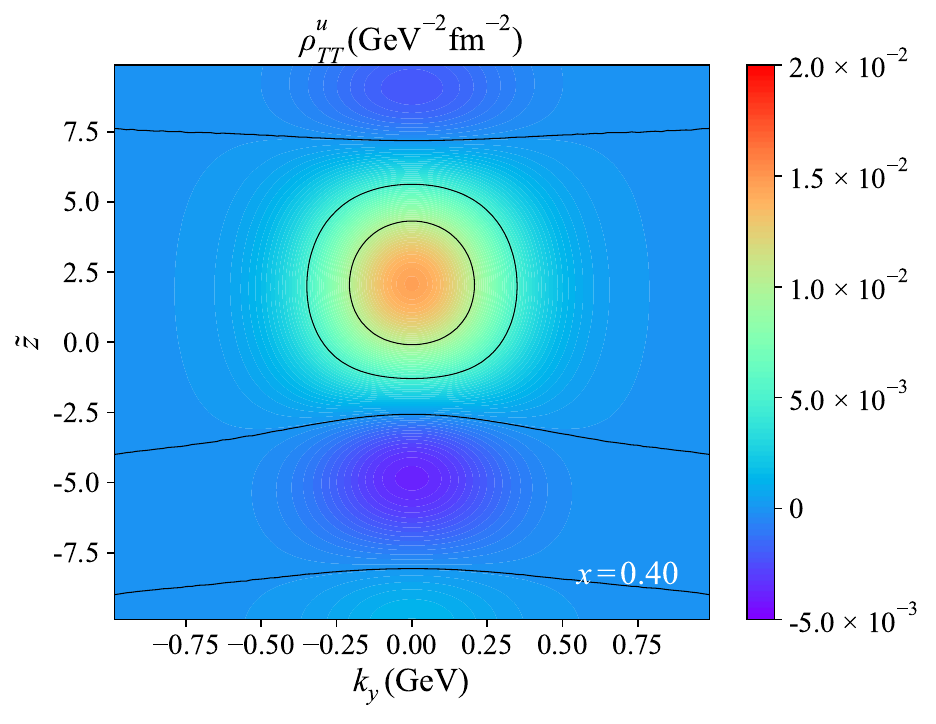}
	}\\
	\subfloat{
		\includegraphics[width=0.31\textwidth]{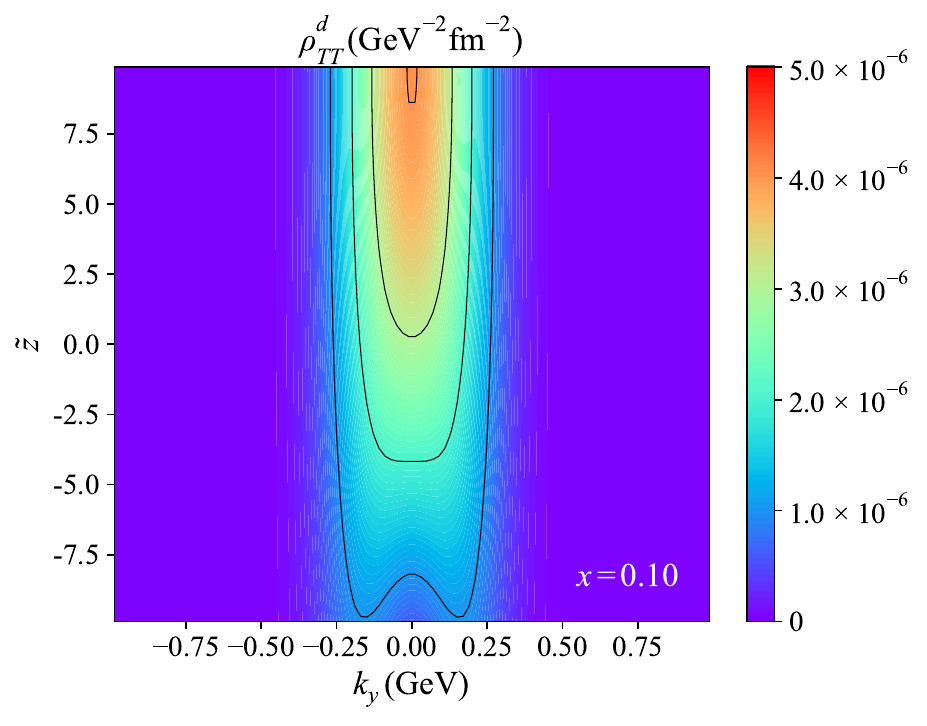}
	}
	\subfloat{
		\includegraphics[width=0.31\textwidth]{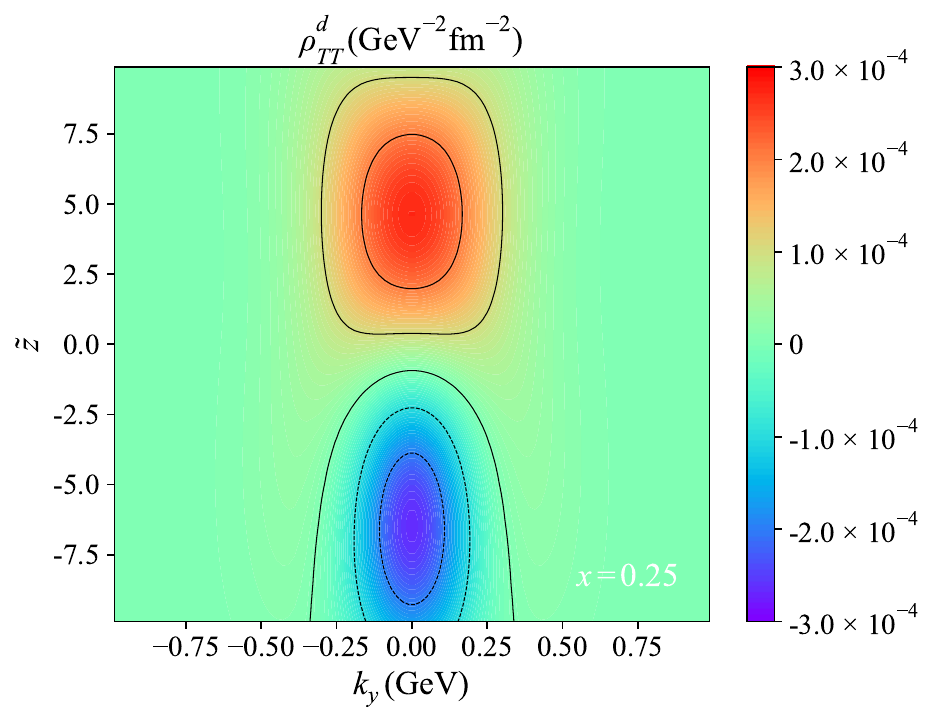}
	}
	\subfloat{
		\includegraphics[width=0.31\textwidth]{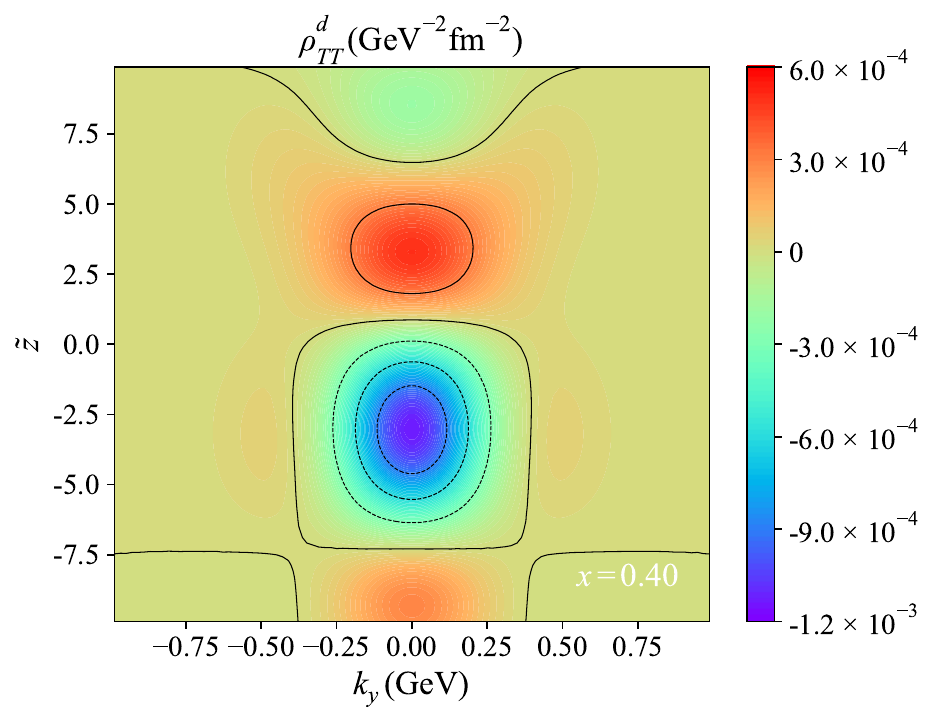}
	}
	\caption{Six-dimensional transverse LF Wigner distribution $\rho_{\mathrm{TT}}\left(\tilde{z},x,\boldsymbol{b}_{\perp}, \boldsymbol{k}_{\perp}\right)$ for $u$ quark (upper panels) and $d$ quark (lower panels). The figure presents the Wigner distributions in the $\tilde{z}-k_y$ plane, with the transverse coordinate fixed at $\boldsymbol{b}_{\perp}=0.4\,\mathrm{GeV}^{-1}\boldsymbol{\hat{e}}_x$ (where $\boldsymbol{\hat{e}}_x$ is the unit vector along the $x$-axis) and the transverse momentum component fixed at $k_x=0.3\,\mathrm{GeV}$. The three columns correspond to $x=0.10$, $x=0.25$, and $x=0.40$.}
	\label{6DProtonTTudzky}
\end{figure}

In Figs.~\ref{6DProtonTTudzbx}--\ref{6DProtonTTudzky}, % In Fig.~\ref{6DProtonTTudzbx}, Fig.~\ref{6DProtonTTudzby}, Fig.~\ref{6DProtonTTudzkx} and Fig.~\ref{6DProtonTTudzky}, 
we plot the six-dimensional transverse light-front Wigner distribution $\rho_{\mathrm{TT}}\left(\tilde{z},x,\boldsymbol{b}_{\perp}, \boldsymbol{k}_{\perp}\right)$ for the $u$ and $d$ quarks of the proton, displayed in the $\tilde{z}-b_x$, $\tilde{z}-b_y$, $\tilde{z}-k_x$, and $\tilde{z}-k_y$ subspaces, respectively. These six-dimensional transverse light-front Wigner distributions correspond to the scenario where both quark and proton are transverse-polarized in the same direction (referred to as the $x$ direction). The numerical results are shown for fixed values of the transverse momentum $\boldsymbol{k}_{\perp}$ or the transverse coordinate $\boldsymbol{b}_{\perp}$, and the longitudinal momentum fraction $x$ is set at $x = 0.10$, $x = 0.25$, and $x = 0.40$ in the first, second, and third columns, respectively.

Upon integrating over the three-dimensional transverse phase space, the six-dimensional transverse light-front Wigner distribution function can be reduced to the transverse distributions. In Fig.~\ref{6DProtonTTudzbx} and Fig.~\ref{6DProtonTTudzkx}, the distributions exhibit centrosymmetry about the origin in the $\tilde{z}-b_x$ and $\tilde{z}-k_x$ subspaces, with the maximum values located at the origin for each fixed $x$. In contrast, Fig.~\ref{6DProtonTTudzby} and Fig.~\ref{6DProtonTTudzky} exhibit an axisymmetric pattern with respect to $b_x=0$. It is noteworthy that the figure within the $\tilde{z}-b_x$ plane exhibits a dipole symmetry, which is particularly conspicuous for substantial values of $x$. \new{Besides, the flavor decomposition shows striking differences: $u$ quark distributions display about three times stronger modulation amplitudes than $d$ quarks, directly reflecting their dominant coupling to scalar diquarks in the proton wavefunction. Notably, the $d$ quark distribution develops a sign inversion in its quadrupole structure at $x = 0.40$ in Fig.~\ref{6DProtonTTudzky}, suggesting competing spin-orbit coupling mechanisms between flavors. These features emerge naturally from the Melosh-Wigner rotation formalism in our light-front model, which properly accounts for relativistic spin transformations.}

%\subsubsection{u Quark}

%\subsubsection{d Quark}

\subsection{Pretzelous Wigner distribution}
\label{sub:TTt}

\begin{figure}[htbp]
	\centering
	\subfloat{
		\includegraphics[width=0.31\textwidth]{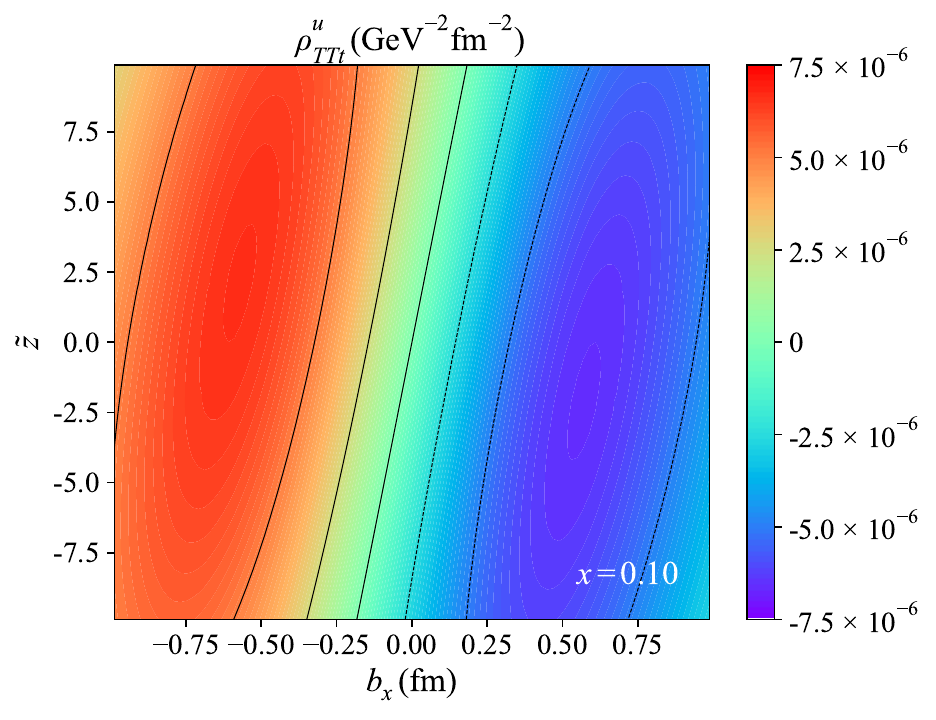}
	}
	\subfloat{
		\includegraphics[width=0.31\textwidth]{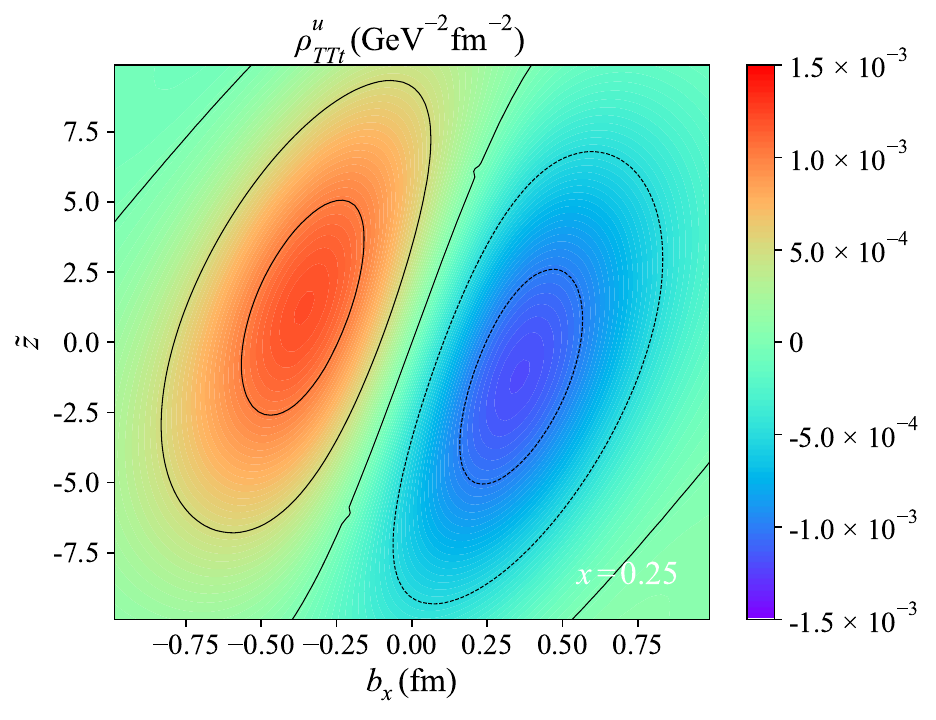}
	}
	\subfloat{
		\includegraphics[width=0.31\textwidth]{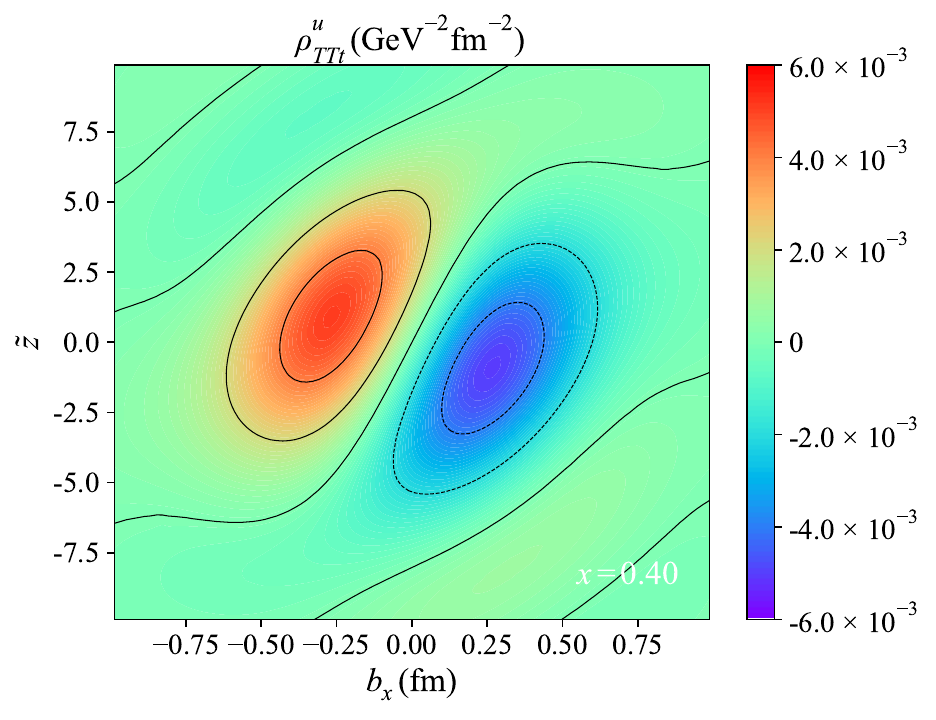}
	}\\
	\subfloat{
		\includegraphics[width=0.31\textwidth]{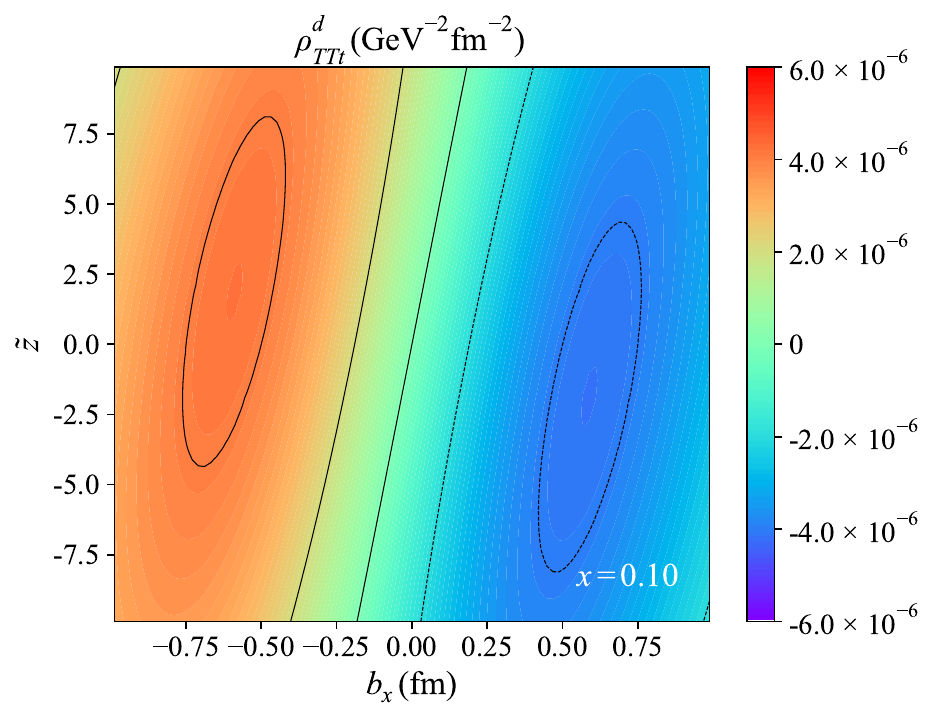}
	}
	\subfloat{
		\includegraphics[width=0.31\textwidth]{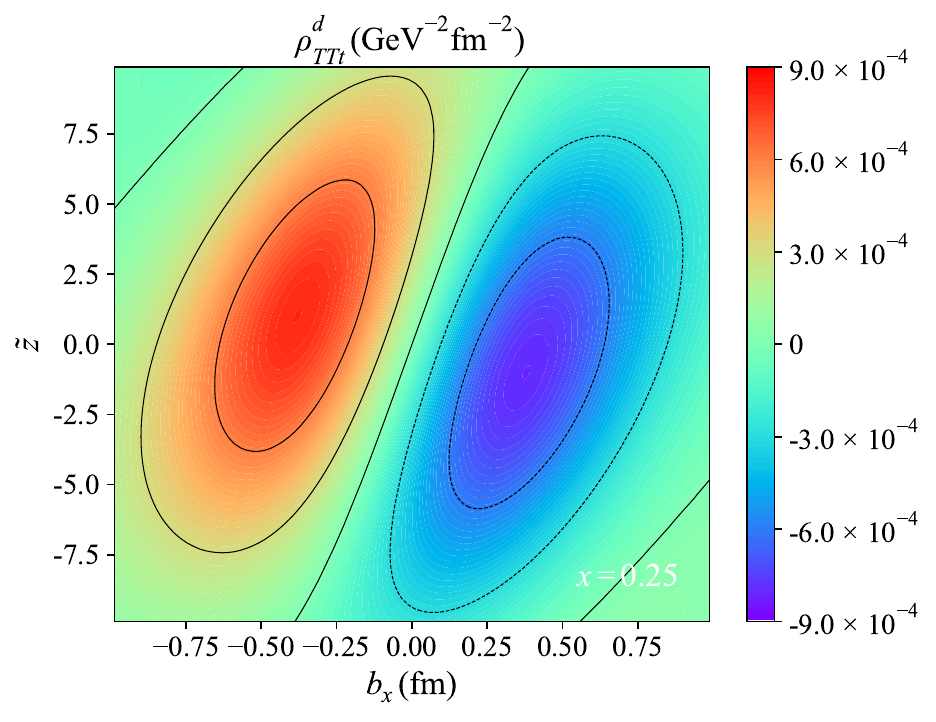}
	}
	\subfloat{
		\includegraphics[width=0.31\textwidth]{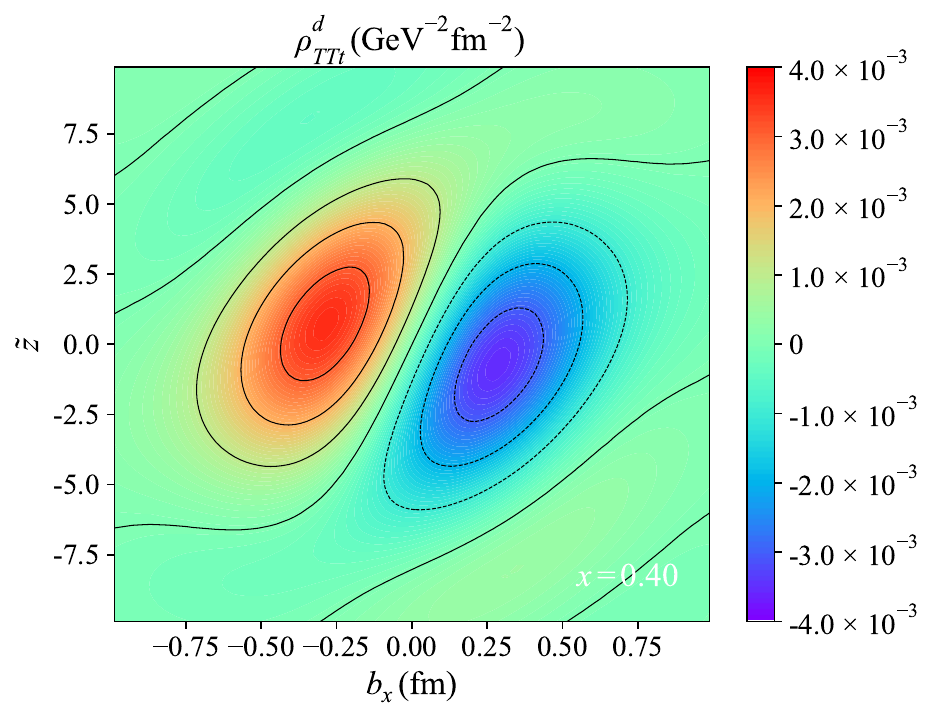}
	}
	\caption{Six-dimensional pretzelous light-front Wigner distribution $\rho_{\mathrm{TTt}}\left(\tilde{z},x,\boldsymbol{b}_{\perp}, \boldsymbol{k}_{\perp}\right)$ for $u$ quark (upper panels) and $d$ quark (lower panels). The figure presents the Wigner distribution in the $\tilde{z}-b_x$ plane, with the transverse momentum fixed at $\boldsymbol{k}_{\perp}=0.3\,\mathrm{GeV}\boldsymbol{\hat{e}}_x$ (where $\boldsymbol{\hat{e}}_x$ is the unit vector in the $x$-direction) and the transverse coordinate component fixed at $b_y=0.4\,\mathrm{GeV}^{-1}$. The three columns correspond to $x=0.10$, $x=0.25$, and $x=0.40$.}
	\label{6DProtonTTtudzbx}
\end{figure}

\begin{figure}[htbp]
	\centering
	\subfloat{
		\includegraphics[width=0.31\textwidth]{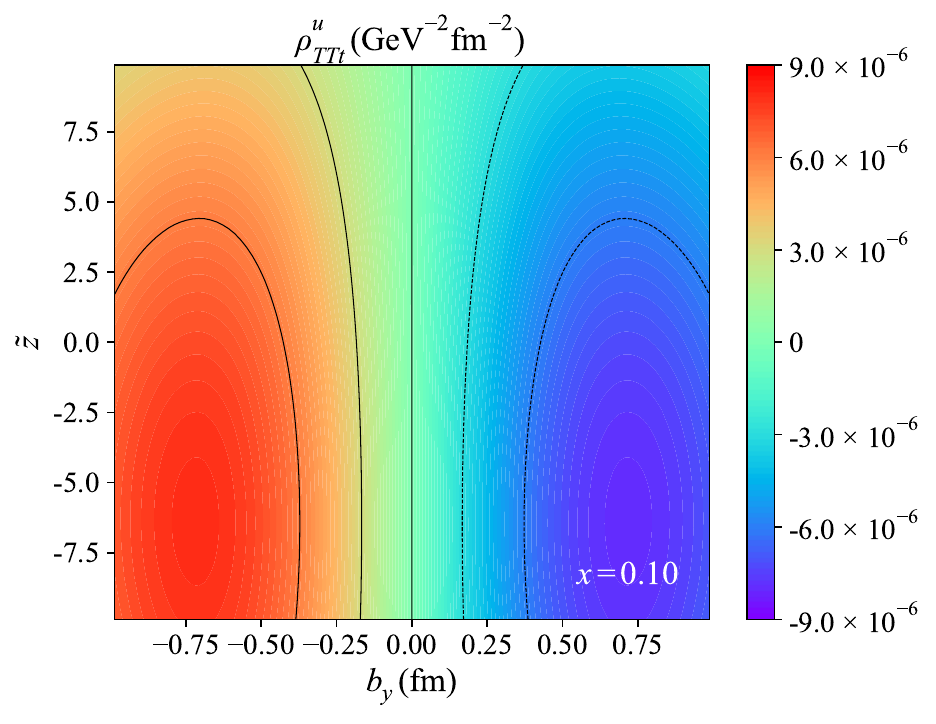}
	}
	\subfloat{
		\includegraphics[width=0.31\textwidth]{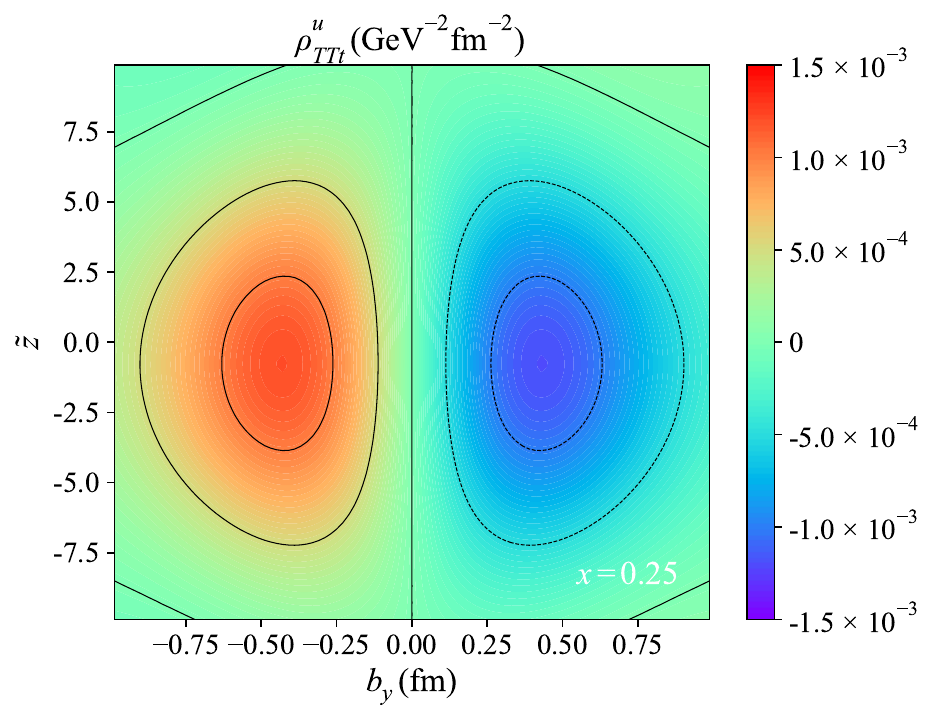}
	}
	\subfloat{
		\includegraphics[width=0.31\textwidth]{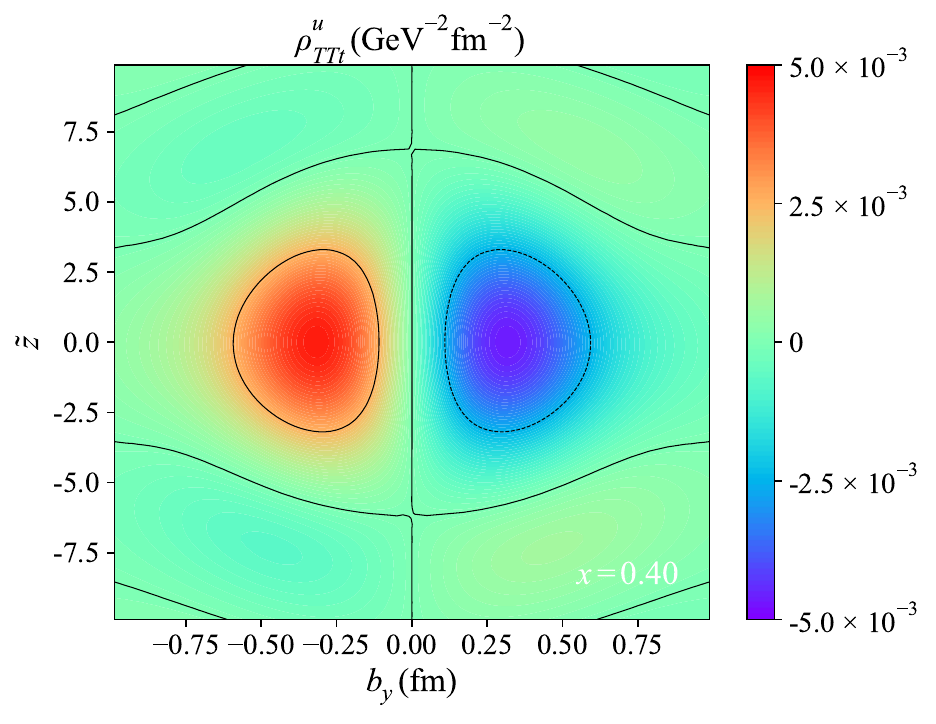}
	}\\
	\subfloat{
		\includegraphics[width=0.31\textwidth]{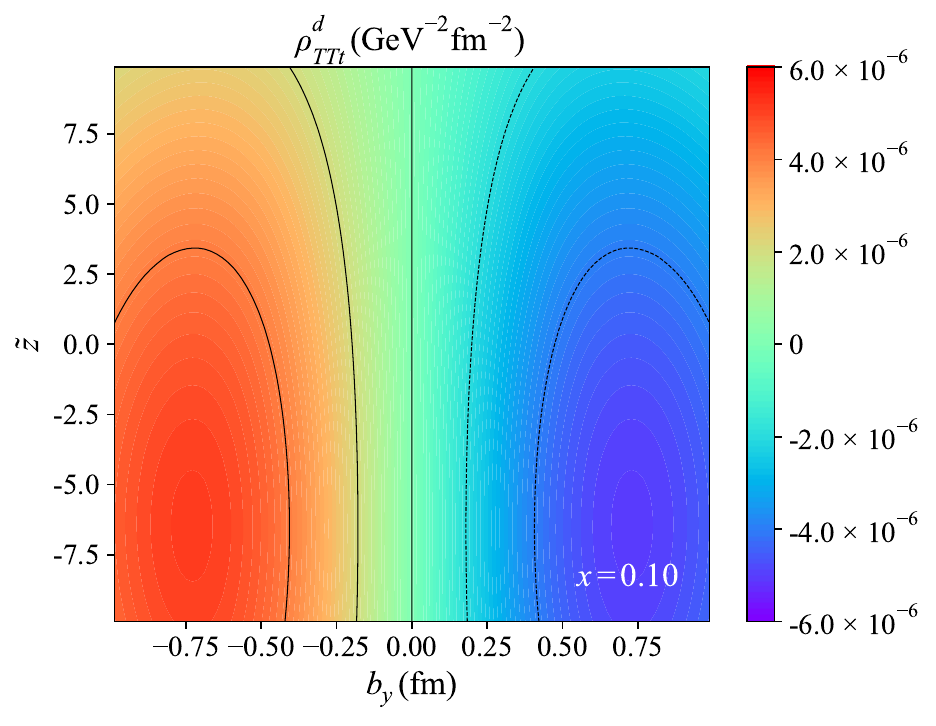}
	}
	\subfloat{
		\includegraphics[width=0.31\textwidth]{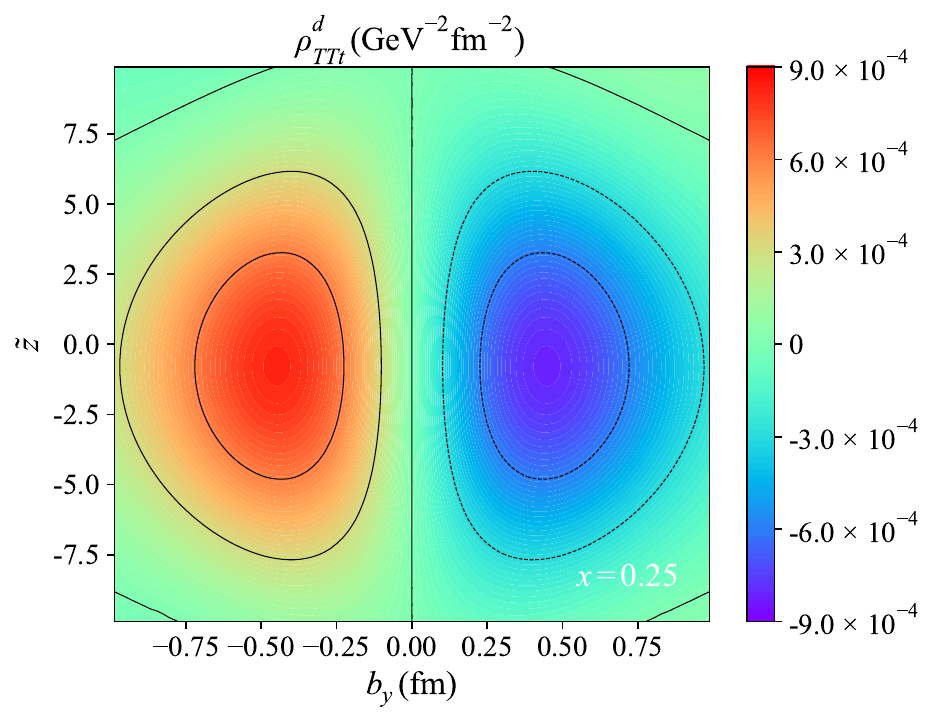}
	}
	\subfloat{
		\includegraphics[width=0.31\textwidth]{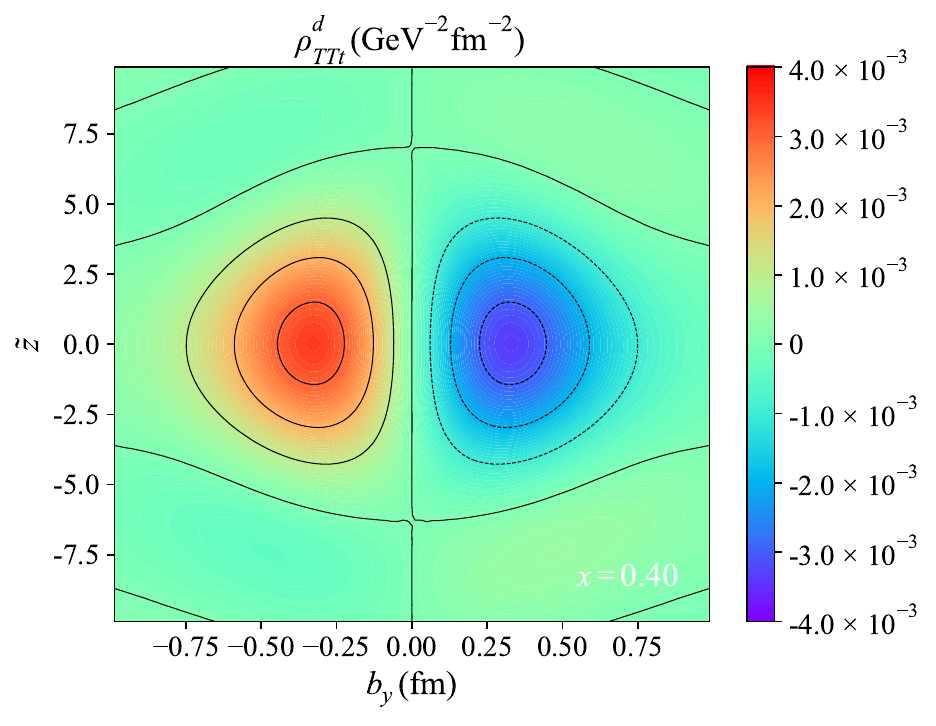}
	}
	\caption{Six-dimensional pretzelous light-front Wigner distribution $\rho_{\mathrm{TTt}}\left(\tilde{z},x,\boldsymbol{b}_{\perp}, \boldsymbol{k}_{\perp}\right)$ for $u$ quark (upper panels) and $d$ quark (lower panels). The figure presents the Wigner distributions in the $\tilde{z}-b_y$ plane, with the transverse momentum fixed at $\boldsymbol{k}_{\perp}=0.3\,\mathrm{GeV}\boldsymbol{\hat{e}}_x$ (where $\boldsymbol{\hat{e}}_x$ is the unit vector in the $x$-direction) and the transverse coordinate component fixed at $b_x=0.4\,\mathrm{GeV}^{-1}$. The three columns correspond to $x=0.10$, $x=0.25$, and $x=0.40$.}
	\label{6DProtonTTtudzby}
\end{figure}

\begin{figure}[htbp]
	\centering
	\subfloat{
		\includegraphics[width=0.31\textwidth]{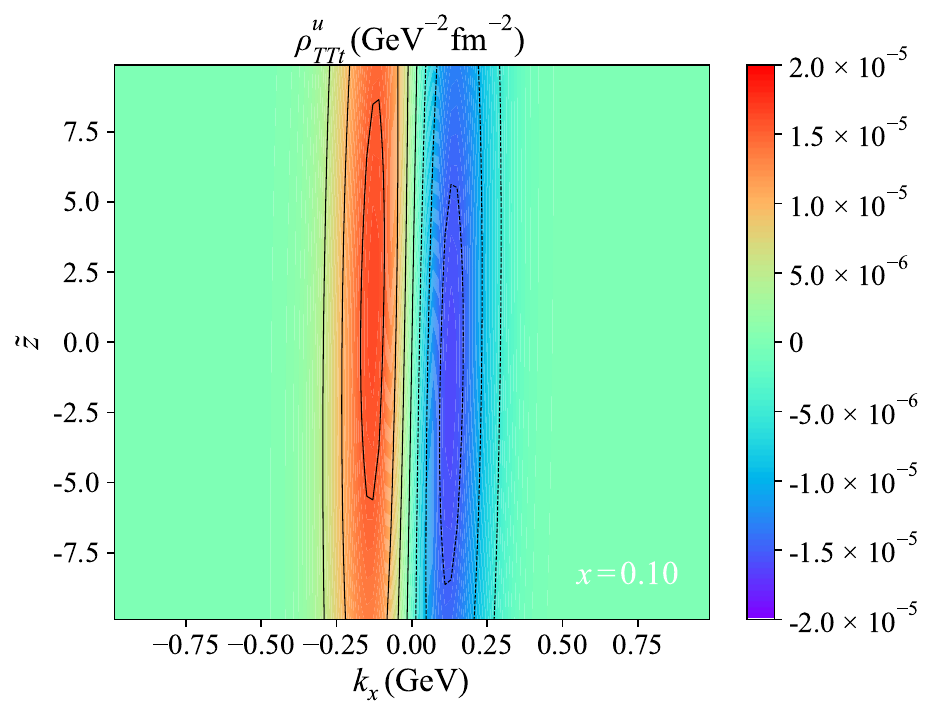}
	}
	\subfloat{
		\includegraphics[width=0.31\textwidth]{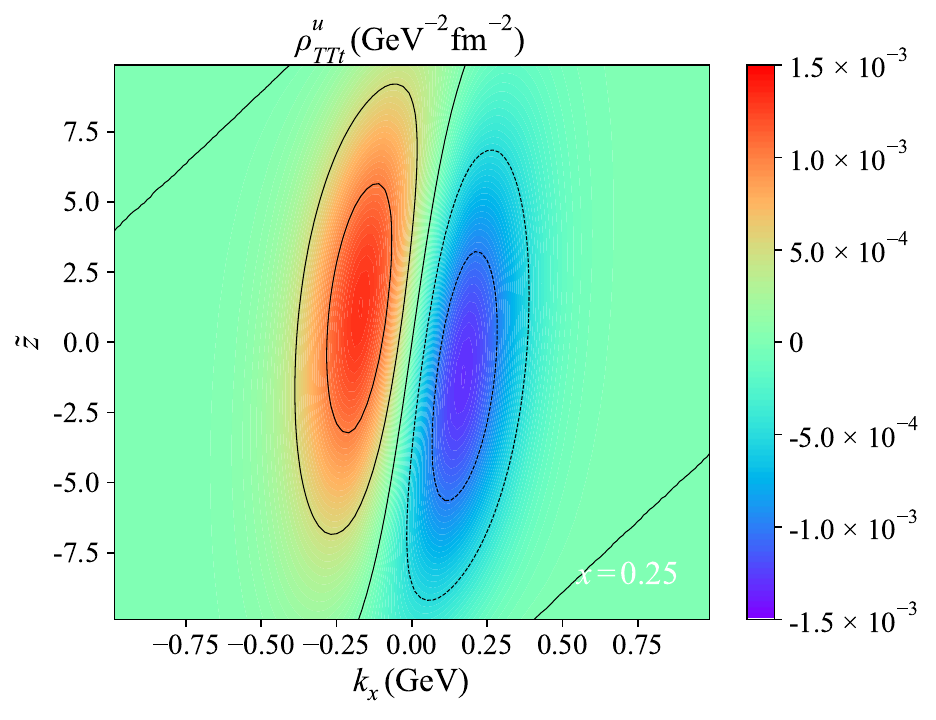}
	}
	\subfloat{
		\includegraphics[width=0.31\textwidth]{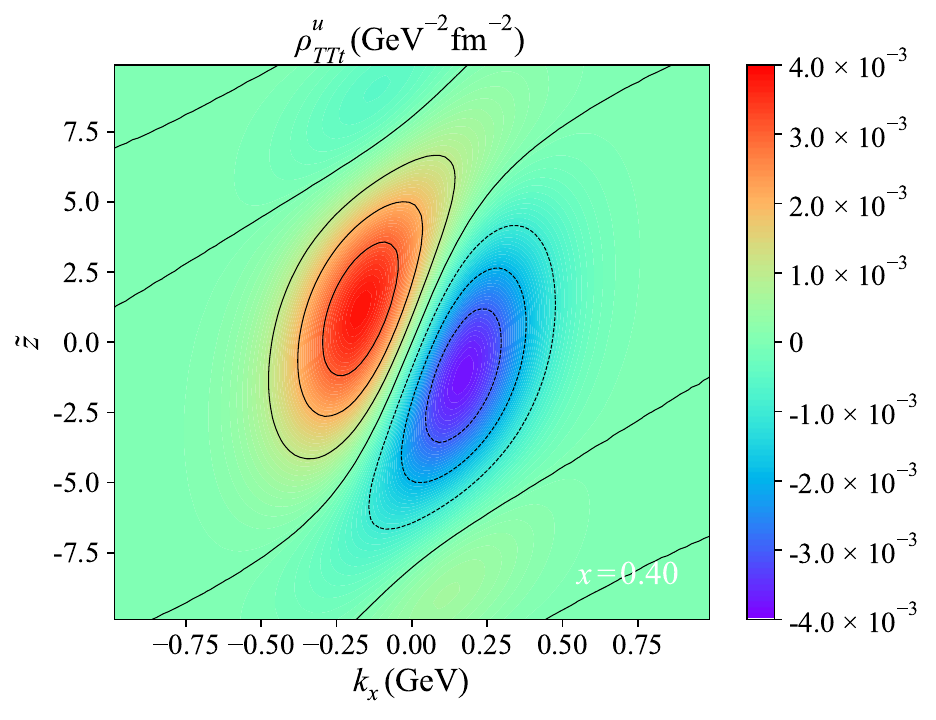}
	}\\
	\subfloat{
		\includegraphics[width=0.31\textwidth]{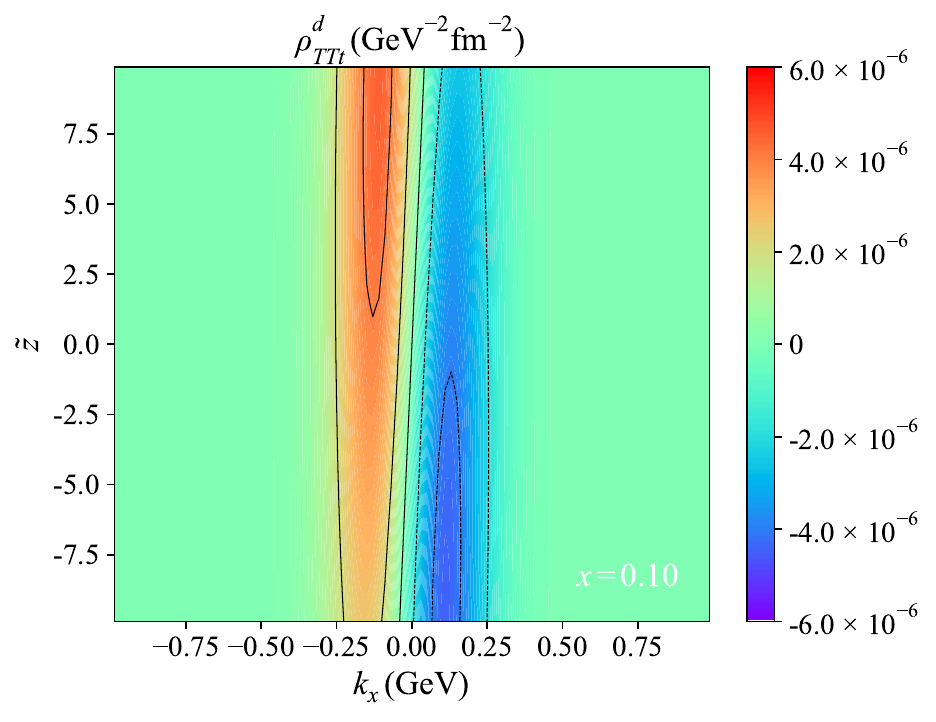}
	}
	\subfloat{
		\includegraphics[width=0.31\textwidth]{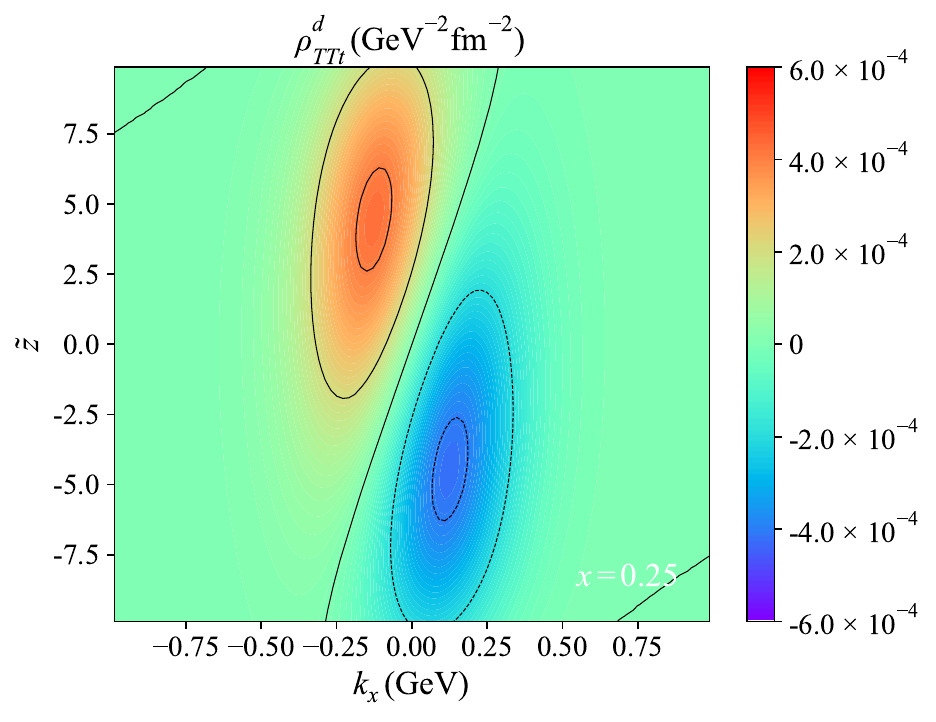}
	}
	\subfloat{
		\includegraphics[width=0.31\textwidth]{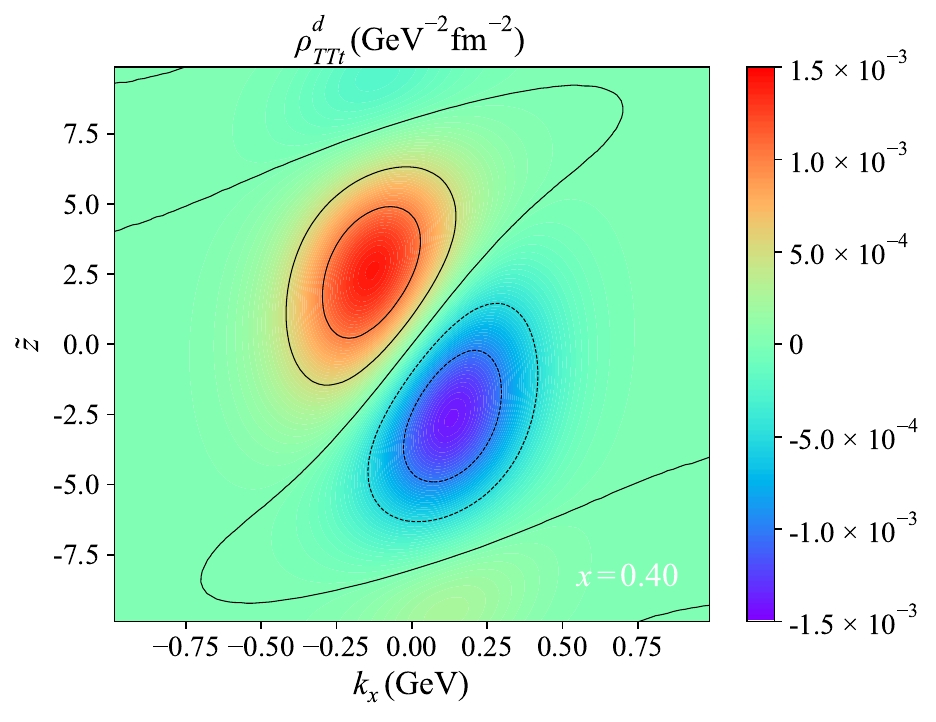}
	}
	\caption{Six-dimensional pretzelous light-front Wigner distribution $\rho_{\mathrm{TTt}}\left(\tilde{z},x,\boldsymbol{b}_{\perp}, \boldsymbol{k}_{\perp}\right)$ for $u$ quark (upper panels) and $d$ quark (lower panels). The figure presents the Wigner distributions in the $\tilde{z}-k_x$ plane, with the transverse coordinate fixed at $\boldsymbol{b}_{\perp}=0.4\,\mathrm{GeV}^{-1}\boldsymbol{\hat{e}}_x$ (where $\boldsymbol{\hat{e}}_x$ is the unit vector along the $x$-axis) and the transverse momentum component fixed at $k_y=0.3\,\mathrm{GeV}$. The three columns correspond to $x=0.10$, $x=0.25$, and $x=0.40$.}
	\label{6DProtonTTtudzkx}
\end{figure}

\begin{figure}[htbp]
	\centering
	\subfloat{
		\includegraphics[width=0.31\textwidth]{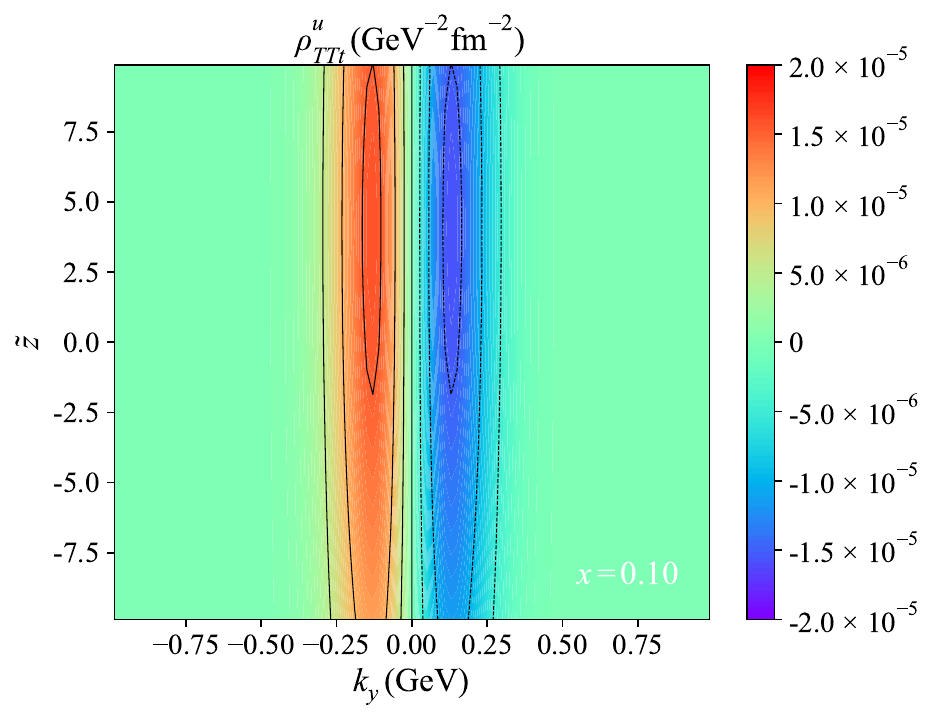}
	}
	\subfloat{
		\includegraphics[width=0.31\textwidth]{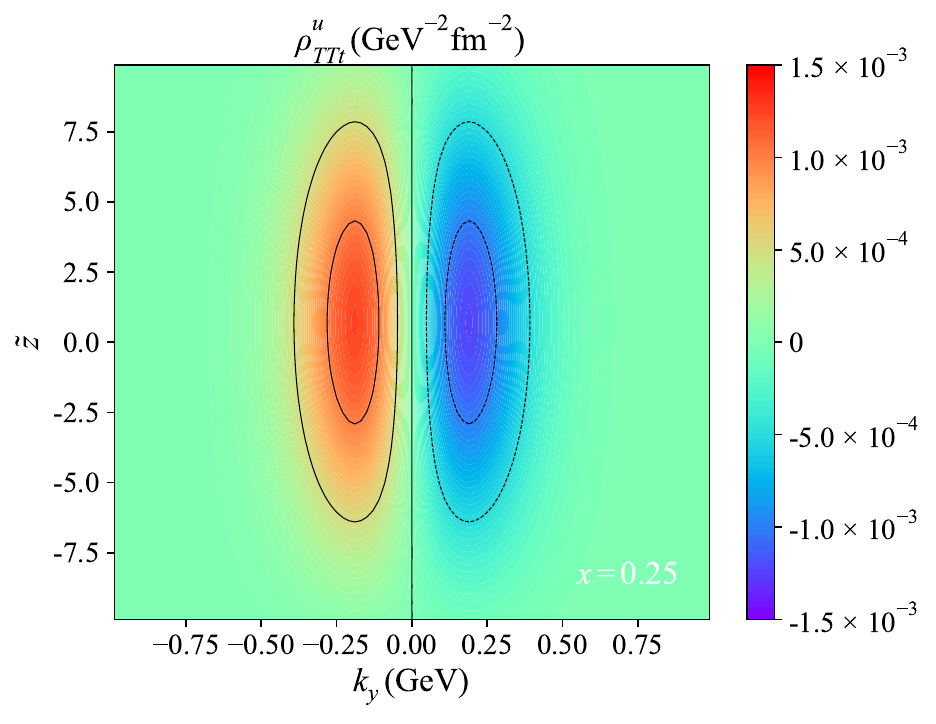}
	}
	\subfloat{
		\includegraphics[width=0.31\textwidth]{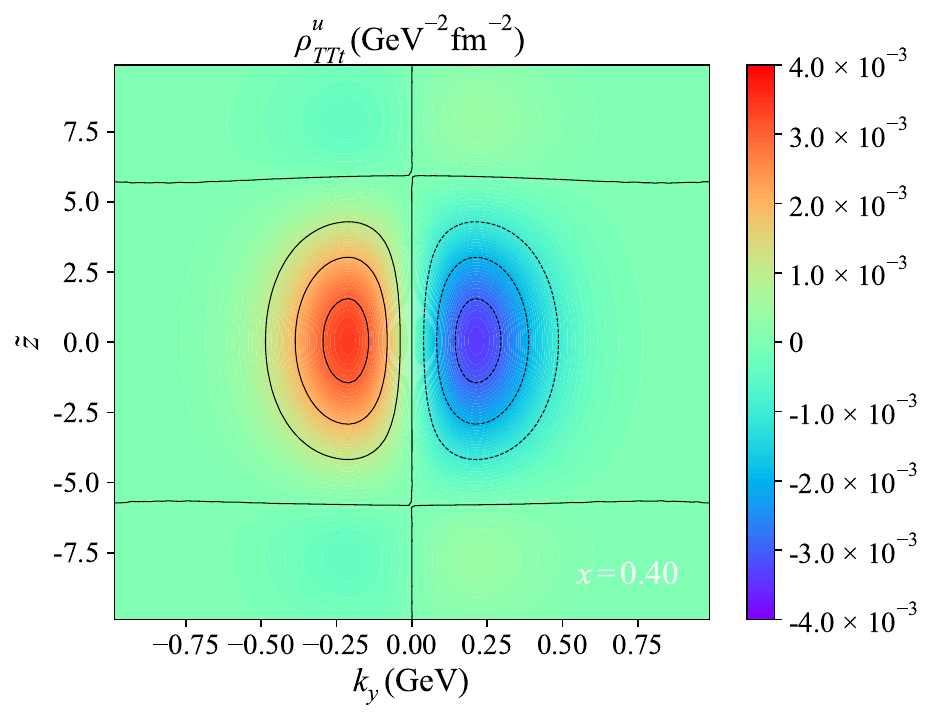}
	}\\
	\subfloat{
		\includegraphics[width=0.31\textwidth]{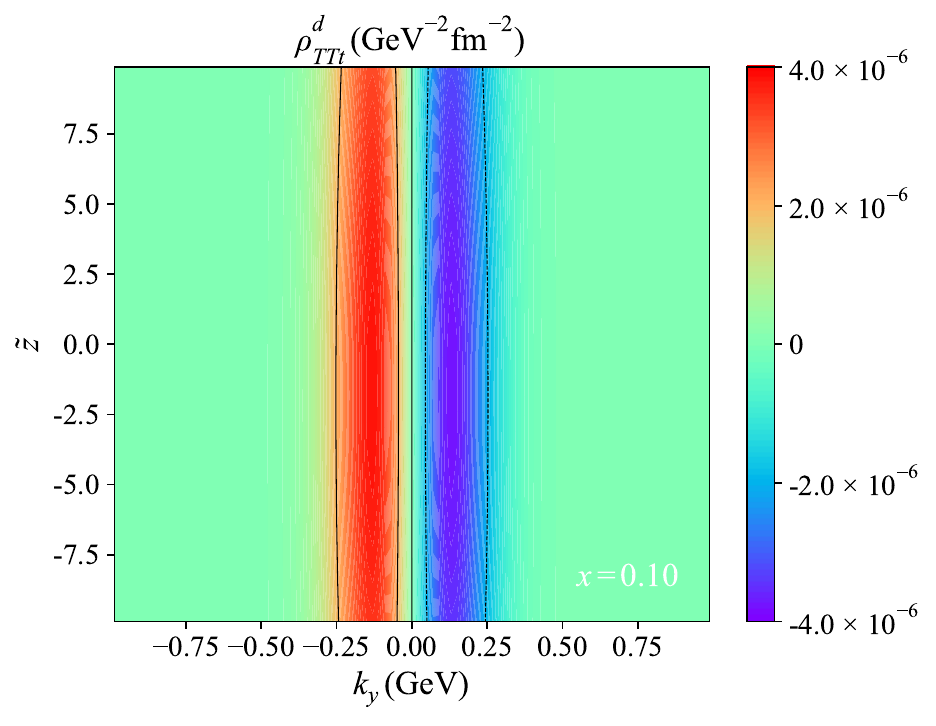}
	}
	\subfloat{
		\includegraphics[width=0.31\textwidth]{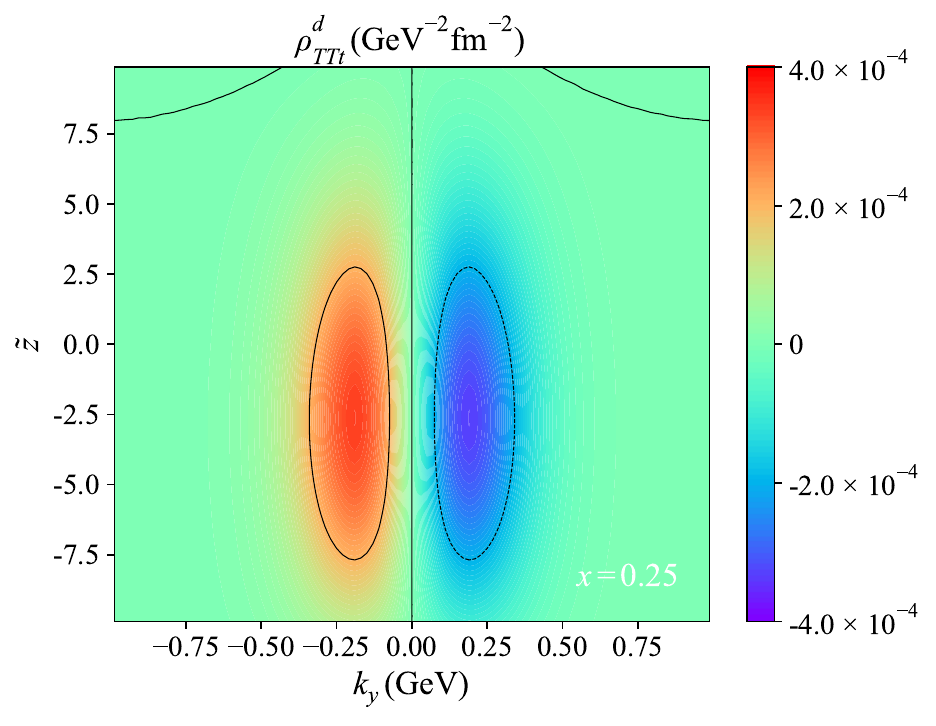}
	}
	\subfloat{
		\includegraphics[width=0.31\textwidth]{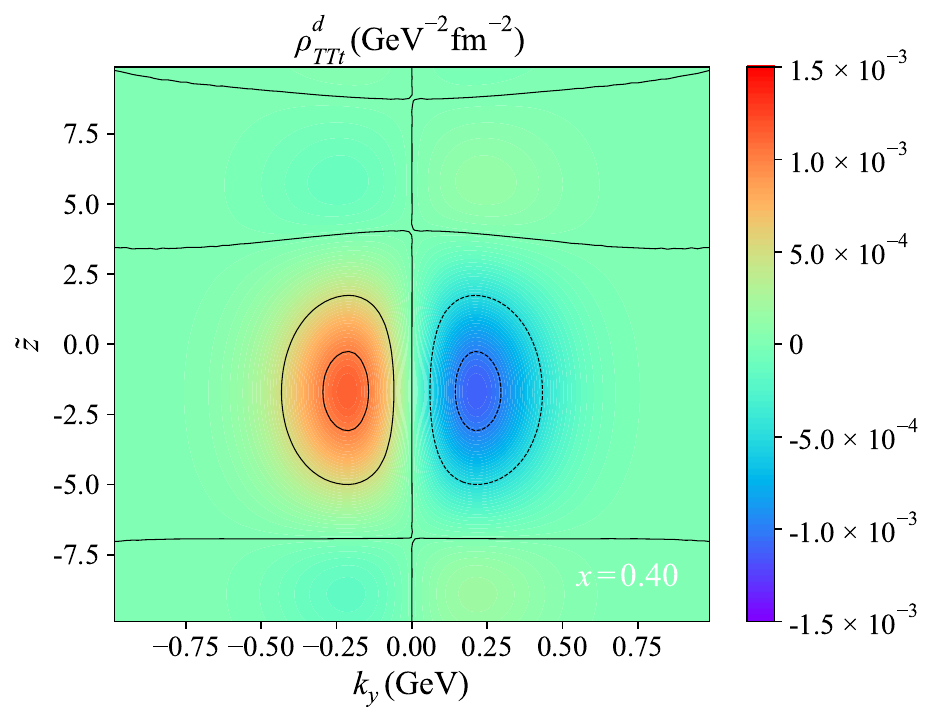}
	}
	\caption{Six-dimensional pretzelous light-front Wigner distribution $\rho_{\mathrm{TTt}}\left(\tilde{z},x,\boldsymbol{b}_{\perp}, \boldsymbol{k}_{\perp}\right)$ for $u$ quark (upper panels) and $d$ quark (lower panels). The figure presents the Wigner distributions in the $\tilde{z}-k_y$ plane, with the transverse coordinate fixed at $\boldsymbol{b}_{\perp}=0.4\,\mathrm{GeV}^{-1}\boldsymbol{\hat{e}}_x$ (where $\boldsymbol{\hat{e}}_x$ is the unit vector along the $x$-axis) and the transverse momentum component fixed at $k_x=0.3\,\mathrm{GeV}$. The three columns correspond to $x=0.10$, $x=0.25$, and $x=0.40$.}
	\label{6DProtonTTtudzky}
\end{figure}

In this subsection, we discuss a different case of transverse polarization. In Figs.~\ref{6DProtonTTtudzbx}--\ref{6DProtonTTtudzky}, % In Fig.~\ref{6DProtonTTtudzbx}, Fig.~\ref{6DProtonTTtudzby}, Fig.~\ref{6DProtonTTtudzkx} and Fig.~\ref{6DProtonTTtudzky}, 
we plot the pretzelous light-front Wigner distribution $\rho_{\mathrm{TT}}^{\perp}\left(\tilde{z},x,\boldsymbol{b}_{\perp}, \boldsymbol{k}_{\perp}\right)$ for the $u$ and $d$ quarks of the proton, displayed in the $\tilde{z}-b_x$, $\tilde{z}-b_y$, $\tilde{z}-k_x$, and $\tilde{z}-k_y$ subspaces, respectively. \new{These six-dimensional pretzelous light-front Wigner distributions provide a unique window into the nucleon spin structure by examining the correlation between quarks polarized along one transverse direction ($\hat{x}$) within a proton polarized along the orthogonal direction ($\hat{y}$). This particular configuration reveals novel quantum interference effects between the quark orbital motion and the proton spin structure that are inaccessible in parallel spin alignments.} The numerical results, which show the relationship between the longitudinal coordinate and the transverse coordinates or transverse momentum, are presented for fixed values of the transverse momentum $\boldsymbol{k}_{\perp}$ or the transverse coordinate $\boldsymbol{b}_{\perp}$, and the longitudinal momentum fraction $x$ is set at $x = 0.10$, $x = 0.25$, and $x = 0.40$ in the first, second, and third columns, respectively.

In Fig.~\ref{6DProtonTTtudzbx} and Fig.~\ref{6DProtonTTtudzkx}, the six-dimensional pretzelous light-front Wigner distribution exhibits centrosymmetry about the coordinate origin in the $\tilde{z}-b_x$ and $\tilde{z}-k_x$ subspaces, with the maximum values located at the origin for every fixed value of $x$. In contrast, Fig.~\ref{6DProtonTTtudzby} and Fig.~\ref{6DProtonTTtudzky} reveal a dipole-symmetric shape about $b_x=0$. In addition, the distribution retains its special non-positive properties. For larger values of $x$, the six-dimensional pretzelous light-front Wigner distribution approximates a quadrupole structure, a behavior that is intrinsically tied to the spin structure of the system.

%\subsubsection{u Quark}

%\bibliographystyle{unsrt}
%\bibliography{references}{}

\begin{thebibliography}{}


%\cite{Feynman:1969ej}
\bibitem{Feynman:1969ej}
R.~P.~Feynman,
Very high-energy collisions of hadrons,
\href{https://doi.org/10.1103/PhysRevLett.23.1415}{Phys. Rev. Lett. \textbf{23}, 1415--1417 (1969)}.

%\cite{Mulders:1995dh}
\bibitem{Mulders:1995dh}
P.~J.~Mulders and R.~D.~Tangerman,
The complete tree level result up to order 1/Q for polarized deep inelastic leptoproduction,
\href{https://doi.org/10.1016/0550-3213(95)00632-X}{Nucl. Phys. B \textbf{461}, 197--237 (1996)}
[\href{https://doi.org/10.48550/arXiv.hep-ph/9510301}{\tt arXiv:hep-ph/9510301}].

%\cite{Goeke:2005hb}
\bibitem{Goeke:2005hb}
K.~Goeke, A.~Metz and M.~Schlegel,
Parameterization of the quark-quark correlator of a spin-1/2 hadron,
\href{https://doi.org/10.1016/j.physletb.2005.05.037}{Phys. Lett. B \textbf{618}, 90--96 (2005)}
[\href{https://doi.org/10.48550/arXiv.hep-ph/0504130}{\tt arXiv:hep-ph/0504130}].

%\cite{Bacchetta:2006tn}
\bibitem{Bacchetta:2006tn}
A.~Bacchetta, M.~Diehl, K.~Goeke, A.~Metz, P.~J.~Mulders and M.~Schlegel,
Semi-inclusive deep inelastic scattering at small transverse momentum,
\href{https://doi.org/10.1088/1126-6708/2007/02/093}{J. High Energy Phys. 02 (2007) 093}
[\href{https://doi.org/10.48550/arXiv.hep-ph/0611265}{\tt arXiv:hep-ph/0611265}].


%\cite{Brambilla:2014jmp}
\bibitem{Brambilla:2014jmp}
N.~Brambilla and others,
QCD and Strongly Coupled Gauge Theories: Challenges and Perspectives,
\href{https://doi.org/10.1140/epjc/s10052-014-2981-5}{Eur. Phys. J. C \textbf{74}, 2981 (2014)}
[\href{https://doi.org/10.48550/arXiv.1404.3723}{\tt arXiv:1404.3723}].

%\cite{Gross:1973id}
\bibitem{Gross:1973id}
D.~J.~Gross and F.~Wilczek,
Ultraviolet Behavior of Nonabelian Gauge Theories,
\href{https://doi.org/10.1103/PhysRevLett.30.1343}{Phys. Rev. Lett. \textbf{30}, 1343--1346 (1973)}.

%\cite{Politzer:1973fx}
\bibitem{Politzer:1973fx}
H.~D.~Politzer,
Reliable Perturbative Results for Strong Interactions?,
\href{https://doi.org/10.1103/PhysRevLett.30.1346}{Phys. Rev. Lett. \textbf{30}, 1346--1349 (1973)}.

%\cite{Collins:1989gx}
\bibitem{Collins:1989gx}
J.~C.~Collins, D.~E.~Soper and G.~F.~Sterman,
Factorization of Hard Processes in QCD,
\href{https://doi.org/10.1142/9789814503266_0001}{Adv. Ser. Direct. High Energy Phys. \textbf{5}, 1--91 (1989)}
[\href{https://doi.org/10.48550/arXiv.hep-ph/0409313}{\tt arXiv:hep-ph/0409313}].

%\cite{Collins:1981uw}
\bibitem{Collins:1981uw}
J.~C.~Collins and D.~E.~Soper,
Parton Distribution and Decay Functions,
\href{https://doi.org/10.1016/0550-3213(82)90021-9}{Nucl. Phys. B \textbf{194}, 445--492 (1982)}.

%\cite{Martin:1998sq}
\bibitem{Martin:1998sq}
A.~D.~Martin, R.~G.~Roberts, W.~J.~Stirling and R.~S.~Thorne,
Parton distributions: a new global analysis,
\href{https://doi.org/10.1007/s100520050220}{Eur. Phys. J. C \textbf{4}, 463--496 (1998)}
[\href{https://doi.org/10.48550/arXiv.hep-ph/9803445}{\tt arXiv:hep-ph/9803445}].

%\cite{Gluck:1994uf}
\bibitem{Gluck:1994uf}
M.~Gl\"uck, E.~Reya and A.~Vogt,
Dynamical parton distributions of the proton and small x physics,
\href{https://doi.org/10.1007/BF01624586}{Z. Phys. C \textbf{67}, 433--448 (1995)}
[\href{https://doi.org/10.48550/arXiv.hep-ph/9403227}{\tt arXiv:hep-ph/9403227}].

%\cite{Gluck:1998xa}
\bibitem{Gluck:1998xa}
M.~Gl\"uck, E.~Reya and A.~Vogt,
Dynamical parton distributions revisited,
\href{https://doi.org/10.1007/s100520050289}{Eur. Phys. J. C \textbf{5}, 461--470 (1998)}
[\href{https://doi.org/10.48550/arXiv.hep-ph/9806404}{\tt arXiv:hep-ph/9806404}].


%\cite{Barone:2001sp}
\bibitem{Barone:2001sp}
V.~Barone, A.~Drago and P.~G.~Ratcliffe,
Transverse polarisation of quarks in hadrons,
\href{https://doi.org/10.1016/S0370-1573(01)00051-5}{Phys. Rept. \textbf{359}, 1--168 (2002)}
[\href{https://doi.org/10.48550/arXiv.hep-ph/0104283}{\tt arXiv:hep-ph/0104283}].

%\cite{Brodsky:2002cx}
\bibitem{Brodsky:2002cx}
S.~J.~Brodsky, D.~S.~Hwang and I.~Schmidt,
Final state interactions and single spin asymmetries in semiinclusive deep inelastic scattering,
\href{https://doi.org/10.1016/S0370-2693(02)01320-5}{Phys. Lett. B \textbf{530}, 99--107 (2002)}
[\href{https://doi.org/10.48550/arXiv.hep-ph/0201296}{\tt arXiv:hep-ph/0201296}].

%\cite{Ji:1996nm}
\bibitem{Ji:1996nm}
X.~D.~Ji,
Deeply virtual Compton scattering,
\href{https://doi.org/10.1103/PhysRevD.55.7114}{Phys. Rev. D \textbf{55}, 7114--7125 (1997)}
[\href{https://doi.org/10.48550/arXiv.hep-ph/9609381}{\tt arXiv:hep-ph/9609381}].

%\cite{Diehl:2003ny}
\bibitem{Diehl:2003ny}
M.~Diehl,
Generalized parton distributions,
\href{https://doi.org/10.1016/j.physrep.2003.08.002}{Phys. Rept. \textbf{388}, 41--277 (2003)}
[\href{https://doi.org/10.48550/arXiv.hep-ph/0307382}{\tt arXiv:hep-ph/0307382}].

%\cite{Belitsky:2005qn}
\bibitem{Belitsky:2005qn}
A.~V.~Belitsky and A.~V.~Radyushkin,
Unraveling hadron structure with generalized parton distributions,
\href{https://doi.org/10.1016/j.physrep.2005.06.002}{Phys. Rept. \textbf{418}, 1--387 (2005)}
[\href{https://doi.org/10.48550/arXiv.hep-ph/0504030}{\tt arXiv:hep-ph/0504030}].

%\cite{Goeke:2001tz}
\bibitem{Goeke:2001tz}
K.~Goeke, M.~V.~Polyakov and M.~Vanderhaeghen,
Hard exclusive reactions and the structure of hadrons,
\href{https://doi.org/10.1016/S0146-6410(01)00158-2}{Prog. Part. Nucl. Phys. \textbf{47}, 401--515 (2001)}.

%\cite{Ji:1996ek}
\bibitem{Ji:1996ek}
X.~D.~Ji,
Gauge-Invariant Decomposition of Nucleon Spin,
\href{https://doi.org/10.1103/PhysRevLett.78.610}{Phys. Rev. Lett. \textbf{78}, 610--613 (1997)}
[\href{https://doi.org/10.48550/arXiv.hep-ph/9603249}{\tt arXiv:hep-ph/9603249}].

%\cite{Avakian:2008dz}
\bibitem{Avakian:2008dz}
H.~Avakian, A.~V.~Efremov, P.~Schweitzer and F.~Yuan,
Transverse momentum dependent distribution function $h_{1T}^\perp$ and the single spin asymmetry $A_{UT}^{\sin(3\phi-\phi_S)}$,
\href{https://doi.org/10.1103/PhysRevD.78.114024}{Phys. Rev. D \textbf{78}, 114024 (2008)}
[\href{https://doi.org/10.48550/arXiv.0805.3355}{\tt arXiv:0805.3355}].

%\cite{Avakian:2009jt}
\bibitem{Avakian:2009jt}
H.~Avakian, A.~V.~Efremov, P.~Schweitzer, O.~V.~Teryaev, F.~Yuan and P.~Zavada,
Insights on non-perturbative aspects of TMDs from models,
\href{https://doi.org/10.1142/S0217732309001200}{Mod. Phys. Lett. A \textbf{24}, 2995--3004 (2009)}
[\href{https://doi.org/10.48550/arXiv.0910.3181}{\tt arXiv:0910.3181}].

%\cite{She:2009jq}
\bibitem{She:2009jq}
J.~She, J.~Zhu and B.-Q.~Ma,
Pretzelosity h(1T)**perpendicular and quark orbital angular momentum,
\href{https://doi.org/10.1103/PhysRevD.79.054008}{Phys. Rev. D \textbf{79}, 054008 (2009)}
[\href{https://doi.org/10.48550/arXiv.0902.3718}{\tt arXiv:0902.3718}].

%\cite{Avakian:2010br}
\bibitem{Avakian:2010br}
H.~Avakian, A.~V.~Efremov, P.~Schweitzer and F.~Yuan,
The transverse momentum dependent distribution functions in the bag model,
\href{https://doi.org/10.1103/PhysRevD.81.074035}{Phys. Rev. D \textbf{81}, 074035 (2010)}
[\href{https://doi.org/10.48550/arXiv.1001.5467}{\tt arXiv:1001.5467}].

%\cite{Accardi:2012qut}
\bibitem{Accardi:2012qut}
A.~Accardi and others,
Electron Ion Collider: The Next QCD Frontier: Understanding the glue that binds us all,
\href{https://doi.org/10.1140/epja/i2016-16268-9}{Eur. Phys. J. A \textbf{52}, 268 (2016)}
[\href{https://doi.org/10.48550/arXiv.1212.1701}{\tt arXiv:1212.1701}].

%\cite{Anderle:2021wcy}
\bibitem{Anderle:2021wcy}
D.~P.~Anderle and others,
Electron-ion collider in China,
\href{https://doi.org/10.1007/s11467-021-1062-0}{Front. Phys. (Beijing) \textbf{16}, 64701 (2021)}
[\href{https://doi.org/10.48550/arXiv.2102.09222}{\tt arXiv:2102.09222}].

%\cite{Boussarie:2023izj}
\bibitem{Boussarie:2023izj}
R.~Boussarie, M.~Burkardt, M.~Constantinou, W.~Detmold, M.~Ebert, M.~Engelhardt, S.~Fleming, L.~Gamberg, X.~Ji, Z.~Kang and others,
TMD handbook,
\href{https://doi.org/10.48550/arXiv.2304.03302}{\tt arXiv:2304.03302}.

%\cite{Wigner:1932eb}
\bibitem{Wigner:1932eb}
E.~P.~Wigner,
{On the quantum correction for thermodynamic equilibrium},
\href{https://doi.org/10.1103/PhysRev.40.749}{Phys. Rev. \textbf{40}, 749--760 (1932)}.

%\cite{Hillery:1983ms}
\bibitem{Hillery:1983ms}
M.~Hillery, R.~F.~O'Connell, M.~O.~Scully and E.~P.~Wigner,
{Distribution functions in physics: Fundamentals},
\href{https://doi.org/10.1016/0370-1573(84)90160-1}{Phys. Rept. \textbf{106}, 121--167 (1984)}.

%\cite{Hiley2015}
\bibitem{Hiley2015}
I.~Licata,
{Beyond peaceful coexistence. The Emergence of Space, Time and Quantum},
\href{https://doi.org/10.13140/RG.2.1.1704.7765}{(2015)}.

%\cite{Ji:2003ak}
\bibitem{Ji:2003ak}
X.~D.~Ji,
Viewing the proton through 'color' filters,
\href{https://doi.org/10.1103/PhysRevLett.91.062001}{Phys. Rev. Lett. \textbf{91}, 062001 (2003)}
[\href{https://doi.org/10.48550/arXiv.hep-ph/0304037}{\tt arXiv:hep-ph/0304037}].

%\cite{Belitsky:2003nz}
\bibitem{Belitsky:2003nz}
A.~V.~Belitsky, X.~D.~Ji and F.~Yuan,
Quark imaging in the proton via quantum phase space distributions,
\href{https://doi.org/10.1103/PhysRevD.69.074014}{Phys. Rev. D \textbf{69}, 074014 (2004)}
[\href{https://doi.org/10.48550/arXiv.hep-ph/0307383}{\tt arXiv:hep-ph/0307383}].

%\cite{Lorce:2011kd}
\bibitem{Lorce:2011kd}
C.~Lorce and B.~Pasquini,
Quark Wigner Distributions and Orbital Angular Momentum,
\href{https://doi.org/10.1103/PhysRevD.84.014015}{Phys. Rev. D \textbf{84}, 014015 (2011)}
[\href{https://doi.org/10.48550/arXiv.1106.0139}{\tt arXiv:1106.0139}].

%\cite{Lorce:2015sqe}
\bibitem{Lorce:2015sqe}
C.~Lorc'e and B.~Pasquini,
Multipole decomposition of the nucleon transverse phase space,
\href{https://doi.org/10.1103/PhysRevD.93.034040}{Phys. Rev. D \textbf{93}, 034040 (2016)}
[\href{https://doi.org/10.48550/arXiv.1512.06744}{\tt arXiv:1512.06744}].

%\cite{Burkardt:2002hr}
\bibitem{Burkardt:2002hr}
M.~Burkardt,
Impact parameter space interpretation for generalized parton distributions,
\href{https://doi.org/10.1142/S0217751X03012370}{Int. J. Mod. Phys. A \textbf{18}, 173--208 (2003)}
[\href{https://doi.org/10.48550/arXiv.hep-ph/0207047}{\tt arXiv:hep-ph/0207047}].

%\cite{Burkardt:2000za}
\bibitem{Burkardt:2000za}
M.~Burkardt,
Impact parameter dependent parton distributions and off forward parton distributions for zeta ---\ensuremath{>} 0,
\href{https://doi.org/10.1103/PhysRevD.62.071503}{Phys. Rev. D \textbf{62}, 071503 (2000)}
[\href{https://doi.org/10.48550/arXiv.hep-ph/0005108}{\tt arXiv:hep-ph/0005108}].
Note: [Erratum: Phys.Rev.D 66, 119903 (2002)]

% spin-0
%\cite{Ma:2018ysi}
\bibitem{Ma:2018ysi}
Z.~L.~Ma and Z.~Lun,
Quark Wigner distribution of the pion meson in light-cone quark model,
\href{https://doi.org/10.1103/PhysRevD.98.054024}{Phys. Rev. D \textbf{98}, 054024 (2018)}
[\href{https://doi.org/10.48550/arXiv.1808.00140}{\tt arXiv:1808.00140}].

%\cite{Kaur:2019jow}
\bibitem{Kaur:2019jow}
S.~Kaur and H.~Dahiya,
Study of kaon structure using the light-cone quark model,
\href{https://doi.org/10.1103/PhysRevD.100.074008}{Phys. Rev. D \textbf{100}, 074008 (2019)}
[\href{https://doi.org/10.48550/arXiv.1908.01939}{\tt arXiv:1908.01939}].

%\cite{Kaur:2019kpi}
\bibitem{Kaur:2019kpi}
N.~Kaur and H.~Dahiya,
Quark Wigner Distributions and GTMDs of Pion in the Light-Front Holographic Model,
\href{https://doi.org/10.1140/epja/s10050-020-00179-9}{Eur. Phys. J. A \textbf{56}, 172 (2020)}
[\href{https://doi.org/10.48550/arXiv.1909.10146}{\tt arXiv:1909.10146}].

%\cite{Zhang:2021tnr}
\bibitem{Zhang:2021tnr}
J.~L.~Zhang and J.~L.~Ping,
Kaon generalized parton distributions and light-front wave functions in the Nambu\textendash{}Jona-Lasinio model,
\href{https://doi.org/10.1140/epjc/s10052-021-09600-z}{Eur. Phys. J. C \textbf{81}, 814 (2021)}.

%\cite{Liu:2015eqa}
\bibitem{Liu:2015eqa}
T.~Liu and B.-Q.~Ma,
Quark Wigner distributions in a light-cone spectator model,
\href{https://doi.org/10.1103/PhysRevD.91.034019}{Phys. Rev. D \textbf{91}, 034019 (2015)}
[\href{https://doi.org/10.48550/arXiv.1501.07690}{\tt arXiv:1501.07690}].

% spin-1/2
%\cite{Mukherjee:2014nya}
\bibitem{Mukherjee:2014nya}
A.~Mukherjee, S.~Nair and V.~K.~Ojha,
Quark Wigner Distributions and Orbital Angular Momentum in Light-front Dressed Quark Model,
\href{https://doi.org/10.1103/PhysRevD.90.014024}{Phys. Rev. D \textbf{90}, 014024 (2014)}
[\href{https://doi.org/10.48550/arXiv.1403.6233}{\tt arXiv:1403.6233}].

%\cite{Mukherjee:2015aja}
\bibitem{Mukherjee:2015aja}
A.~Mukherjee, S.~Nair and V.~K.~Ojha,
Wigner distributions for gluons in a light-front dressed quark model,
\href{https://doi.org/10.1103/PhysRevD.91.054018}{Phys. Rev. D \textbf{91}, 054018 (2015)}
[\href{https://doi.org/10.48550/arXiv.1501.03728}{\tt arXiv:1501.03728}].

%\cite{More:2017zqq}
\bibitem{More:2017zqq}
J.~More, A.~Mukherjee and S.~Nair,
Quark Wigner Distributions Using Light-Front Wave Functions,
\href{https://doi.org/10.1103/PhysRevD.95.074039}{Phys. Rev. D \textbf{95}, 074039 (2017)}
[\href{https://doi.org/10.48550/arXiv.1701.00339}{\tt arXiv:1701.00339}].

%\cite{Chakrabarti:2016yuw}
\bibitem{Chakrabarti:2016yuw}
D.~Chakrabarti, T.~Maji, C.~Mondal and A.~Mukherjee,
{Wigner distributions and orbital angular momentum of a proton},
\href{https://doi.org/10.1140/epjc/s10052-016-4258-7}{Eur. Phys. J. C \textbf{76}, 409 (2016)}
[\href{https://doi.org/10.48550/arXiv.1601.03217}{\tt arXiv:1601.03217}].

%\cite{Chakrabarti:2017teq}
\bibitem{Chakrabarti:2017teq}
D.~Chakrabarti, T.~Maji, C.~Mondal and A.~Mukherjee,
Quark Wigner distributions and spin-spin correlations,
\href{https://doi.org/10.1103/PhysRevD.95.074028}{Phys. Rev. D \textbf{95}, 074028 (2017)}
[\href{https://doi.org/10.48550/arXiv.1701.08551}{\tt arXiv:1701.08551}].

%\cite{Chakrabarti:2019wjx}
\bibitem{Chakrabarti:2019wjx}
D.~Chakrabarti, N.~Kumar, T.~Maji and A.~Mukherjee,
Sivers and Boer\textendash{}Mulders GTMDs in light-front holographic quark\textendash{}diquark model,
\href{https://doi.org/10.1140/epjp/s13360-020-00470-0}{Eur. Phys. J. Plus \textbf{135}, 496 (2020)}
[\href{https://doi.org/10.48550/arXiv.1902.07051}{\tt arXiv:1902.07051}].

%\cite{Kaur:2019lox}
\bibitem{Kaur:2019lox}
S.~Kaur and H.~Dahiya,
Study of spin-spin correlations between quark and a spin-$\frac{1}{2}$ particle,
\href{https://doi.org/10.1155/2020/9429631}{Adv. High Energy Phys. \textbf{2020}, 9429631 (2020)}
[\href{https://doi.org/10.48550/arXiv.1906.04662}{\tt arXiv:1906.04662}].

%\cite{Kumar:2017xcm}
\bibitem{Kumar:2017xcm}
N.~Kumar and C.~Mondal,
{Wigner distributions for an electron},
\href{https://doi.org/10.1016/j.nuclphysb.2018.04.014}{Nucl. Phys. B \textbf{931}, 226--249 (2018)}
[\href{https://doi.org/10.48550/arXiv.1705.03183}{\tt arXiv:1705.03183}].

%\cite{Han:2022tlh}
\bibitem{Han:2022tlh}
Y.~Han, T.~Liu and B.-Q.~Ma,
Six-dimensional light-front Wigner distribution of hadrons,
\href{https://doi.org/10.1016/j.physletb.2022.137127}{Phys. Lett. B \textbf{830}, 137127 (2022)}
[\href{https://doi.org/10.48550/arXiv.2202.10359}{\tt arXiv:2202.10359}].

%\cite{Brodsky:2006ku}
\bibitem{Brodsky:2006ku}
S.~J.~Brodsky, D.~Chakrabarti, A.~Harindranath, A.~Mukherjee and J.~P.~Vary,
{Hadron optics in three-dimensional invariant coordinate space from deeply virtual compton scattering},
\href{https://doi.org/10.1103/PhysRevD.75.014003}{Phys. Rev. D \textbf{75}, 014003 (2020)}
[\href{https://doi.org/10.48550/arXiv.hep-ph/0611159}{\tt arXiv:hep-ph/0611159}].

%\cite{Brodsky:2006in}
\bibitem{Brodsky:2006in}
S.~J.~Brodsky, D.~Chakrabarti, A.~Harindranath, A.~Mukherjee and J.~P.~Vary,
{Hadron optics: Diffraction patterns in deeply virtual Compton scattering},
\href{https://doi.org/10.1016/j.physletb.2006.08.061}{Phys. Lett. B \textbf{641}, 440--446 (2006)}
[\href{https://doi.org/10.48550/arXiv.hep-ph/0604262}{\tt arXiv:hep-ph/0604262}].

%\cite{Miller:2019ysh}
\bibitem {Miller:2019ysh}
G.~A.~Miller and S.~J.~Brodsky,
{Frame-independent spatial coordinate $\tilde{z}$: Implications for light-front wave functions, deep inelastic scattering, light-front holography, and lattice QCD calculations},
\href{https://doi.org/10.1103/PhysRevC.102.022201}{Phys. Rev. C \textbf{102}, 022201 (2020)}
[\href{https://doi.org/10.48550/arXiv.1912.08911}{\tt arXiv:1912.08911}].

%\cite{Han:2023fav}
\bibitem{Han:2023fav}
Y.~Han, T.~Liu and B.-Q.~Ma,
Six-dimensional light-front Wigner distribution of the pion,
\href{https://doi.org/10.1016/j.nuclphysa.2023.122757}{Nucl. Phys. A \textbf{1040}, 122757 (2023)}.

%\cite{Field:1976ve}
\bibitem{Field:1976ve}
R.~D.~Field and R.~P.~Feynman,
{Quark Elastic Scattering as a Source of High Transverse Momentum Mesons},
\href{https://doi.org/10.1103/PhysRevD.15.2590}{Phys. Rev. D \textbf{15}, 2590--2616 (1977)}[Report Number: CALT-68-565].

%\cite{Close:1973422}
\bibitem{Close:1973422}
F.~E.~Close,
{$\nu$W2 at small $\omega^{\prime}$ and resonance form factors in a quark model with broken SU (6)},
\href{https://doi.org/10.1016/0370-2693(73)90389-4}{Phys. Lett. B \textbf{43}, 422--426 (1973)}.

%\cite{PhysRev.179.1547}
\bibitem{PhysRev.179.1547}
J.~D.~Bjorken,
{Asymptotic Sum Rules at Infinite Momentum},
\href{https://doi.org/10.1103/PhysRev.179.1547}{Phys. Rev. \textbf{179}, 1547--1553 (1969)}.

%\cite{feynman2018photon}
\bibitem{feynman2018photon}
R.~P.~Feynman,
{Photon-hadron interactions},
\href{https://doi.org/}{(2018)}
[Publisher: CRC Press].

%\cite{Ma:1991xq}
\bibitem{Ma:1991xq}
B.-Q.~Ma,
{Melosh rotation: Source for the proton's missing spin},
\href{https://doi.org/10.1088/0954-3899/17/5/001}{J. Phys. G \textbf{17}, L53--L58 (1991)}
[\href{https://doi.org/10.48550/arXiv.0711.2335}{\tt arXiv:0711.2335}].

%\cite{Ma:1992sj}
\bibitem{Ma:1992sj}
B.-Q.~Ma and Q.~R.~Zhang,
{The proton spin and the Wigner rotation},
\href{https://doi.org/10.1007/BF01557707}{Z. Phys. C \textbf{58}, 479--482 (1993)}
[\href{https://doi.org/10.48550/arXiv.hep-ph/9306241}{\tt arXiv:hep-ph/9306241}].

%\cite{Wigner:1939cj}
\bibitem{Wigner:1939cj}
E.~P.~Wigner,
{On Unitary Representations of the Inhomogeneous Lorentz Group},
\href{https://doi.org/10.2307/1968551}{Annals Math. \textbf{40}, 149--204 (1939)}
[Editor: Y.~S.~Kim and W.~W.~Zachary].

%\cite{Melosh:1974cu}
\bibitem{Melosh:1974cu}
H.~J.~Melosh,
{Quarks: Currents and constituents},
\href{https://doi.org/10.1103/PhysRevD.9.1095}{Phys. Rev. D \textbf{9}, 1095 (1974)}.

%\cite{Buccella:1974bz}
\bibitem{Buccella:1974bz}
F.~Buccella, C.~A.~Savoy and P.~Sorba,
{Current Quarks, Constituent Quarks and the Poincare Group},
\href{https://doi.org/10.1007/BF02816884}{Lett. Nuovo Cim. \textbf{10}, 455 (1974)}
[Report Number: INFN-ROME-546].

%\cite{Ma:1996np}
\bibitem{Ma:1996np}
B.-Q.~Ma,
{The x dependent helicity distributions for valence quarks in nucleons},
\href{https://doi.org/10.1016/0370-2693(96)00208-0}{Phys. Lett. B \textbf{375}, 320--326 (1996)}
[\href{https://doi.org/10.48550/arXiv.hep-ph/9604423}{\tt arXiv:hep-ph/9604423}].
Note: [Erratum: Phys.Lett.B 380, 494 (1996)]

%\cite{Ma:1997gy}
\bibitem{Ma:1997gy}
B.-Q.~Ma, I.~Schmidt and J.~Soffer,
{The Quark spin distributions of the nucleon},
\href{https://doi.org/10.1016/S0370-2693(98)01158-7}{Phys. Lett. B \textbf{441}, 461--467 (1998)}
[\href{https://doi.org/10.48550/arXiv.hep-ph/9710247}{\tt arXiv:hep-ph/9710247}].

%\cite{Ma:2002ir}
\bibitem{Ma:2002ir}
B.-Q.~Ma, D.~Qing and I.~Schmidt,
{Electromagnetic form-factors of nucleons in a light cone diquark model},
\href{https://doi.org/10.1103/PhysRevC.65.035205}{Phys. Rev. C \textbf{65}, 035205 (2002)}
[\href{https://doi.org/10.48550/arXiv.hep-ph/0202015}{\tt arXiv:hep-ph/0202015}].

%\cite{Liu:2014npa}
\bibitem{Liu:2014npa}
T.~Liu and B.-Q.~Ma,
{Generalized form factors of the nucleon in a light-cone spectator-diquark model},
\href{https://doi.org/10.1103/PhysRevC.89.055202}{Phys. Rev. C \textbf{89}, 055202 (2014)}
[\href{https://doi.org/10.48550/arXiv.1408.4873}{\tt arXiv:1408.4873}].

%\cite{Lu:2004au}
\bibitem{Lu:2004au}
Z.~Lu and B.-Q.~Ma,
{Sivers function in light-cone quark model and azimuthal spin asymmetries in pion electroproduction},
\href{https://doi.org/10.1016/j.nuclphysa.2004.06.006}{Nucl. Phys. A \textbf{741}, 200--214 (2004)}
[\href{https://doi.org/10.48550/arXiv.hep-ph/0406171}{\tt arXiv:hep-ph/0406171}].

%\cite{Bacchetta:2008af}
\bibitem{Bacchetta:2008af}
A.~Bacchetta, F.~Conti and M.~Radici,
{Transverse-momentum distributions in a diquark spectator model},
\href{https://doi.org/10.1103/PhysRevD.78.074010}{Phys. Rev. D \textbf{78}, 074010 (2008)}
[\href{https://doi.org/10.48550/arXiv.0807.0323}{\tt arXiv:0807.0323}].

%\cite{Lu:2012gu}
\bibitem{Lu:2012gu}
Z.~Lu and I.~Schmidt,
{T-odd quark-gluon-quark correlation function in the diquark model},
\href{https://doi.org/10.1016/j.physletb.2012.05.023}{Phys. Lett. B \textbf{712}, 451--455 (2012)}
[\href{https://doi.org/10.48550/arXiv.1202.0700}{\tt arXiv:1202.0700}].

% GPD
%\cite{Burkardt:2003je}
\bibitem{Burkardt:2003je}
M.~Burkardt and D.~S.~Hwang,
{Sivers asymmetry and generalized parton distributions in impact parameter space},
\href{https://doi.org/10.1103/PhysRevD.69.074032}{Phys. Rev. D \textbf{69}, 074032 (2004)}
[\href{https://doi.org/10.48550/arXiv.hep-ph/0309072}{\tt arXiv:hep-ph/0309072}].

%\cite{Chakrabarti:2005zm}
\bibitem{Chakrabarti:2005zm}
D.~Chakrabarti and A.~Mukherjee,
{Generalized parton distributions in the impact parameter space with non-zero skewedness},
\href{https://doi.org/10.1103/PhysRevD.72.034013}{Phys. Rev. D \textbf{72}, 034013 (2005)}
[\href{https://doi.org/10.48550/arXiv.hep-ph/0506006}{\tt arXiv:hep-ph/0506006}].

%\cite{Hwang:2007tb}
\bibitem{Hwang:2007tb}
D.~S.~Hwang and D.~Mueller,
{Implication of the overlap representation for modelling generalized parton distributions},
\href{https://doi.org/10.1016/j.physletb.2008.01.014}{Phys. Lett. B \textbf{660}, 350--359 (2008)}
[\href{https://doi.org/10.48550/arXiv.0710.1567}{\tt arXiv:0710.1567}].

% liu5D
%\cite{Liu:2014vwa}
\bibitem{Liu:2014vwa}
T.~Liu,
{Quark orbital motions from Wigner distributions},
\href{https://doi.org/}{arXiv:1406.7709}.

% spin
%\cite{Zhu:2010ewp}
\bibitem{Zhu:2010ewp}
J.~Zhu and B.-Q.~Ma,
{Probing the leading-twist transverse-momentum-dependent parton distribution function $h^{\perp}_{1T}$ via the polarized proton-antiproton Drell-Yan process},
\href{https://doi.org/10.1103/PhysRevD.82.114022}{Phys. Rev. D \textbf{82}, 114022 (2010)}
[\href{https://doi.org/10.48550/arXiv.1103.4201}{\tt arXiv:1103.4201}].

%\cite{Zhu:2011zza}
\bibitem{Zhu:2011zza}
J.~Zhu and B.-Q.~Ma,
{Proposal for measuring new transverse momentum dependent parton distributions $g_{1T}$ and $h_{1L}^\perp$ through semi-inclusive deep inelastic scattering},
\href{https://doi.org/10.1016/j.physletb.2010.12.036}{Phys. Lett. B \textbf{696}, 246--251 (2011)}
[\href{https://doi.org/10.48550/arXiv.1104.4564}{\tt arXiv:1104.4564}].

%\cite{Lu:2011cw}
\bibitem{Lu:2011cw}
Z.~Lu, B.-Q.~Ma and J.~Zhu,
{Azimuthal asymmetries in single polarized proton-proton Drell-Yan processes},
\href{https://doi.org/10.1103/PhysRevD.84.074036}{Phys. Rev. D \textbf{84}, 074036 (2011)}
[\href{https://doi.org/10.48550/arXiv.1108.4974}{\tt arXiv:1108.4974}].

% transition
%\cite{Xiao:2003wf}
\bibitem{Xiao:2003wf}
B.~W.~Xiao and B.-Q.~Ma,
{Pion photon and photon pion transition form-factors in the light cone formalism},
\href{https://doi.org/10.1103/PhysRevD.68.034020}{Phys. Rev. D \textbf{68}, 034020 (2003)}
[\href{https://doi.org/10.48550/arXiv.hep-ph/0312162}{\tt arXiv:hep-ph/0312162}].

%\cite{Ahluwalia:1993xa}
\bibitem{Ahluwalia:1993xa}
D.~V.~Ahluwalia and M.~Sawicki,
{Front form spinors in the Weinberg-Soper formalism and generalized Melosh transformations for any spin},
\href{https://doi.org/10.1103/PhysRevD.47.5161}{Phys. Rev. D \textbf{47}, 5161--5168 (1993)}
[\href{https://doi.org/10.48550/arXiv.nucl-th/9603019}{\tt arXiv:nucl-th/9603019}].

% wave functions
%\cite{Brodsky:1980vj}
\bibitem{Brodsky:1980vj}
S.~J.~Brodsky, T.~Huang and G.~P.~Lepage,
{SLAC-PUB-2540, published in the Proceedings of the XXth International Conference on High Energy Physics, Madison, Wisconsin (1980)}.

%\cite{brodsky1981slac}
\bibitem{brodsky1981slac}
S.~J.~Brodsky,
{SJ Brodsky, T. Huang and GP Lepage, Conf. Proc. C \textbf{810816}, 143 (1981)}.


%\cite{Brodsky:1982nx}
\bibitem{Brodsky:1982nx}
S.~J.~Brodsky, T.~Huang and G.~P.~Lepage,
{Hadronic and nuclear interactions in QCD},
[Report Number: SLAC-PUB-2868, Journal: Springer Tracts Mod. Phys., Volume: 100, Pages: 81--144, Year: 1982].

%\cite{Terentev:1976jk}
\bibitem{Terentev:1976jk}
M.~V.~Terentev,
{On the Structure of Wave Functions of Mesons as Bound States of Relativistic Quarks},
[Report Number: ITEP-5-1976, Journal: Sov. J. Nucl. Phys., Volume: 24, Pages: 106, Year: 1976].

%\cite{Karmanov:1979if}
\bibitem{Karmanov:1979if}
V.~A.~Karmanov,
{LIGHT FRONT WAVE FUNCTION OF RELATIVISTIC COMPOSITE SYSTEM IN EXPLICITLY SOLVABLE MODEL},
\href{https://doi.org/10.1016/0550-3213(80)90204-7}{Nucl. Phys. B \textbf{166}, 378--398 (1980)}
[Report Number: ITEP-55-1979].

%\cite{Chung:1988mu}
\bibitem{Chung:1988mu}
P.~L.~Chung, F.~Coester and W.~N.~Polyzou,
{Charge Form-Factors of Quark Model Pions},
\href{https://doi.org/10.1016/0370-2693(88)90995-1}{Phys. Lett. B \textbf{205}, 545--548 (1988)}.

\bibitem{Vega:2013bxa}
A.~Vega, I.~Schmidt, T.~Gutsche and V.~E.~Lyubovitskij,
{Nucleon GPDs in a light-front quark model derived from soft-wall AdS/QCD},
\href{https://doi.org/10.48550/arXiv.1306.1597}
[\href{https://doi.org/10.48550/arXiv.1306.1597}{\tt arXiv:1306.1597}].

%\cite{brodsky1981quantum}
\bibitem{brodsky1981quantum}
S.~J.~Brodsky, T.~Huang and G.~P.~Lepage,
{Quantum chromodynamics and hadronic interactions at short distances},
\href{https://www.osti.gov/biblio/6299168}{(1981)}
[Institution: Stanford Linear Accelerator Center].

%\cite{Wigner:1932eb}
\bibitem{Huang:1994dy}
T.~Huang, B.-Q.~Ma and Q.~X.~Shen,
{Analysis of the pion wave function in light cone formalism},
\href{https://doi.org/10.1103/PhysRevD.49.1490}{Phys. Rev. D \textbf{49}, 1490--1499 (1994)}
[\href{https://doi.org/10.48550/arXiv.hep-ph/9402285}{\tt arXiv:hep-ph/9402285}].


%\cite {Lorce:2011dv}
\bibitem {Lorce:2011dv}
A.~Accardi, A.~Bacchetta, W.~Melnitchouk and M.~Schlegel,
What can break the Wandzura - Wilczek relation?,
\href {https://doi.org/10.1088/1126-6708/2009/11/093}{J. High Energy Phys. \textbf{2009(11)}, 093}
[\href{https://doi.org/10.48550/arXiv.0907.2942}{\tt arXiv:0907.2942 [hep-ph]}].




%\cite{Hofstadter:1953zjy}
\bibitem{Hofstadter:1953zjy}
R.~Hofstadter, H.~R.~Fechter and J.~A.~McIntyre,
High - Energy Electron Scattering and Nuclear Structure Determinations,
\href{https://doi.org/10.1103/PhysRev.92.978}{Phys. Rev. \textbf{92}, 978 (1953)}.




%\cite{Pasquini:2013uja}
\bibitem{Pasquini:2013uja}
B.~Pasquini and C.~Lorce,
The partonic structure of the nucleon from generalized transverse momentum - dependent parton distributions,
in {Advanced Studies Institute on Symmetries and Spin},
\href{https://arxiv.org/abs/1304.1479}{\tt arXiv:1304.1479 [hep-ph]}, 2013.


%\cite{Hatta:2016dxp}
\bibitem{Hatta:2016dxp}
Y.~Hatta, B.-W.~Xiao and F.~Yuan,
Probing the Small-
x
Gluon Tomography in Correlated Hard Diffractive Dijet Production in Deep Inelastic Scattering,
\href{https://doi.org/10.1103/PhysRevLett.116.202301}{Phys. Rev. Lett. \textbf{116}, 202301 (2016)}
[\href{https://arxiv.org/abs/1601.01585}{\tt arXiv:1601.01585 [hep-ph]}].


%\cite{Bhattacharya:2017bvs}
\bibitem{Bhattacharya:2017bvs}
S.~Bhattacharya, A.~Metz and J.~Zhou,
Generalized TMDs and the exclusive double Drell–Yan process,
\href{https://doi.org/10.1016/j.physletb.2017.05.081}{Phys. Lett. B \textbf{771}, 396--400 (2017)}
[\href{https://arxiv.org/abs/1702.04387}{\tt arXiv:1702.04387 [hep-ph]}];
[Erratum: Phys. Lett. B \textbf{810}, 135866 (2020)].


%\cite{Ji:2004xq}
\bibitem{Ji:2004xq}
X.-d.~Ji, J.-P.~Ma and F.~Yuan,
QCD factorization for spin - dependent cross sections in DIS and Drell - Yan processes at low transverse momentum,
\href{https://doi.org/10.1016/j.physletb.2004.07.026}{Phys. Lett. B \textbf{597}, 299--308 (2004)}
[\href{https://arxiv.org/abs/hep-ph/0405085}{\tt arXiv:hep-ph/0405085}].


% tansform
% cite{Mantovani:2018qxy}

%\cite{Schlumpf:1993rm}
\bibitem {Schlumpf:1993rm}
F.~Schlumpf,
{Magnetic moments of the baryon decuplet in a relativistic quark model},
\href {https://doi.org/10.1103/PhysRevD.48.4478}{Phys. Rev. D \textbf{48}, 4478--4480 (1993)}
[\href{https://doi.org/10.48550/arXiv.hep-ph/9305293}{\tt arXiv:hep-ph/9305293}].

%\cite{Weber:1990fx}
\bibitem{Weber:1990fx}
H.~J.~Weber,
{Electromagnetic $N$-$\Delta$ Transition in a Light Cone Quark Model},
\href{https://doi.org/10.1016/0003-4916(91)90064-F}{Annals Phys. \textbf{207}, 417--427 (1991)}
[Report Number: UVa-INPP-90-5].


%\cite{Maris:2002yu}
\bibitem{Maris:2002yu}
P.~Maris,
{Effective masses of diquarks},
\href{https://doi.org/10.1007/s00601-002-0111-7}{Few Body Syst. \textbf{32}, 41--52 (2002)}
[\href{https://doi.org/10.48550/arXiv.nucl-th/0204020}{\tt arXiv:nucl-th/0204020}].

%\cite{Maris:2004bp}
\bibitem{Maris:2004bp}
P.~Maris,
{Electromagnetic properties of diquarks},
\href{https://doi.org/10.1007/s00601-004-0064-0}{Few Body Syst. \textbf{35}, 117--127 (2004)}
[\href{https://doi.org/10.48550/arXiv.nucl-th/0409008}{\tt arXiv:nucl-th/0409008}].

%\cite{Eichmann:2008ef}
\bibitem{Eichmann:2008ef}
G.~Eichmann, I.~C.~Cloet, R.~Alkofer, A.~Krassnigg and C.~D.~Roberts,
{Toward unifying the description of meson and baryon properties},
\href{https://doi.org/10.1103/PhysRevC.79.012202}{Phys. Rev. C \textbf{79}, 012202 (2009)}
[\href{https://doi.org/10.48550/arXiv.0810.1222}{\tt arXiv:0810.1222}].

%\cite{Eichmann:2009qa}
\bibitem{Eichmann:2009qa}
G.~Eichmann, R.~Alkofer, A.~Krassnigg and D.~Nicmorus,
{Nucleon mass from a covariant three-quark Faddeev equation},
\href{https://doi.org/10.1103/PhysRevLett.104.201601}{Phys. Rev. Lett. \textbf{104}, 201601 (2010)}
[\href{https://doi.org/10.48550/arXiv.0912.2246}{\tt arXiv:0912.2246}].

%\cite{Eichmann:2011ec}
\bibitem{Eichmann:2011ec}
G.~Eichmann and C.~S.~Fischer,
{Unified description of hadron-photon and hadron-meson scattering in the Dyson-Schwinger approach},
\href{https://doi.org/10.1103/PhysRevD.85.034015}{Phys. Rev. D \textbf{85}, 034015 (2012)}
[\href{https://doi.org/10.48550/arXiv.1111.0197}{\tt arXiv:1111.0197}].

%\cite{Cloet:2013gva}
\bibitem{Cloet:2013gva}
I.~C.~Cloet, C.~D.~Roberts and A.~W.~Thomas,
{Revealing dressed-quarks via the proton's charge distribution},
\href{https://doi.org/10.1103/PhysRevLett.111.101803}{Phys. Rev. Lett. \textbf{111}, 101803 (2013)}
[\href{https://doi.org/10.48550/arXiv.1304.0855}{\tt arXiv:1304.0855}].

% Exp



%\cite {H1:2012xnw}
\bibitem {H1:2012xnw}
H.~Abramowicz \textit {et al.} [H1, ZEUS Collaboration],
Combination and QCD Analysis of Charm Production Cross Section Measurements in Deep-Inelastic ep Scattering at HERA,
\href {https://doi.org/10.1140/epjc/s10052-013-2311-3}{Eur. Phys. J. C \textbf{73}, 2311 (2013)}
[\href{https://arxiv.org/abs/1211.1182}{\tt arXiv:1211.1182 [hep-ex]}].


%% LL

%\cite{LHPC:2010jcs}
\bibitem{LHPC:2010jcs}
J.~D.~Bratt and others (LHPC collaboration),
Nucleon structure from mixed action calculations using 2+1 flavors of asqtad sea and domain wall valence fermions,
\href{https://doi.org/10.1103/PhysRevD.82.094502}{Phys. Rev. D \textbf{82}, 094502 (2010)}
[\href{https://doi.org/10.48550/arXiv.1001.3620}{\tt arXiv:1001.3620 [hep-lat]}].


%\cite{Alexandrou:2020sml}
\bibitem{Alexandrou:2020sml}
C.~Alexandrou, S.~Bacchio, M.~Constantinou, J.~Finkenrath, K.~Hadjiyiannakou, K.~Jansen, G.~Koutsou, H.~Panagopoulos, and G.~Spanoudes,
Complete flavor decomposition of the spin and momentum fraction of the proton using lattice QCD simulations at physical pion mass,
\href{https://doi.org/10.1103/PhysRevD.101.094513}{Phys. Rev. D \textbf{101}, 094513 (2020)}
[\href{https://doi.org/10.48550/arXiv.2003.08486}{\tt arXiv:2003.08486 [hep-lat]}].


%\cite{HERMES:2006jyl}
\bibitem{HERMES:2006jyl}
A.~Airapetian and others (HERMES collaboration),
Precise determination of the spin structure function g(1) of the proton, deuteron and neutron,
\href{https://doi.org/10.1103/PhysRevD.75.012007}{Phys. Rev. D \textbf{75}, 012007 (2007)}
[\href{https://doi.org/10.48550/arXiv.hep-ex/0609039}{\tt arXiv:hep-ex/0609039}].

%\cite{Lorce:2011ni}
\bibitem{Lorce:2011ni}
C.~Lorce, B.~Pasquini, X.~Xiong and F.~Yuan,
{The quark orbital angular momentum from Wigner distributions and light-cone wave functions},
\href{https://doi.org/10.1103/PhysRevD.85.114006}{Phys. Rev. D \textbf{85}, 114006 (2012)}
[\href{https://doi.org/10.48550/arXiv.1111.4827}{\tt arXiv:1111.4827}].


%\cite{Diehl:2013xca,CLAS:2007clm,Kumericki:2016ehc}


%\cite {Diehl:2013xca}
\bibitem {Diehl:2013xca}
M.~Diehl and P.~Kroll,
Nucleon form factors, generalized parton distributions and quark angular momentum,
\href {https://doi.org/10.1140/epjc/s10052-013-2397-7}{Eur. Phys. J. C \textbf{73}, 2397 (2013)}
[\href{https://doi.org/10.48550/arXiv.1302.4604}{\tt arXiv:1302.4604 [hep-ph]}].

%\cite{CLAS:2007clm}
\bibitem{CLAS:2007clm}
F.~X.~Girod and others (CLAS collaboration),
Measurement of Deeply virtual Compton scattering beam - spin asymmetries,
\href{https://doi.org/10.1103/PhysRevLett.100.162002}{Phys. Rev. Lett. \textbf{100}, 162002 (2008)}
[\href{https://doi.org/10.48550/arXiv.0711.4805}{\tt arXiv:0711.4805 [hep-ex]}].

%\cite{Kumericki:2016ehc}
\bibitem{Kumericki:2016ehc}
Kresimir Kumericki, Simonetta Liuti, and Herve Moutarde,
GPD phenomenology and DVCS fitting: Entering the high-precision era,
\href{https://doi.org/10.1140/epja/i2016-16157-3}{Eur. Phys. J. A \textbf{52}, 157 (2016)}
[\href{https://doi.org/10.48550/arXiv.1602.02763}{\tt arXiv:1602.02763 [hep-ph]}].






%\cite{AbdulKhalek:2021gbh}
\bibitem{AbdulKhalek:2021gbh}
R.~Abdul Khalek and others,
{Science Requirements and Detector Concepts for the Electron - Ion Collider}: {EIC Yellow Report},
\href{https://doi.org/10.1016/j.nuclphysa.2022.122447}{Nucl. Phys. A \textbf{1026}, 122447 (2022)}
[\href{https://doi.org/10.48550/arXiv.2103.05419}{\tt arXiv:2103.05419 [physics.ins-det]}].

%\cite{Dudek:2012vr}
\bibitem{Dudek:2012vr}
J.~Dudek and others,
{Physics Opportunities with the 12 GeV Upgrade at Jefferson Lab},
\href{https://doi.org/10.1140/epja/i2012-12187-1}{Eur. Phys. J. A \textbf{48}, 187 (2012)}
[\href{https://doi.org/10.48550/arXiv.1208.1244}{\tt arXiv:1208.1244 [hep - ex]}].


%\cite{Ji:2020ena}
\bibitem{Ji:2020ena}
X.~Ji, F.~Yuan, and Y.~Zhao,
{What we know and what we don't know about the proton spin after 30 years},
\href{https://doi.org/10.1038/s42254-020-00248-4}{Nature Rev. Phys. \textbf{3}, 27--38 (2021)}
[\href{https://doi.org/10.48550/arXiv.2009.01291}{\tt arXiv:2009.01291 [hep-ph]}].


%\cite{COMPASS:2017jbv}
\bibitem{COMPASS:2017jbv}
M.~Aghasyan and others (COMPASS collaboration),
{First measurement of transverse - spin - dependent azimuthal asymmetries in the Drell - Yan process},
\href{https://doi.org/10.1103/PhysRevLett.119.112002}{Phys. Rev. Lett. \textbf{119}, 112002 (2017)}
[\href{https://doi.org/10.48550/arXiv.1704.00488}{\tt arXiv:1704.00488 [hep-ex]}].

%\cite{COMPASS:2012ozz}
\bibitem{COMPASS:2012ozz}
C.~Adolph and others (COMPASS collaboration),
{Experimental investigation of transverse spin asymmetries in muon - p SIDIS processes: Collins asymmetries},
\href{https://doi.org/10.1016/j.physletb.2012.09.055}{Phys. Lett. B \textbf{717}, 376--382 (2012)}
[\href{https://doi.org/10.48550/arXiv.1205.5121}{\tt arXiv:1205.5121 [hep-ex]}].


%\cite{Zhou:2016rnt}
\bibitem{Zhou:2016rnt}
J.~Zhou,
{Elliptic gluon generalized transverse-momentum-dependent distribution inside a large nucleus},
\href{https://doi.org/10.1103/PhysRevD.94.114017}{Phys. Rev. D \textbf{94}, 114017 (2016)}
[\href{https://doi.org/10.48550/arXiv.1611.02397}{\tt arXiv:1611.02397}].

%\cite{Bhattacharya:2023hbq}
\bibitem{Bhattacharya:2023hbq}
S.~Bhattacharya, D.~Zheng and J.~Zhou,
{Probing the Quark Orbital Angular Momentum at Electron-Ion Colliders Using Exclusive \ensuremath{\pi}0 Production},
\href{https://doi.org/10.1103/PhysRevLett.133.051901}{Phys. Rev. Lett. \textbf{133}, 051901 (2024)}
[\href{https://doi.org/10.48550/arXiv.2312.01309}{\tt arXiv:2312.01309}].


%\cite{Bacchetta:2011gx}
\bibitem{Bacchetta:2011gx}
A.~Bacchetta and M.~Radici,
{Constraining quark angular momentum through semi-inclusive measurements},
\href{https://doi.org/10.1103/PhysRevLett.107.212001}{Phys. Rev. Lett. \textbf{107}, 212001 (2011)}
[\href{https://doi.org/10.48550/arXiv.1107.5755}{\tt arXiv:1107.5755}].

%\cite{Jaffe:1989jz}
\bibitem{Jaffe:1989jz}
R.~L.~Jaffe and A.~Manohar,
{The g1 Problem: Fact and Fantasy on the Spin of the Proton},
\href{https://doi.org/10.1016/0550-3213(90)90506-9}{Nucl. Phys. B \textbf{337}, 509--546 (1990)}.



%\cite{Jaffe:1996zw}
\bibitem{Jaffe:1996zw}
R.~L.~Jaffe,
{Spin, twist and hadron structure in deep inelastic processes},
in {Ettore Majorana International School of Nucleon Structure: 1st Course: The Spin Structure of the Nucleon},
\href{https://doi.org/10.48550/arXiv.hep - ph/9602236}{\tt arXiv:hep-ph/9602236}
[MIT - CTP - 2506], pp. 42--129, Jan. 1996.


%\cite{JeffersonLabHallA:2013mjr}
\bibitem{JeffersonLabHallA:2013mjr}
K.~Allada et al. (Jefferson Lab Hall A Collaboration),
{Single spin asymmetries of inclusive hadrons produced in electron scattering from a transversely polarized $^3$He target},
\href{https://doi.org/10.1103/PhysRevC.89.042201}{Phys. Rev. C \textbf{89}, 042201 (2014)}
[\href{https://doi.org/10.48550/arXiv.1311.1866}{\tt arXiv:1311.1866}].






\end{thebibliography}
%\nocite{*}

%\end{document}

\end{document}